\documentclass{aa}

\usepackage{natbib}
\usepackage{graphicx}
\usepackage{txfonts}

\begin{document}

\setlength{\voffset}{0cm}

\title{
  Axisymmetric simulations of magneto--rotational core collapse:\\
  dynamics and gravitational wave signal
}

\titlerunning{
  Magneto--rotational core collapse
}

\author{
  M. Obergaulinger \and M.A. Aloy \and E. M\"uller
}

\institute{
  Max-Planck-Institut f\"ur Astrophysik, 
  Karl-Schwarzschild-Str.\,1, 
  D-85741 Garching bei M\"unchen, Germany
}

\offprints{M. Obergaulinger}

\date{Received / Accepted }


\abstract
  {}
  {
    We have performed a comprehensive parameter study of the collapse of
    rotating, strongly magnetized stellar cores in axisymmetry to
    determine their gravitational wave signature based on the Einstein
    quadrupole formula.
  }
%
{
  We use a Newtonian explicit magnetohydrodynamic Eulerian code based on
  the relaxing-TVD method for the solution of the ideal MHD equations,
  and apply the constraint-transport method to guarantee a
  divergence--free evolution of the magnetic field.  We neglect effects
  due to neutrino transport and employ a simplified equation of state.
  The initial models are polytropes in rotational equilibrium with a
  prescribed degree of differential rotation and rotational energy. The
  initial magnetic fields are purely poloidal the field strength ranging
  from $10^{10}\mathrm{G}$ to $10^{13}\mathrm{G}$.  The evolution of
  the core is followed until a few ten milliseconds past core bounce.
}
%
{The initial magnetic fields are amplified mainly by the differential
  rotation of the core giving rise to a strong toroidal field component
  with an energy comparable to the rotational energy. The poloidal field
  component grows by compression during collapse, but does not change
  significantly after core bounce.  In large parts of the simulated
  cores the growth time of the magneto--rotational instability (MRI) is
  of the order of a few milliseconds. The saturation field strengths
  that can be reached both via a pure $\Omega$ dynamo or the MRI are of
  the order of $10^{15}\mathrm{G}$ at the surface of the core.
  Sheet-like circulation flows which produce a strong poloidal field
  component transporting angular momentum outwards develop due to MRI,
  provided the initial field is not too weak. Weak initial magnetic
  fields ($\la 10^{11}\mathrm{G}$) have no significant effect on the
  dynamics of the core and the gravitational wave signal.  Strong
  initial fields ($\ga 10^{12}\mathrm{G}$) cause considerable angular
  momentum transport whereby rotational energy is extracted from the
  collapsed core which loses centrifugal support and enters a phase of
  secular contraction. The gravitational wave amplitude at bounce
  changes by up to a few ten percent compared to the corresponding
  non-magnetic model. If the angular momentum losses are large, the
  post--bounce model. If the angular momentum losses are large the
  post--bounce equilibrium state of the core changes from a
  centrifugally to a pressure supported one.  This transition imprints
  in the gravitational wave signal a reduction of the amplitude of the
  large--scale oscillations characteristic of cores bouncing due to
  centrifugal forces.
  
  In some models the quasi-periodic large-scale oscillations are
  replaced by higher frequency irregular oscillations. This pattern
  defines a new signal type which we call a type IV gravitational wave
  signal.  Collimated bipolar outflows give rise to a unique feature
  that may allow their detection by means of gravitational wave
  astronomy: a large positive quadrupole wave amplitude of similar
  size as that of the bounce signal.}
%
{}


\maketitle


\section{Introduction}
\label{Sek:Intro}

The gravitational binding energy liberated by the collapse of the iron
core of a massive ($ M \ga 8 \ M_{\odot}$) star to a neutron star is the
commonly accepted energy source of type Ib/c and type II supernovae, as
a few percent of this energy are sufficient to unbind and rapidly eject
the stellar envelope and to create the supernova outburst.  However,
which physical processes turn the central implosion into the explosion
of the stellar layers surrounding the forming neutron star is still
debated in spite of many efforts over more than three decades. Heating
of stellar gas just outside the proto--neutron star (PNS) by neutrinos
diffusing and being advected out of its interior is thought to play a
crucial role in the explosion mechanism.  However, as current
neutrino--driven supernova models produce (weak) explosions only for low
mass progenitors (for a recent review, see
e.g.\,\citealp{Janka_etal_astroph_2004__CCSN_Modelling}), there may be a
need to include additional physics in the models in order to make them
work successfully for more massive progenitors, too.

On this account magneto--rotational core collapse, which has been
studied by a few authors in the past
\citep[]{LeBlanc_Wilson_ApJ_1970__MHD_SN,
  Bisnovatyi-Kogan_Popov_Samokhin_APSS_1976__MHD_SN,
  Meier_etal_ApJ_1976__MHD_SN, Mueller_Hillebrandt_AAP_1979__MHD_SN,
  Ohnishi_TRIAE_1983__MHD_Collapse, Symbalisty_ApJ_1984__MHD_SN}, has
become an active research field in recent years
\citep[]{Wheeler_Meier_Wilson_ApJ_2002__MHD_SN,
  Akiyama_etal_ApJ_2003__MRI_SN, Kotake_etal_ApJ_2004__SN,
  Kotake_etal_PRD_2004__MagCollapse_GW, Takiwaki_etal_ApJ_2004__MHD_SN,
  Wheeler_Akiyama_astroph_04__MHD_SN,
  Yamada_Sawai_ApJ_2004__MHDCollapse,
  Ardeljan_Bisnovatyi-Kogan_Moiseenko_MNRAS_2005__MHD_SN,
  Kotake_Yamada_Sato_ApJ_2005__MHDCollapse_Nu,
  Sawai_Kotake_Yamada_astroph_2005__MHDSN}.  Further reasons for this
activity are the availability of sufficient computational power for the
necessarily multi--dimensional magneto-hydrodynamic (MHD) simulations,
observations indicating very asymmetric explosions
\citep[]{Wang_etal_ApJ_1996__SN_Polar, Wang_etal_ApJ_2001__Bipolar_SN,
  Leonard_etal_ApJ_2001__SN_Spectropol}, and the interpretation of
Anomalous X-Ray Pulsars and Soft Gamma-Ray Repeaters as magnetars,
i.e.\,very strongly magnetized neutron stars
\citep[]{Duncan_Thompson_ApJL_1992__Magnetars,
  Thompson_Duncan_ApJ_1996__Magnetars,
  Kouveliotou_etal_ApJL_1999__SGR1900+14}.

Concerning the initial conditions for magneto--rotational core collapse
the up to now most advanced evolutionary calculations of {\it rotating}
massive stars \citep{Heger_Woosley_Spruit_2005} predict that the initial
rotation rates are more than an order of magnitude smaller than (i) the
minimum ones used in past (parameter) studies of magneto--rotational
core collapse, and (ii) those predicted by previous evolutionary
calculations (see,
e.g.\,\citealp{Woosley_Heger_Weaver_RMP_2002,Hirschi_etal_astroph_2003__PreSN})
which lead to neutron stars rotating very rapidly ($\sim 1\,$ms) at
birth. The latter studies ignored the torques exerted in differentially
rotating regions by the magnetic fields that thread them. Thus, the
stars end up with 30 to 50 times more angular momentum than in the
models by \citet{Heger_Woosley_Spruit_2005} in that part of their core
destined to collapse to a neutron star.

The strength (and distribution) of the initial magnetic field in the
stellar core is unknown. If weak initially, several possible
amplification mechanisms exist that may amplify the magnetic field of
the collapsing progenitor to a dynamically important strength.  Linear
amplification of the field by means of differential rotation will occur
\citet{Meier_etal_ApJ_1976__MHD_SN}, which transforms rotational energy
into magnetic energy by winding up any seed polodial field into a
toroidal magnetic field.  This process can be accompanied by the action
of meridional (e.g.\,convective) motions that transform toroidal into
poloidal fields.  Both processes together lead to the so--called
$\alpha$-$\Omega$ dynamo.  Recently, the magneto-rotational instability
(MRI) \citep[see][]{Balbus_Hawley_RMP_1998__MRI} has received a lot of
interest in the context of supernova collapse and explosion
\citep{Akiyama_etal_ApJ_2003__MRI_SN, Kotake_etal_ApJ_2004__SN,
  Yamada_Sawai_ApJ_2004__MHDCollapse,
  Sawai_Kotake_Yamada_astroph_2005__MHDSN}.  Unlike linear wrapping, the
MRI will give rise to an exponential growth of the field strength while
working on the same time scale (see however
\citealp{Sawai_Kotake_Yamada_astroph_2005__MHDSN}).  The MRI saturation
field is independent of the initial field, i.e.\,even quite small
initial fields can be amplified to dynamically important strengths.  The
MRI will occur if the radial gradient of the angular velocity is
negative, a condition arising quite naturally in core collapse
situations.

A major effect of magnetic fields on the collapse dynamics is the
transport of angular momentum.  Due to its very low (fluid and $\nu$
shear) viscosity (see e.g.\,\citealp{Keil96}) a collapsing
non--magnetized stellar core maintains its Lagrangian angular momentum
profile $\vec j(m)$, $m$ being the Lagrangian mass coordinate, on time
scales of $\sim 1\,$s, but magnetic fields can significantly
redistribute the angular momentum \citep{Meier_etal_ApJ_1976__MHD_SN}.
This can slow down the forming neutron star and thus counteract the
effects of rotation.  In some cases, even retrograde rotation may result
in some parts of the core \citep{Mueller_Hillebrandt_AAP_1979__MHD_SN}.
Angular momentum transfer can also destabilize the rotational
equilibrium the core resides in after a centrifugal bounce at
sub--nuclear densities, and lead to a subsequent (second) collapse to
nuclear densities and beyond that releases large amounts of
gravitational binding energy \citep{Symbalisty_ApJ_1984__MHD_SN}.
Conversely, the violent convective flow both inside the neutrino sphere
and between the neutrino sphere and the shock will transport and amplify
magnetic fields in the collapsed core of a supernova
\citep{Thompson_Murray_ApJ_2001__MHD_conv_SN}

Analytic considerations \citep{Meier_etal_ApJ_1976__MHD_SN,
  Wheeler_Meier_Wilson_ApJ_2002__MHD_SN} and numerical simulations
\citep{LeBlanc_Wilson_ApJ_1970__MHD_SN, Symbalisty_ApJ_1984__MHD_SN,
  Akiyama_etal_ApJ_2003__MRI_SN, Kotake_etal_PRD_2004__MagCollapse_GW,
  Kotake_etal_ApJ_2004__SN, Yamada_Sawai_ApJ_2004__MHDCollapse,
  Ardeljan_Bisnovatyi-Kogan_Moiseenko_MNRAS_2005__MHD_SN,
  Sawai_Kotake_Yamada_astroph_2005__MHDSN} show that magneto-rotational
core collapse might lead to jet--like explosions.  Though the magnetic
stress will remain below equipartition strength in most regions of the
star, it might affect the dynamics of the core through its anisotropic
components.  Magnetic stresses can assist in pushing the stalled shock,
or may even drive a mildly relativistic outflow in form of a jet along
the rotation axis, which is powered by the rotational energy transfered
to the jet by the magnetic stresses.

Observations of gravitational waves (GWs) will allow one to learn more
about the supernova mechanism as they provide pristine information
directly from the stellar interior, in particular about the amount and
distribution of the angular momentum, and the strength and topology of
the magnetic field.  Simplified Newtonian
\citep{Mueller_AAP_1982__RotCollapse_GW, Moenchmeyer_1991,
  Yamada_Sato_ApJ_1994__RotCollapse, ZM97,
  Kotake_Yamada_Sato_PRD_2003__CCGW,
  Kotake_etal_PRD_2004__MagCollapse_GW,
  Fryer_Holz_Hughes_ApJ_2004__Collapse_GW, Ott_etal_ApJ_2004__GW,
  Yamada_Sawai_ApJ_2004__MHDCollapse} and general--relativistic
\citep{DFM1, DFM2} calculations of {\it rotational} core collapse
predict the emission of a strong signal around core bounce, and that the
magnitude of the bounce signal as well as the post--bounce gravitational
radiation depend sensitively on the initial rotation rate and rotation
profile.  Newtonian hydrodynamic simulations using more sophisticated
micro- and transport--physics, as well as state--of--the--art rotating
progenitors \citep{Heger_Woosley_Spruit_2005} show that the
gravitational wave signal at core bounce is small compared to the signal
produced by convective motions in the post--bounce core and by aspheric
neutrino emission \citet{Mueller_etal_ApJ_2004__CCSN_GW}. {\it
  Magneto-rotational} effects on the gravitational wave signature were
first investigated in detail by
\citet{Kotake_etal_PRD_2004__MagCollapse_GW} and
\citet{Yamada_Sawai_ApJ_2004__MHDCollapse} who found differences from
the signature of purely hydrodynamic models only in the case of very
strong initial fields ($|\vec B| \ga 10^{12} \mathrm{G}$).
 
In the following we present a comprehensive parameter study of the
axisymmetric Newtonian core collapse of rotating magnetized polytropes
and of their gravitational wave signature. Our study extends the work of
\citet{ZM97} (hereafter ZM) and the complementary work of \citet{DFM1,
  DFM2} (hereafter DFM), who investigated the hydrodynamic collapse of a
large set of rotating polytropes in Newtonian and general relativistic
gravity, respectively. The simulations have been performed with the
recently developed MHD difference scheme of
\citet{Pen_Arras_Wong_APJS_2003__rTVD_MHD_Code}. They incorporate
neither neutrino transport nor nuclear burning processes.  Due to the
reduced complexity of our models we could explore many of them covering
a large region in parameter space, the focus being the gravitational
wave signal emitted by magnetized stellar cores and their dynamic
evolution.  Because of our assumptions and approximations the validity
of our models is limited to the stage of core collapse and to the first
few ten milliseconds of their post--bounce evolution.

Several related but less comprehensive numerical MHD studies have been
performed in the past few years:
\citet{Yamada_Sawai_ApJ_2004__MHDCollapse} used the ZEUS-2D code,
employed the parametric equation of state of
\citet{Yamada_Sato_ApJ_1994__RotCollapse}, considered no neutrinos, and
followed the evolution of initially rapidly rotating and very strongly
magnetized ($|\vec B| \ga 10^{12} \mathrm{G}$ cores with a purely
homogeneous poloidal field.  Contrary to
\citet{LeBlanc_Wilson_ApJ_1970__MHD_SN} and
\citet{Symbalisty_ApJ_1984__MHD_SN} they find that the magnetic field
becomes strongest behind the shock wave and not in the inner core, and
thus is the main driving factor of the observed jet outflow along the
rotation axis.  Besides a field amplification by differential rotation,
they also observe the possible action of the MRI. They calculate the
gravitational wave signal in the quadrupole approximation finding no
substantial difference between the bounce signal of magnetized and
non--magnetized models.  \citet{Kotake_etal_PRD_2004__MagCollapse_GW}
also use the ZEUS-2D code to which they add an approximate neutrino
cooling with a leakage scheme.  They assume an initially predominantly
toroidal magnetic field in their investigated 14 models of which all but
one are very strongly magnetized $|\vec B| > 10^{11} \mathrm{G}$.
Besides the simplified equation of state of
\citet{Yamada_Sato_ApJ_1994__RotCollapse} they also consider two
realistic equations of state.
\citet{Kotake_etal_PRD_2004__MagCollapse_GW} focus their study on the
effect of the magnetic field on the gravitational wave signal, and find
that the gravitational wave amplitudes are lowered by $\sim 10\%$ for
models with the strongest initial magnetic fields ($|\vec B| \sim
10^{14} \mathrm{G}$).  \citet{Kotake_etal_ApJ_2004__SN},
\citet{Takiwaki_etal_ApJ_2004__MHD_SN}, and
\citet{Kotake_Yamada_Sato_ApJ_2005__MHDCollapse_Nu} all using the same
input physics and numerics as
\citet{Kotake_etal_PRD_2004__MagCollapse_GW} are concerned with the
effects of the magnetic fields on the anisotropic neutrino radiation and
convection, on the propagation of the shock wave, and on the
rotation--induced anisotropic neutrino heating through parity--violating
effects, respectively.  \citet{Kotake_etal_ApJ_2004__SN} find that the
aspherical shapes of the shock and of the neutrino sphere (oblate or
prolate depending on the initial rotation law) are enhanced in the
magnetized models, and that the MRI is expected to develop on the prompt
shock propagation time scale. \citet{Takiwaki_etal_ApJ_2004__MHD_SN}
observe the formation of a tightly collimated shock wave along the
rotational axis for strongly magnetized models.
\citet{Kotake_Yamada_Sato_ApJ_2005__MHDCollapse_Nu} find an at most
0.5\% change of the neutrino heating rates even in their most strongly
magnetized models ($|\vec B| \ga 10^{13} \mathrm{G}$).
\citet{Ardeljan_Bisnovatyi-Kogan_Moiseenko_MNRAS_2005__MHD_SN} employ a
2D implicit Lagrangian code, a simplified equation of state, and
consider energy losses by neutrinos and iron dissociation. They add a
magnetic field of quadrupole--like symmetry with an energy of $10^{-6}$
of the core's gravitational binding energy to the collapsed,
post--bounce differentially rotating, stationary core.  The toroidal
field component of this seed field first grows linearly due to
differential rotation, but then starts to amplify exponentially due to
the action of the MRI.  The resulting drastic increase of the magnetic
pressure eventually causes an explosion with an energy of
$0.6\,10^{51}\,$erg. Finally,
\citet{Sawai_Kotake_Yamada_astroph_2005__MHDSN} extend the work of
\citet{Yamada_Sawai_ApJ_2004__MHDCollapse} by considering
inhomogeneously magnetized cores mainly in the very strong field regime
($|\vec B| \sim 10^{12} \mathrm{G} \ldots 10^{13} \mathrm{G}$), which
may produce magnetars. They find that poloidal magnetic fields which are
initially concentrated toward the rotation axis produce more energetic
explosions and more prolate shocks than cores with an initially uniform
field. A core with an initially quadrupolar field
\citep{Ardeljan_Bisnovatyi-Kogan_Moiseenko_MNRAS_2005__MHD_SN} gives
rise to a collimated fast jet ($v \la c/2$), while a core with a pure
toroidal fields shows no sign of an explosion.

The paper is organized as follows: we will describe the physics included
in our models in Sect.\,\ref{Sek:Modelle}, and briefly introduce our
numerical method in Sect.\,\ref{Sek:Num}.  Our results will be discussed
in Sect.\,\ref{Sek:Ergebnissse}, and a summary and conclusions of our
work will be presented in Sect.\,\ref{Sek:Schluss}.  In the appendices
we provide
a brief discussion of the relaxing TVD scheme employed in our numerical
code (appendix \ref{Sek:rTVD}),
a presentation of the quadrupole formula used for the extraction of the
GW signal (appendix \ref{Sek:GW}),
and a compilation of some characteristic properties of all models
(appendix \ref{Sek:Syn}).

\section{Physics of our models}
\label{Sek:Modelle}

\subsection{Evolution equations}
\label{Suk:Modelle:Glgn}

We evolve the density $\rho$, the velocity $\vec v$, the total energy
density $e_{\star} \equiv \varepsilon + e_{\mathrm{kin}} +
e_{\mathrm{mag}}$ ($\varepsilon$, $e_{\mathrm{kin}} \equiv \frac{1}{2}
\rho \vec v^2$, and $e_{\mathrm{mag}} \equiv \frac{1}{2} \frac{\vec
  B^2}{4\pi}$ are the internal, kinetic, and magnetic energy density,
respectively), and the magnetic field $\vec B$ of our models using the
equations of Newtonian ideal magnetohydrodynamics (MHD):
\begin{eqnarray}
  \label{Gl:Konti}
  \partial_t \rho + \nabla_m (\rho v^m ) & =  & 0, \\
  \partial_t \left(\rho v_n\right) + \nabla_m \left(\rho v_n v^m 
             + P_{\star} - b_n b^m\right) & =  & f_n \label{Gl:Euler}, \\
  \partial_t e_{\star} + \nabla_m \left(\left(e_{\star} + P_{\star}\right)v^m 
             - b^m b_n v^n\right) & =  & q.
  \label{Gl:Energie}
\end{eqnarray}
Here, Latin indices run from $1$ to $3$, and Einstein's sum convention
applies.  $P_{\star} \equiv P_{\mathrm{gas}} + \vec b^2 / 2$ is the
total pressure, which is the sum of the gas pressure $P_{\mathrm{gas}}$
and the isotropic magnetic pressure $P_{\mathrm{mag}} \equiv \vec b^2 /
2$ with $\vec b = \vec B / \sqrt{4\pi}$.

Using the \emph{ideal} MHD equations, we neglect effects due to the
viscosity and the finite conductivity of the gas.  This is normally a
very good approximation for the stellar interior.  However, during core
collapse interesting hydrodynamic effects might arise from the inclusion
of viscosity, in particular in the case of MHD
\citep{Thompson_Quataert_Burrows_ApJ_2004__Vis_Rot_SN}. Non--ideal terms
in the induction equation might lead to reconnection of field lines,
thus possibly affecting the topology of the field, and might prove
important for several kinds of hydromagnetic instabilities
\citep{Spruit_AAP_1999__MHD_star}.

For a Newtonian self--gravitating fluid the source terms $f_n$ and $q$
in the MHD momentum (\ref{Gl:Euler}) and energy (\ref{Gl:Energie})
equations are given by
\begin{eqnarray}
  f_n & = & - \rho \nabla_n \Phi,
  \label{Gl:Grimp} \\
  q   & = & - \rho \vec v \, \vec \nabla \Phi,
  \label{Gl:Grerg}
\end{eqnarray}
where the gravitational potential \(\Phi\) obeys the Poisson equation
\begin{equation}
  \bigtriangleup \Phi = 4 \pi G \rho,
  \label{Gl:Poisson}
\end{equation}
with $G$ being the gravitational constant.  The Poisson equation is
solved in every time step using the solver of
\citet{Mueller_Steinmetz_CPC_1995__Self_grav_flow}, which is based on
the integral form of Poisson's equation and on an expansion of the
density distribution into spherical harmonics.

We integrate the MHD equations in spherical coordinates assuming
axisymmetry, i.e.\,we cannot simulate non--axisymmetric instabilities
which can occur if the rotation rate exceeds a critical value during
core collapse due to angular momentum conservation
\citep{Tassoul_Book__Theo_rotating_stars}.  Axisymmetry also inhibits
the growth of various MHD instabilities
\citep{Spruit_AAP_1999__MHD_star}. We further assume equatorial symmetry
in order to reduce the computational costs of a simulation.

\subsection{Microphysics}
\label{Suk:Modelle:MikPhy}

We do neither consider nuclear reactions nor neutrino transport, and use
a simplified equation of state (EOS).  Since neutrinos are thought to
play a major role in the revival of the stalled prompt shock, our
approach is limited to the collapse, bounce, and shock formation phases
when neutrinos are not yet dynamically important.  This limitation
allows us to focus on a specific, not yet comprehensively studied part
of core collapse, namely the influence of MHD effects.

We have used the approximate, analytic EOS of
\citet{Janka_Zwerger_Moenchmeyer_AAP_1993__Art_vis_SN} in our parameter
study. It is based on a decomposition of the gas pressure $P$ into the
sum of a polytropic part $P_{\mathrm{p}}$ and a thermal part
$P_{\mathrm{th}}$:
\begin{equation}
  P = P_{\mathrm{p}} + P_{\mathrm{th}}.
  \label{Gl:HybEOS}
\end{equation}
The polytropic part is given by 
\begin{equation}
  \label{Gl:PPol}
  P_{\mathrm{p}} = \kappa_{\mathrm{p}} \cdot \rho^{\Gamma_{\mathrm{p}}}
\end{equation}
where the adiabatic index
\begin{equation}
  \label{Gl:Gamma_p}
  \Gamma_{\mathrm{p}} = 
  \left\{ 
    \begin{array}{ccc} 
      \Gamma_1 & \mathrm{for} & \rho \le \rho_{\mathrm{nuc}} \\
      \Gamma_2 & \mathrm{for} & \rho > \rho_{\mathrm{nuc}}
    \end{array}
  \right.
\end{equation}
describes cold matter that undergoes a phase transition at nuclear
density $\rho_{\mathrm{nuc}} = 2\cdot 10^{14} \ \mathrm{cm \, s}^{-1}$.
Below this threshold density, the pressure is dominated by a
relativistic degenerate electron gas with $\Gamma_{\mathrm{p}} =
\Gamma_1 \la 4/3$.  At $\rho_{\mathrm{nuc}}$, the EOS stiffens
considerably, and the adiabatic index jumps to a value
$\Gamma_\mathrm{p} = \Gamma_2 \sim 2.5$, which mimics the phase
transition to incompressible nuclear matter.  Continuity of the pressure
at nuclear density implies
\begin{equation}
  \label{Gl:Kappas}
  \kappa_2 = \kappa_1 \cdot \rho_{\mathrm{nuc}}^{\Gamma_1-\Gamma_2},
\end{equation}
where $\kappa_1 = 4.897\, \cdot 10^{14}\,$cgs--units.

The polytropic pressure changes only due to adiabatic processes,
i.e.\,it cannot describe dissipation of kinetic into thermal energy in
shocks. This shock heating is treated by the thermal part of the EOS
which has the form of an ideal gas EOS:
\begin{equation}
  \label{Gl:PThe}
  P_{\mathrm{th}} = (\gamma-1) \varepsilon_{\mathrm{th}}.
\end{equation}
In our simulations, we took $\gamma = 1.5$ corresponding to a gas
composed of a mixture of a relativistic and a non--relativistic
component.

\subsection{Initial models}
\label{Suk:Init}

\subsubsection{Equilibrium models for rotating polytropes}
\label{Suk:Modelle:RotPol}

Concerning rotation and magnetic field, the conditions in the stellar
core at the onset of collapse are not well constrained.  Therefore, we
have investigated the evolution of a sufficiently broad set of
simplified stellar models.  Except for the magnetic field, the initial
models are the same as those studied by ZM and DFM.

The initial models are rotating polytropes in hydrostatic equilibrium.
constructed with the method of
\citet{Eriguchi_Mueller_AAP_1985__rotating_eql}.  They rotate according
to the so--called $j$--constant law
\citep{Eriguchi_Mueller_AAP_1985__rotating_eql}
\begin{equation}
  \label{Gl:jconst}
  \Omega(\varpi) = \frac{\Omega_0}{1+\left(\frac{\varpi}{A}\right)^2},
\end{equation}
where the angular velocity $\Omega$ is given as a function of the
cylindrical radial coordinate $\varpi$.  The parameter $\Omega_0$ is the
maximum angular velocity, and the parameter $A$, having the dimension of
a length, determines the degree of differential rotation of the core.
Deviations from rigid body rotation are significant only for radii
$\varpi \gg A$, where the rotation law approaches that of a
configuration with constant specific angular momentum $j$.

\subsubsection{The hydrodynamic parameter space}
\label{Suk:Modelle:HParameterraum}

The initial models are calculated using the hybrid EoS (\ref{Gl:HybEOS})
in the ``cold'' limit, i.e.\,$P_{\mathrm{th}} = 0$, an adiabatic index
$\Gamma_1 = 4/3$, the j-constant rotation law (\ref{Gl:jconst}), and
with a central density $\rho_{\mathrm{c}} = \cdot 10^{10} \ \mathrm{cm
  \, s}^{-1}$.  Instead of $\Omega_0$ we use the ratio of rotational to
gravitational binding energy, $\beta_{\mathrm{rot}} \equiv |
E_{\mathrm{rot}} / E_{\mathrm{grav}} |$ (initially $E_{\mathrm{grav}}
\sim 5.5\cdot 10^{51} \ \mathrm{erg}$) to parametrize the models.  The
two--dimensional parameter space $(A, \beta_{\mathrm{rot}})$ is covered
by $4 \times 5$ initial models according to Table\,\ref{Tab:InitPars}.

As $\Gamma_1 = 4/3$ for our initial models, they are only marginally
stable against collapse, which can be triggered by either lowering the
coefficient $\kappa_1$ or the adiabatic index $\Gamma_1$.  The former
approach mimics the energy consumption due to photo--disintegration of
nuclei, while the reduction of $\Gamma_1$ models the softening of the
EOS due to deleptonisation.  We apply the latter method to initiate the
collapse.

For realistic equations of state $1.28 \la \Gamma_1 \la 1.325$
and $2.4 \la \Gamma_2 \la 3$, respectively.  Following ZM and
DFM we constructed models with five different values of the sub-nuclear
adiabatic index $\Gamma_1$, and set $\Gamma_2 = 2.5$
(Table\,\ref{Tab:InitPars}). The nomenclature of the models also follows
that of ZM and DFM; i.e.\,model A1B3G5 has a rotation parameter $A =
5\cdot 10^{9}\ \mathrm{cm}$ (A1), a fractional rotational energy
$\beta=0.9\%$ (B3), and a sub-nuclear adiabatic index $\Gamma_1=1.28$
(G5; see Table\,\ref{Tab:InitPars}).

\begin{table}[ht]
  \caption[Initial Models] 
  {
    Initial models and their parametrisation:
    $A$ and $\beta_{\mathrm{rot}}$ are the rotation law parameter
    (equation \ref{Gl:jconst}) and the ratio of rotational to
    gravitational energy, respectively.  Larger values of $A$ correspond
    to more rigidly rotating cores.  $\Gamma_1$ is the sub-nuclear
    adiabatic index of our hybrid equation of state (see
    Sect.\,\ref{Suk:Modelle:MikPhy}).  
  }
  \label{Tab:InitPars}
  \centering
  \begin{tabular}{cl|cl|cl}
    \hline \hline
    Model & $A [\mathrm{cm}]$ & Model & $\beta_{\mathrm{rot}} [\%]$ 
                              & Model & $\Gamma_1$ \\
    \hline
    A1 &  $5\cdot 10^{9}$  & B1 & $\approx 0.25$ & G1 & $1.325$\\
    A2 &  $1\cdot 10^{8}$  & B2 & $\approx 0.45$ & G2 & $1.32$ \\
    A3 &  $5\cdot 10^{7}$  & B3 & $\approx 0.9$  & G3 & $1.31$ \\
    A4 &  $1\cdot 10^{7}$  & B4 & $\approx 1.8$  & G4 & $1.30$ \\
       &               & B5 & $\approx 4.0$  & G5 & $1.28$ \\
    \hline \hline
  \end{tabular}
\end{table}

\subsubsection{Magnetic field configuration}
\label{Suk:Modelle:IniMag}

The magnetic fields of our initial models are calculated from the vector
potential of a circular current loop of radius $r_{\mathrm{mag}}$ in the
equatorial plane \citep{Jackson_Book_1962__Edyn}.  The only
non-vanishing component of the magnetic vector potential is $A_{\phi}$.
It is given by
\begin{eqnarray}
  \label{Gl:Dipol-Init:A_phi}
  A_{\phi} ( \vec r ) \propto
  \frac{1}{r_{\mathrm{mag}}} 
  & \int & 
  \frac{\mathrm{d} \ r'^3}{3} \, \mathrm{d} (-\cos \theta') \, 
  \mathrm{d} \phi' \, \nonumber \\
  & & \frac
  { \sin \theta' \cos \phi' \delta (\cos \theta') 
    \delta  ( r - r_{\mathrm{mag}} ) }
  { |\vec r - \vec r'| }
  ,
\end{eqnarray}
which can be expanded in terms of Legendre-Polynomials $P_n^m$, yielding
\begin{equation}
  \label{Gl:Dipol-Init:A_phi-Legendre}
  A_{\phi} \propto
  \sum _ {n = 0} ^ {\infty}
  \frac { (-1)^n (2n-1)!! r_{<}^{2n+1} }
  {2^n ( n+1)! r_{>}^{2n+2}}
  P_{2n+1}^1 ( \cos \theta).
\end{equation}
Here, $r_{<} = \mathrm{min} ( r, r_{\mathrm{mag}})$ and $r_{>} =
\mathrm{max} ( r, r_{\mathrm{mag}})$ and $(2n-1)!!$ = $1 \cdot 3 \cdot 5
\cdot ... \cdot (2n+1)$.  The constant of proportionality in these
expressions is fixed by demanding that the magnetic field strength in
the core is equal to a given value.

\begin{figure}
  \resizebox{\hsize}{!}{\includegraphics{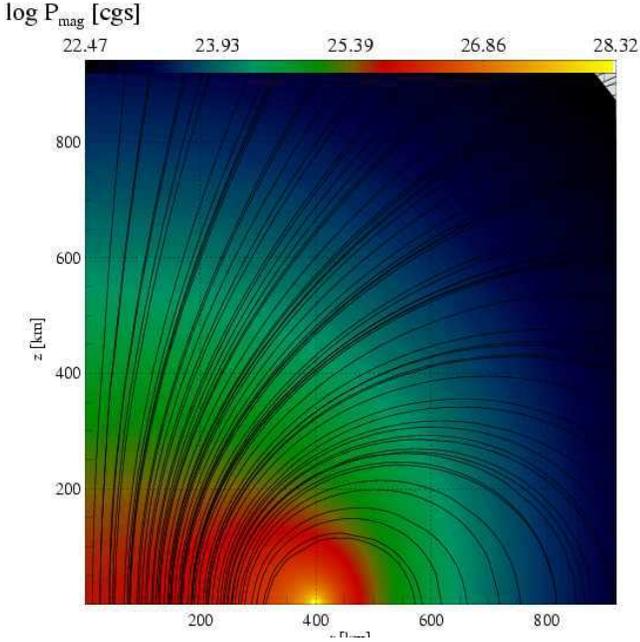}}
  \caption[Initial Magnetic Field] {The initial magnetic field
    configuration of the models of series AaBbGg-D3M13.  Besides the
    field lines the distribution of the magnetic pressure (color coded)
    is displayed.  $P_{\mathrm{mag}}$ is largest near the center of the
    field generating current loop located at a radius of
    $r_{\mathrm{mag}}^{D3} = 400\ \mathrm{km}$ in the equatorial plane,
    and drops rapidly with increasing radius for $r >
    r_{\mathrm{mag}}^{D3}$.  }
  \label{Fig:MagInit}
\end{figure}

The magnetic field strength in the core's center is normalized by $B_{0}
= \sqrt{4\pi}b_0$.  For very small radii ($r \ll r_{\mathrm{mag}}$) the
field resembles a uniform field parallel to the rotation axis, whereas
at large radii the field lines bend towards the equatorial plane. The
field strength is largest in the interior of the current loop
(Fig.\,\ref{Fig:MagInit}).

Table\,\ref{Tab:MinitPars} provides an overview of the parametrisation
of the magnetic field.  The models are denoted as follows: Model
A3B3G3-D3M12 is the hydrodynamic initial model A3B3G3
(Sec.\,\ref{Suk:Modelle:HParameterraum}) endowed with a magnetic field
generated by a current loop located at a radius of $r_{\mathrm{mag}} =
400\ \mathrm{km}$ and a maximum field strength of $b_0 = 10^{12} \ 
\mathrm{G}$ (see Table\,\ref{Tab:MinitPars}).  The initial magnetic
energy $E_{\mathrm{mag}}$ for models AaBbGg-DdM12 is given in
Table\,\ref{Tab:MinitErg}.

Most simulations were performed using models AaBbGg-D3Mm ($m = 10, 11,
12, 13$). Their magnetic field configuration is shown in
Fig.\,\ref{Fig:MagInit}.  Since the radius of the current loop
($r_{\mathrm{mag}} = 400\ \mathrm{km}$) that generates this field
configuration is small compared to the radius of the stellar core
($r_{\mathrm{core}} \sim 1500 \ldots 2100\ \mathrm{km}$), the magnetic
energy is highly concentrated in the center of the stellar core.  This
is different for models AaBbGg-D0Mm, which posses a homogeneous field
directed along the rotational axis. The magnetic energy of these
``uniform--field'' models is much larger than that of the corresponding
''current--loop'' models AaBbGg-DdMm ($d = 1, 2, 3, 4$) due to the
contributions of the outer layers of the core.

The magnetic field strengths of our initial models are -- as in most
studies of magneto-rotational collapse -- much higher than those
estimated to exist in realistic stellar cores.  Magnetic field strengths
in iron cores probably do not exceed $10^{9} \mathrm{G}$, and the
toroidal field component is expected to be much stronger than the
poloidal one \citep{Heger_Woosley_Spruit_2005}.  However, as such
``weak'' initial fields do not give rise to important dynamic effects on
the time scales under consideration here (unless MRI amplification would
take place; see Sect.~\ref{Sek:Ergebnissse}), and as we want to
investigate the principal effects of magnetic fields on the core
collapse, we consider stronger fields in our parameter study.  Weaker
initial fields may lead to similar effects after a longer amplification
phase.  Our initial fields are purely toroidal, but they rapidly (within
a fraction of the collapse time scale) develop a strong toroidal
component.

\begin{table}[ht]
  \caption[Initial magnetic fields] {Parametrisation of the initial
  magnetic fields for the models of series AaBbGg-DdMm by the radius
  of the field generating current loop centered at $r_{\mathrm{mag}}$
  (parametrized by $d = 1, 2, 3, 4, 0$) and the field strength in the
  core's center $B_0=\sqrt{4\pi}b_0$ (parametrized by $ m = 10, 11,
  12, 13$).  For models AaBbGg-D0Mm the field generating current loop
  is located at infinity yielding a uniform magnetic field throughout
  the entire core.  }
  \label{Tab:MinitPars}
  \centering
  \begin{tabular}{cc|cl}
    \hline\hline
    Model & $r_{\mathrm{mag}} \ [\mathrm{km}]$ & Model 
          & $b_0  \ [\textrm{G}]$ \\
    \hline
    D1 & $100$   & M10 & $10^{10}$ \\
    D2 & $200$   & M11 & $10^{11}$ \\
    D3 & $400$   & M12 & $10^{12}$ \\
    D4 & $800$   & M13 & $10^{13}$ \\
    D0 & $\infty$  & & \\
    \hline\hline
  \end{tabular}
\end{table}

\begin{table}[ht]
  \caption[Initial magnetic fields] {Initial magnetic energy
  $E_{\mathrm{mag}}$ and typical values of the ratio of magnetic to
  gravitational energy $\beta_{\mathrm{mag}}$ (the exact values depend
  also on the hydrodynamic initial model and its gravitational energy)
  for the models of series AaBbGg-DdM12. The magnetic energy of models
  with $b_0 \ne 10^{12}\mathrm{G}$ can be obtained by a simple
  scaling relation, e.g.\,$E_{\mathrm{mag}}^{AaBbGg-D2M10} =
  E_{\mathrm{mag}}^{AaBbGg-D2M12}\, (10^{10} / 10^{12})^2$. }
  \label{Tab:MinitErg}
  \centering
  \begin{tabular}{cll}
    \hline \hline
    Model & $E_{\mathrm{mag}} \ [\mathrm{erg}]$ 
          & $\log \beta_{\mathrm{mag}}$\\
    \hline 
    D0M12 & $1 \cdot 10^{49}$ & -2.7\\
    D4M12 & $6 \cdot 10^{48}$ & -3.0\\
    D3M12 & $7 \cdot 10^{47}$ & -3.9\\
    D2M12 & $7 \cdot 10^{46}$ & -4.9\\
    D1M12 & $1 \cdot 10^{46}$ & -5.8\\
    \hline \hline
  \end{tabular}
\end{table}

\subsection{Gravitational-wave emission}
\label{Suk:Modelle:GW}

During collapse, bounce, and explosion the rapid infall of matter and in
particular its more or less abrupt slowdown give rise to strong
variations of the matter--density quadrupole moment of any aspheric
core.  This causes the emission of gravitational radiation.

We calculate the gravitational wave amplitude of the core using the
quadrupole formula in spherical coordinates, and applying the extension
of the formulation of \citet[ MSMK, hereafter]{Moenchmeyer_1991} to the
MHD case due to \citet{Kotake_etal_PRD_2004__MagCollapse_GW}. Our
treatment includes the hydrodynamic, gravitational, and magnetic forces
acting on the fluid.  We calculate the quadrupole amplitude
$A^{\mathrm{E}2}_{20}$ according to the formula (see Appendix
\ref{Sek:GW}):
\begin{eqnarray}
  \label{Gl:AE220_1}
  A^{\mathrm{E}2}_{20} & = &
  \frac{G}{c^4} \frac{32\pi^{\frac{3}{2}}}{\sqrt{15}}
  \int_{0}^{1} \mathrm{d} z \int_{0}^{\infty} 
  \mathrm{d} \frac{r^3}{3} \nonumber \\
  && 
  \left[
    f_{rr} (3z^2-1) + f_{\theta\theta}(2-3z^2)  - f_{\phi\phi}
    - 6 f_{r\theta} z\sqrt{1-z^2} - \right. \nonumber \\
  && \left. - r \partial_r\Phi (3z^2-1) + 3 \partial_{\theta}\Phi z\sqrt{1-z^2}
  \right],
\end{eqnarray}
where the components of $f_{ij}$ are given by
\begin{equation}
  \label{Gl:GWLumHM}
  f_{ij}  = \rho v_i v_j - b_i b_j.
\end{equation}
In the following, we will refer to the various parts of the total
amplitude as follows: $A^{\mathrm{E}2}_{20;v_iv_j}$,
$A^{\mathrm{E}2}_{20;b_ib_j}$, and $A^{\mathrm{E}2}_{20;\mathrm{G}_i}$
denote the contributions of the terms involving $v_i v_j$, $b_i b_j$,
and $\partial_i \Phi$ in Eq.\,\ref{Gl:AE220_1}, respectively.
Furthermore $A^{\mathrm{E}2}_{20;\mathrm{hyd}}$,
$A^{\mathrm{E}2}_{20;\mathrm{mag}}$, and
$A^{\mathrm{E}2}_{20;\mathrm{grav}}$ are the sums over all components of
$A^{\mathrm{E}2}_{20;v_iv_j}$, $A^{\mathrm{E}2}_{20;b_ib_j}$, and
$A^{\mathrm{E}2}_{20;\mathrm{G}_i}$, respectively.

The radiative quadrupole moment $M^{\mathrm{E}2}_{20}$
(Eq.\,\ref{Gl:ME220}) is a measure of the asphericity of the core's
density distribution.  It is positive for a very prolate core, and
negative in the limit of very oblate cores.  Its first time derivative
$N^{\mathrm{E}2}_{20}$ (Eq.\,\ref{Gl:NE220}) measures the asphericity of
the mass--flux and the momentum distribution of the core, and its second
time derivative, the quadrupole amplitude $A^{\mathrm{E}2}_{20}$, is a
measure of the asphericity of the forces acting on the fluid.  As a rule
of thumb, a prolate mass--flux or a prolate momentum distribution
(e.g.\,a bipolar jet--like outflow along the rotational axis) gives rise
to a positive value of $N^{\mathrm{E}2}_{20}$. Forces that act on the
core in a way to make it more oblate, such as the centrifugal force that
has its manifestation in the $A^{\mathrm{E}2}_{20;v_{\phi}v_{\phi}}$
part of the amplitude, will give rise to a negative contribution to the
total amplitude (negative sign of the $A^{\mathrm{E}2}_{20;
  v_{\phi}v_{\phi}}$ term in Eq.\,\ref{Gl:AE220_1}).

The different signs of the hydrodynamic and the magnetic contributions
to the amplitude (Eq.\,\ref{Gl:GWLumHM}) resulting from the different
signs of the hydrodynamic (Reynolds) and the magnetic (Maxwell) stresses
in the MHD flux terms, will -- for suited topologies of field and flow
-- lead to a more or less prominent phase shift between the hydrodynamic
and the magnetic amplitude.  If the gravitational wave amplitude
$A^{\mathrm{E}2}_{20;\mathrm{G}_i}$ is in phase with the hydrodynamic
amplitude $A^{\mathrm{E}2}_{20;\mathrm{hyd}}$ (which holds well for many
models, in particular for those with relatively long oscillation
periods; see Sect.\,\ref{Sek:Ergebnissse}), the magnetic amplitude
$A^{\mathrm{E}2}_{20;\mathrm{mag}}$ may be phase shifted with respect to
$A^{\mathrm{E}2}_{20;\mathrm{hyd}} + A^{\mathrm{E}2}_{20;\mathrm{G}_i}$.
Such a phase shift was observed by \citet{Yamada_Sawai_ApJ_2004__MHDCollapse}.

\section{Numerical method}
\label{Sek:Num}

The MHD equations are integrated using a newly developed Eulerian,
finite volume code based on the algorithm devised by
\citet{Pen_Arras_Wong_APJS_2003__rTVD_MHD_Code}.  This code employs the
relaxing TVD method of \citet{Jin_Xin_CPAM_1995__rTVD} for the solution
of the advection equations and the constraint--transport formulation of
\citet{Evans_Hawley_ApJ_1988__CTM} to deal with the divergence
constraint of the magnetic field.

We have rewritten the original code of Pen et\,al.\,in order to adjust
for the simulations of stellar core collapse.  This included the
transformation of the equations from Cartesian to spherical coordinates,
the calculation of the gravitational potential, the implementation of
the gravitational source terms in the momentum and energy equations, and
the implementation of an approximate equation of state for iron core
matter.  The integration of the fluid equations and of the induction
equation is based on a second-order (piecewise--linear) relaxing TVD
method.  For a short summary of this method see Appendix \ref{Sek:rTVD}.
For the time evolution we use an operator--split approach based on a
method of lines \citep{LeVeque_Book_1992__Conservation_Laws}.

The simulations were performed on a grid of 380 logarithmically spaced
radial zones up to $R_{\star} \ \mathrm{km}$, where $R_{\star} \sim
1700...2000 \ \mathrm{km}$ is the radius of the initial stellar model.
The central resolution was $(\Delta_r)_{\mathrm{c}} \approx 300 \ 
\mathrm{m}$.  The angular grid consisted of 60 equidistant zones in the
domain $0 \leq \theta \leq \frac{\pi}{2}$.  This grid resolution has
been chosen after obtaining converged results when running several
models at different resolutions.  The numerical convergence of our
simulations is demonstrated in the Appendix
(Sect.\,\ref{Sek:Convergence}).

The stellar models used by us (polytropes) are quite compact
configurations of matter characterized by a sharp transition from a high
density interior to a low density surface layer.  Note that this feature
is also seen in sophisticated stellar evolution calculations, which
predict a steep density gradient at the outer edge of the iron core. In
rotating models the transition layer is aspherical and must completely
be contained within the spherical boundary of the numerical grid,
i.e.\,its numerical treatment requires special care. Grid zones outside
the core are filled with an ''atmosphere'' fluid of some prescribed
density $\rho_{\mathrm{atmo}}$ at rest.  During collapse the infall of
matter creates a region near the edge of the core where the density
might become so low that numerical problems arise.  To overcome these
difficulties, we follow the approach of DFM1 and set the hydrodynamic
variables $\rho,\vec v, e$ equal to some prescribed values
$\rho_{\mathrm{atmo}}, \vec 0 , e_{\mathrm{atmo}}$ in all zones where
the fluid density falls short of a given threshold
$\rho_{\mathrm{cut}}$.  In this way the atmosphere can adjust to the
(non--spherical) varying shape of the star.  We used
$\rho_{\mathrm{cut}} = 10^{5}\ \mathrm{cm \, s}^{-1}$, and set
$\rho_{\mathrm{atmo}} = 10^{3}\ \mathrm{cm \, s}^{-1}$.  From the
atmospheric density and velocity one can calculate the energy density
assuming zero thermal energy.

The evolution of the magnetic field is turned off for zones marked as
atmosphere, i.e.\,$\vec b$ remains constant in the atmosphere consistent
with the assumed zero velocity of the atmosphere gas.

\section{Results}
\label{Sek:Ergebnissse}

The results of our simulations show that magneto--rotational collapse
can be categorized in essence into two limiting cases depending on the
strength of the initial magnetic field. If the initial magnetic field is
weak, its influence on the dynamics and the gravitational wave emission
is negligible during the time scales of our simulations (see
Sect.\,\ref{Suk:Dis:SchwachFeld}).  The results of the hydrodynamic
simulations of ZM and DFM apply in this case without any modification.
On the other hand, and not unexpectedly, initially strongly magnetized
cores evolve quite differently, as will be discussed in
Sect.\,\ref{Suk:Dis:StarkFeld}.  Note that since the MRI acts
independently of the strength of the initial magnetic field, the
distinction between dynamically negligible and dynamically important
fields may be an artificial one resulting from our inability to simulate
the MRI for magnetic fields below a certain threshold (see
Sect.\,\ref{Suk:MRI}). Hence, also initially weak magnetic fields may
cause similar dynamical effects as strong ones.  The dependence of our
results on the initial magnetic field configuration will be discussed in
Sect.\,\ref{Suk:Dis:FC}, and further information about the temporal
evolution of all models is provided in appendix \ref{Sek:Syn}.

\subsection{Weak initial fields}
\label{Suk:Dis:SchwachFeld}

\subsubsection{GW signal and dynamics}
\label{Suk:Dis:SchwachFeld:GW}

\begin{figure*}[!htbp]
  \centering
  \includegraphics[width=5.6cm]{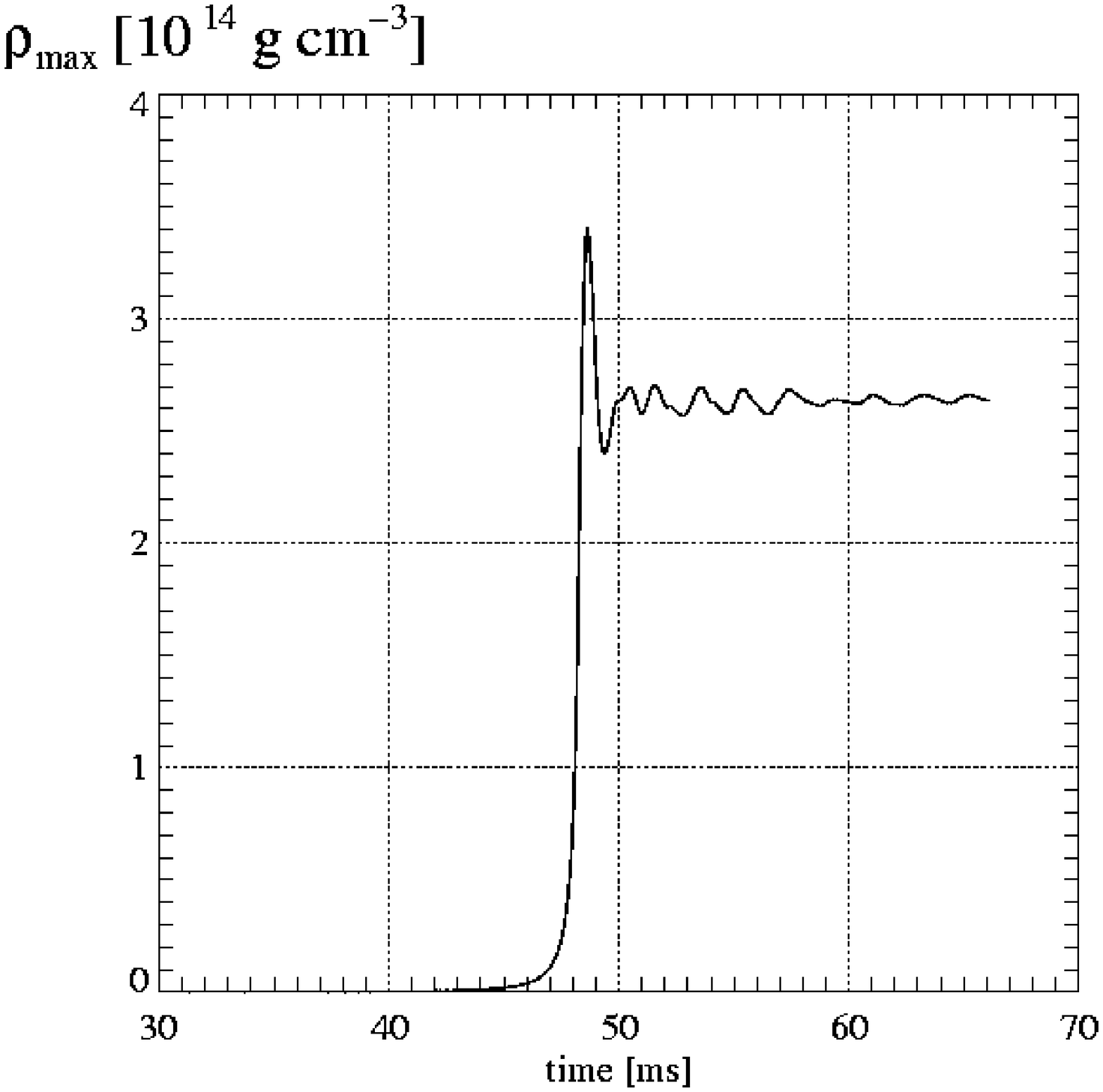}
  \includegraphics[width=5.6cm]{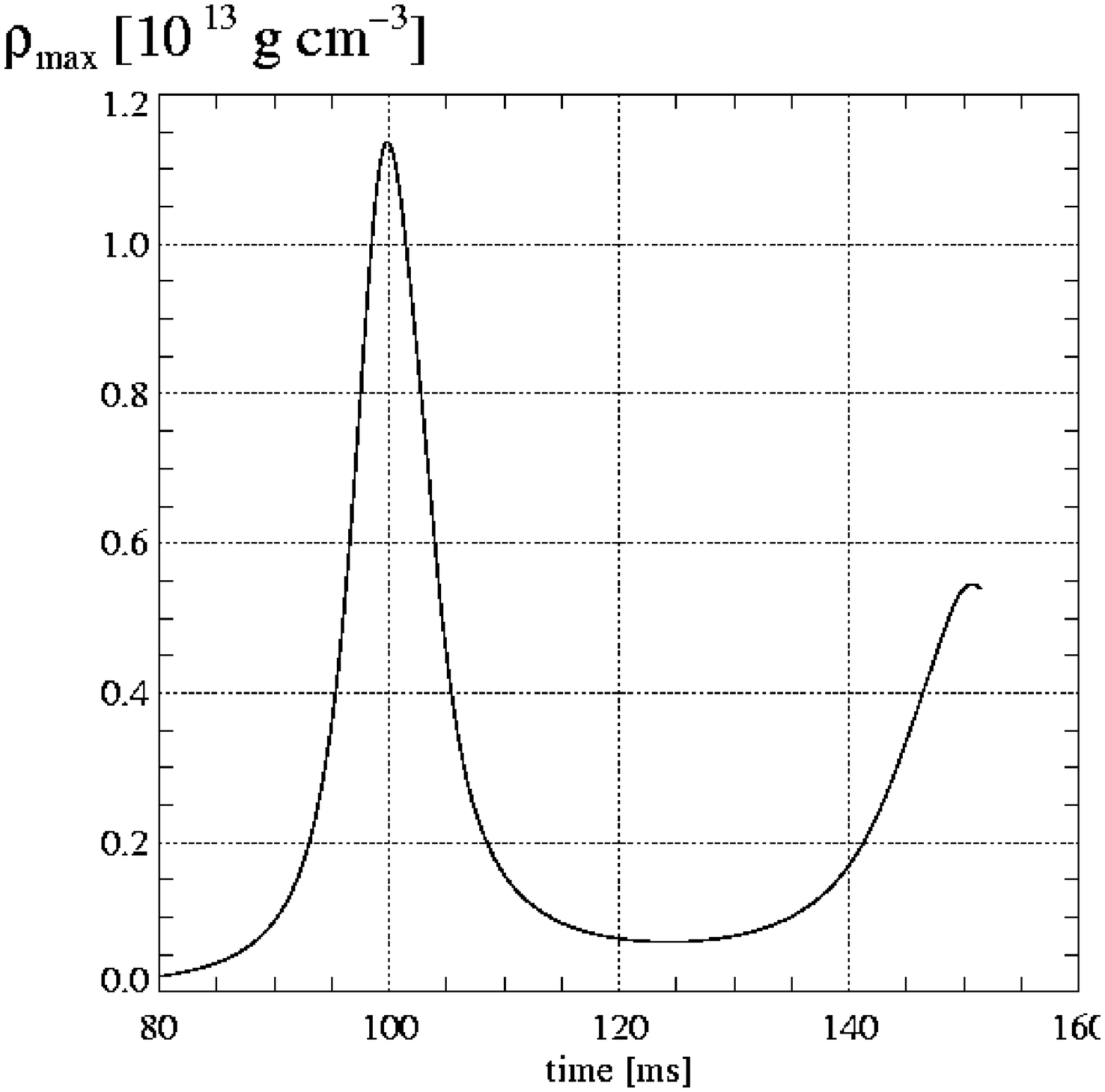}
  \includegraphics[width=5.6cm]{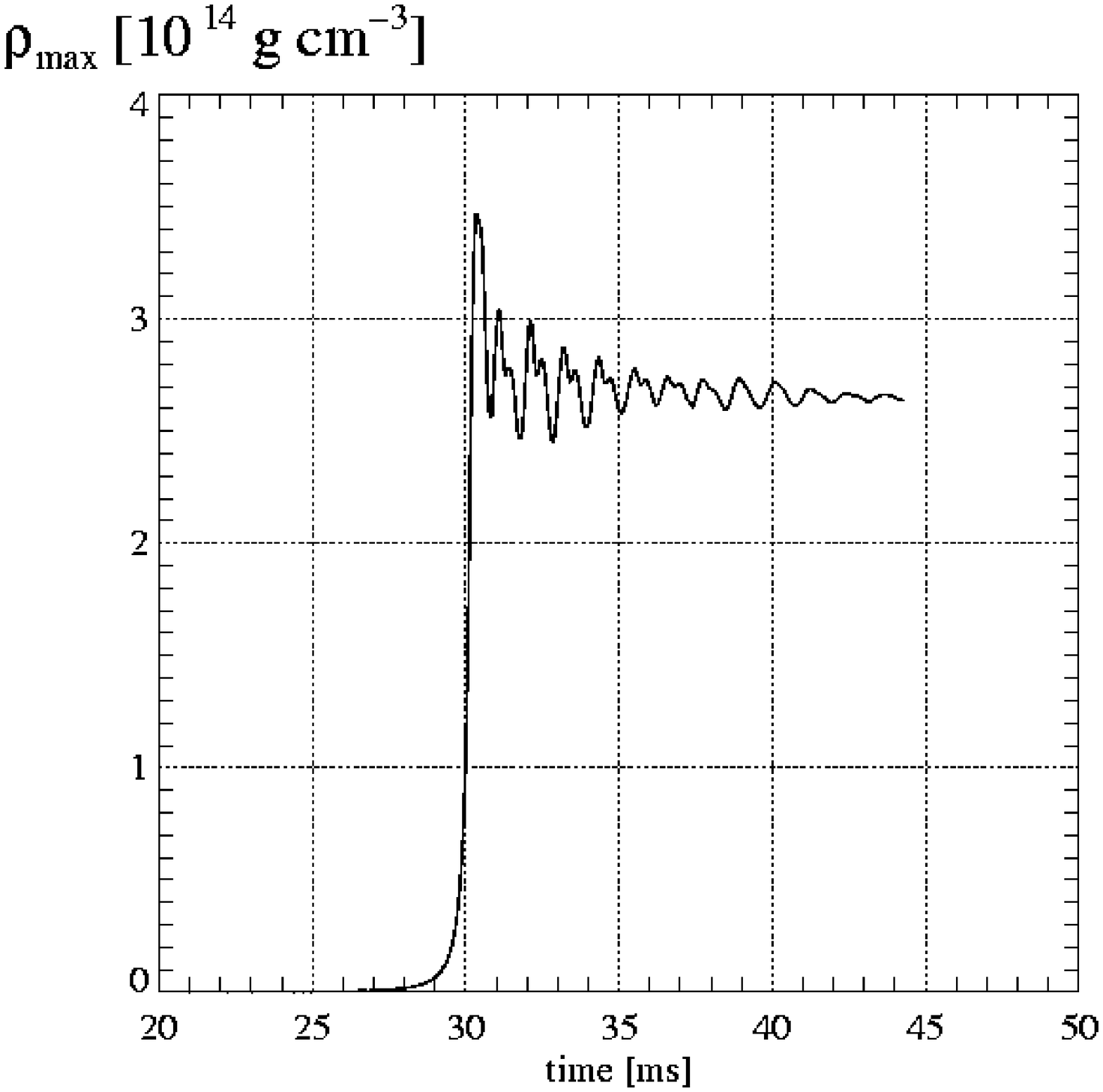}
  \includegraphics[width=5.6cm]{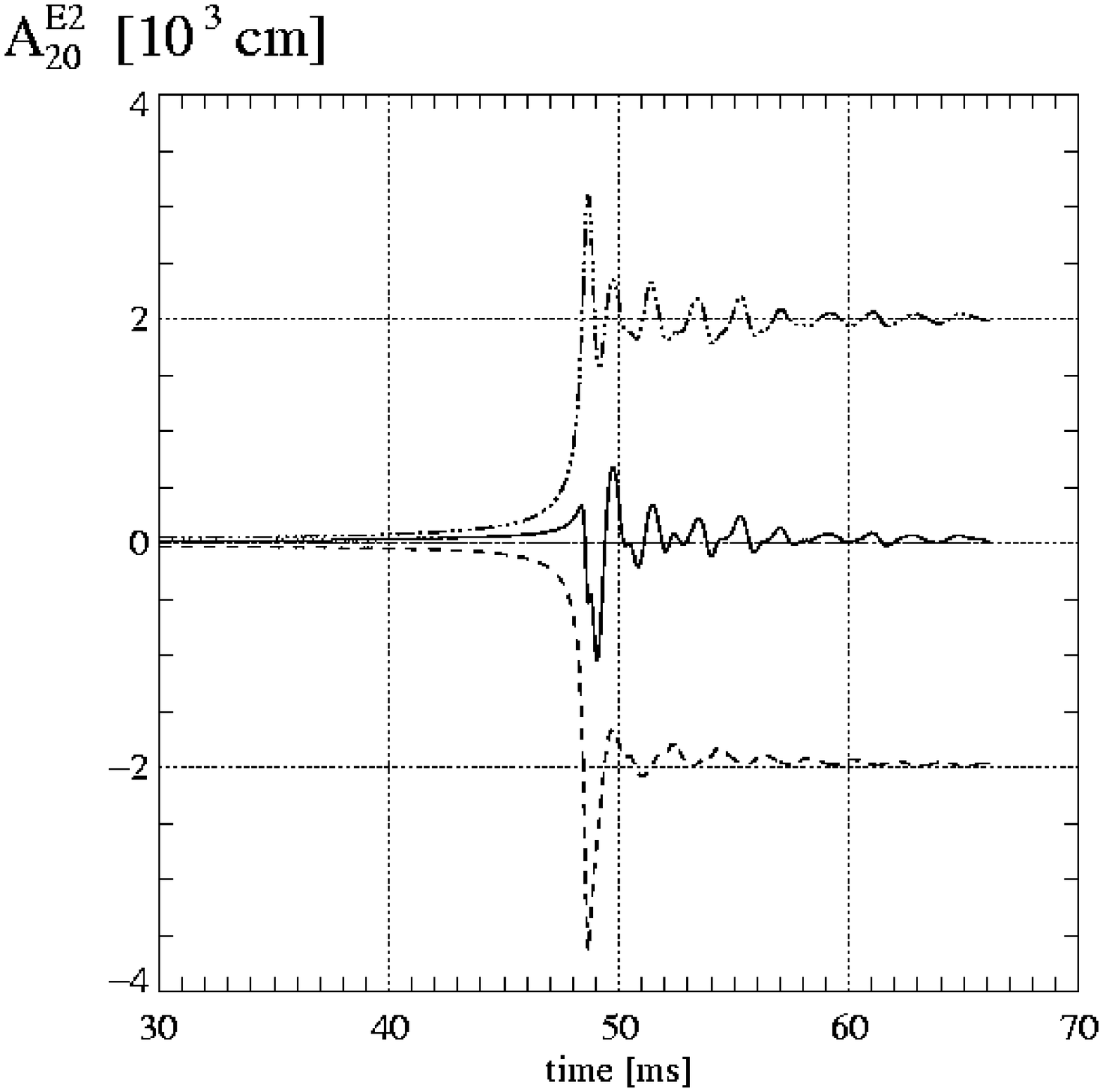}
  \includegraphics[width=5.6cm]{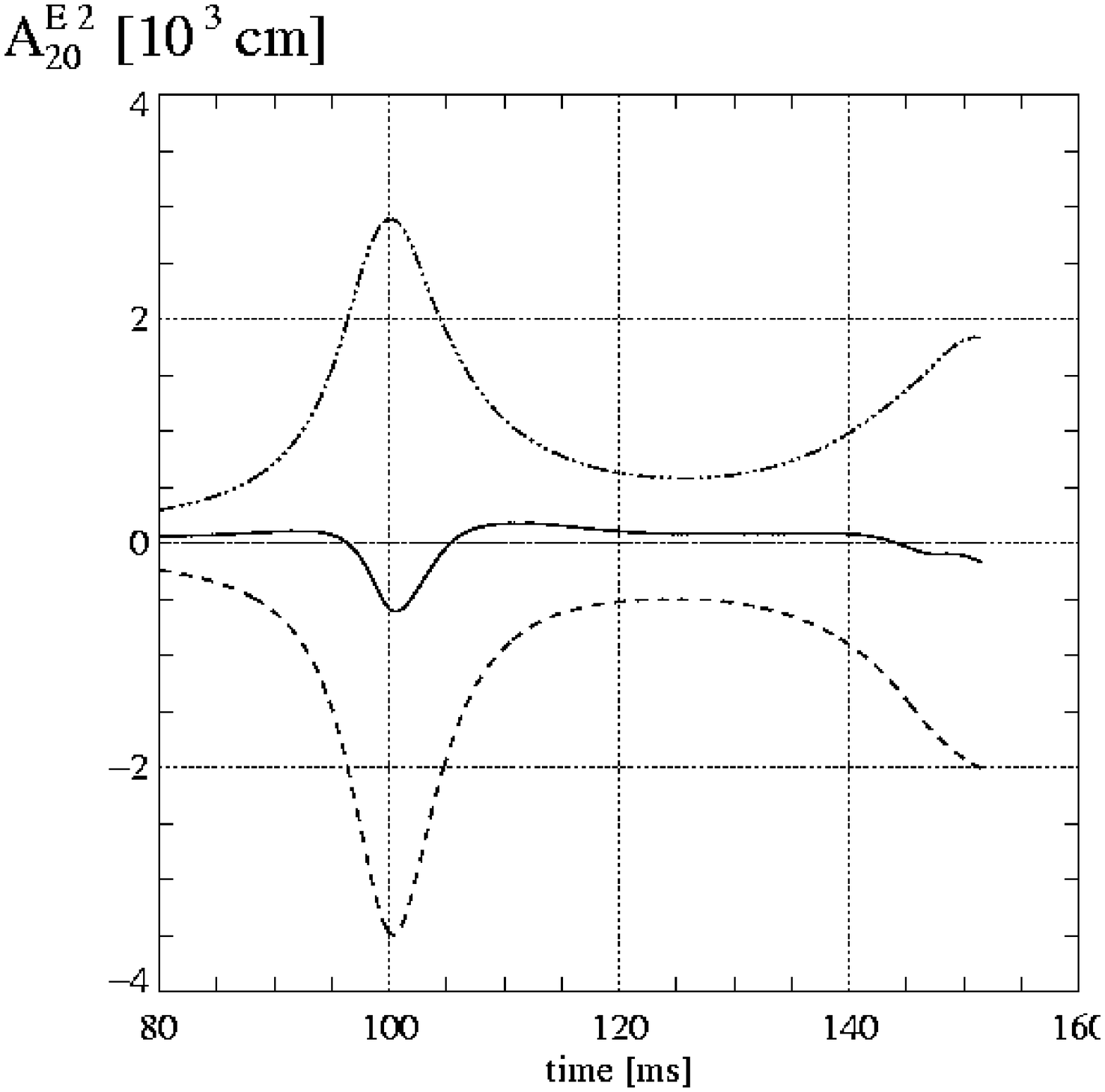}
  \includegraphics[width=5.6cm]{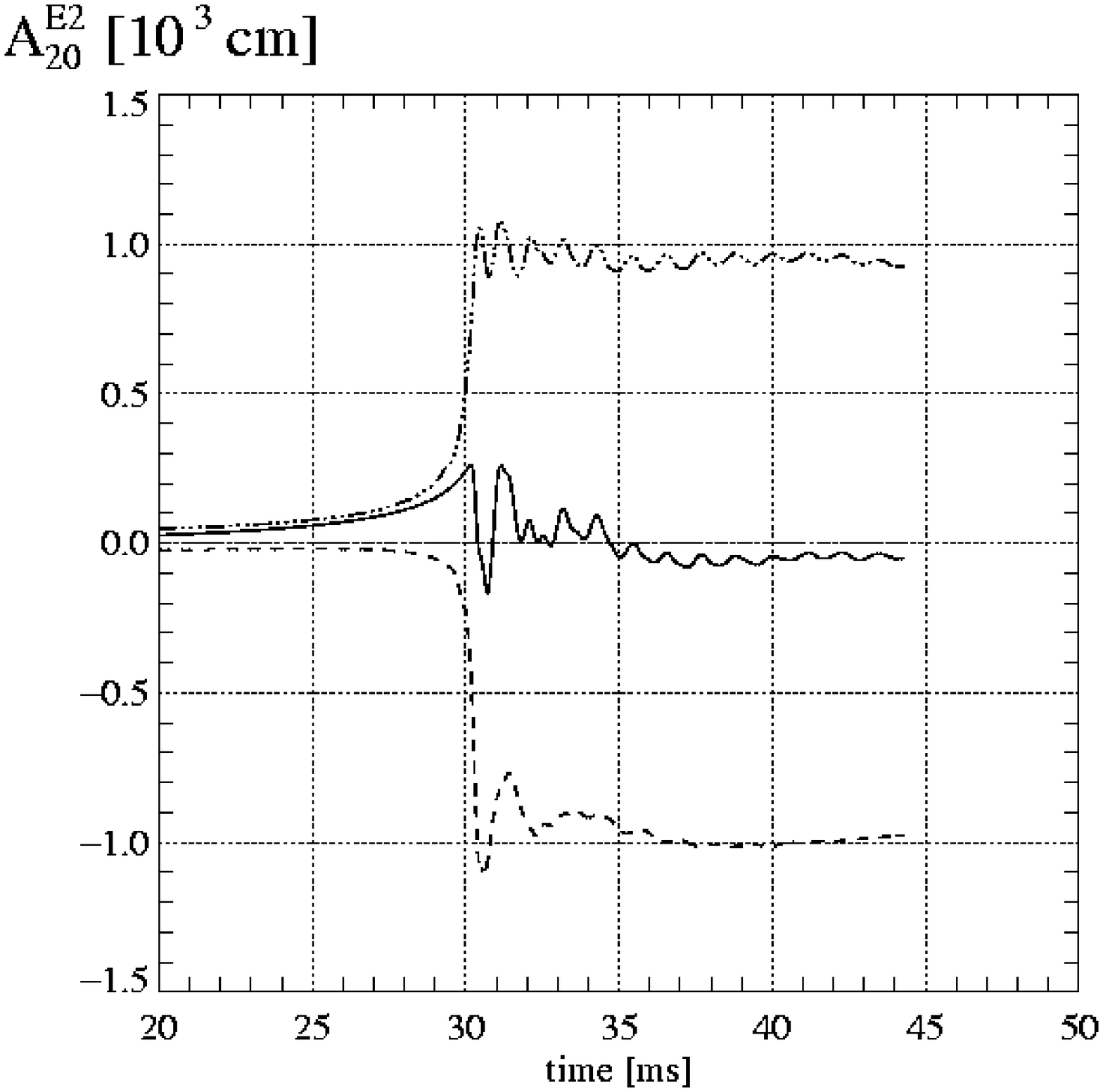}
  \caption[GW-signal types: non-magnetic case] 
  {The evolution of the maximum density (top panels) and the GW
    amplitude (bottom panels) of three weak--field models: the standard
    type-I model A1B3G3-D3M10 (left panels), the type-II model
    A2B4G1-D3M10 (middle panels), and the type-III model A3B3G5-D3M10
    (right panels).  The bottom panels show the total GW amplitude
    $A^{\mathrm{E}2}_{20}$ (solid lines) together with the partial
    amplitudes $A^{\mathrm{E}2}_{20; \mathrm{hyd}}$ (dashed lines),
    $A^{\mathrm{E}2}_{20; \mathrm{mag}}$ (dash-dotted lines; here almost
    zero), and $A^{\mathrm{E}2}_{20; \mathrm{grav}}$ (dash-dot-dot-dot
    line). }
\label{Fig:sFF_rho_und_GW}
\end{figure*}

The gravitational wave signals of initially weakly magnetized cores do
not differ from those obtained by MM, ZM and DFM for the corresponding
non--magnetized initial models, because the magnetic fields never become
dynamically important during the simulations.  Consequently, both the
magnetic force contribution to the GW amplitude (see
Sect.\,\ref{Suk:Modelle:GW}) and the back--reaction of the magnetic
field on the flow are negligible. This ''weak--field'' behavior holds
for most models with initial fields of $\la 10^{11} \ \mathrm{G}$.
However, in some models bouncing at relative low central density due to
centrifugally forces (e.g.\,models A4B5G5-D3M12 and A2B4G1-D3M12) even a
field of $10^{12} \ \mathrm{G}$ does not affect the evolution of the
core significantly.  We note that these numbers may change when the
field amplification by the MRI, which works independently of the seed
field strength, is properly simulated (see Sect.\,\ref{Suk:MRI}).

Three types of GW signals are distinguished by MM and ZM which they call
type-I, type-II, and type-III, respectively. The evolution of three
representative weak--field models that emit these different signal types
is illustrated in Fig.\,\ref{Fig:sFF_rho_und_GW}.  Although the GW
signals are very similar to those of the corresponding non--magnetized
cores of ZM, we will discuss them and the underlying dynamics here in
some detail for reason of comparison with the GW signals of strongly
magnetized cores (see Sect.\,\ref{Suk:Dis:StarkFeld}).

In the case of a \emph{type-I} (or standard--type) GW signal (model
A1B3G3-D3M10; Fig.\,\ref{Fig:sFF_rho_und_GW}, left panels), the core
bounces due to the stiffening of the nuclear EOS at supra--nuclear
densities.  Pressure waves crossing the inner core stop the infall and
lead to the formation of an outward moving shock wave.  Hence, the
typical time scale is roughly given by the sound crossing time of the
inner core at bounce ($\sim 1\,$ms).  After bounce the core exhibits
damped oscillations on roughly the same time scale.  During collapse,
the GW amplitude is increasingly positive due to the gravity
contribution $A^{\mathrm{E}2}_{20;\mathrm{grav}}>0$ which dominates the
hydrodynamic one $A^{\mathrm{E}2}_{20;\mathrm{hyd}} < 0$.  At bounce the
GW amplitude decreases strongly assuming large negative values, because
the modulus of the centrifugal contribution
$A^{\mathrm{E}2}_{20;v_{\phi}v_{\phi}}$ becomes very large when the
rotational energy approaches its maximum.  The oscillations of the core
produce the oscillations of $A^{\mathrm{E}2}_{20;\mathrm{grav}}$ and
hence of the total GW amplitude on the same time scale.  As the shock
wave is almost spherical (except for initially very rapidly rotating
cores), the post--bounce GW amplitude is predominantly produced by the
central core with only a modest contribution of the outer layers.  The
hydrodynamic contribution $A^{\mathrm{E}2}_{20;\mathrm{hyd}}$ is
dominated by $A^{\mathrm{E}2}_{20;v_{\phi}v_{\phi}}$.

A \emph{type-II} signal (model A2B4G1-D3M10;
Fig.\,\ref{Fig:sFF_rho_und_GW}, middle panels) is emitted, if, for an
initially sufficiently fast rotation and a sufficiently large degree of
differential rotation, the core suffers a bounce due to centrifugal
forces before nuclear density is reached.  After bounce the core
exhibits several large amplitude oscillations, which are considerably
less damped than in the case of a pressure dominated bounce.  The
oscillations are reflected in the GW signal: each time the density
reaches a maximum the GW amplitude becomes strongly negative, while
being positive otherwise.  The negative GW amplitude again results from
the enhanced rotational energy during this collapse phase when the inner
core also becomes strongly oblate.  The contribution of the
centrifugal-force amplitude $A^{\mathrm{E}2}_{20;v_{\phi}v_{\phi}}$ to
the hydrodynamic one $A^{\mathrm{E}2}_{20;\mathrm{hyd}}$ is negligible.
Since much of the dynamics happens at relatively large radii, the GW
signal is produced by the whole core.  The contribution of the central
core is typically negative because of its rotationally flattened, oblate
shape, whereas the outer layers contribute to the overall amplitude with
a positive signal due to the prolate shape of the outward propagating
shock wave.

The signals of \emph{type III} (model A3B3G5-D3M10;
Fig.\,\ref{Fig:sFF_rho_und_GW}, right panels) that are emitted by cores
collapsing very rapidly due to a very soft sub--nuclear EOS are
characterized by the lack of a strongly negative GW amplitude at bounce.
This results from the only modest rise of the rotational energy at
bounce implying a relatively small centrifugal contribution to the GW
signal $A^{\mathrm{E}2}_{20;v_{\phi}v_{\phi}}$, and from the relatively
large positive contribution of the still collapsing layers outside to
the shock wave.  Past bounce the signal exhibits rapid (on the
hydrodynamic time scale $t_{\mathrm{hyd}} \approx 1,$ms) oscillations
like the maximum density.  The contributions from the outer layers of
the core are also responsible that the signal then remains positive for
an extended period of time (i.e.\,several oscillation periods) during
which the amplitude gradually decreases.

\subsubsection{Field amplification}
\label{Suk:Dis:SchwachFeld:Feld}

\begin{figure}[!htbp]
  \centering 
  \includegraphics[width=4cm]{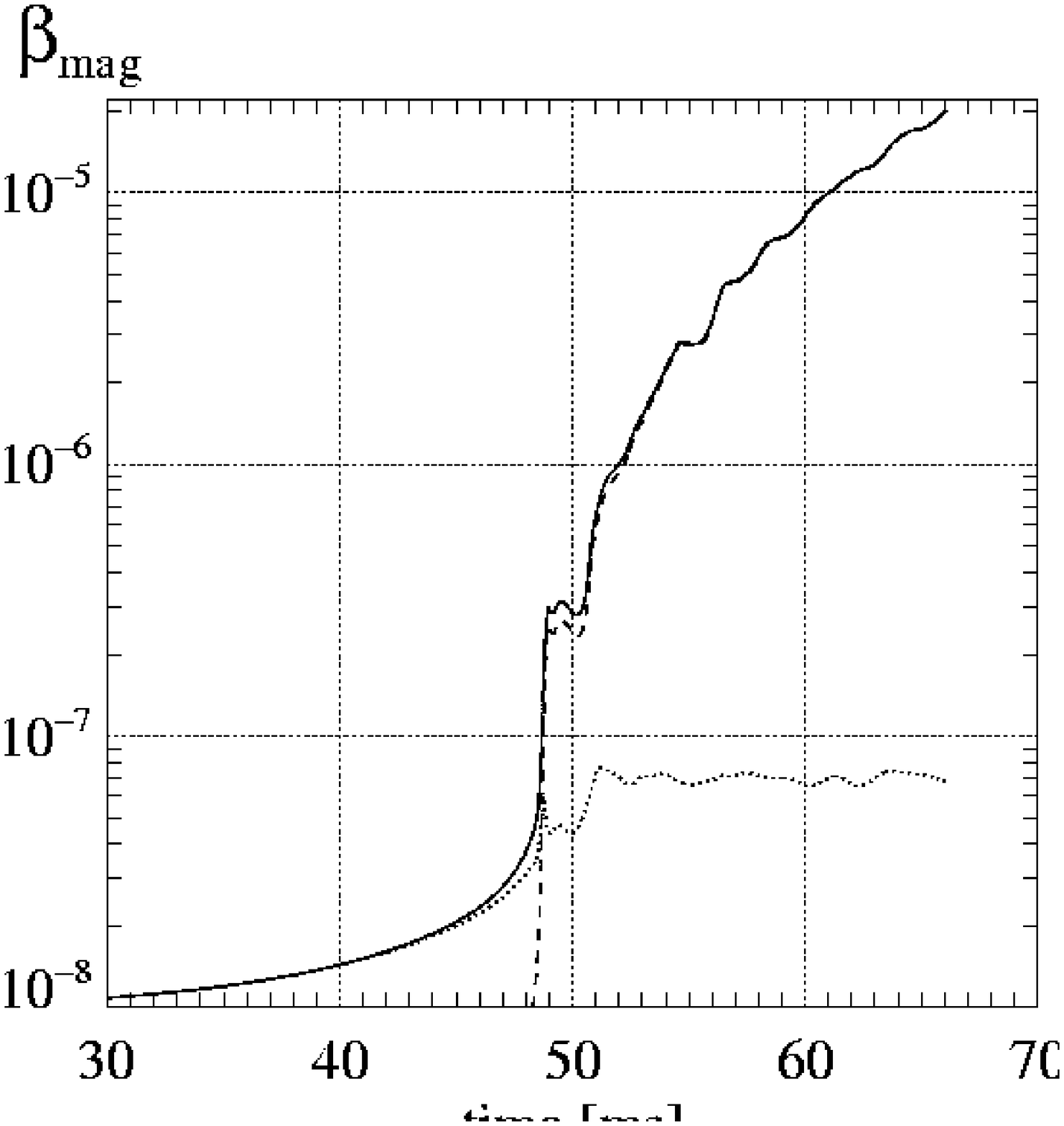}
  \includegraphics[width=4cm]{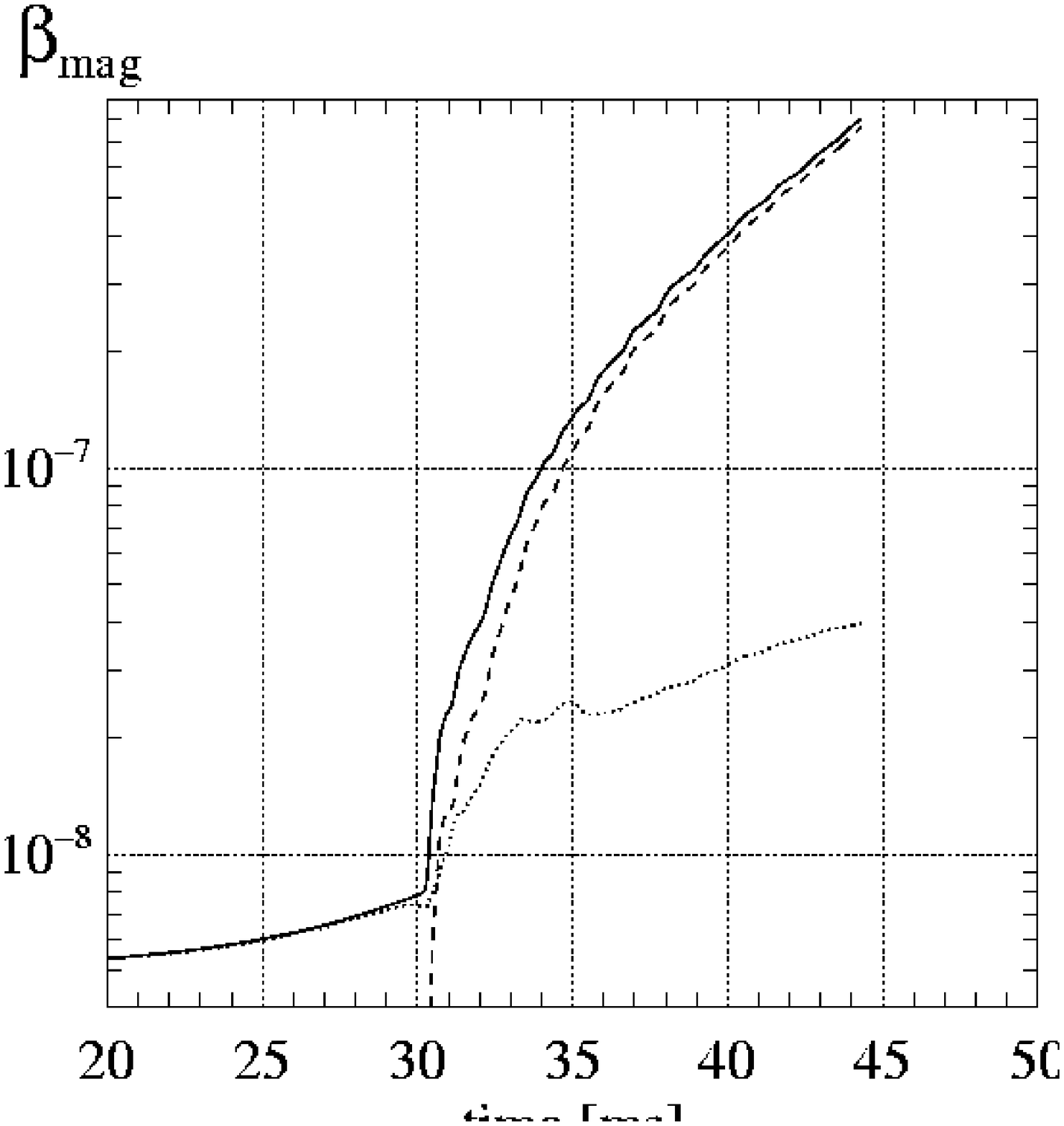} \par
  \includegraphics[width=4cm]{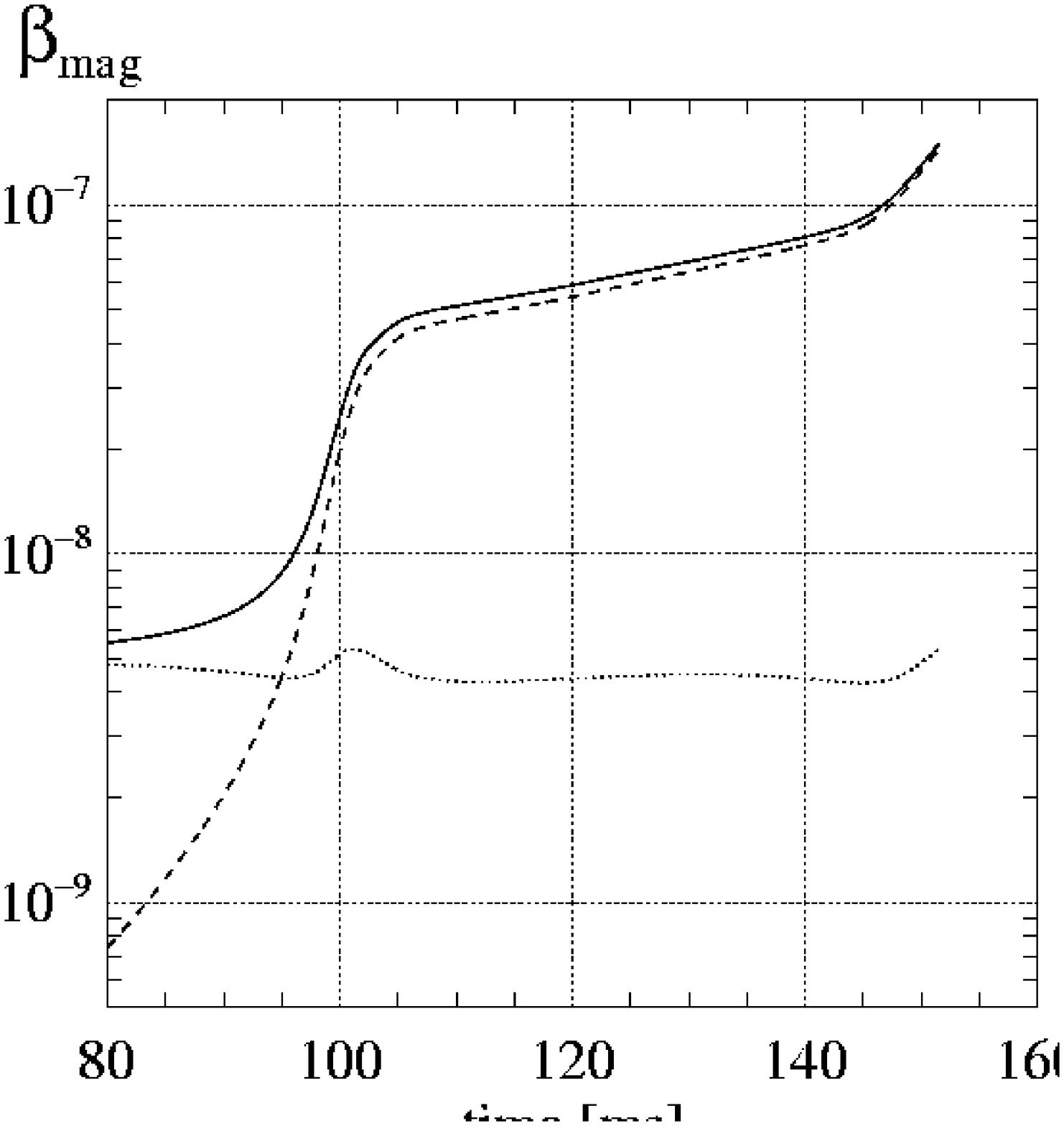}
  \includegraphics[width=4cm]{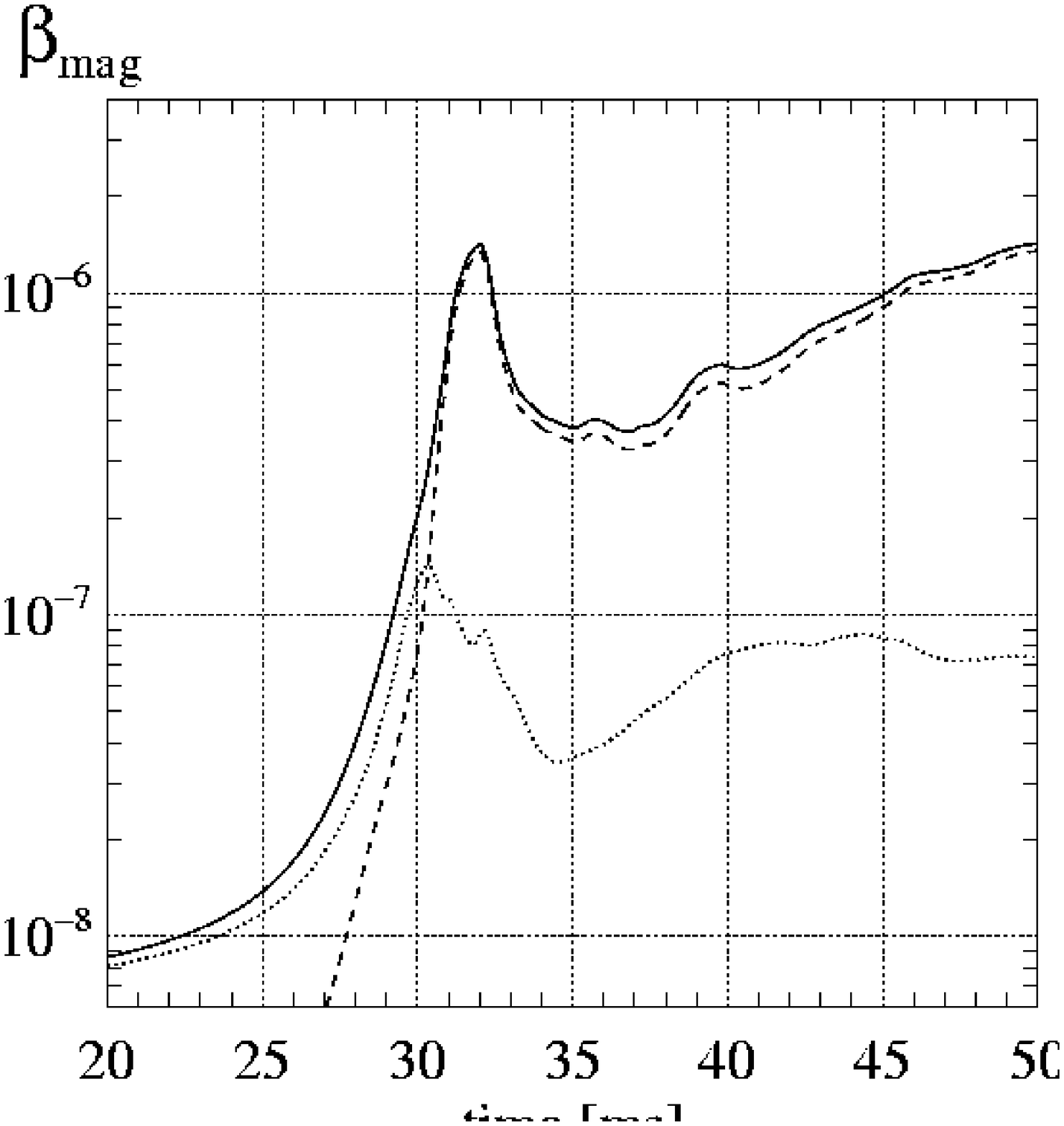}
  \caption[Evolution of the Magnetic Field: Weak-Field limit.]  
  { The evolution of the magnetic energy parameters
    $\beta_{\mathrm{mag}}$ (total energy, solid line), $\beta_{\phi}$
    (toroidal field energy, dashed line), and
    $\beta_{\mathrm{mag}}-\beta_{\phi}$ (poloidal field energy, dotted
    line) of models A1B3G3-D3M10 (upper left panel), A3B3G5-D3M10 (upper
    right panel), A2B4G1-D3M10 (lower left panel), and A4B5G5-D3M10
    (lower right panel), respectively.  Note that in the upper left
    panel $\beta_{\mathrm{mag}} \approx \beta_{\phi}$ and, thus, both
    lines are almost indistinguishable.  }
\label{Fig:Mag1}
\end{figure}

The evolution of the magnetic energy is illustrated for four selected
models in Fig.\,\ref{Fig:Mag1} showing the ratio of the field energy and
the gravitational energy for the total ($\beta_{\mathrm{mag}}$),
toroidal ($\beta_{\phi}$), and poloidal
($\beta_{\mathrm{mag}}-\beta_{\phi}$) magnetic field, respectively.
Both the total and the toroidal magnetic energies rise sharply at bounce
by a factor of $10^{2}$ to $10^{3}$.  The magnetic energy is mostly
stored in form of the toroidal magnetic field $B_{\phi}$, which is
created from the poloidal component by the action of differential
rotation. The amplification process extends beyond the end of our
simulations.

As our study is restricted to axisymmetric simulations, the main field
amplification mechanism is the conversion of rotational into magnetic
energy via the $\Omega$--dynamo.  Axisymmetry suppresses most of the
instabilities of the (toroidal) field that are necessary to close the
dynamo loop of a full $\alpha-\Omega$ dynamo, where a poloidal field is
converted into a toroidal one and amplified by differential rotation,
and then converted back into a poloidal field by some instability of the
toroidal field
\citep{Spruit_AAP_1999__MHD_star,Spruit_AAP_2002__Dynamo}.

Before we discuss the properties of some models in greater detail, we
summarize a few general trends:
\begin{itemize}
\item 
  The larger the initial rotational energy (for a given degree of
  differential rotation and a given EOS), the larger is the rate of
  field amplification.
\item
  The higher the degree of differential rotation (for a given
  $\beta_{\mathrm{rot}}$), the larger is the rate of field
  amplification.
\item
  Among a series of models with the same initial configuration, the
  magnetic field is amplified more efficiently for models whose EOS
  has a larger sub--nuclear adiabatic index $\Gamma_1$ and do not
  suffer a centrifugal bounce.  The time scale for the amplification
  of the magnetic field is set by the rotation period of the core.
\item
  The collapse of models with a relatively stiff sub--nuclear EOS
  (AaBbG1-DdMm) leads to a very extended post--bounce core having a
  rotational frequency much less than that of a corresponding type-I
  model even though the rotational energy and $\beta_{\mathrm{rot}}$
  may be larger in the former model.  Thus, the field is amplified
  more slowly in models suffering a sub--nuclear bounce due to
  centrifugal forces, and their cores experience several phases of
  expansion during which they slow down. This leads to a significantly
  less efficient amplification or even a weakening of the magnetic
  field.  We thus find that type-I models are most efficient in terms
  of magnetic field amplification.
\item Due to the efficient field amplification most of our weak--field
  models reach magnetic field strengths of the order of
  $10^{13}\mathrm{G}$ to $10^{15}\mathrm{G}$ within a few tens of
  milliseconds after core bounce.  This is in the range of field
  strengths observed in magnetars.  From the definition of
  $\beta_{\mathrm{mag}}$ follows, that
  \begin{equation}
    \label{Gl:field-beta}
    \beta_{\mathrm{mag}} 
    \sim
    \frac{\frac{4 \pi}{3} b^2/2 \ R^4}{G M^2}
    ,
  \end{equation}
  where $b$ is a typical value of the field within the core, and $R$
  and $M$ are its mass and radius, respectively.  Thus, a typical
  post--bounce field corresponds to a magnetic energy parameter of
  \begin{equation}
    \label{Gl:Ampl-field-beta}
    \beta_{\mathrm{mag}} 
    \sim 
    0.05 \ \% \left( \frac {b}{10^{15} \ \mathrm{G}}\right)^2 
    \left( \frac{R}{20 \ \mathrm{km}}\right)^{4}
    \left( \frac{M}{0.5 \ M_{\sun}}\right)^{-2}
    .
  \end{equation}
\end{itemize}

The maximum energy that can go into the magnetic field is limited by the
energy that is contained in the differential rotation of the core.  None
of our weak--field cores is evolved long enough for the field to reach
saturation strength.  However, examining the field amplification process
for models with strong ($\ga 10^{12} \mathrm{G}$) initial fields,
which reach dynamically important values not too early, we conclude that
efficient field amplification ceases when the magnetic energy of the
collapsed core approaches a significant fraction of its rotational
energy, i.e.\,some ten percent.  As all of our cores initially have
similar rotational energies ($\beta_{\mathrm{rot}}$ about a few
percent), we expect them to reach (surface) field strengths of the same
order if there is sufficient time for amplification.  From our models
with initially stronger fields we can estimate the saturation field
strengths to be of the order of $b \ga 10^{15}\ \mathrm{G}$.  Using
the estimate for $\beta_{\mathrm{mag}}$ given above
(\ref{Gl:Ampl-field-beta}), these fields correspond to
$\beta_{\mathrm{mag}} \sim ( 0.05 ... 5 ) \, \%$. Such values we have
found in our simulations.  The magnetic field in most post--bounce cores
is strongly concentrated in the inner core, reflecting the density
structure. Thus, for usual standard--type cores, the field is strongest
a few kilometres from the center, and drops rapidly towards larger
radii. When a strong shear layer is present at the surface of the inner
core, the field may also be strongest there. This is the case for many
of the weak-field cores, but does not necessarily hold for the
strong-field ones. In the latter case, the simulations tend to give
values of $\beta_{\mathrm{mag}}$ in excess of the previous estimate.

The field amplification via the $\Omega$--dynamo is most efficient, if
the poloidal field component is large compared to the toroidal one.  A
characteristic growth time for the generation of $b_{\phi}$ from a
radial component $b_{\varpi}$ is given by
\citep{Meier_etal_ApJ_1976__MHD_SN}
\begin{equation}
  \label{Gl:Omega-Zeit}
  \tau_{\Omega}= \frac{b_{\phi}}{b_{\varpi} \varpi 
                 \partial_{\varpi} \Omega} \,.
\end{equation}
In our models the poloidal component does not grow after bounce (apart
from the effects of compression and expansion during the large--scale
oscillations of type-II models) unless there is an instability
(Fig.\,\ref{Fig:Mag1}; Sect.\,\ref{Suk:MRI}).

The rate of field amplification is similar for the fast collapsing cores
A1B3G5-D3M10, A2B4G5-D3M10 and A3B3G5-D3M10, which have a soft
sub--nuclear EOS and which do not suffer a centrifugal bounce.  A
slightly more efficient amplification is observed for the initially more
differentially and faster rotating cores A2B4G5-D3M10 and A3B3G5-D3M10
than for the initially rigidly rotating model A1B3G5-D3M10.  An even
faster amplification is observed for the slower collapsing model
A1B3G3-D3M10, where 10\,ms after bounce $\beta_{\mathrm{mag}} = 8\cdot
10^{-5}$ compared to $\beta_{\mathrm{mag}} = 8\cdot 10^{-6}$ for model
A3B3G5-D3M10 (Fig.\,\ref{Fig:Mag1}, upper two panels).  This reflects
the different infall profiles of the cores whereby the regions of
strongest magnetic field (initially around and interior to the field
generating current loop at $r_{\mathrm{mag}} = 400\ \mathrm{km}$) are
swept along to different positions in the core.  In case of the model
A1B3G3-D3M10 this region is much closer ($r \approx 40\ \mathrm{km}$) to
the center at the time of bounce than for model A3B3G5-D3M10, where it
is located at $r \approx 120\ \mathrm{km}$.  Therefore, the fraction of
the magnetic field that is available in the regions where the
$\Omega$--dynamo acts most efficiently is larger for the former model,
i.e.\,the field amplification can proceed more efficiently.

\begin{figure}[!htbp]
  \resizebox{\hsize}{!}{\includegraphics{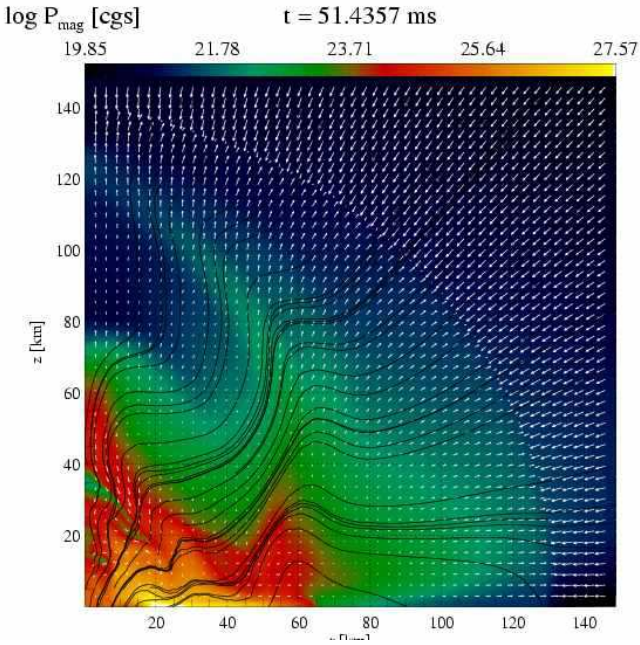}}
  \caption[Model A1B3G3-D3M10:] 
  { Magnetic pressure (color shaded), velocity field (arrows), and
    magnetic field lines of model A1B3G3-D3M10 at $t=51.44\ \mathrm{ms}$.  }
  \label{Fig:133-310:Magnet_Bounce}
\end{figure}

\begin{figure}[!htbp]
  \resizebox{\hsize}{!}{\includegraphics{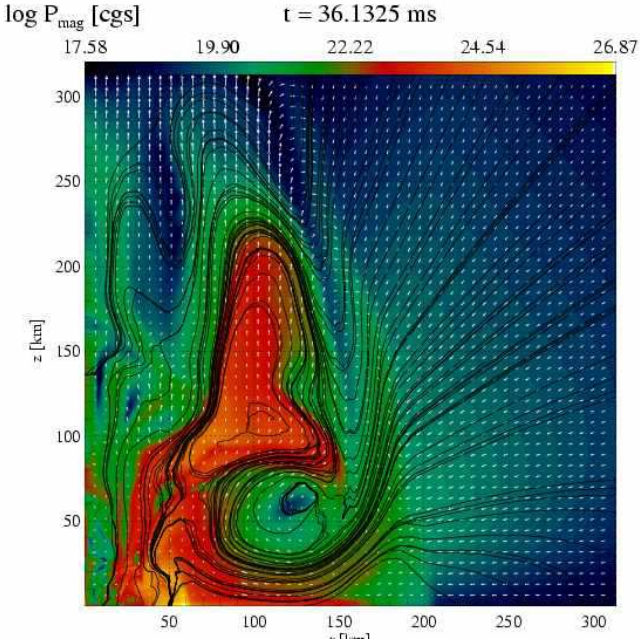}}
  \caption[Model A4B5G5-D3M10:] 
  { Same as Fig.\,\ref{Fig:455-310:Magnet_Bounce}, but for model
    A4B5G5-D3M10 at $t=36.13\ \mathrm{ms}$. }
  \label{Fig:455-310:Magnet_Bounce}
\end{figure}

During collapse and to an even larger extent during and after core
bounce the magnetic field configuration of model A1B3G3-D3M10 is
considerably distorted.  The field lines initially surrounding the
off--center current loop (Fig.\,\ref{Fig:MagInit}) are pulled towards
the center.  Apart from an overall compression the field geometry
remains basically the same as in the initial model.  A complex pattern
of ``filamentary'' regions of high and low magnetic fields develops as
the field lines get entangled by the fluid flow around and after bounce.
No simple classification of the magnetic field as a dipole, quadrupole,
etc., is possible.  In the subsequent evolution the tangled fields
follow the outward propagating shock front which is almost spherically
symmetric in this initially rigidly rotating model
(Fig.\,\ref{Fig:133-310:Magnet_Bounce}).

Multiple--bounce models that experience several phases of collapse and
expansion exhibit a similar evolution of the magnetic field.  During a
contraction phase angular momentum conservation yields a more efficient
amplification of the magnetic field energy, whereas
$\beta_{\mathrm{mag}}$ grows much slower in an expansion phase.  This is
the case for model A2B4G1-D3M10 (Fig.\,\ref{Fig:Mag1}, lower left
panel).  The magnetic energy parameter $\beta_{\mathrm{mag}}$ rises by a
factor $\approx 3$ during $10\,$ms around core bounce ($t_{\mathrm{b}}
\approx 100\ \mathrm{ms}$), but then requires $40\,$ms to amplify the
field by another factor of $\approx 1.6$.  The amplification rate
increases again strongly during the subsequent contraction phase.  For
model A4B5G5-D3M10, the amplification factor during core bounce is about
$20$, but later the field energy gets greatly reduced
($\beta_{\mathrm{mag}}$ drops by a factor $\approx 3.5$) during the
extremely violent expansion that follows the bounce in this model
(Fig.\,\ref{Fig:Mag1}, lower right panel; the evolution of the maximum
density is shown in Fig.\,\ref{Fig:455:Dyn}).  Afterwards, differential
rotation again starts to increase the magnetic field, but on a longer
time scale.

Model A4B5G5-D3M10 possesses a torus shaped initial density
distribution, and maintains it throughout the entire evolution.  The
field structure of this model tends to become very complex during the
evolution.  In the final model, at $t= 36.13\ \mathrm{ms}$, the field
exhibits both sheet--like regions of strong fields ($\approx 100\ 
\mathrm{km}$ off the axis), and also toroidal (at $r \approx 120\,\ 
\mathrm{km}$, $z \approx 60\ \mathrm{km}$) and cylindrical (e.g.\,
$\approx 50\ \mathrm{km}$ off the axis) weak--field regions
(Fig.\,\ref{Fig:455-310:Magnet_Bounce}).

\subsection{The strong field case}
\label{Suk:Dis:StarkFeld}

For initially strong magnetic fields the collapse shows significant
deviations from the purely hydrodynamic case.  The most striking new
feature is the braking of the rotation of the core by magnetic stresses.
The initially strongest fields are amplified by differential rotation
during collapse to such a level that they can cause considerable
braking, whereas initial fields of the order of $10^{12} \ \mathrm{G}$
or less require the action of an MRI--like instability to reach a level
sufficient for significant angular momentum transport on the collapse
time scale.  In the following we will first discuss the former class of
models divided into models bouncing due to pressure forces
(Sect.\,\ref{Suk:PreBouMo}) or centrifugal forces
(Sect.\,\ref{Suk:CenBouMo}), and later address the issue of the MRI in
Sect.\,\ref{Suk:MRI}.

\subsubsection{Models bouncing due to pressure forces}
\label{Suk:PreBouMo}

\begin{figure*}[!htbp] 
  \centering
  \includegraphics[width=5.6cm]{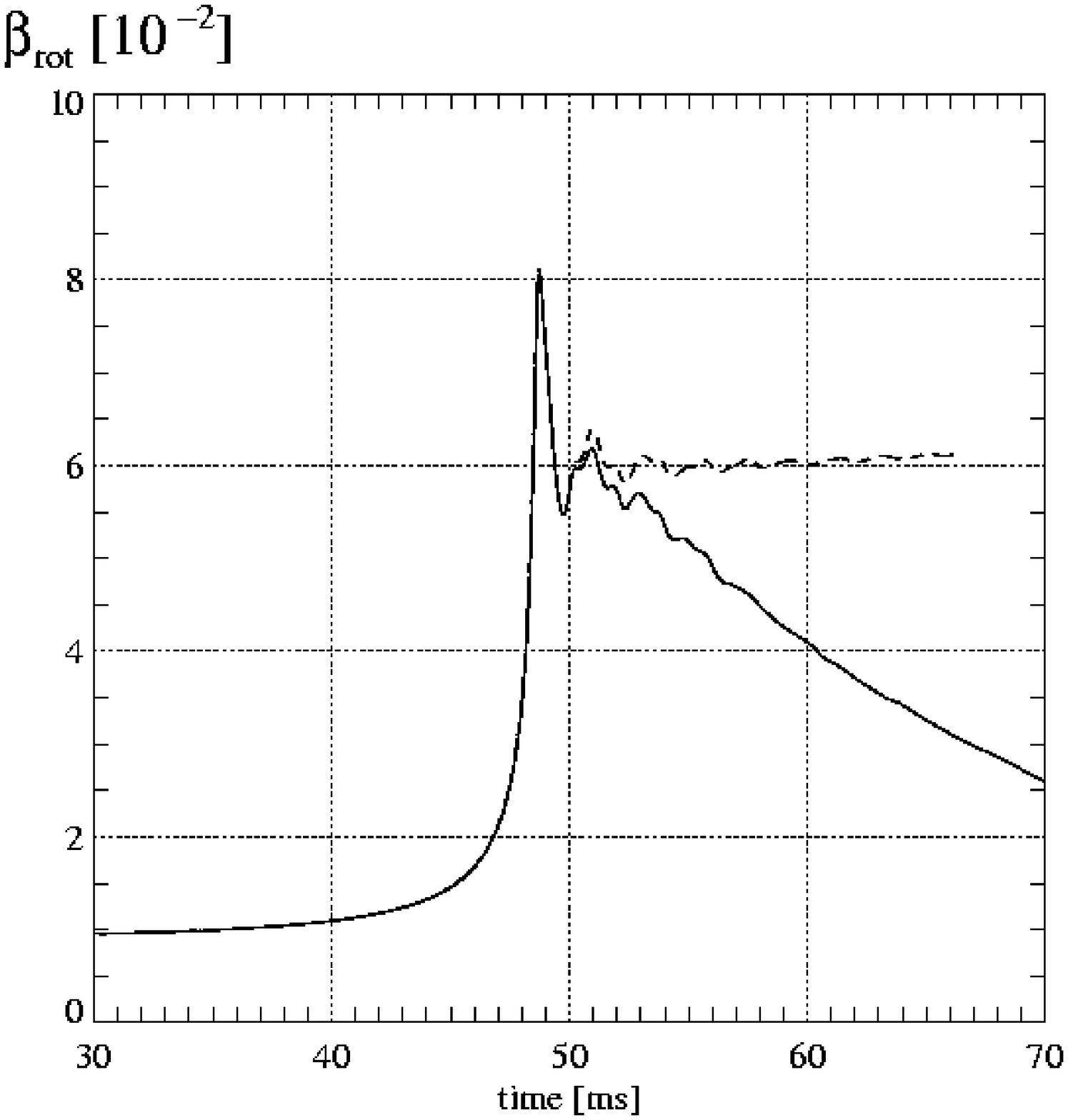}
  \includegraphics[width=5.6cm]{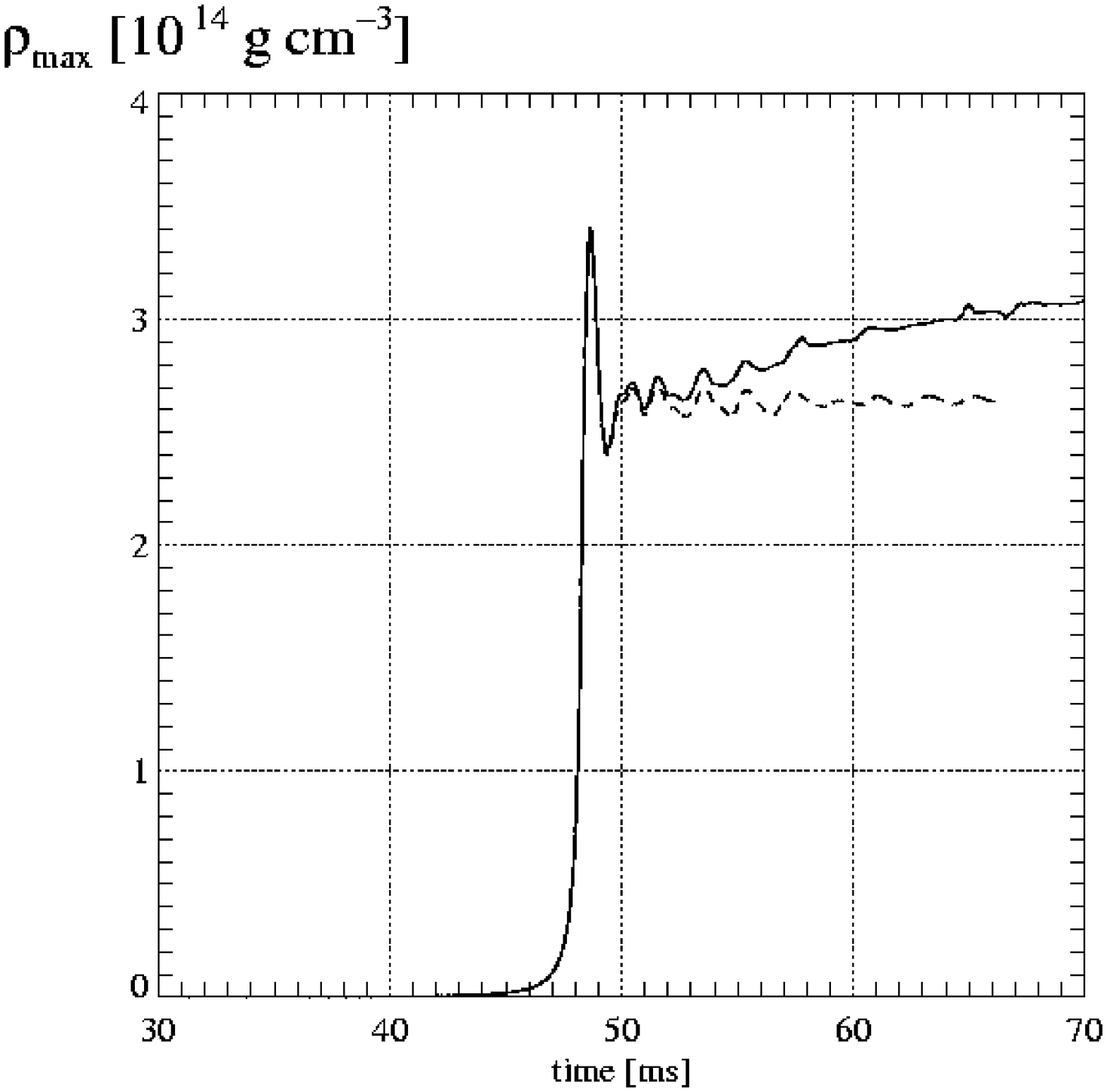}
  \includegraphics[width=5.6cm]{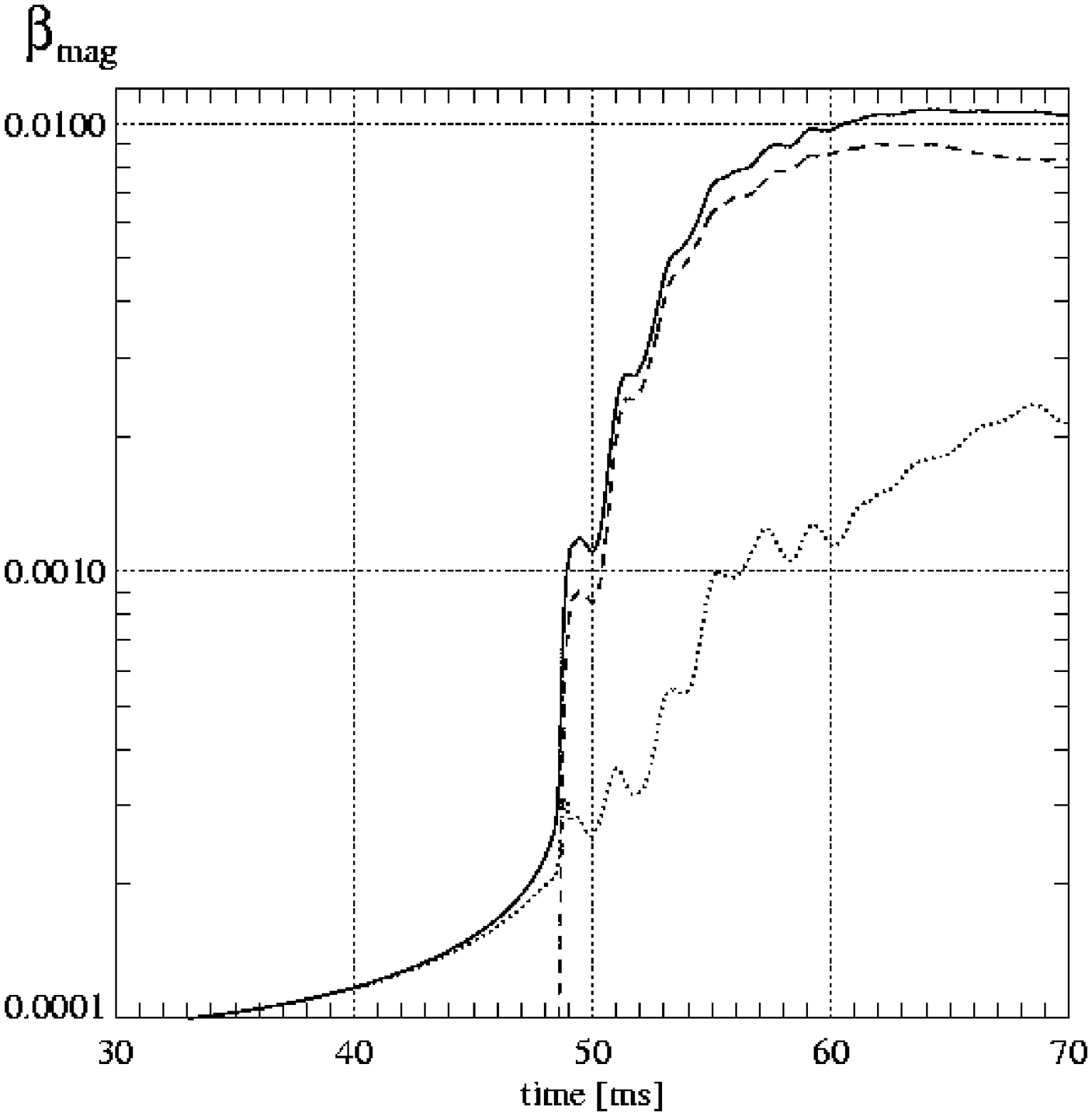}
  \includegraphics[width=5.6cm]{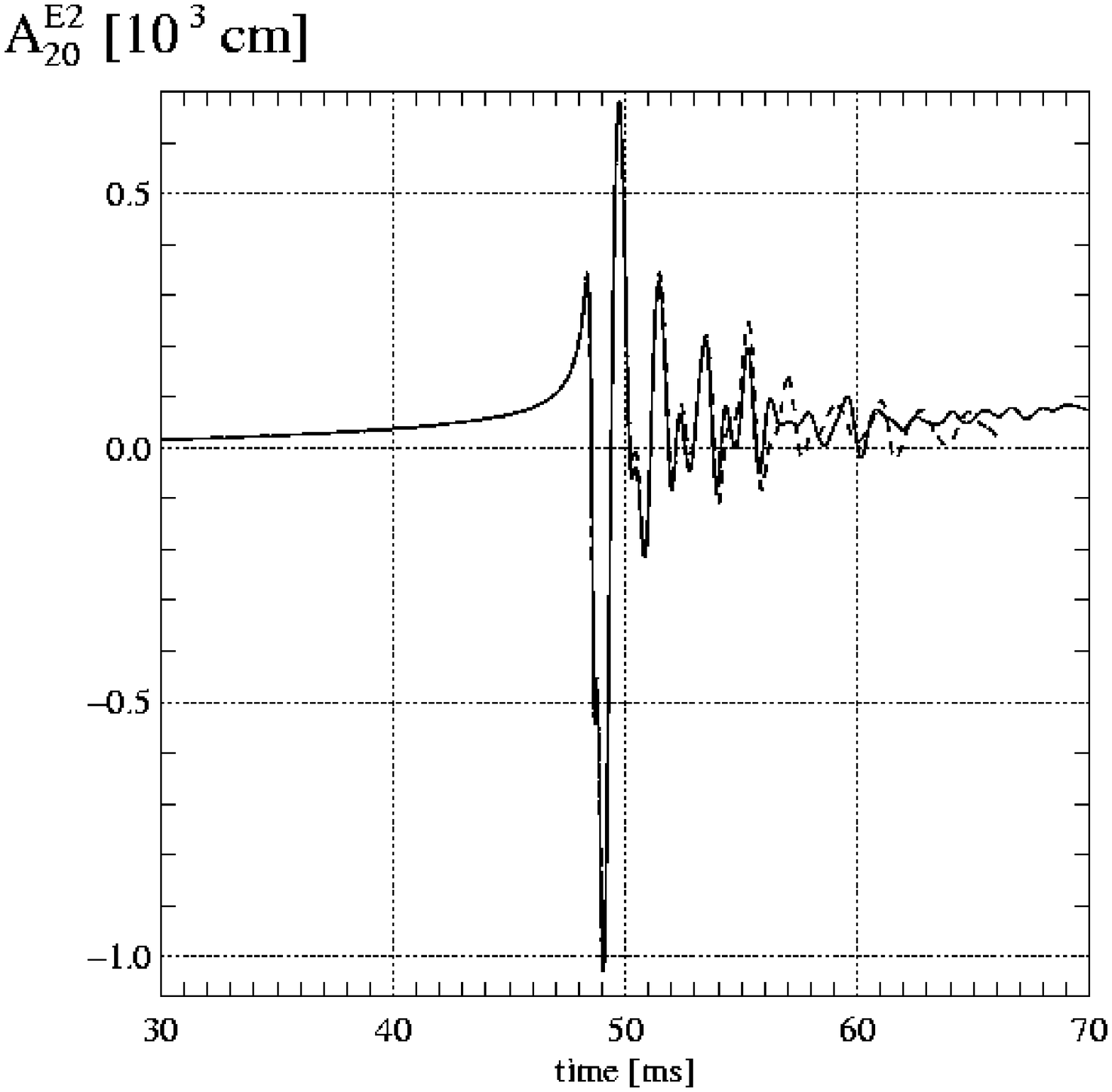}
  \includegraphics[width=5.6cm]{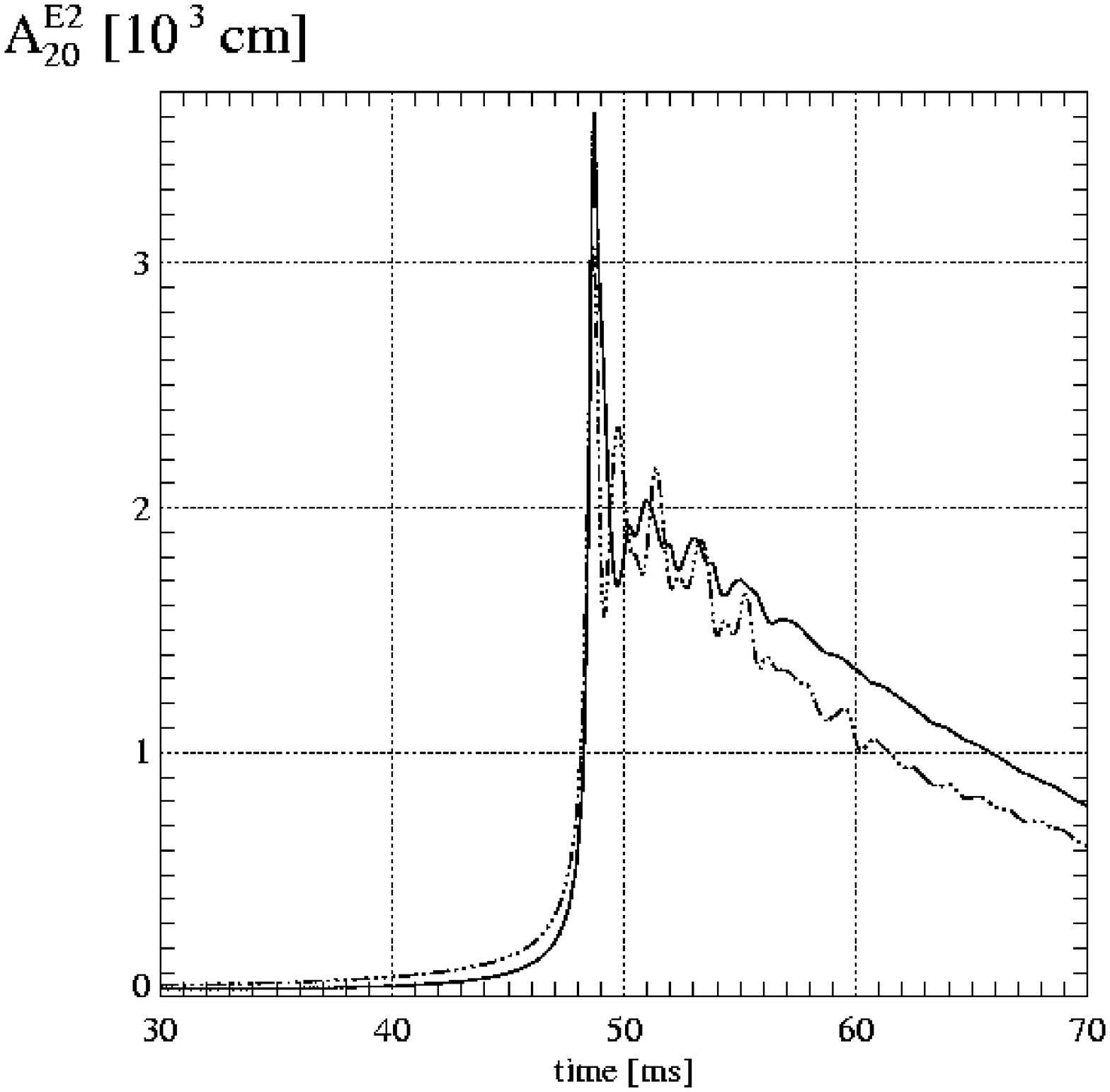}
  \includegraphics[width=5.6cm]{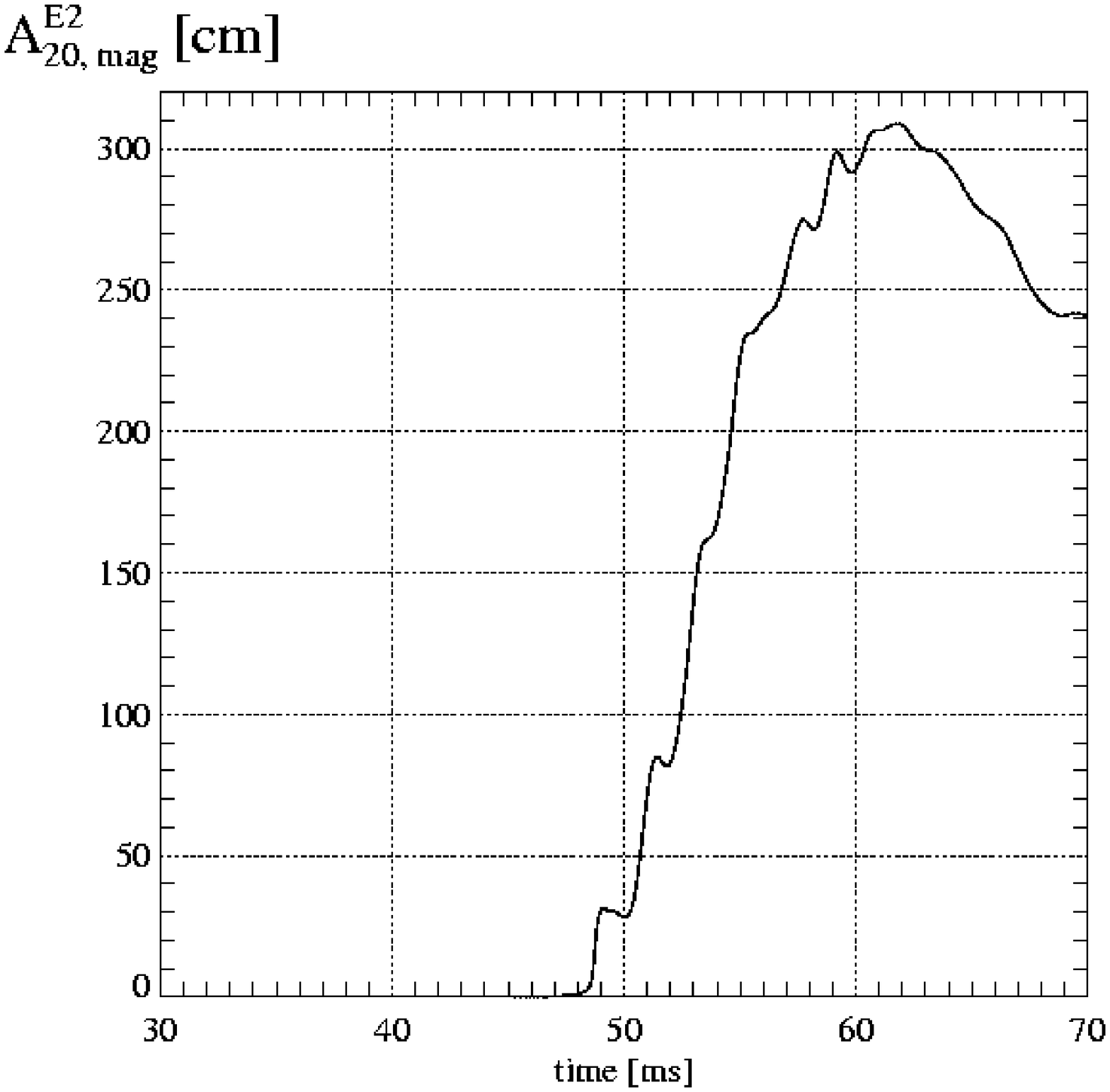}
  \caption[Evolution of model A1B3G3-D3M12]
  { The dynamics and the GW signal of model A1B3G3-D3M12.
    The upper left and middle panels provide a comparison of the
    evolution of the rotational energy parameter $\beta_{\mathrm{rot}}$
    and the maximum density of this model (solid lines) with those of
    model A1B3G3-D3M10 (dashed lines).  The upper right panel displays
    the evolution of $\beta_{\mathrm{mag}}$ (solid line), $\beta_{\phi}$
    (dashed line), and $\beta_{\mathrm{mag}} - \beta_{\phi}$ (dotted
    line), respectively.  The GW signal is shown in the lower panels:
    total amplitude (solid line, left; the dashed line gives the total
    amplitude of model A1B3G3-D3M10),
    $-A^{\mathrm{E}2}_{20;\mathrm{hyd}}$ (solid line, middle),
    $A^{\mathrm{E}2}_{20;\mathrm{grav}}$ (dashed line, middle), and
    $A^{\mathrm{E}2}_{20;\mathrm{mag}}$ (right).  The corresponding
    variables of model A1B3G3-D3M10 are shown in
    Fig.\,\ref{Fig:sFF_rho_und_GW}. 
  }
  \label{Fig:133-312:Dyn}
\end{figure*}

\begin{figure}[!htbp]
  \resizebox{\hsize}{!}{\includegraphics{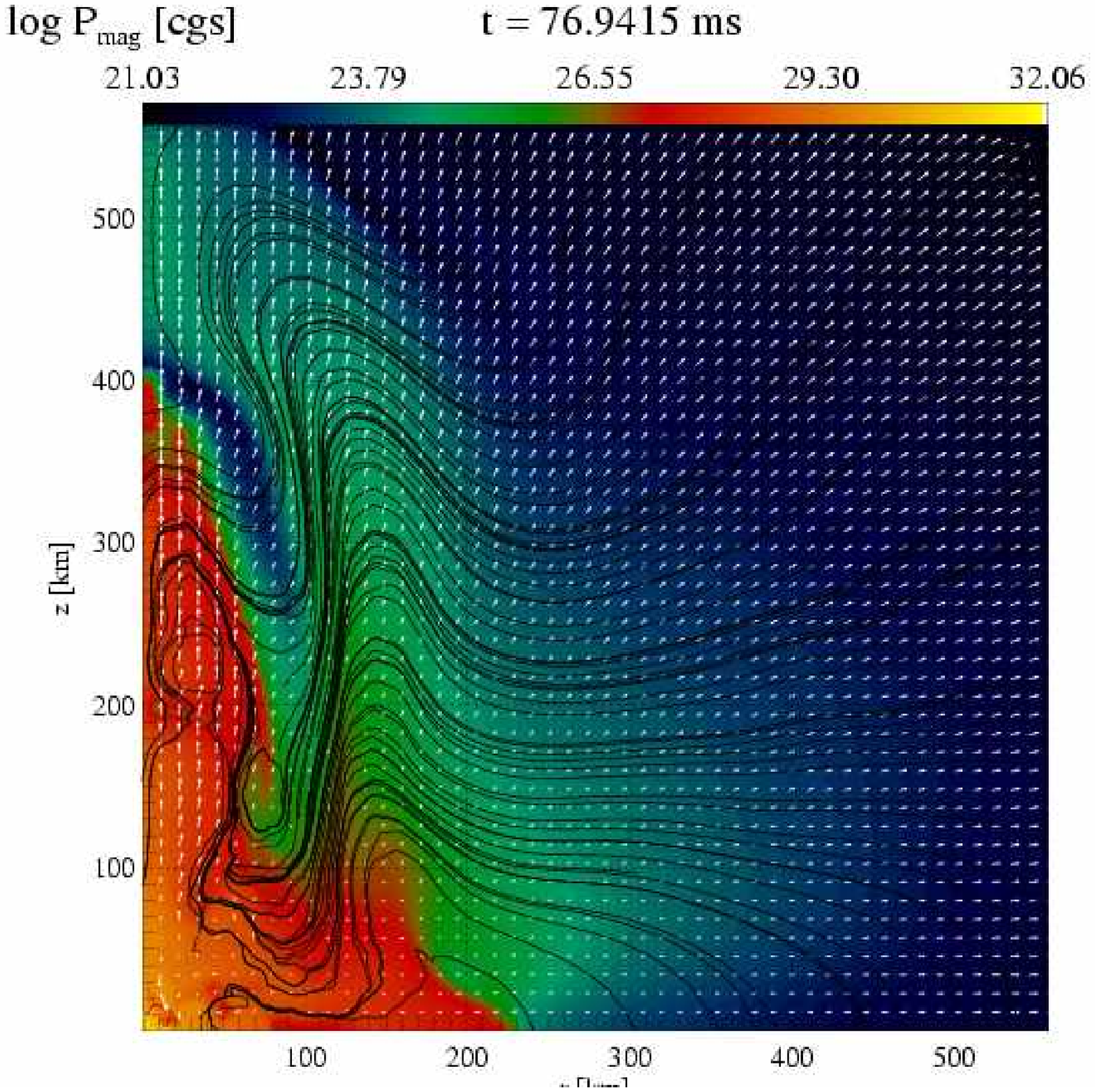}}
  \caption[Model A1B3G3-D3M12: bipolar outflow]
  { Snapshot of the inner regions of model A1B3G3-D3M12 at
    $t=76.94\ \mathrm{ms}$ showing the magnetic pressure (color shaded), the
    bipolar flow field (vectors), and the magnetic field lines,
    respectively.  }
  \label{Fig:133-312:Outflow}
\end{figure}

As in the case of weak--field models, the poloidal field energy of a
single bounce models such as A1B3G3-D3M12 is approximately constant
during the early post--bounce evolution, but it is strongly amplified
afterwards by MRI--like modes in the core which efficiently remove
angular momentum from the inner core and decrease its rotational energy.
The evolution of the rotational energy hence shows a clear deviation
from that of a weak--field model (Fig.\,\ref{Fig:133-312:Dyn}), which
however should not be the case, as the MRI does not depend on the seed
field. We will explain in Sect.\,\ref{Suk:MRI} why this different
behavior arises.

The loss of rotational support also affects the density structure.  The
core of model A1B3G3-D3M12 is slightly less rotationally flattened than
the core of a non--magnetized model such as A1B3G3-D3M10.  The shock
that forms at bounce, when the models hardly differ, is unaffected by
the different evolution of the central region, whereas matter in the
central region experiences an additional acceleration due to the
presence of the strong magnetic field.  The decrease of the rotation
rate, and also of the degree of differential rotation in the
strong--field case limits the amplification of the toroidal field
component by wrapping of field lines (Fig.\,\ref{Fig:133-312:Dyn}, upper
right panel), and the total magnetic energy has a significant poloidal
contribution.  At $t\approx 70\ \mathrm{ms}$, the ratio of toroidal and
total magnetic energy has dropped to a value of about $80\,\%$.  The
poloidal component grows a little during the post--bounce evolution by
the action of meridional motions in the core.  When the magnetic field
becomes dynamically important, magnetic energy is used to accelerate the
fluid, which puts an end to the phase of efficient field amplification.
During the late stages of evolution, the total field energy remains
approximately constant. Its final value corresponds to
$\beta_{\mathrm{mag}} \approx 1 \, \%$.  The field structure of model
A1B3G3-D3M12 differs from that of the weak--field core A1B3G3-D3M10 in
the central parts of the core at late times due to the MRI modes (that
could not be resolved for the weaker field; see Sect.\,\ref{Suk:MRI}),
but farther away from the center the fields have a similar topology.

Around the time of bounce the GW signal of model A1B3G3-D3M12 is very
similar to the one of the less magnetized core A1B3G3-D3M10, and also
the contributions of the individual parts of the total signal are
comparable (Fig.\,\ref{Fig:133-312:Dyn}, lower panels).  In the
post--bounce evolution, both the hydrodynamic and the gravitational part
of the amplitude start to decrease in magnitude, as the rotationally
induced asphericity of the core decreases along with the extraction of
rotational energy.  The magnetic amplitude increases strongly as
magnetic forces act on the core decreasing its rotation.  The
oscillations of the core are still imprinted on the post--bounce GW
amplitude, but their impact on the signal decreases with time.  In the
long run, the amplitude assumes positive values and varies relatively
little.  Dynamically, this phase is characterized by the emergence of a
weak outflow along the polar axis far behind the shock wave created at
core bounce.  This outflow is predominantly driven by magnetic forces
(Fig.\,\ref{Fig:133-312:Outflow}).  The positive long--term GW amplitude
is mainly due to the $A^{\mathrm{E}2}_{20;v_rv_r}$ term which is large
inside the outflow, whereas the inner parts of the model and the almost
spherical shock wave contribute little to the total signal.  Despite the
differences in the post--bounce GW amplitude contributions of the two
models A1B3G3-D3M10 and A1B3G3-D3M12 their total signals are quite
similar, i.e.\,distinguishing the two models observationally would be
very difficult.

\begin{figure*}[!htbp]
  \centering
  \includegraphics[width=5.6cm]{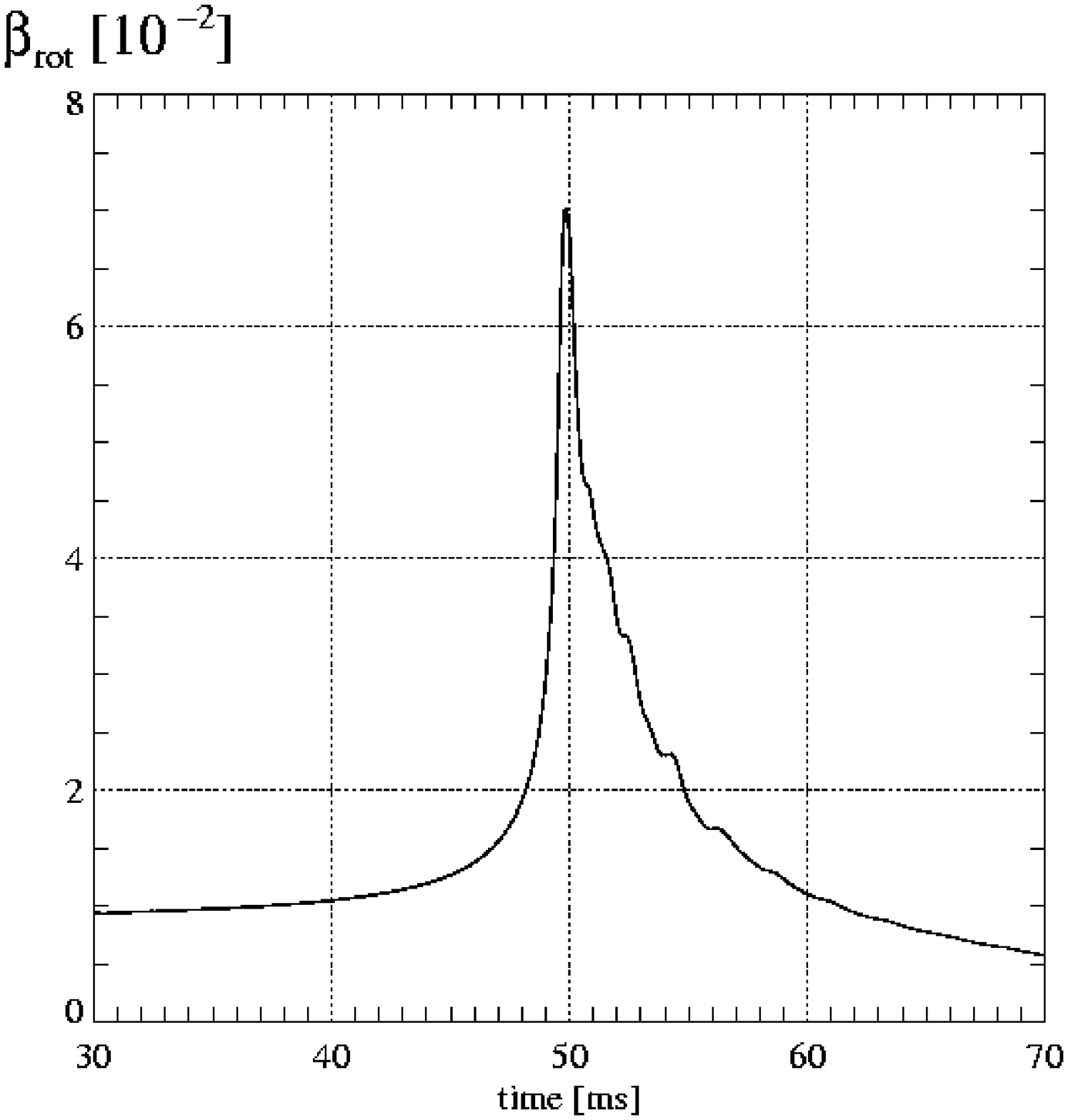}
  \includegraphics[width=5.6cm]{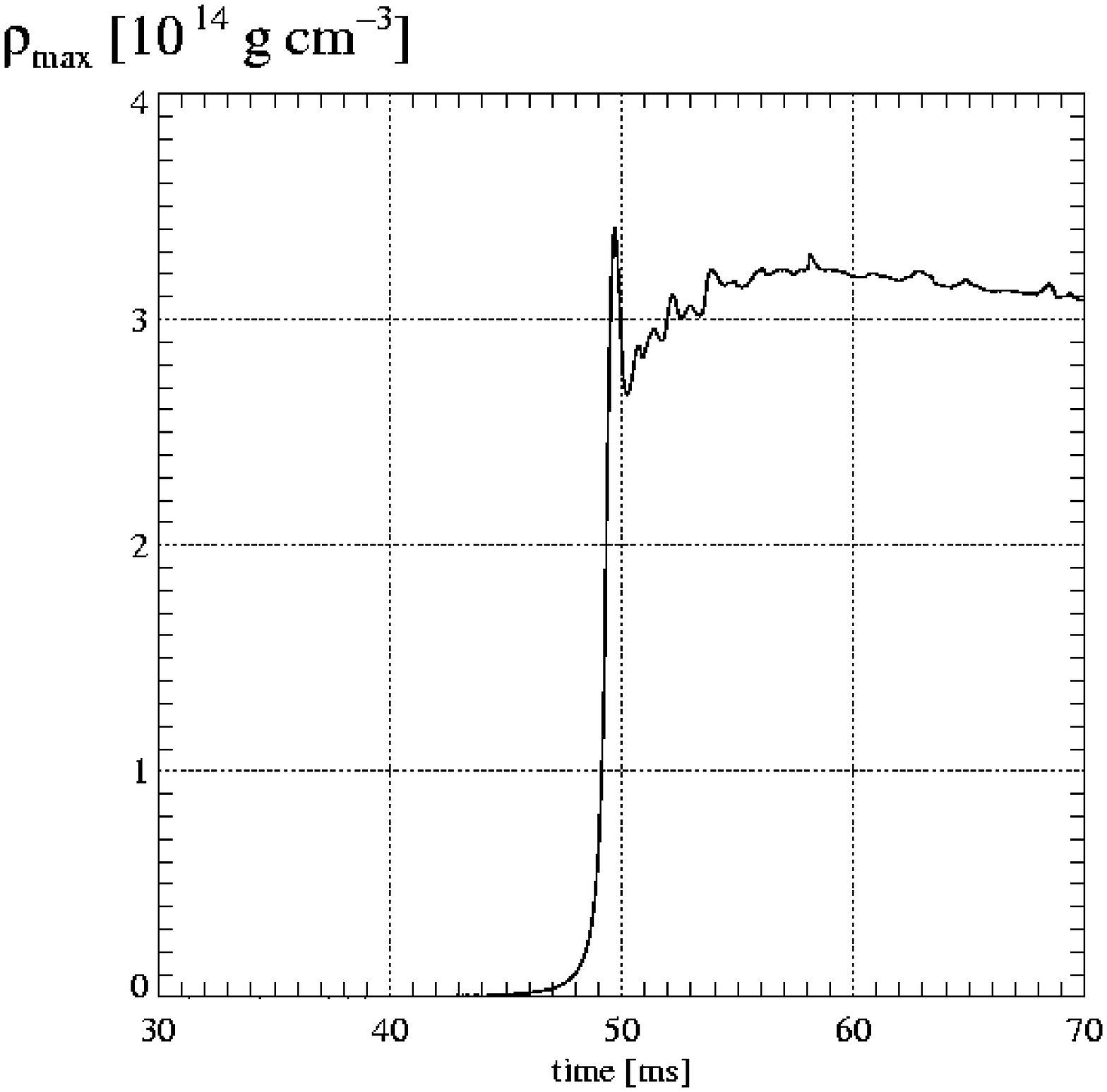}
  \includegraphics[width=5.6cm]{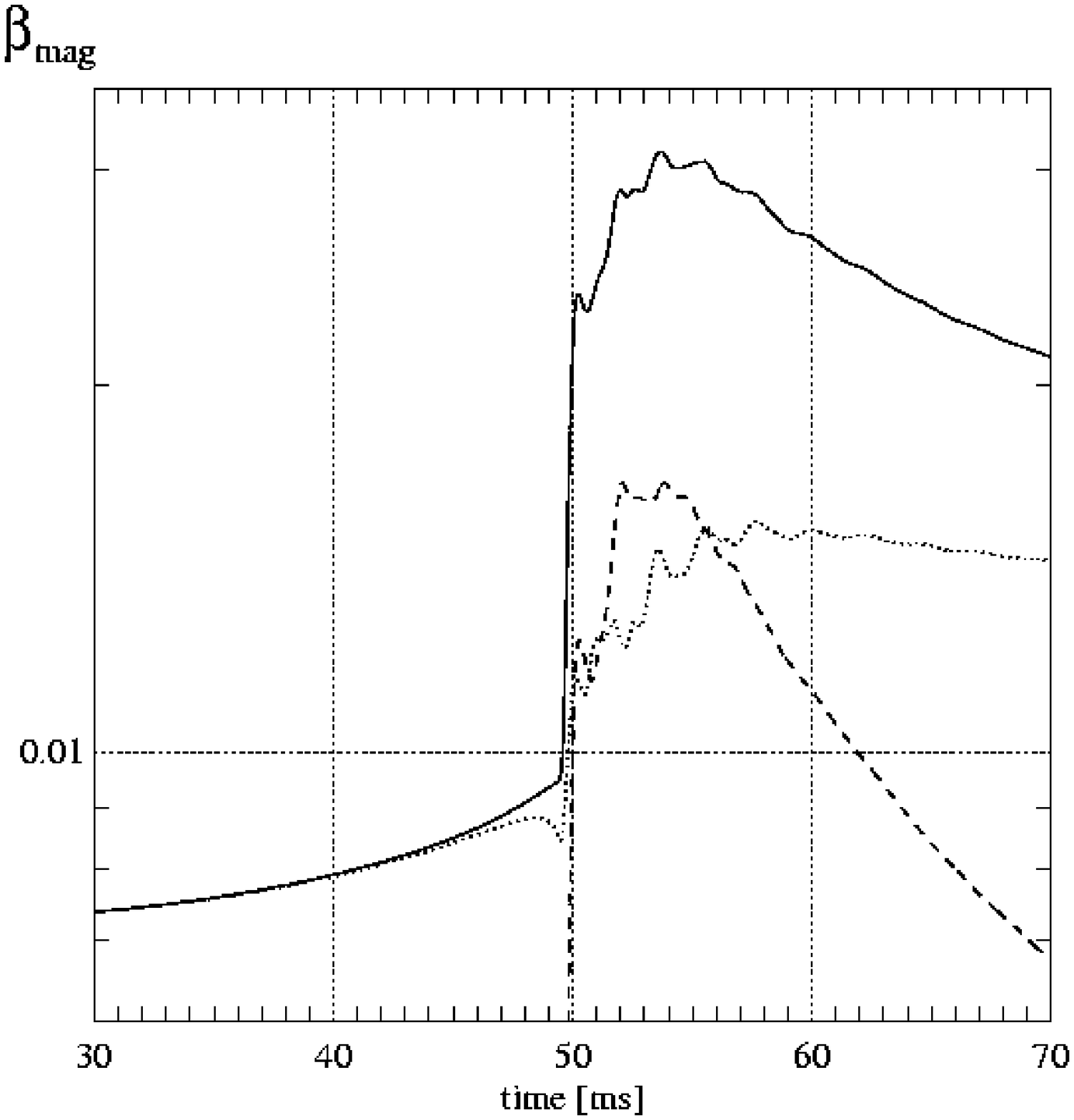}
  \includegraphics[width=5.6cm]{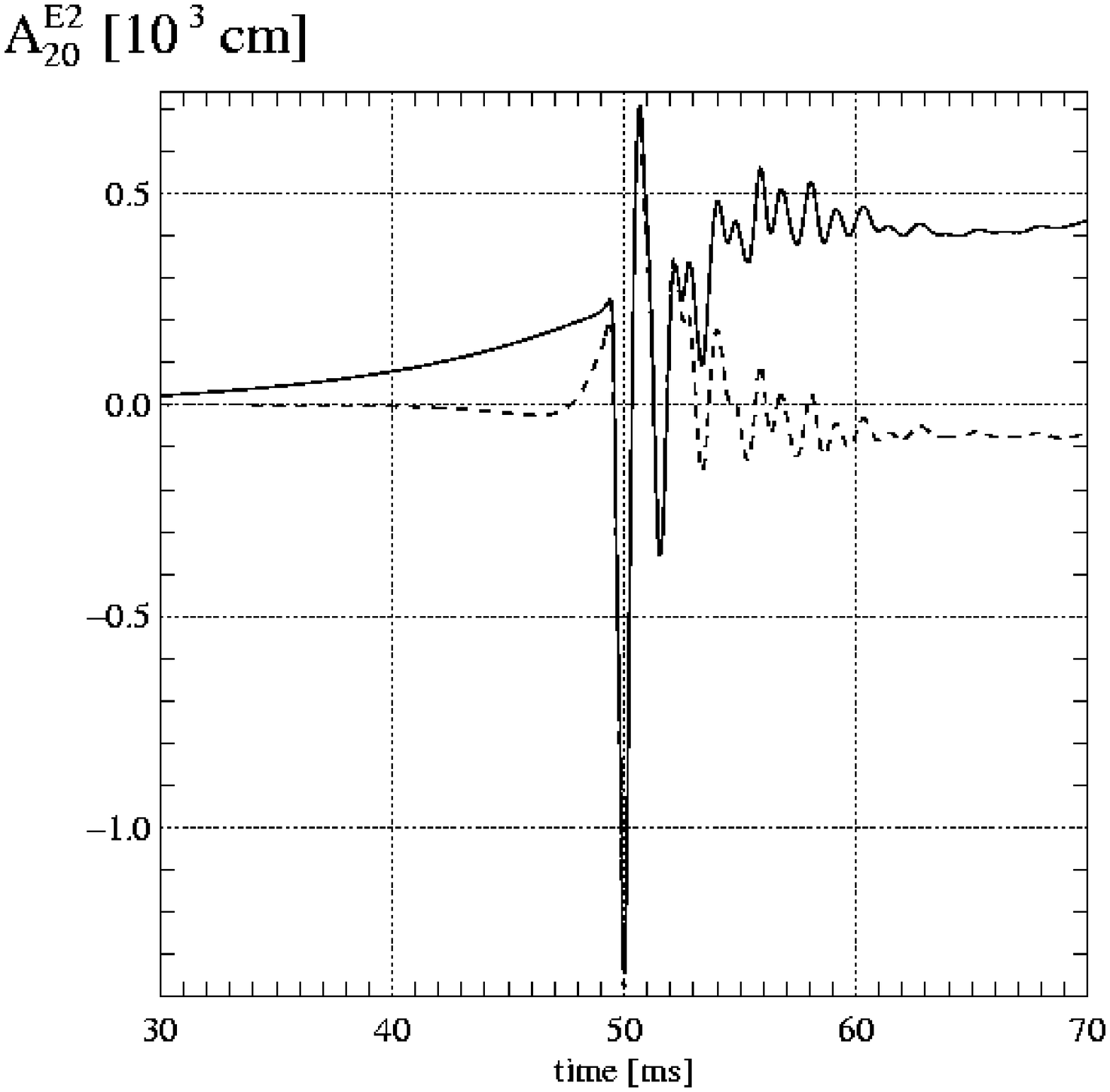}
  \includegraphics[width=5.6cm]{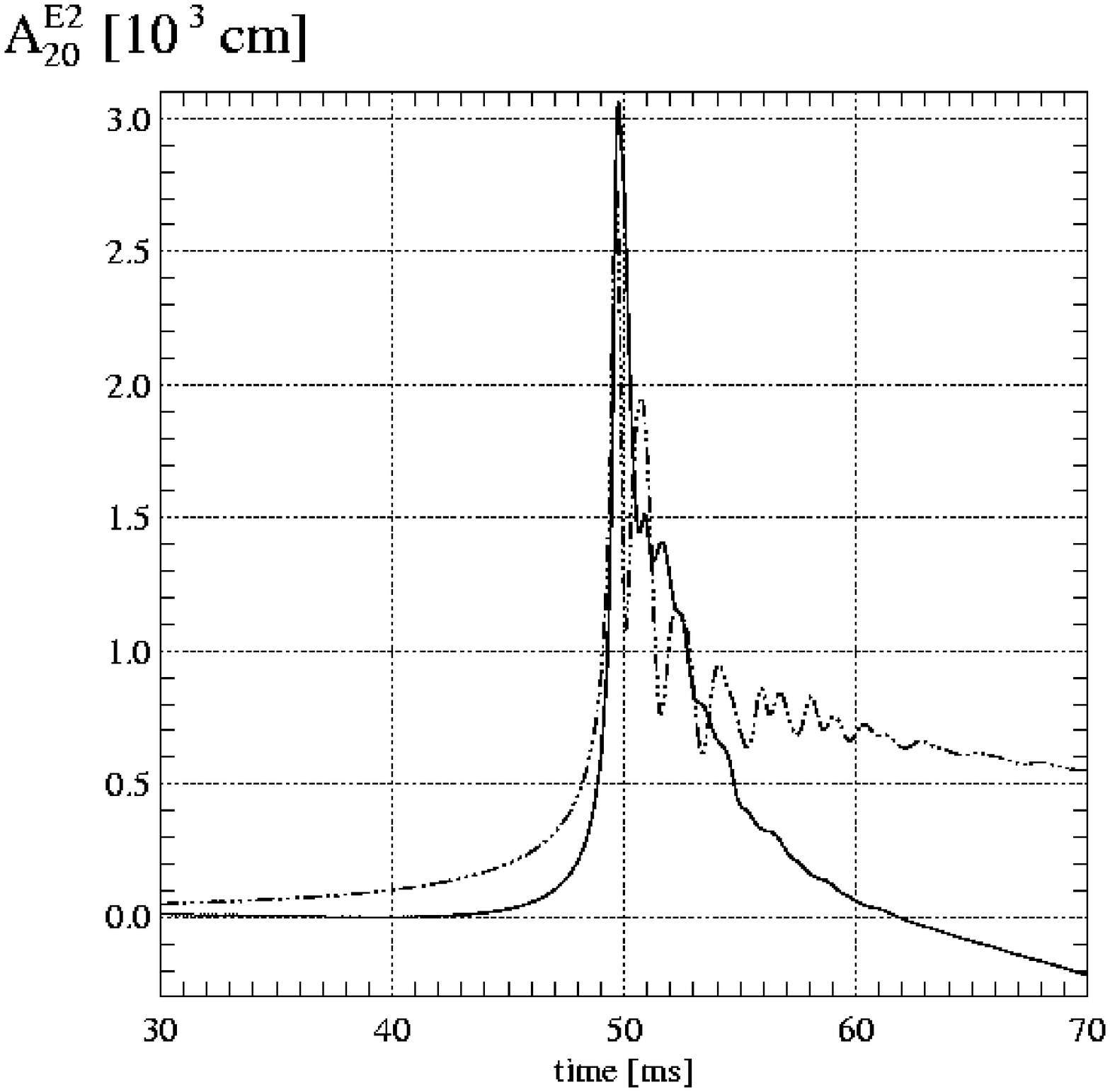}
  \includegraphics[width=5.6cm]{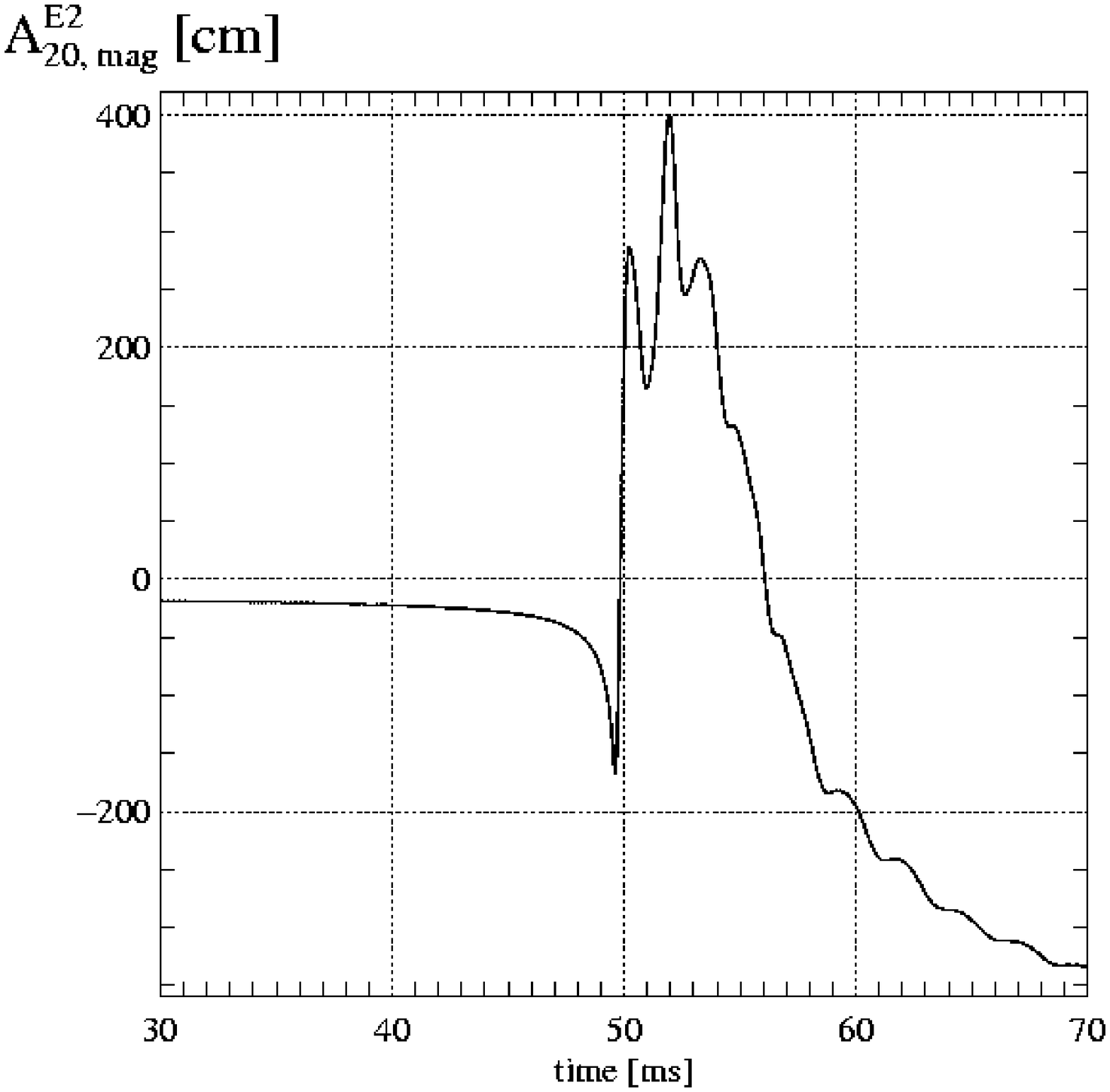}
  \caption[Evolution of model A1B3G3-D3M13]
  { The dynamical evolution and the GW signal of model A1B3G3-D3M13.
    The upper panels display the temporal evolution of the rotational
    energy parameter $\beta_{\mathrm{rot}}$ (left), the maximum
    density $\rho_{\mathrm{max}}$ (middle), and the magnetic energy
    parameters (right; $\beta_{\mathrm{mag}}$ (solid line),
    $\beta_{\phi}$ (dashed line), and $\beta_{\mathrm{mag}} -
    \beta{\phi}$ (dotted line)), respectively.  The GW signal of the
    model is shown in the lower panels: total amplitude (solid line,
    left; the dashed line gives the contribution of the layers with $r
    \le 59.4\,$km), $-A^{\mathrm{E}2}_{20;\mathrm{hyd}}$ (solid line,
    middle), $A^{\mathrm{E}2}_{20;\mathrm{grav}}$ (dashed line,
    middle), and $A^{\mathrm{E}2}_{20;\mathrm{mag}}$ (right).  }
  \label{Fig:133-313:Dyn}
\end{figure*}

\begin{figure*}
  \centering
  \includegraphics[width=16cm]{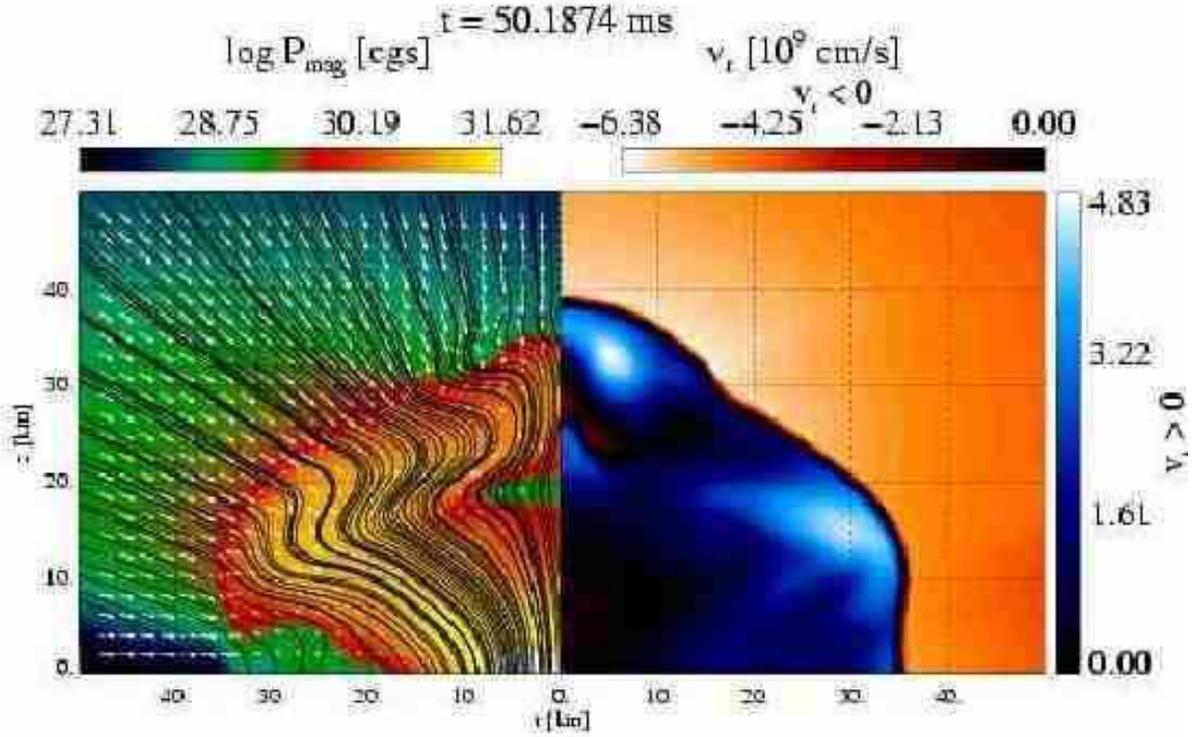}
  \caption[Model A1B3G3-D3M13: First density minimum]
  { Snapshot of model A1B3G3-D3M13 at $t=50.19\ \mathrm{ms}$, when the
    maximum density reaches the first post-bounce minimum.  In the left
    panel the flow field (vectors) and the magnetic field (field lines)
    are displayed together with the magnetic pressure (color--coded).
    The right panel shows the color--coded velocities of in-flowing
    (i.e.\, $v_r>0$) and out-flowing matter (i.e.\, $v_r<0$),
    respectively.  }
\label{Fig:133-313:ErstesMin}
\end{figure*}

\begin{figure*}
  \centering
  \includegraphics[width=16cm]{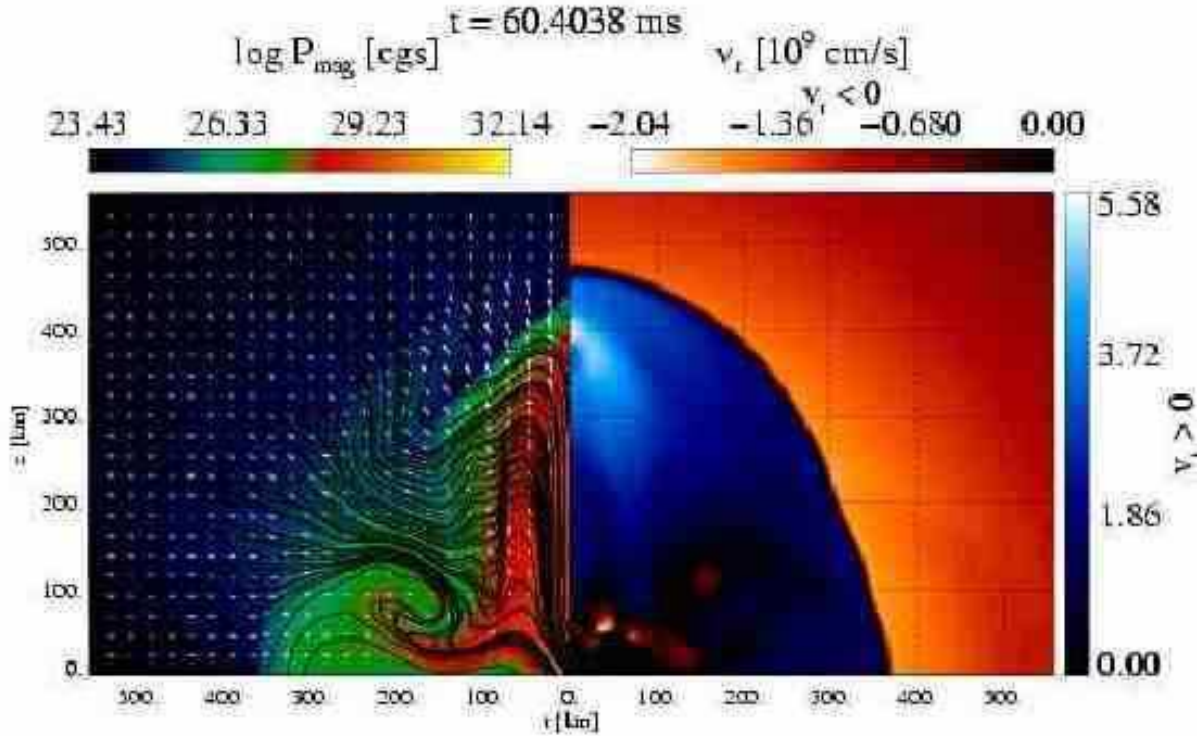}
  \caption[Model A1B3G3-D3M13 at $60\ \mathrm{ms}$] 
  { Same as Fig.\,\ref{Fig:133-313:ErstesMin}, but at $t=60.40\ \mathrm{ms}$.
    Note that the length scales are changed compared to
    Fig.\,\ref{Fig:133-313:ErstesMin}. }
  \label{Fig:133-313:60ms}
\end{figure*}

\begin{figure*}
  \centering
  \includegraphics[width=16cm]{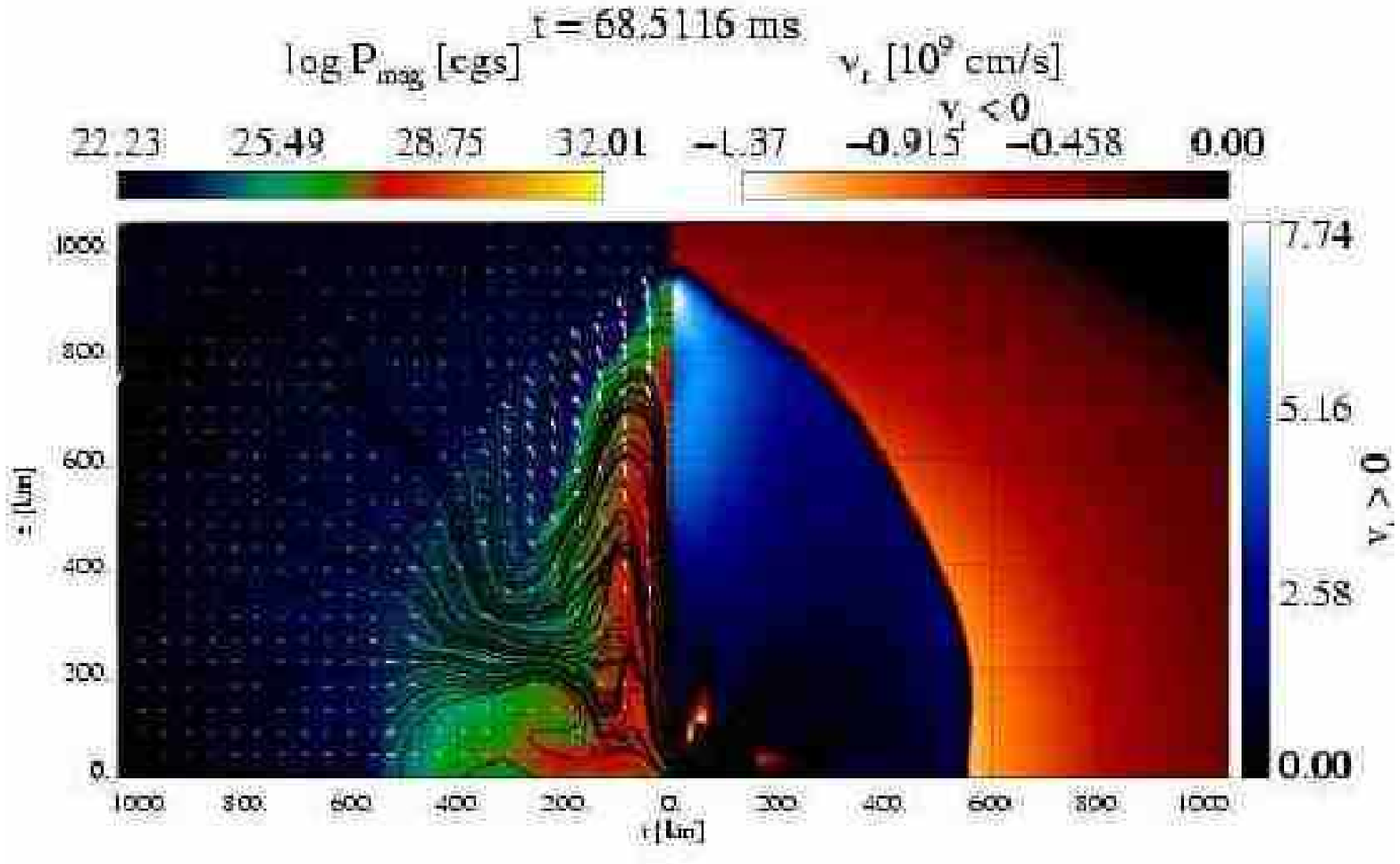}
  \caption[Model A1B3G3-D3M13 at $68\ \mathrm{ms}$]
  { Same as Fig.\,\ref{Fig:133-313:ErstesMin}, but at $t=68.51\ \mathrm{ms}$.
    Note that the length scales of the plot are changed compared to
    Figs.\,\ref{Fig:133-313:ErstesMin} and \ref{Fig:133-313:60ms}. }
  \label{Fig:133-313:68ms}
\end{figure*}

Similar features as those observed in the case of model A1B3G3-D3M12
(Fig.\,\ref{Fig:133-312:Dyn}) are present in the evolution of the
stronger magnetized model A1B3G3-D3M13 (Fig.\,\ref{Fig:133-313:Dyn}),
but they are much more prominent.  Unlike all previously discussed
models, this core shows significant deviations from the purely
hydrodynamic case already during collapse.  For model A1B3G3-D3M13 the
core bounce is delayed by $\approx 1 \ \mathrm{ms}$ compared to the
weaker magnetized models A1B3G3-D3M10...12.  The rotational energy
reaches a maximum value of $\beta_{\mathrm{rot}}^{\mathrm{max}} \approx
7\,\%$ near bounce, which is lower than in the less magnetized models
($\beta_{\mathrm{rot}}^{\mathrm{max}} \approx 8\,\%$).  Within the next
$\approx 6 \ \mathrm{ms}$ the core looses $\approx 80\,\%$ of the
rotational energy it had acquired at bounce.  Subsequently, the rate of
energy loss decreases, and the core contracts
(Fig.\,\ref{Fig:133-313:Dyn}). Its maximum density rises from
$\rho_{\mathrm{max}} = 2.7 \cdot 10^{14}\ \mathrm{cm \, s}^{-1}$ in the
first post--bounce density minimum to $\rho_{\mathrm{max}} = 3.2 \cdot
10^{14}\ \mathrm{cm \, s}^{-1}$ at $t = 58\ \mathrm{ms}$ before it
starts to decreases by a small amount again.  The process of magnetic
braking prevents a further amplification of the field, and it prevents
the transformation of poloidal into toroidal magnetic field.  The ratio
$\beta_{\phi} / \beta_{\mathrm{mag}}$ does not approach unity, as in
case of the models with weaker initial fields, but it remains below
$\approx 60\,\%$ decreasing slowly later.  At $t \approx 71
\mathrm{ms}$, when we stopped the computations, the value of
$\beta_{\mathrm{mag}}$ had dropped to $\beta_{\mathrm{mag}} \approx 2.05
\, \%$, and we expect it to eventually decrease to a value of $\sim 1
\,\%$.

In model A1B3G3-D3M12 the immediate post--bounce core is very similar to
the corresponding non--magnetic one as the magnetic field is not
sufficiently strong to change the dynamics unless an MRI--like
instability sets in.  This is different from the more strongly
magnetized model A1B3G3-D3M13, where without the MRI the combined
amplification due to contraction and differential rotation is sufficient
to cause the braking effect.  This is generally true for all of our
models: very strong initial fields ($> 10^{12} \ \mathrm{G}$) do not
need amplification by an instability to brake the core, whereas weaker
ones heavily depend on it.

After bounce, the shock wave develops a non--spherical, bulb--like shape
, and the twisted field lines give rise to a highly magnetized
post--shock fluid that pushes the shock near the equatorial plane around
$\theta \approx 70\degr$ (Fig.\,\ref{Fig:133-313:ErstesMin}).  Near the
axis, the immediate post-shock matter is only weakly magnetized, the
ratio of gas pressure and magnetic pressure being largest at a
significantly smaller radius ($r_{\mathrm{max, mag}} \approx 27\ 
\mathrm{km}$) than the shock position ($r_{\mathrm{shock}} \approx 40\ 
\mathrm{km}$).  The shock maintains its bulb--like shape also during the
secondary contraction phase that goes along with the extraction of
rotational energy, and fast outflow both at high and low latitudes.  At
later times ($t=60.40\ \mathrm{ms}$, Fig.\,\ref{Fig:133-313:60ms}), the
fluid pattern has changed.  The formerly important outflow near the
equator stalls, and the shock surface becomes more elongated.  The
near--axis outflow velocity has further increased, its maximum value
being $v_{r}^{\mathrm{max}} = 0.19c$ in the strongly magnetized region
well behind the shock ($r_{\rm{max mag}} = 270\ \mathrm{km} <
r_{\mathrm{shock}} = 330\ \mathrm{km}$).  In the outflow along the
rotational axis, a cylindrical shell of very low pressure gas can be
identified (at $\theta \approx 10\degr$ in
Fig.\,\ref{Fig:133-313:60ms}).  The structure of the magnetic field is
dominated by two extended highly magnetized regions, one located near
the equator at $(r \la 200\ \mathrm{km}$, $\theta \approx
80\degr$), and a cylindrical one oriented along the rotational axis
having a width of $\approx 100\,$km (left panel of
Fig.\,\ref{Fig:133-313:60ms}).  At an even later epoch, $t=68.51\ 
\mathrm{ms}$ (Fig.\,\ref{Fig:133-313:68ms}), only little has changed
near the equatorial plane, however at the axis the highly magnetic gas
is about to catch up with the shock wave, which is now significantly
prolate with an axis ratio of about $3:2$.  The outflow has accelerated
further ($v_r = 8\cdot 10^{9} \ \mathrm{cm \, s}^{-1} \sim c/4$), the
maximum velocity corresponding to the highly magnetized regions behind
the shock at $r = 870\ \mathrm{km}$.  Hence, we see the formation of a
jet--like outflow from the core.

In its very interior the core is rotating slower than its non--magnetic
variant at late times.  Furthermore, it has developed a roughly toroidal
region of relatively slow and highly rigid counter--rotation.  This
region grows from a layer at $r \approx 10\ \mathrm{km}$, i.e.\,near the
edge of the inner core, which experiences sufficiently strong magnetic
stresses to reverse its direction of rotation.  Interior to the
retrograde rotating region the angular velocity decreases with time.

Assuming that the total energy density of a fluid element,
$e_{\mathrm{tot}} = \varepsilon+\frac{1}{2}\rho\vec v^2+\frac{1}{2}\vec
b^2-\rho\cdot\Phi$, is converted entirely into kinetic energy
$\frac{1}{2}\rho v_{\infty}^2$, terminal outflow velocities $v_{\infty}$
of up to $1.85\cdot 10^{10}\ \mathrm{cm \, s}^{-1}$ are observed for
model A1B3G3-D3M13 at late times, particularly in the bipolar outflow
along the rotation axis (Fig.\,\ref{Fig:133-313:68ms}).  The large
velocities stretch our \emph{non--relativistic} MHD approach, and imply
that a realistic simulation of the late evolution of the outflow will
require relativistic MHD.

The GW amplitude of model A1B3G3-D3M13 (Fig.\,\ref{Fig:133-313:Dyn},
lower panels) is enhanced by about $10\,\%$ at core bounce.  After
bounce, ringing on the dynamical time scale of the core the average
amplitude shifts to positive values.  The amplitudes and frequencies
of the oscillations are hardly affected by the overall shift of the
average amplitude to positive values implying a different origin of
the oscillations and the overall shift.  A few milliseconds after
bounce $A^{\mathrm{E}2}_{20}$ grows to $\sim 400 \ \mathrm{cm}$, and the rapid
variations of the signal cease. The hydrodynamic and gravitational
contributions to the total amplitude are greatly reduced in magnitude.
The inner core has lost a large amount of its rotational energy and is
of almost spherical shape.  Its contribution to the total signal
amplitude is hence relatively small for epochs well after bounce.

\begin{figure}
  \resizebox{\hsize}{!}{\includegraphics{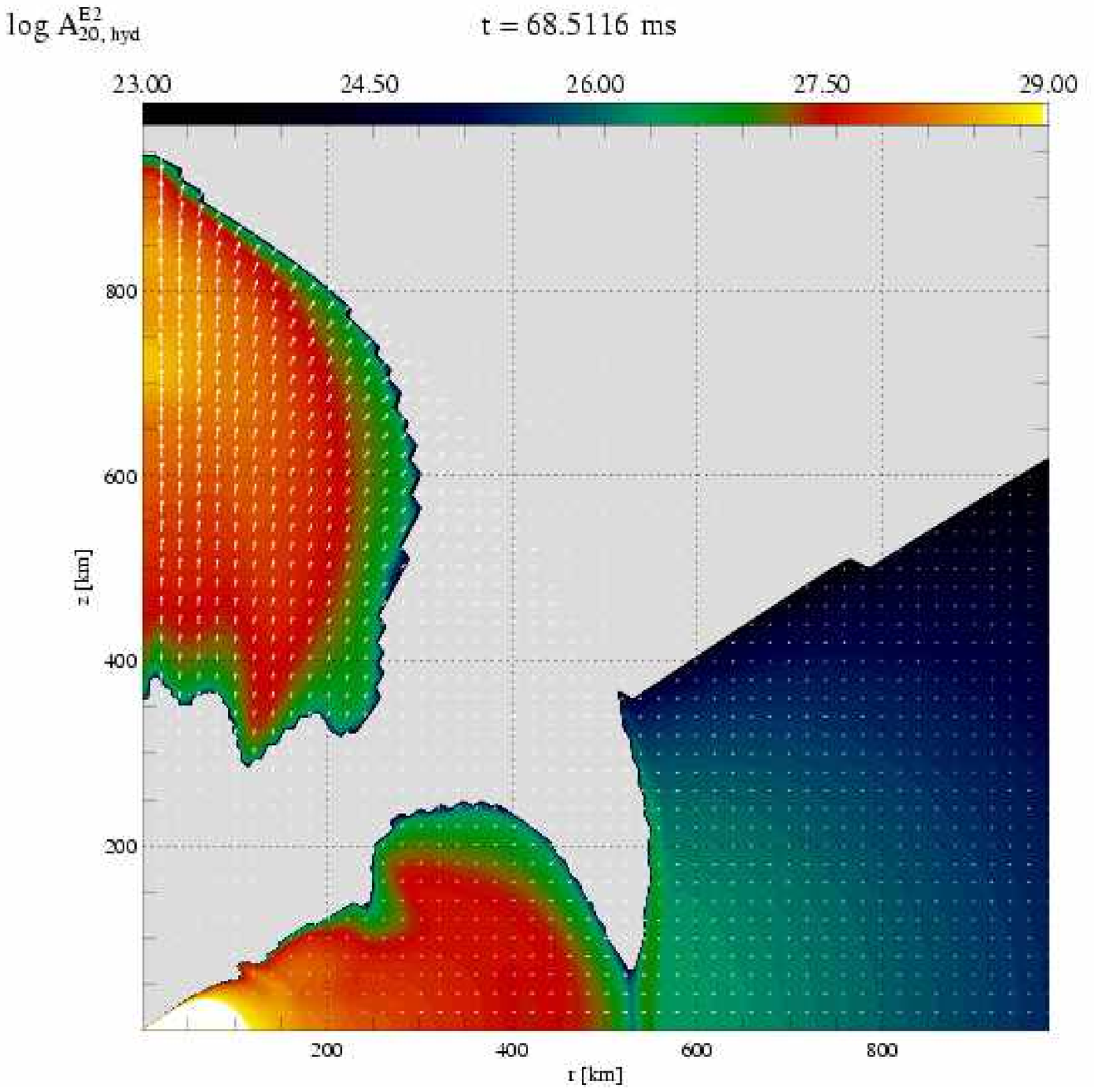}}
  \caption[Model A1B3G3-D3M13: GW hydro part]
  { The hydrodynamic plus gravitational parts of the GW amplitude of
    model A1B3G3-D3M13 at $t=68.5\ \mathrm{ms}$.  Positive contributions
    to the integrand in the formula for the hydrodynamic plus
    gravitational amplitude are displayed by colors. The bipolar outflow
    shows up in the large positive parts near the axis.  
  }
  \label{Fig:133-313:GWLumH}
\end{figure}

While the hydrodynamic amplitude produced by the innermost layers of the
core is very small, the gravitational and magnetic contributions reach
quite large values of several $10^2\,$cm.  However, as both
contributions are of opposite sign they cancel each other almost
completely (Fig.\,\ref{Fig:133-313:Dyn}), i.e.\, only a small net
amplitude results.  The relative smallness of the hydrodynamic amplitude
of the central core indicates that -- once most of the rotational energy
is extracted -- magnetic forces may become more important for the core's
structure than the centrifugal ones.

The very non--spherical shape of the shock wave, and in particular the
appearance of the jet--like outflow that carries a large (radial)
kinetic energy, causes a long--lasting positive GW amplitude the major
contribution being the $A^{\mathrm{E}2}_{20;v_rv_r}$ amplitude of the
outflow (Fig.\,\ref{Fig:133-313:GWLumH}).  Growing larger with time this
contribution eventually makes the hydrodynamic amplitude to become
positive at $t = 61.5 \ \mathrm{ms}$.  Thus, a bipolar motion can be
identified in the GW signature through the appearance of a long--lasting
positive signal.

\begin{figure}
  \resizebox{\hsize}{!}{\includegraphics{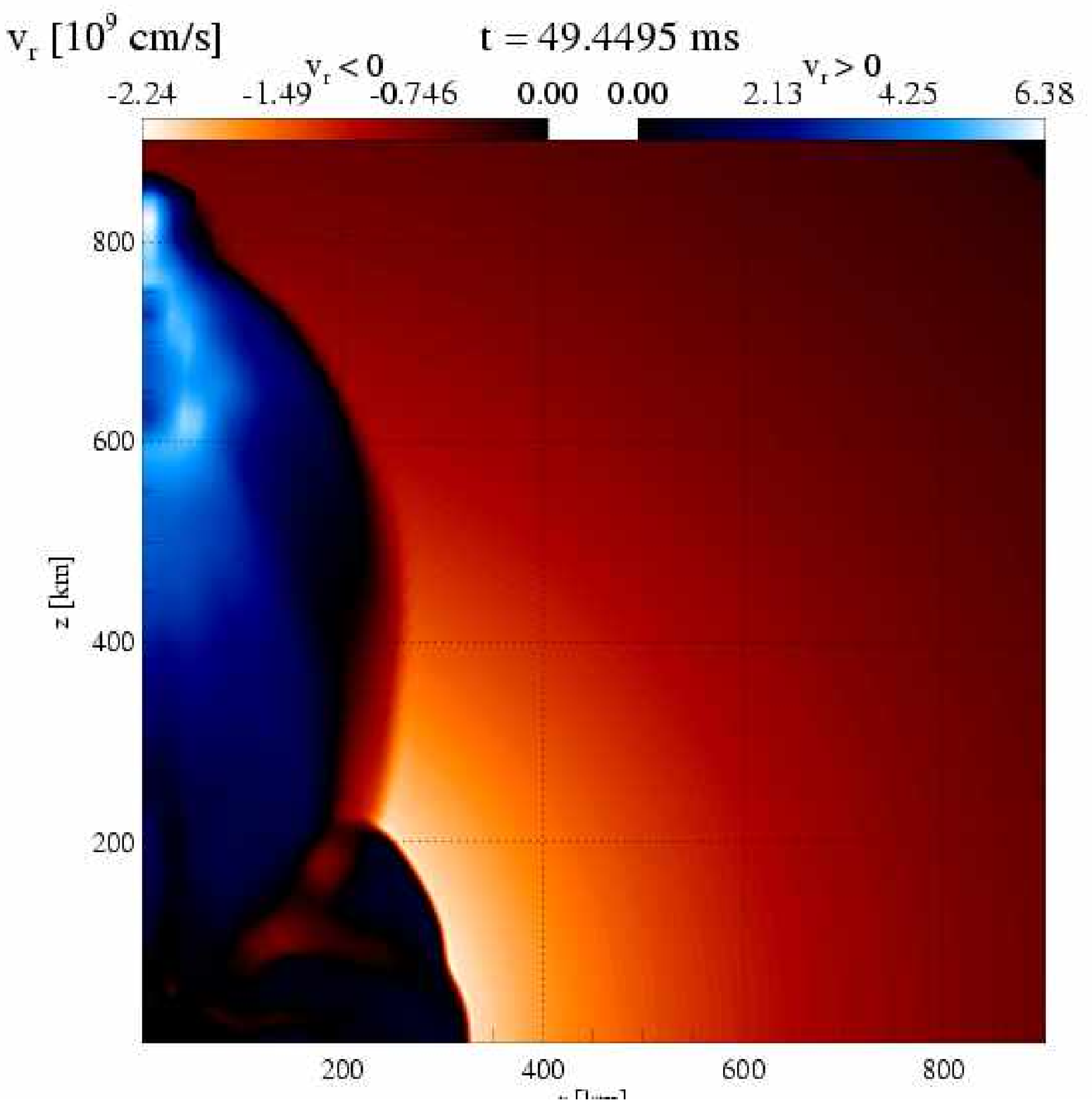}}
  \caption[Model A3B3G5-D3M13 at $50\ \mathrm{ms}$] 
  { Snapshot of model A3B3G5-D3M13 at $t=49.46\ \mathrm{ms}$ showing the
    velocities of inflows (blue colors) and outflows (red
    colors).  }
  \label{Fig:335-313:50ms}
\end{figure}

The behavior just described holds for models A1B3G3/5-D3Mm, A2B4G5-D3Mm,
and A3B3G5-D3Mm, too.  They differ, however, concerning the time scales
and the vigorousness of the phenomena.  Most dramatic is the evolution
of model A3B3G5-D3M13 (Fig.\,\ref{Fig:133-313:Dyn}), where the loss of
rotational energy allows the core to contract to densities
($\rho_{\mathrm{max}} = 4.4\cdot 10^{14}\ \mathrm{cm \, s}^{-1}$) that
exceed the bounce density $\rho_{\mathrm{b}} = 3.46\cdot 10^{14}\ 
\mathrm{cm \, s}^{-1}$ by up to $30\,\%$.  The shock wave is already
strongly prolate when it forms, and the magnetic field is highly
concentrated both towards the equator and -- like in model A1B3G3-D3M13
-- in a cylindrical region oriented along the rotation axis.  The
rapidly moving, highly magnetized outflow deforms the shock wave giving
rise to a shock surface with an axis ratio of about $3:1$.  At $t=49.46\ 
\mathrm{ms}$, magnetic (hoop) stresses in the pinched toroidal field
have accelerated the gas further yielding a so--called ``nose cone''
(well known from simulations of magnetized jets) that is visible in the
outermost part of the jet--like outflow (Fig.\,\ref{Fig:335-313:50ms}).

\begin{figure*}[!htbp]
  \centering
  \includegraphics[width=5.6cm]{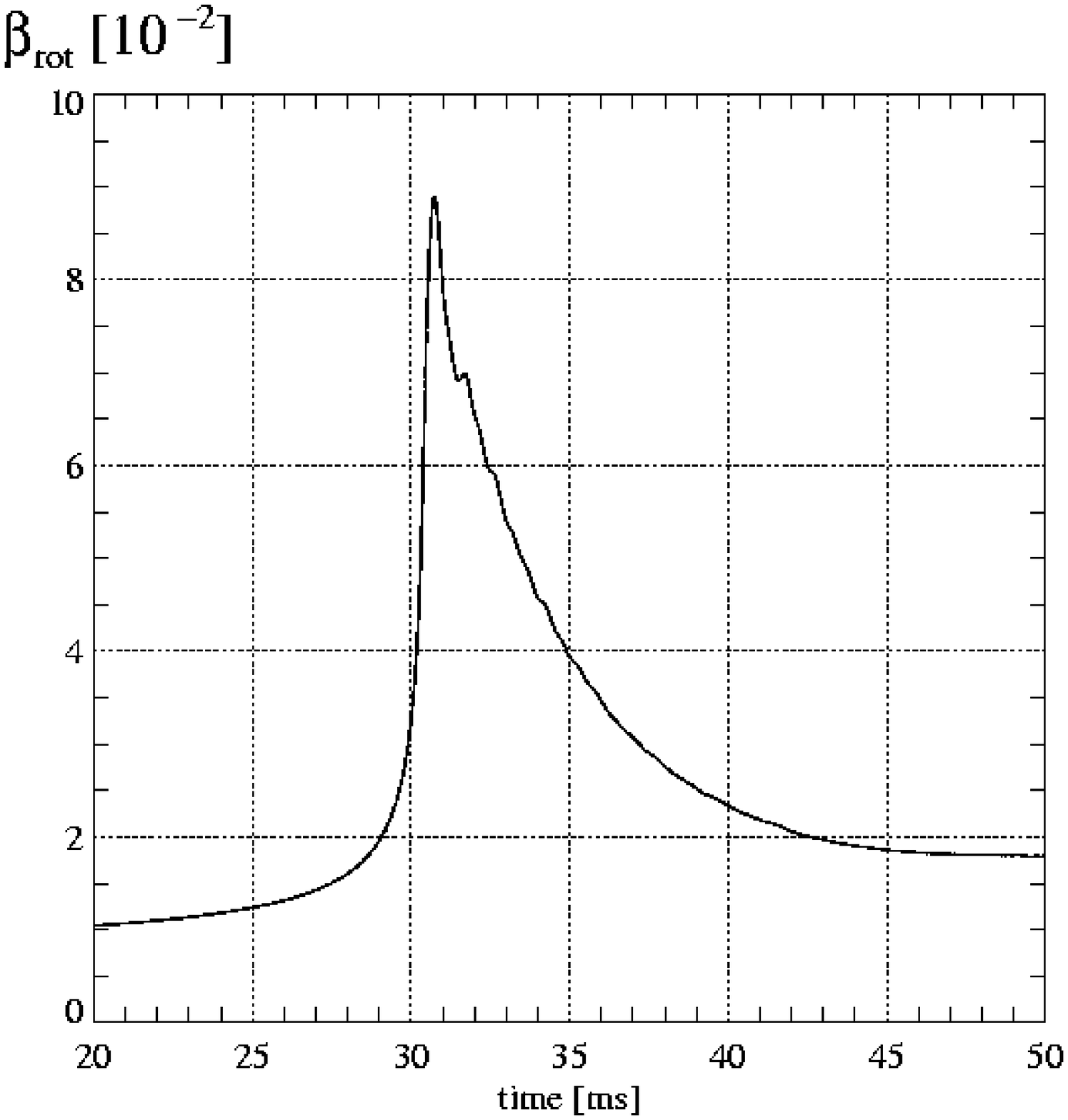}
  \includegraphics[width=5.6cm]{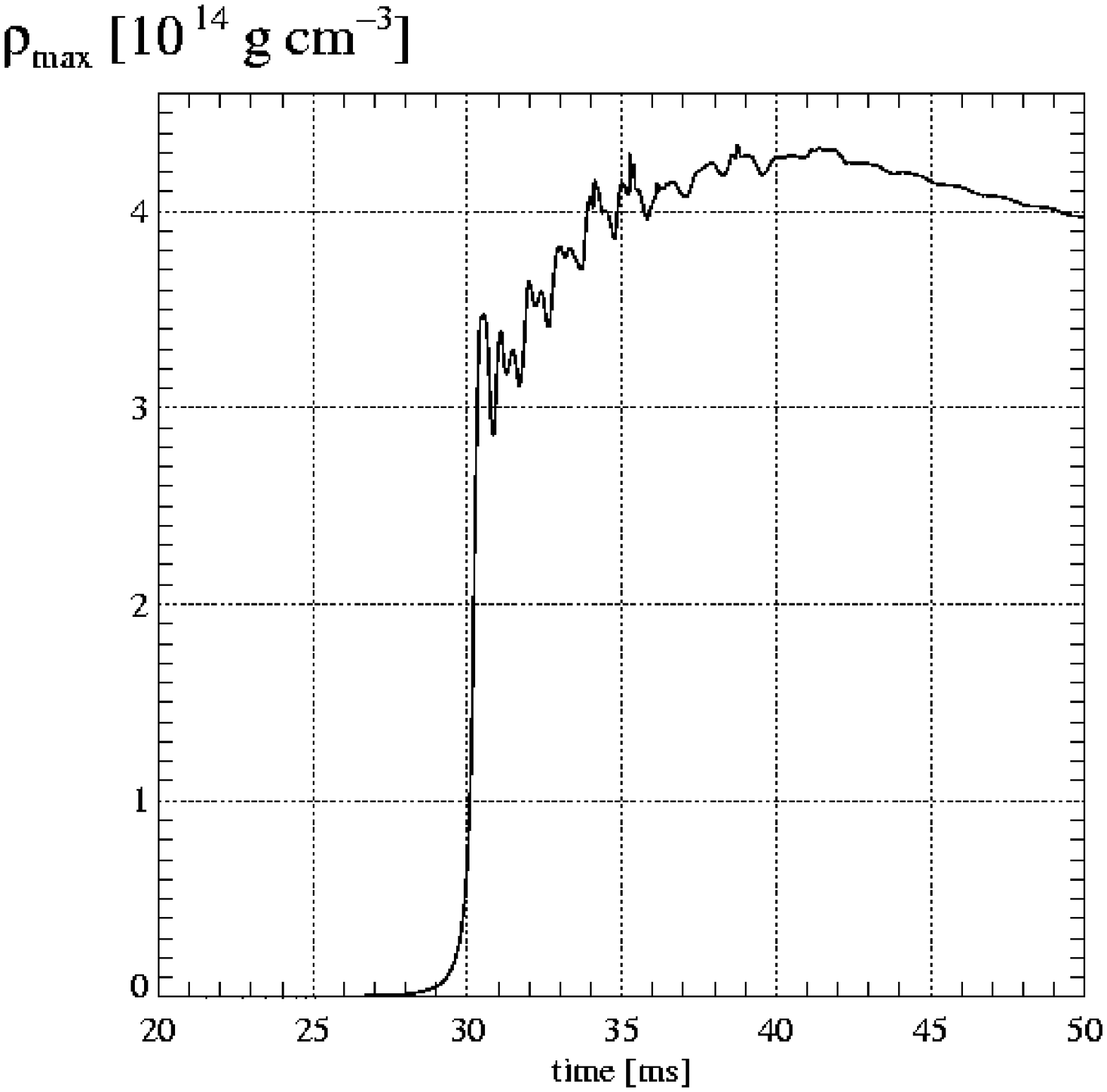}
  \includegraphics[width=5.6cm]{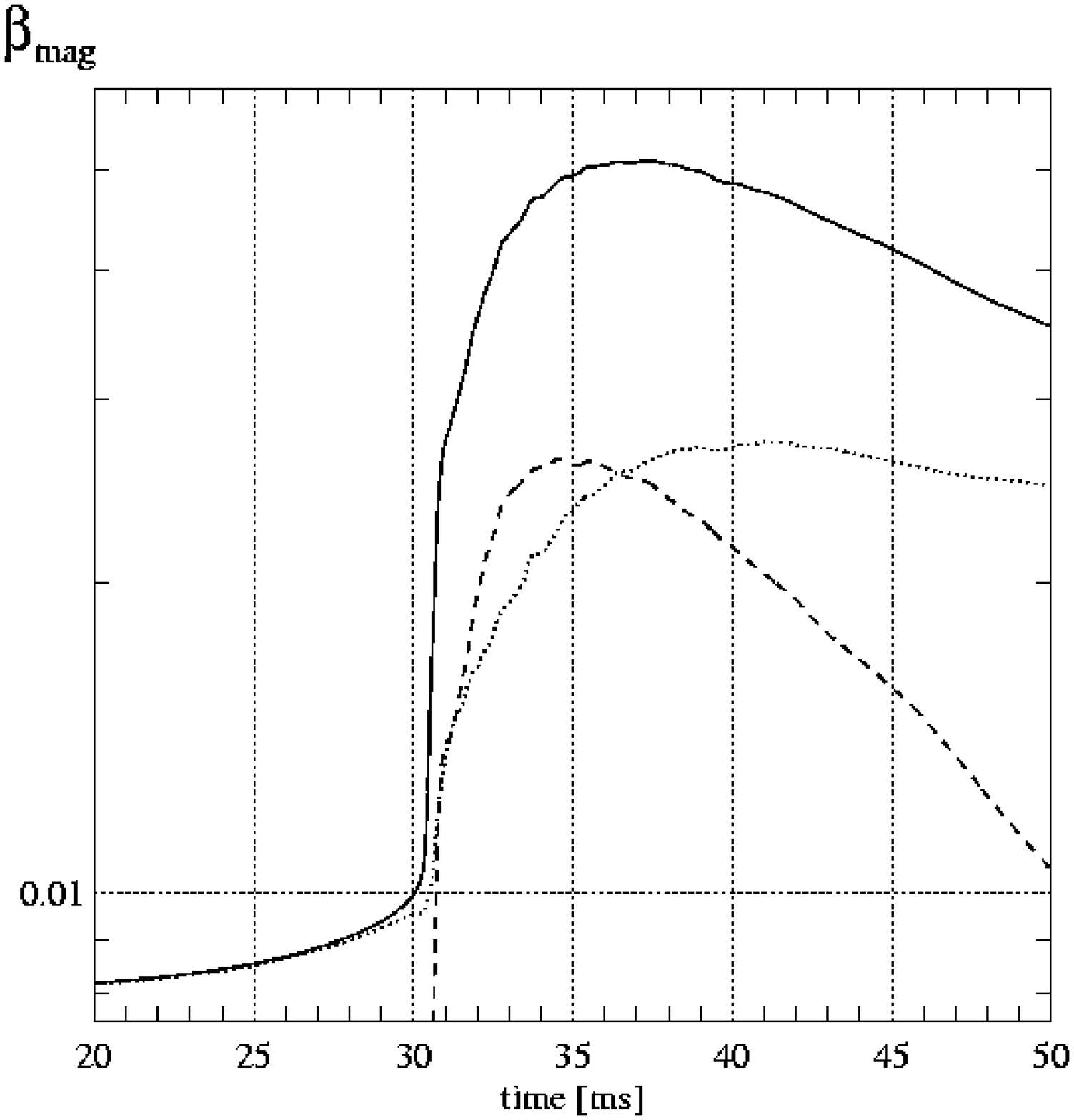}
  \includegraphics[width=5.6cm]{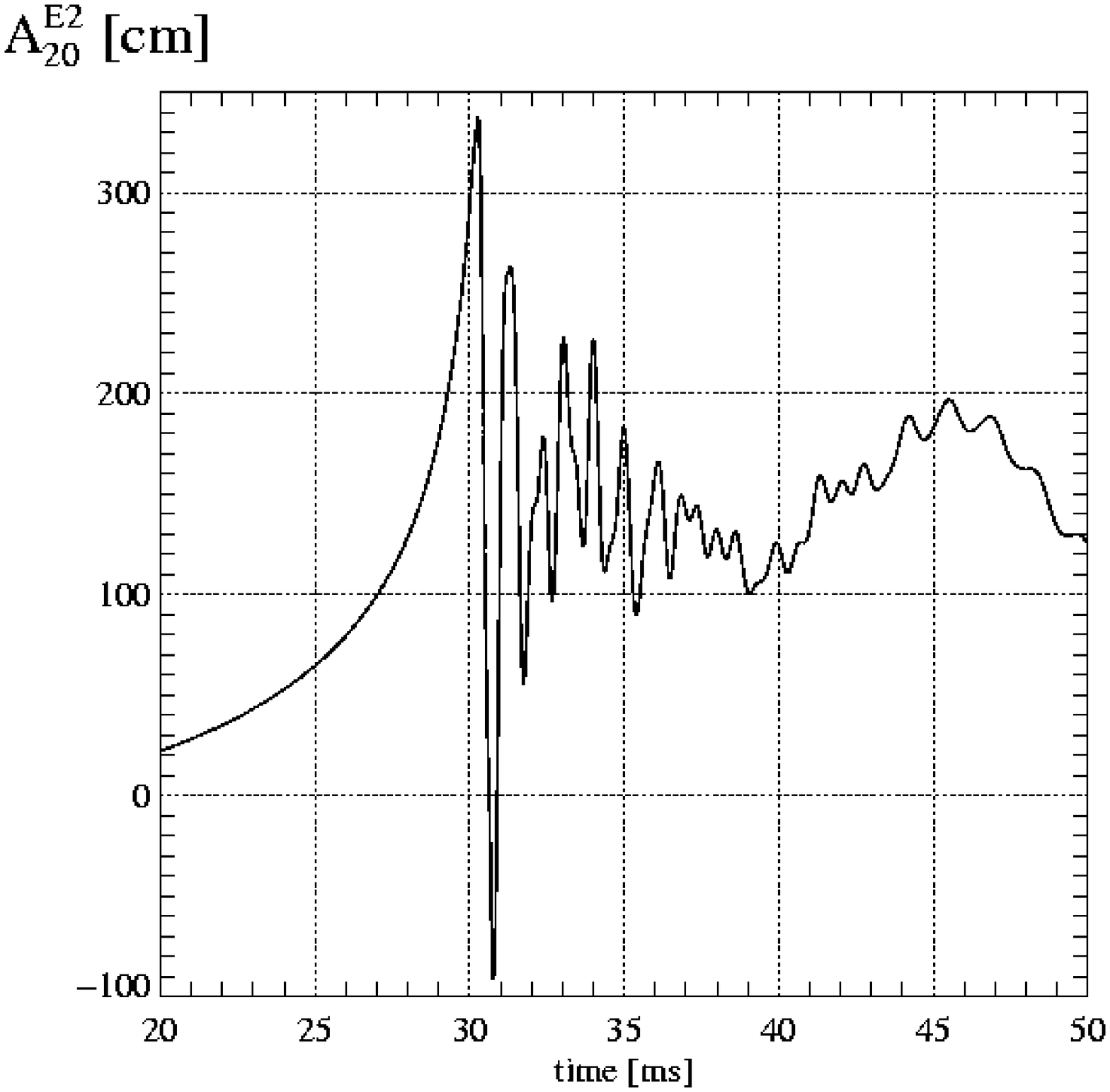}
  \includegraphics[width=5.6cm]{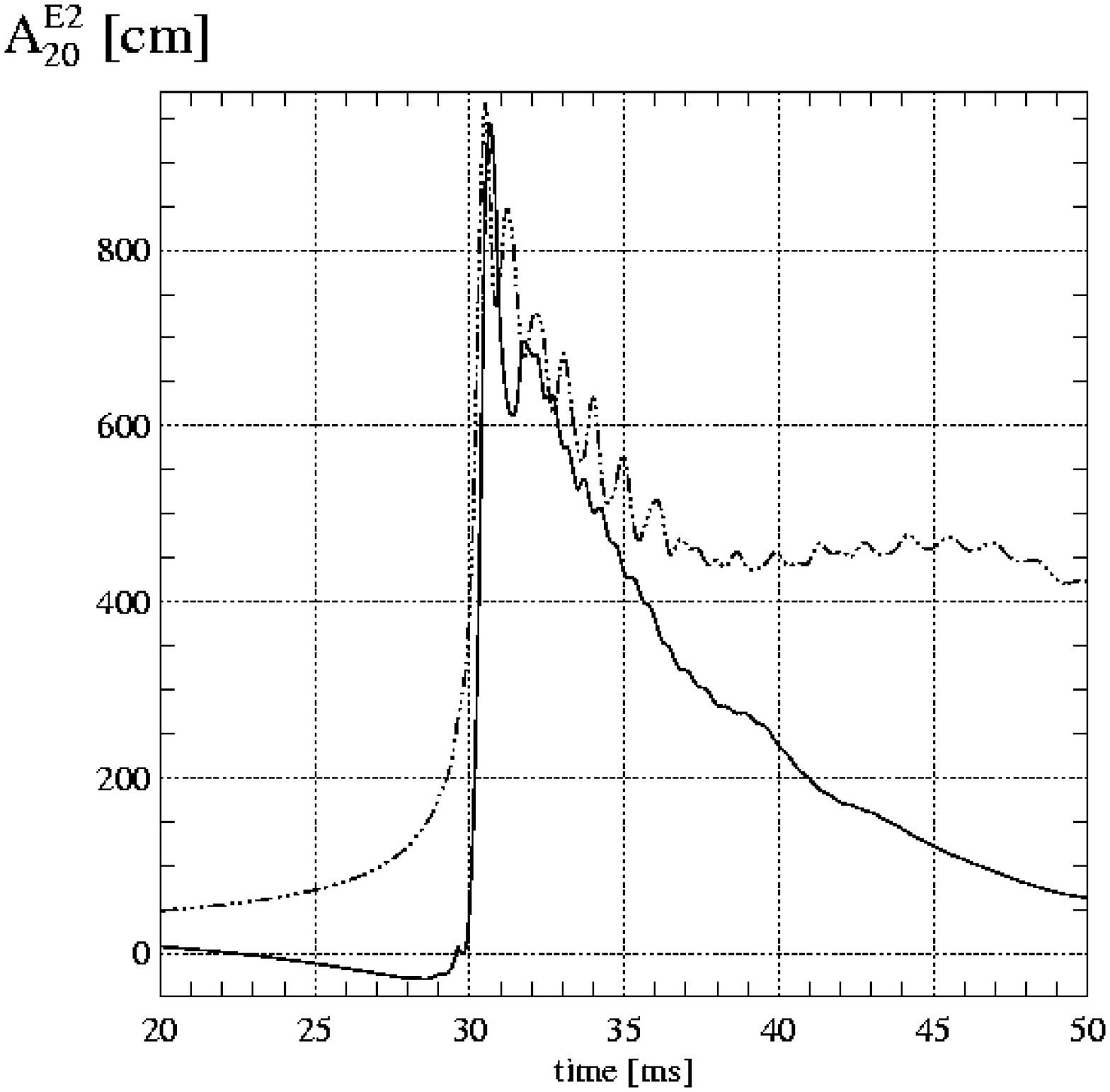}
  \includegraphics[width=5.6cm]{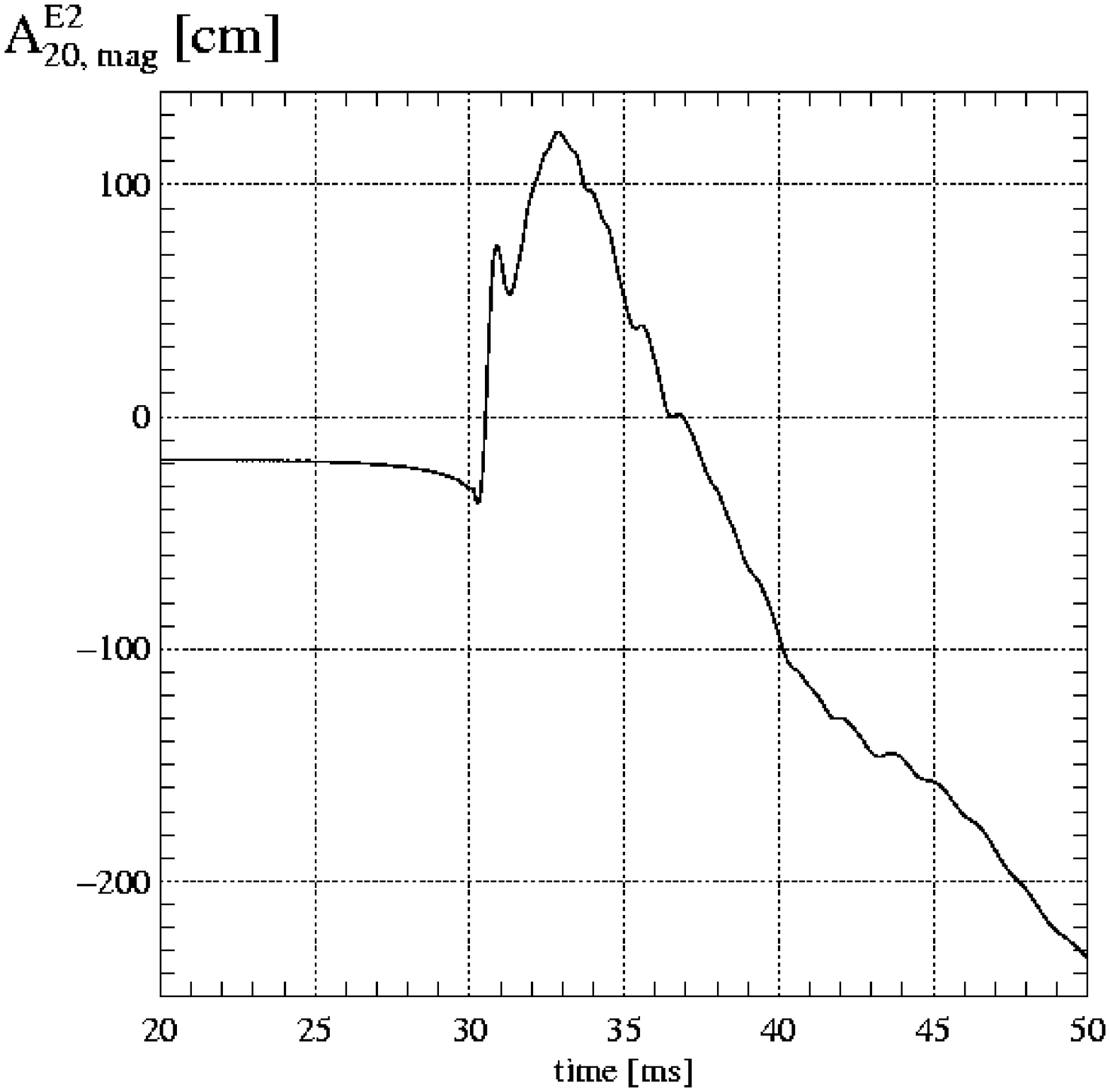}
  \caption[Evolution of model A3B3G5-D3M13]
  { The dynamical evolution and the GW signal of model A3B3G5-D3M13.
    The upper panels display the temporal evolution of the rotational
    energy parameter $\beta_{\mathrm{rot}}$ (left), the maximum
    density $\rho_{\mathrm{max}}$ (middle), and the magnetic energy
    parameters (right; $\beta_{\mathrm{mag}}$ (solid line),
    $\beta_{\phi}$ (dashed line), and $\beta_{\mathrm{mag}} -
    \beta{\phi}$ (dotted line)), respectively.  The GW signal of the
    model is shown in the lower panels: total amplitude (solid line,
    left), $-A^{\mathrm{E}2}_{20;\mathrm{hyd}}$ (solid line, middle),
    $A^{\mathrm{E}2}_{20;\mathrm{grav}}$ (dashed line, middle), and
    $A^{\mathrm{E}2}_{20;\mathrm{mag}}$ (right).  }
  \label{Fig:335-313:Dyn}
\end{figure*}

The GW amplitude of model A3B3G5-D3M13 (Fig.\,\ref{Fig:335-313:Dyn}),
whose weak--field counterpart A3B3G5-D3M10 emits a GW signal of type III
(Fig.\,\ref{Fig:sFF_rho_und_GW}), is modified by the strong initial
field.  Relative to the weak--field model, the pre--bounce maximum is
enhanced, and the size of the (negative) peak at bounce is reduced by
the same amount, whereas the amplitude of the first (positive)
post--bounce peak remains nearly unchanged.  As the core loses angular
momentum, the size of the hydrodynamic and gravitational contributions
to the GW amplitude decreases.  The post--bounce oscillations of the
signal, which are mainly due to the central core, are superimposed on a
nearly constant positive amplitude of half the size of the bounce
amplitude indicating the presence of the collimated bipolar outflow
(Fig.\,\ref{Fig:335-313:50ms}).  The evolution of the magnetic
contributions of models A1B3G3-D3M13 and A3B3G5-D3M13 are similar.  In
both models, a phase of positive amplitude concurrent with the most
efficient braking of the core's rotation is followed by a decrease to
large negative values (Fig.\,\ref{Fig:335-313:Dyn}, lower right panel).
In the latter phase the magnetic amplitude, which is mostly produced in
the central core, provides the main contribution to the total signal.

Increasing the initial magnetic field strength to $\sim
10^{12}\mathrm{G}$ in models which exhibit a type-I GW signal in the
non--magnetic case, the size of the large (negative) bounce amplitude
decreases slightly. This also holds for even stronger magnetic fields in
case of models that are influenced by rotation to a higher degree such
as A3B2G4-D3Mm and A3B3G4-D3Mm, respectively. However, for rigidly and
moderately fast rotating models (A1B1G3-D3Mm and A1B3G3-D3Mm) the size
of the bounce signal grows again strongly when the initial field
strength reaches $10^{13}\mathrm{G}$.

The ratio of the GW amplitudes of the positive peaks immediately prior
to and immediately after the signal minimum at bounce can both increase
and decrease by the effects of a strong magnetic field.  In type-I
models with a small influence of the rotation on the dynamics, a
stronger initial field increases the ratio, and for sufficiently fast
rotating models (A3B2G4-D3Mm) it decreases.  Two models with a type-I GW
signal are close to the transition from a bounce caused by pressure or
centrifugal forces (A2B4G4-DdMm and A3B3G4-DdMm), i.e.\,their cores are
only partially stabilized by the stiffening of the EOS at nuclear matter
density.  For both models the amplitude ratio of the pre- and
post--bounce peaks increases with increasing magnetic field strength,
while it decreases for most type-III models.

\subsubsection{Models bouncing due to centrifugal forces}
\label{Suk:CenBouMo}

\begin{figure*}[!htbp]
  \centering
  \includegraphics[width=5.6cm]{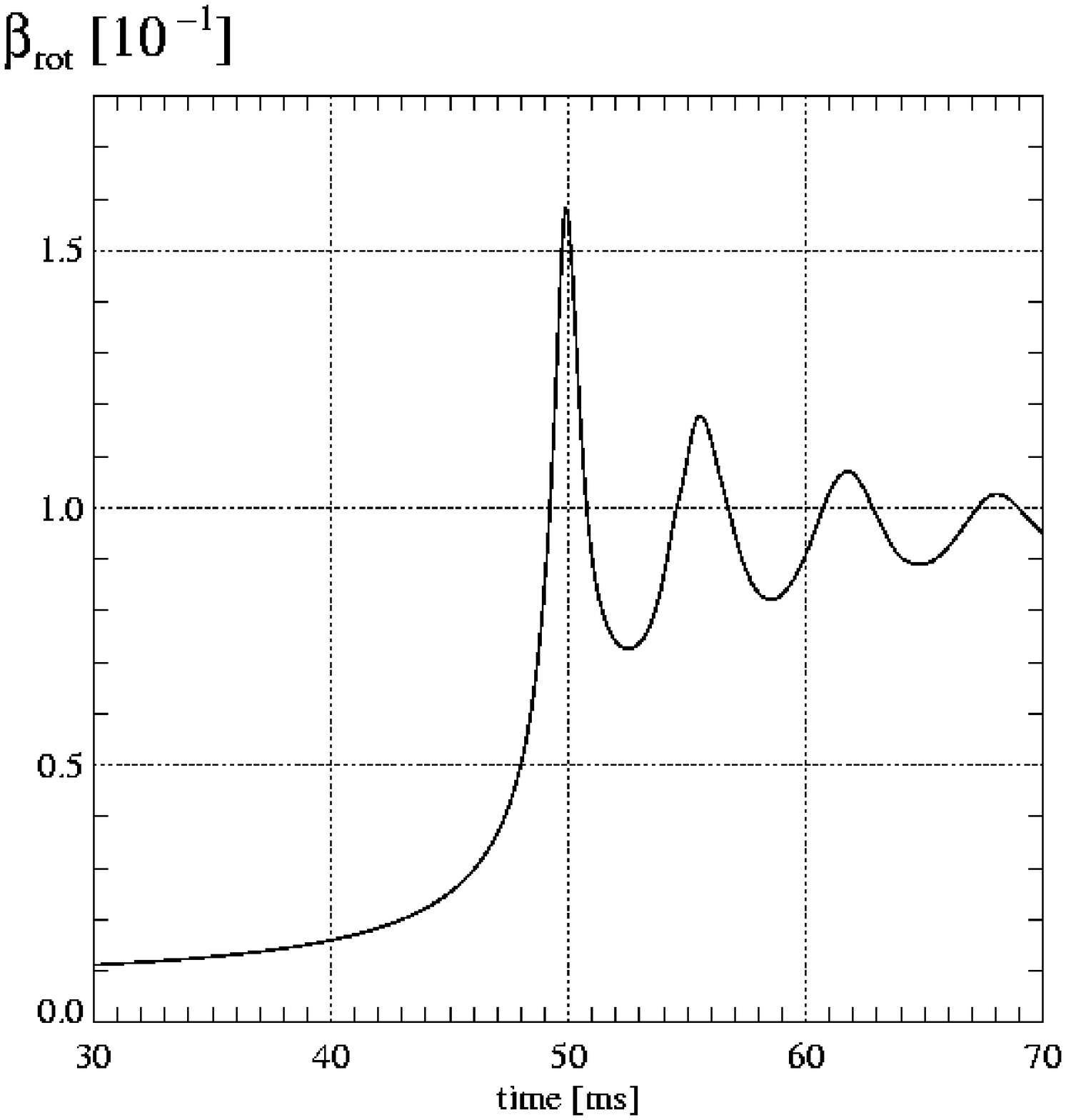}
  \includegraphics[width=5.6cm]{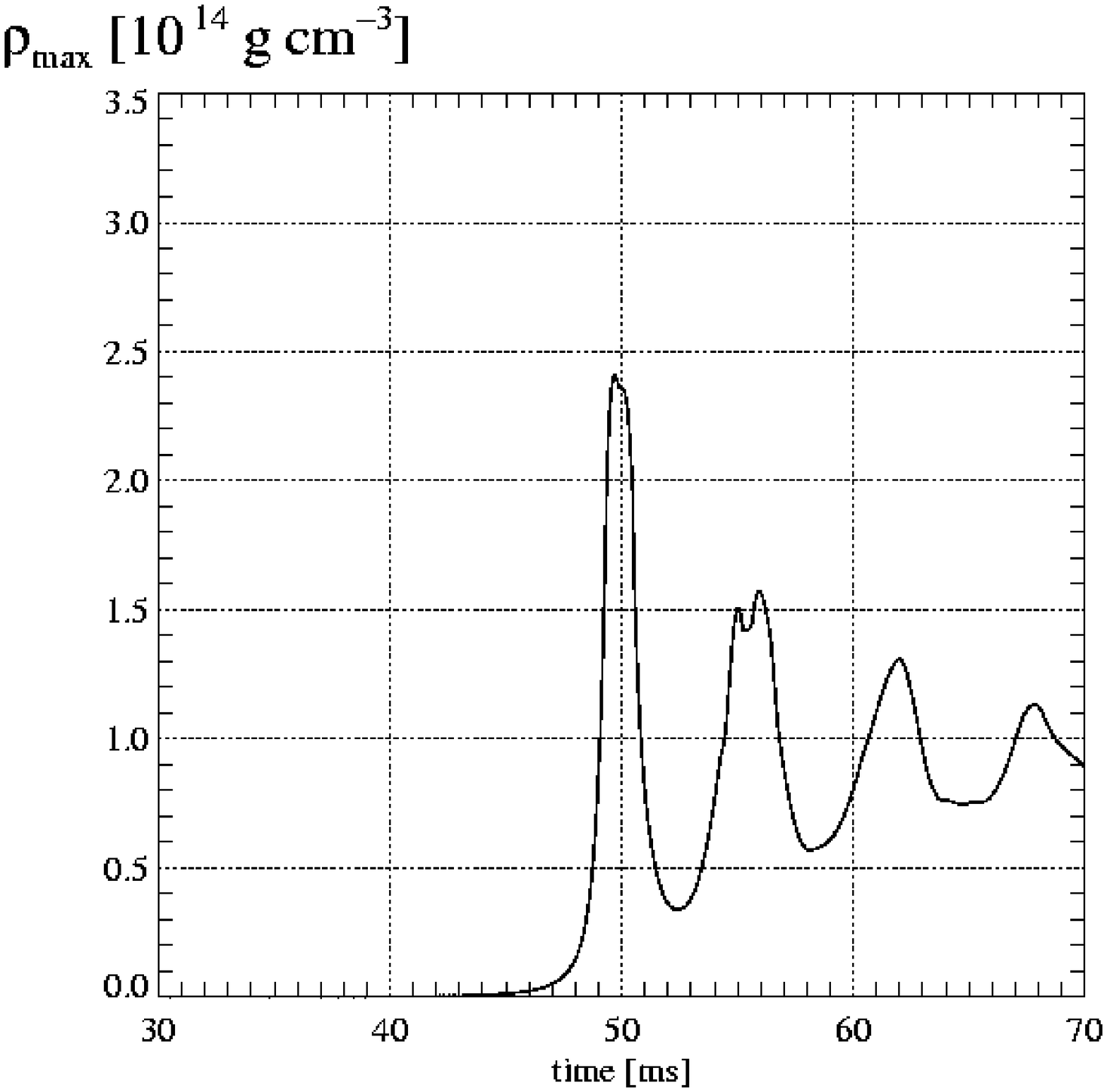}
  \includegraphics[width=5.6cm]{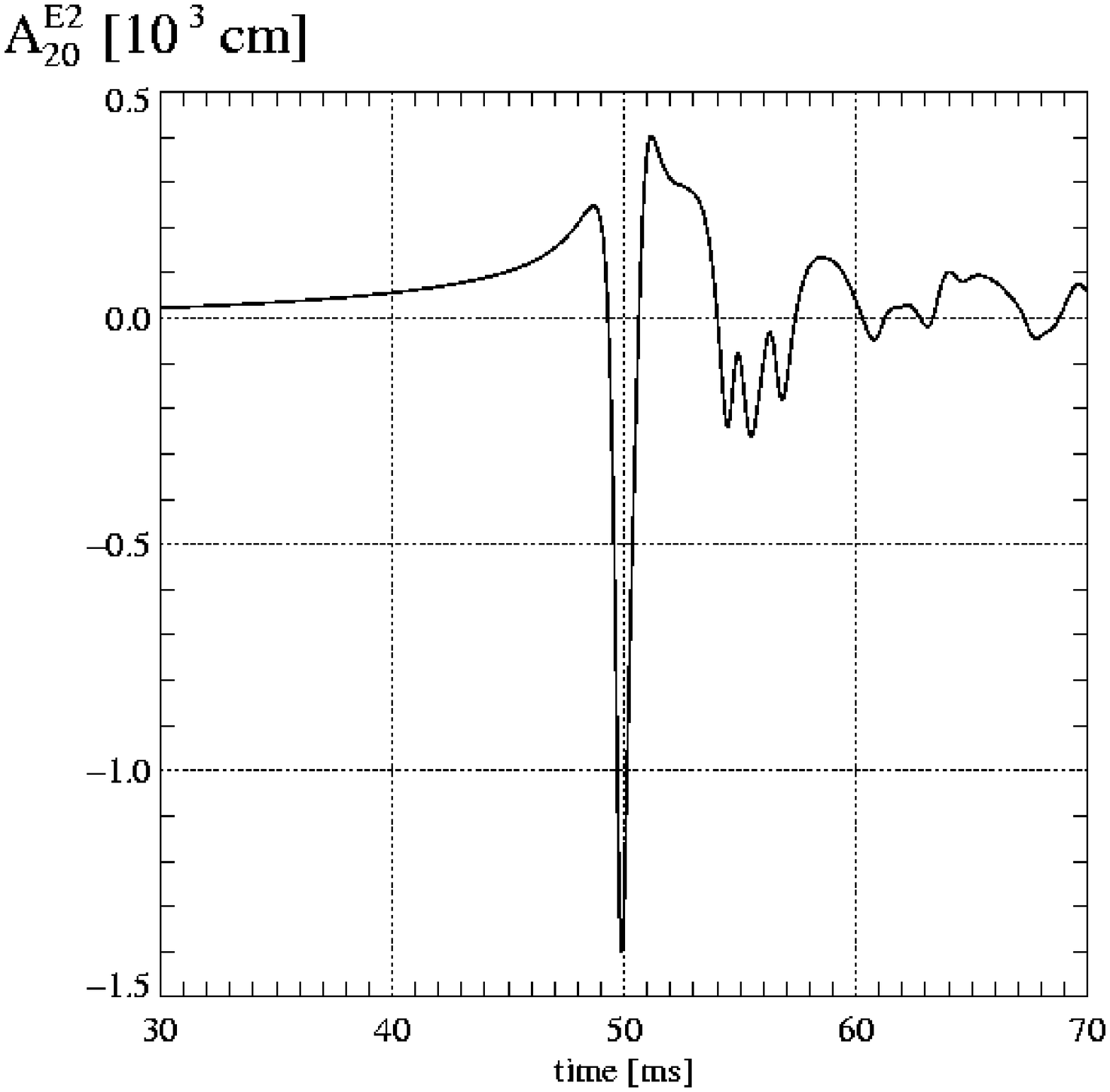}
  \includegraphics[width=5.6cm]{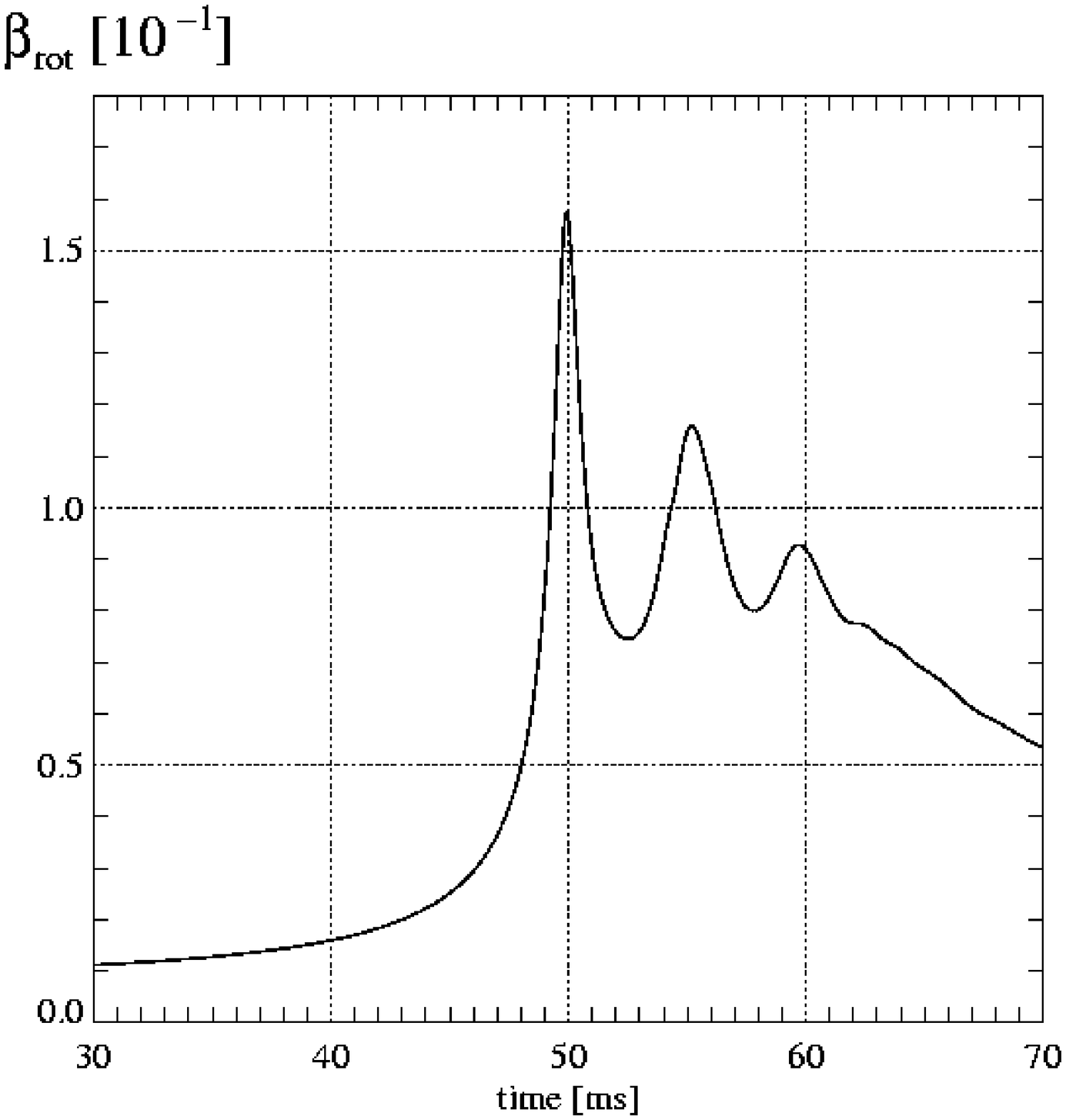}
  \includegraphics[width=5.6cm]{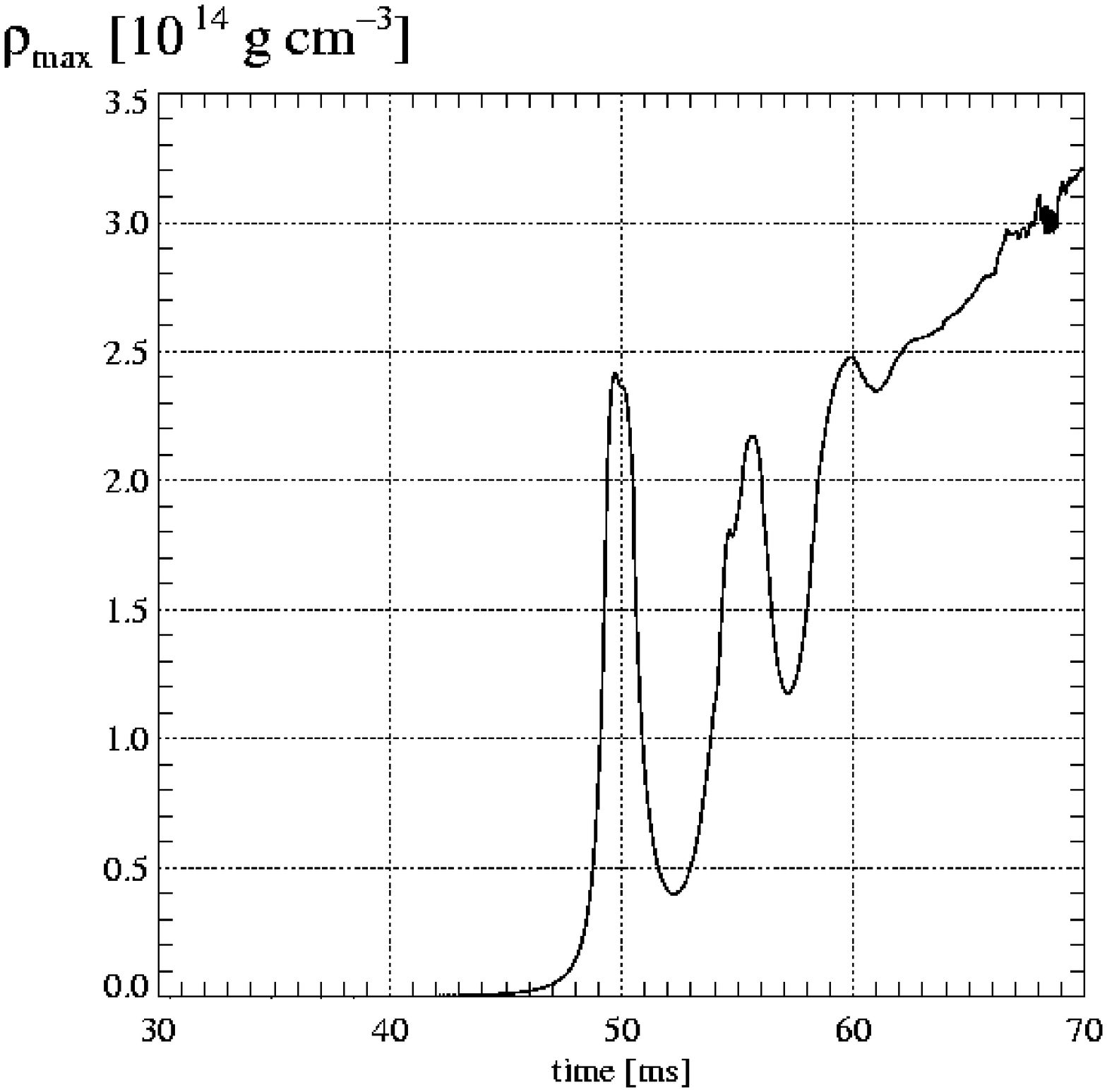}
  \includegraphics[width=5.6cm]{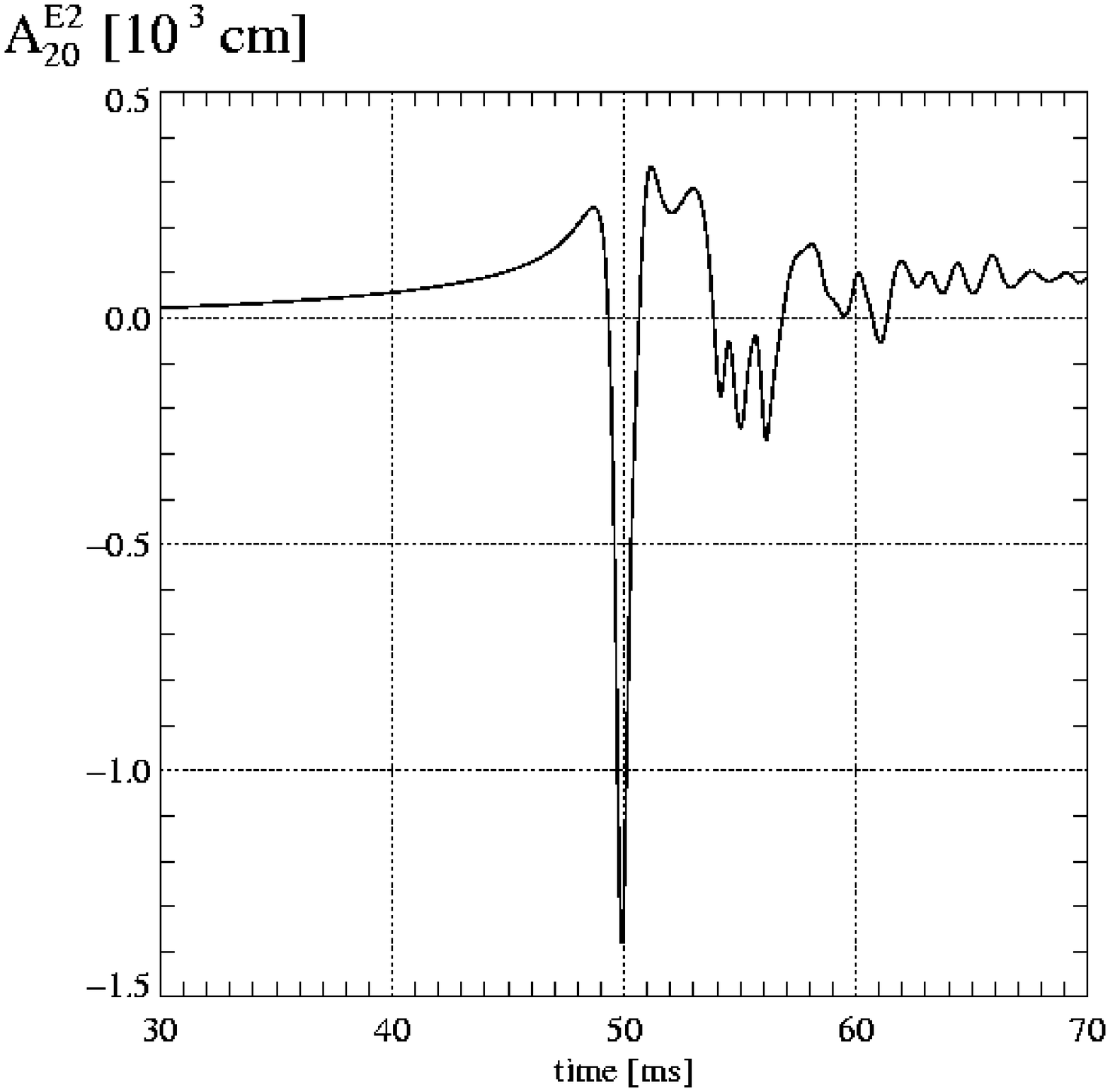}
  \caption[Dynamical evolution of model A3B3G3-D3M12]
  { The evolution of the rotational energy parameter $\beta_{rot}$
    (left panels), of the maximum density (middle panels) and the GW
    signal (right panels) of models A3B3G3-D3M10 (top panels), and
    A3B3G3-D3M12 (bottom panels), respectively.  }
  \label{Fig:333-312:Dyn}
\end{figure*}

\begin{figure*}[!htbp]
  \centering
  \includegraphics[width=5.6cm]{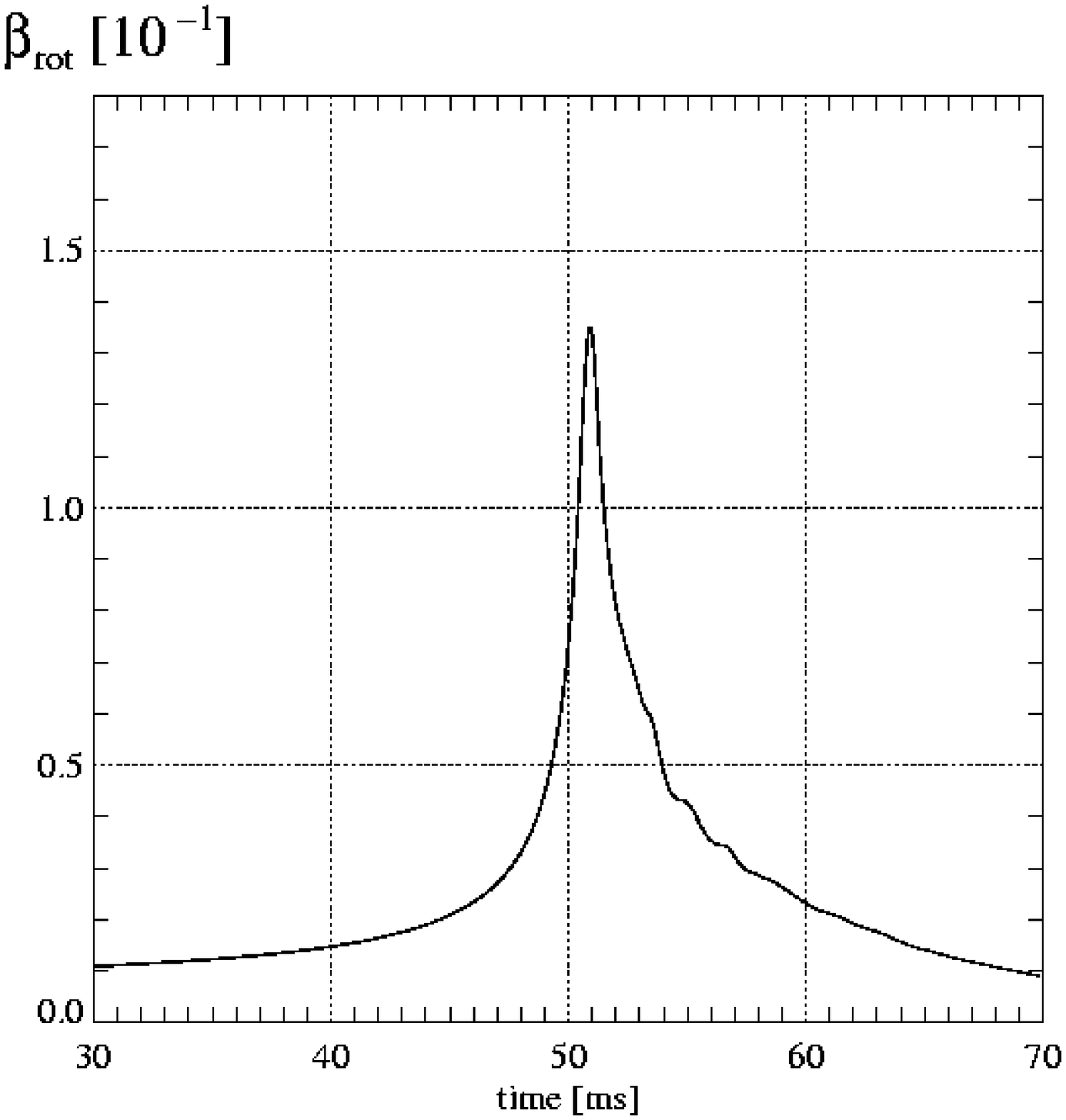} 
  \includegraphics[width=5.6cm]{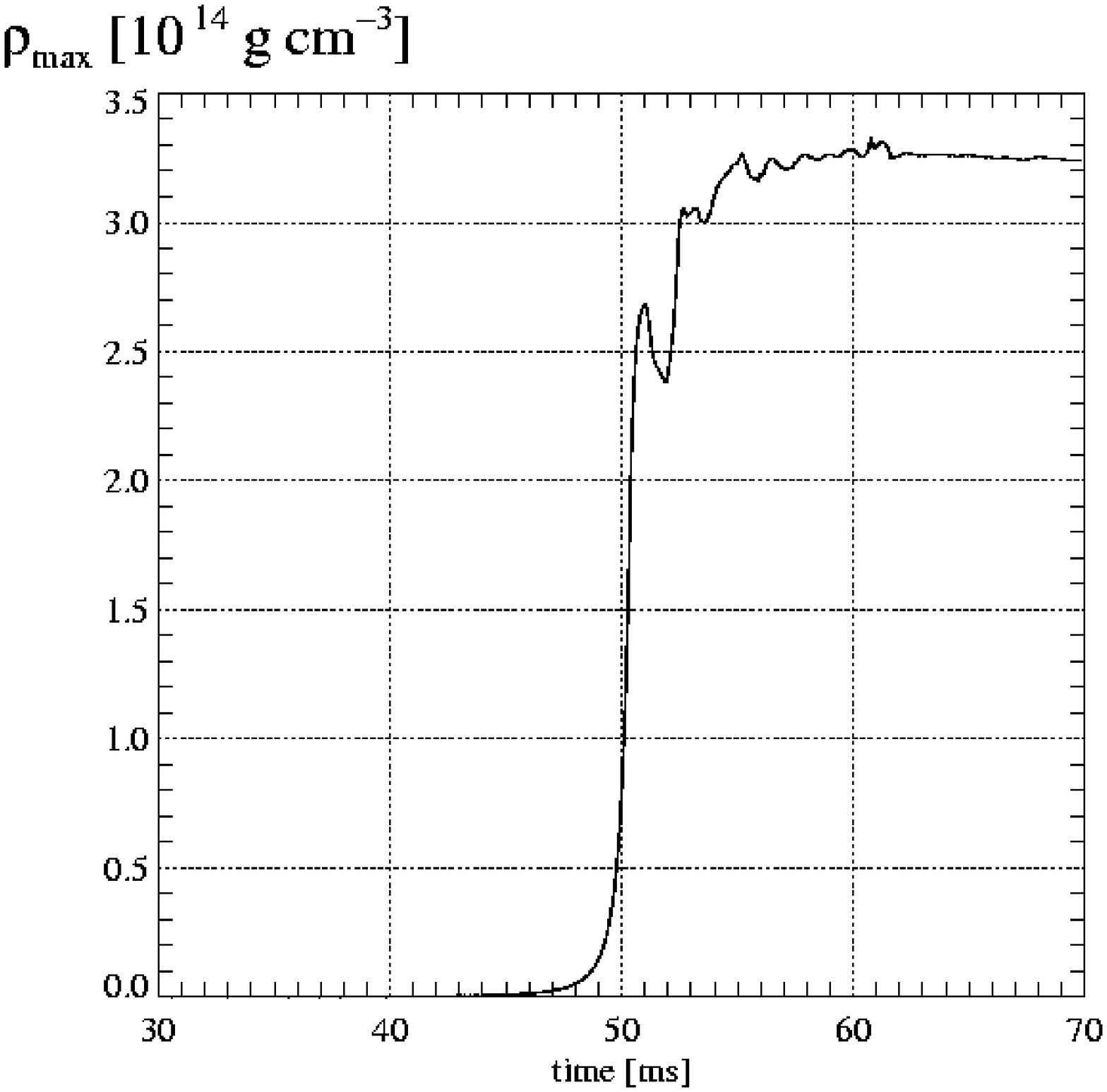} 
  \includegraphics[width=5.6cm]{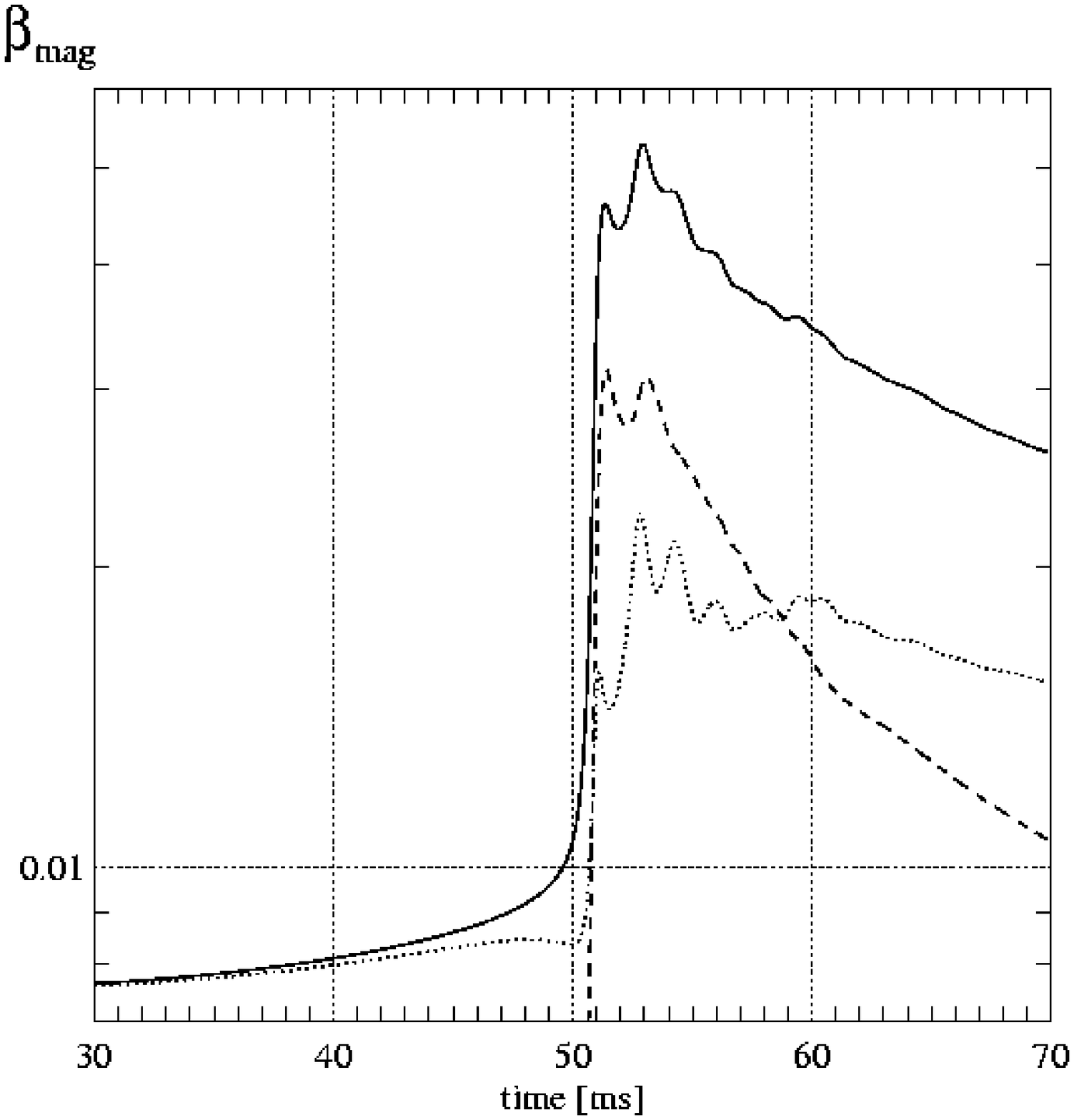} 
  \includegraphics[width=5.6cm]{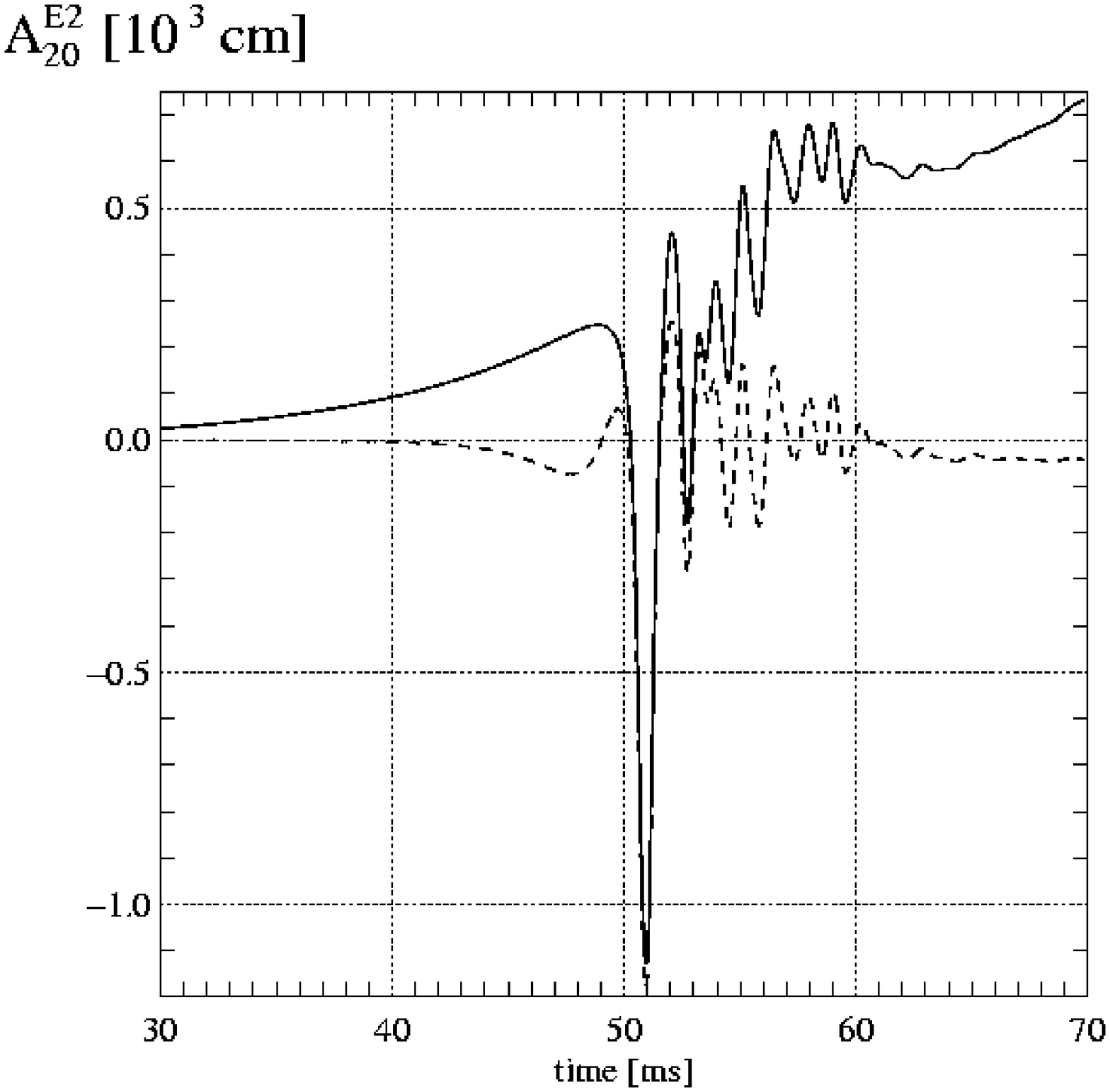} 
  \includegraphics[width=5.6cm]{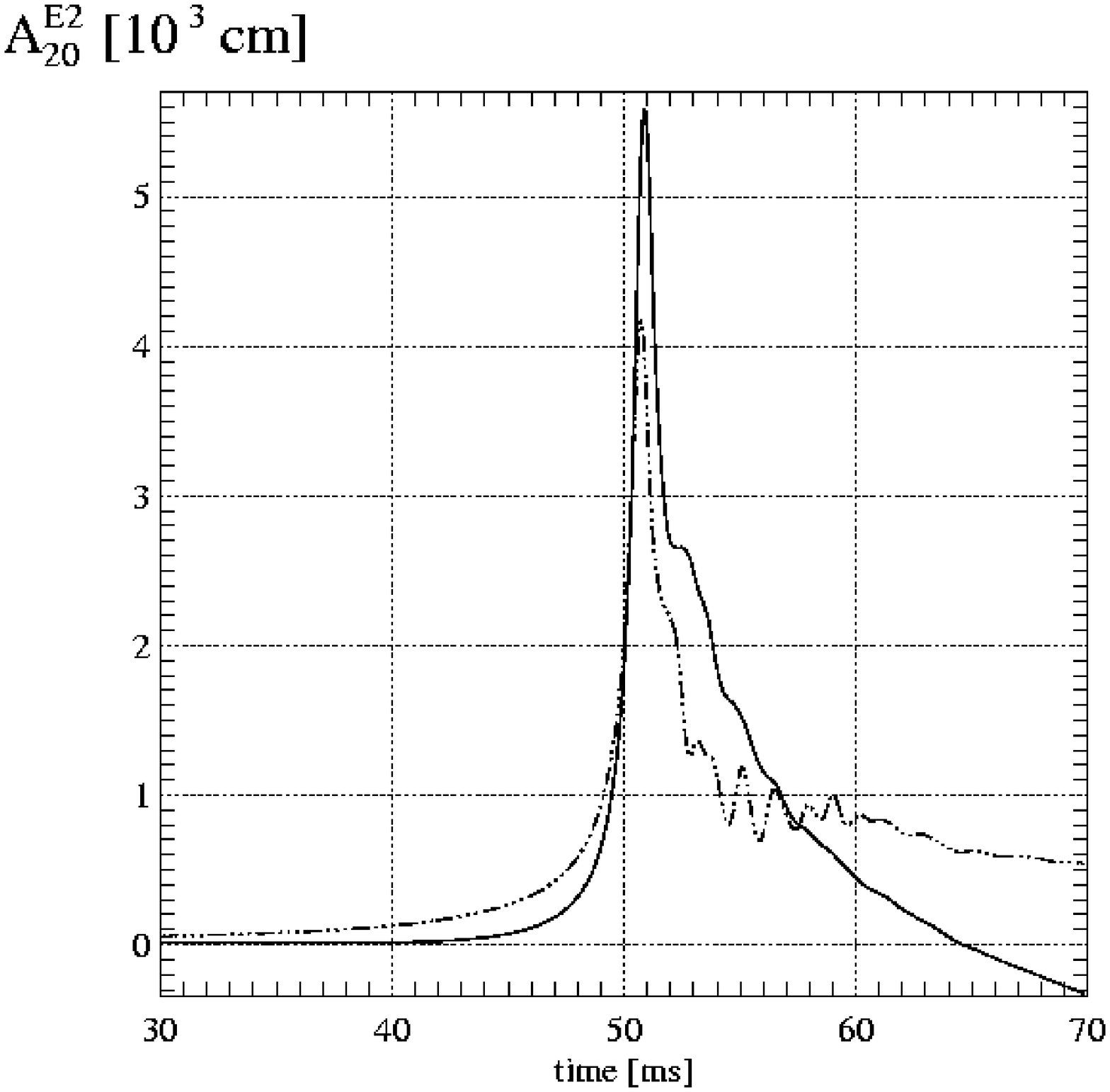} 
  \includegraphics[width=5.6cm]{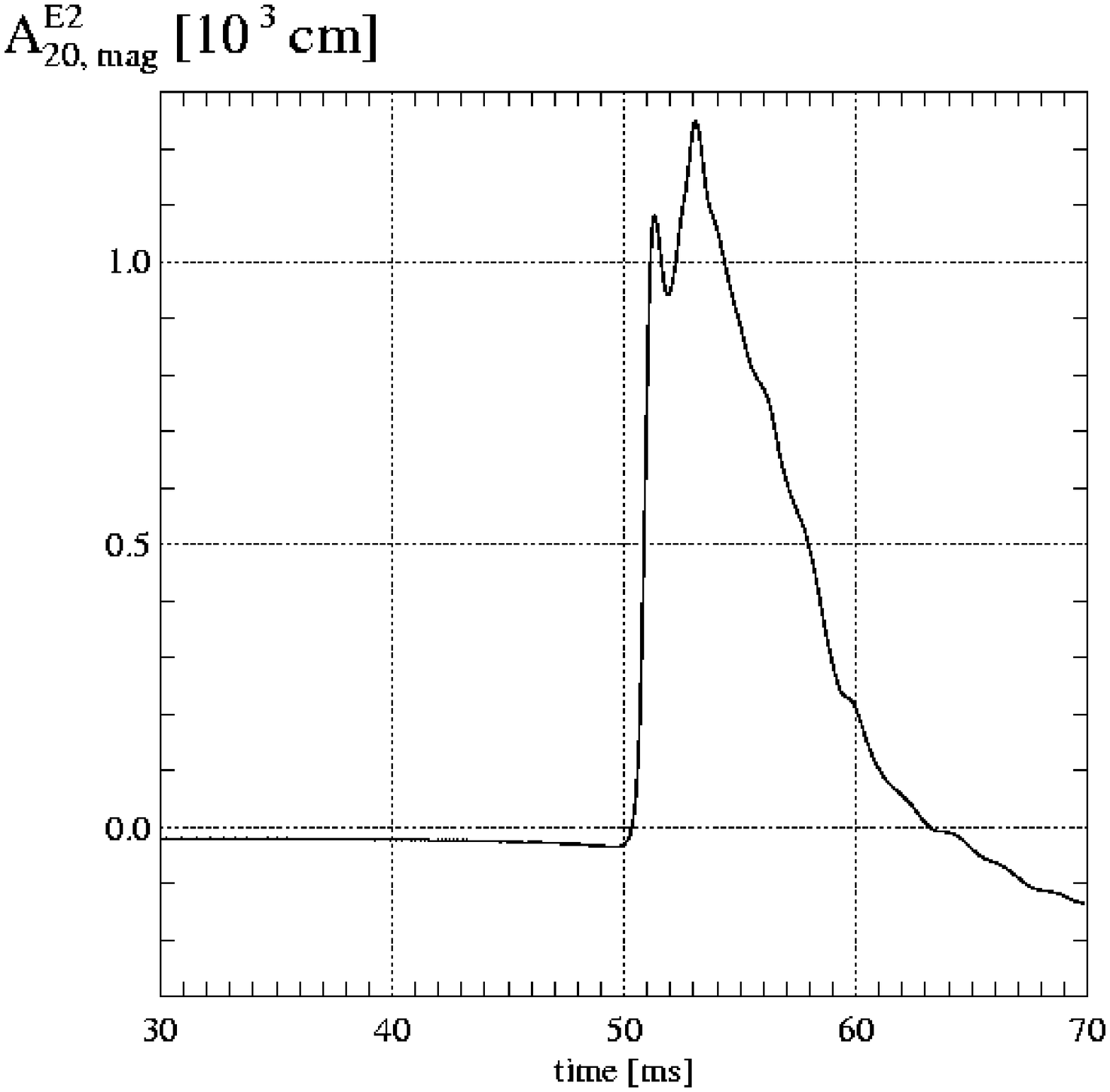} 
  \caption[Evolution of model A3B3G3-D3M13]
  { The dynamical evolution and the GW signal of model A3B3G3-D3M13.
    The upper panels display the temporal evolution of the rotational
    energy parameter $\beta_{\mathrm{rot}}$ (left), the maximum
    density $\rho_{\mathrm{max}}$ (middle), and the magnetic energy
    parameters (right; $\beta_{\mathrm{mag}}$ (solid line),
    $\beta_{\phi}$ (dashed line), and $\beta_{\mathrm{mag}} -
    \beta{\phi}$ (dotted line)), respectively.  The GW signal of the
    model is shown in the lower panels: total amplitude (solid line,
    left; the dashed line gives the contribution of the layers with $r
    \le 55.6\,$km), $-A^{\mathrm{E}2}_{20;\mathrm{hyd}}$ (solid line,
    middle), $A^{\mathrm{E}2}_{20;\mathrm{grav}}$ (dashed line,
    middle), and $A^{\mathrm{E}2}_{20;\mathrm{mag}}$ (right).  
  }
  \label{Fig:333-313:Dyn}
\end{figure*}

The weak--field models of series A3B3G3-D3Mm belong to the transition
class between standard--type and centrifugally bouncing cores.  They
bounce mostly due to centrifugal forces and exhibit large--scale
pulsations after bounce, but their maximum density exceeds nuclear
matter density during bounce. Both the period and the damping of the
pulsations are significantly larger than for purely centrifugally
bouncing type-II models (Fig.\,\ref{Fig:333-312:Dyn}).  Their GW signal
consists of a pronounced peak of negative amplitude at bounce, and
subsequent relatively long--period oscillations.

For the strong--field variants of this model series, e.g.\,model
A3B3G3-D3M12 (Fig.\,\ref{Fig:333-312:Dyn}), the main evolutionary effect
of the magnetic field is the extraction of rotational energy after
bounce that proceeds on time scales much longer than the dynamic time
scale of the inner core.  Model A3B3G3-D3M12 reaches a maximum rotation
parameter $\beta_{rot}^{\mathrm{max}} = 0.16$ at bounce, which is nearly
the same value as in the non--magnetic case.  At the time of the second
density maximum $\beta_{rot}$ is only slightly smaller than in the
weak--field model A3B3G3-D3M10, but at the next (3rd) maximum the
rotational energy has decreased by about $15\,\%$.  Unlike in the
weak--field model no fourth density maximum occurs.  Instead,
$\beta_{rot}$ decreases monotonically.  The rotation rate of model
A3B3G3-D3M12 is sufficiently large for a sufficiently long time to allow
the core to exhibit several centrifugal pulsations before it settles
down in a pressure supported final equilibrium state.  During these
pulsations the time--averaged value of $\rho_{\mathrm{max}}$ increases,
and densities significantly larger than the bounce density $\rho_b =
2.41\cdot 10^{14}\ \mathrm{cm \, s}^{-1}$ are reached.

The GW signal of the strong--field model A3B3G3-D3M12 is similar to that
of the weak--field model A3B3G3-D3M10 during the first $\approx 10 \ 
\mathrm{ms}$ after bounce when the cores of both models still undergo
large--scale pulsations.  Later when the rotational energy has decreased
sufficiently and the pulsations begin to fade away, the signals begin to
differ slightly. The maximum density now increases monotonically.
Rotation is no longer important for the core's stabilization, as it is
supported by pressure forces.  Thus, like in the case of a
standard--type core (e.g.\,A1B3G3-DdMm), the rapid oscillations of the
GW amplitude for $t \ga 60 \mathrm{\ ms}$ occur on the core's
dynamic time scale. Superimposed to the oscillations is a positive mean
amplitude $A^{\mathrm{E}2}_{20} \approx 100 \mathrm{cm}$, which results
from the enhanced asphericity of the shock wave as compared to that of
the weak--field model.

The most strongly magnetized model of series A3B3G3-D3Mm, namely model
A3B3G3-D3M13, behaves very differently from its weak--field variants as
e.g.\,model A3B3G3-D3M10 (Fig.\,\ref{Fig:333-312:Dyn}).  Unlike the
latter model it suffers a single bounce at $\rho_b = 2.69 \cdot 10^{14}\ 
\mathrm{cm \, s}^{-1}$ (a value about $10\,\%$ larger than for the
weak--field models) with a subsequent contraction phase caused by the
decrease of the core's rotational energy.  After bounce the maximum
density reaches a value of $\rho_{\mathrm{max}} = 3.3\cdot 10^{14}\ 
\mathrm{cm \, s}^{-1}$.  Unlike model A3B3G3-D3M12, it does not exhibit
any large--scale pulsations.  Instead, it rapidly approaches a high
density state without an intermediate phase where the density in the
entire core is less that nuclear matter density, as it is the case for
model A3B3G3-D3M12.  The rotational energy ($\beta_{\mathrm{max}} =
0.135$) is less than for models with weaker initial fields
(e.g.\,$\,\beta_{\mathrm{max}} = 0.16$ for model A3B3G3-D3M10), and
decreases by $90\,\%$ within $15\,$ms (Fig.\,\ref{Fig:333-313:Dyn}).

The collapse and post--bounce dynamics of model A3B3G3-D3M13 is similar
to that of strong--field single--bounce models, as e.g.\,model
A1B3G3-D3M13. Many features of the latter model are also present in
model A3B3G3-D3M13: the bulb--like shock surface, the two highly
magnetized post--shock regions (one near the equator and the other one
along the polar axis), and a region of retrograde rotation near the
equatorial plane at the edge of the inner core. The ratio of magnetic
and gas pressure is largest at the axis at $r=175\ \mathrm{km}$ well
behind the shock located at $r_{\mathrm{shock}} \approx 200\ 
\mathrm{km}$.

The GW signal of model A3B3G3-D3M13 is qualitatively similar to that of
its less strongly magnetized variant A3B3G3-D3M12, but the features
caused by the magnetic field are more pronounced
(Fig.\,\ref{Fig:333-313:Dyn}).  The peak amplitude at bounce
$A^{\mathrm{E}2}_{20} = -1128 \ \mathrm{cm}$ is, contrary to model
A3B3G3-D3M12, considerably less negative than in the weak-field model
A3B3G3-D3M10 ($A^{\mathrm{E}2}_{20} = -1401 \ \mathrm{cm}$).
Immediately after bounce the GW signal shows certain similarities with
that of model A1B3G3-D3M13 (Fig.\,\ref{Fig:133-313:Dyn}): a large
positive peak is followed by a series of oscillations occurring on the
local dynamic time scale superimposed on an increase of the mean
amplitude to a value of $A^{\mathrm{E}2}_{20} \approx 600 \ \mathrm{cm}$
within $t \approx 5\ \mathrm{ms}$, which is the time it takes to turn
the initially roughly spherical shock into a relatively wide bipolar
outflow directed along the rotation axis.

Similar to cores bouncing due to pressure forces, the extraction of
rotational energy from cores bouncing due to centrifugal forces proceeds
qualitatively differently for models with an initial magnetic field
strengths of $10^{12} \ \mathrm{G}$ and $10^{13}\ \mathrm{G}$,
respectively.  For the weaker field the extraction process relies on an
instability--driven angular momentum transport, which is not required
for the stronger field models.

\begin{figure*}[!htbp]
  \centering
  \includegraphics[width=5.6cm]{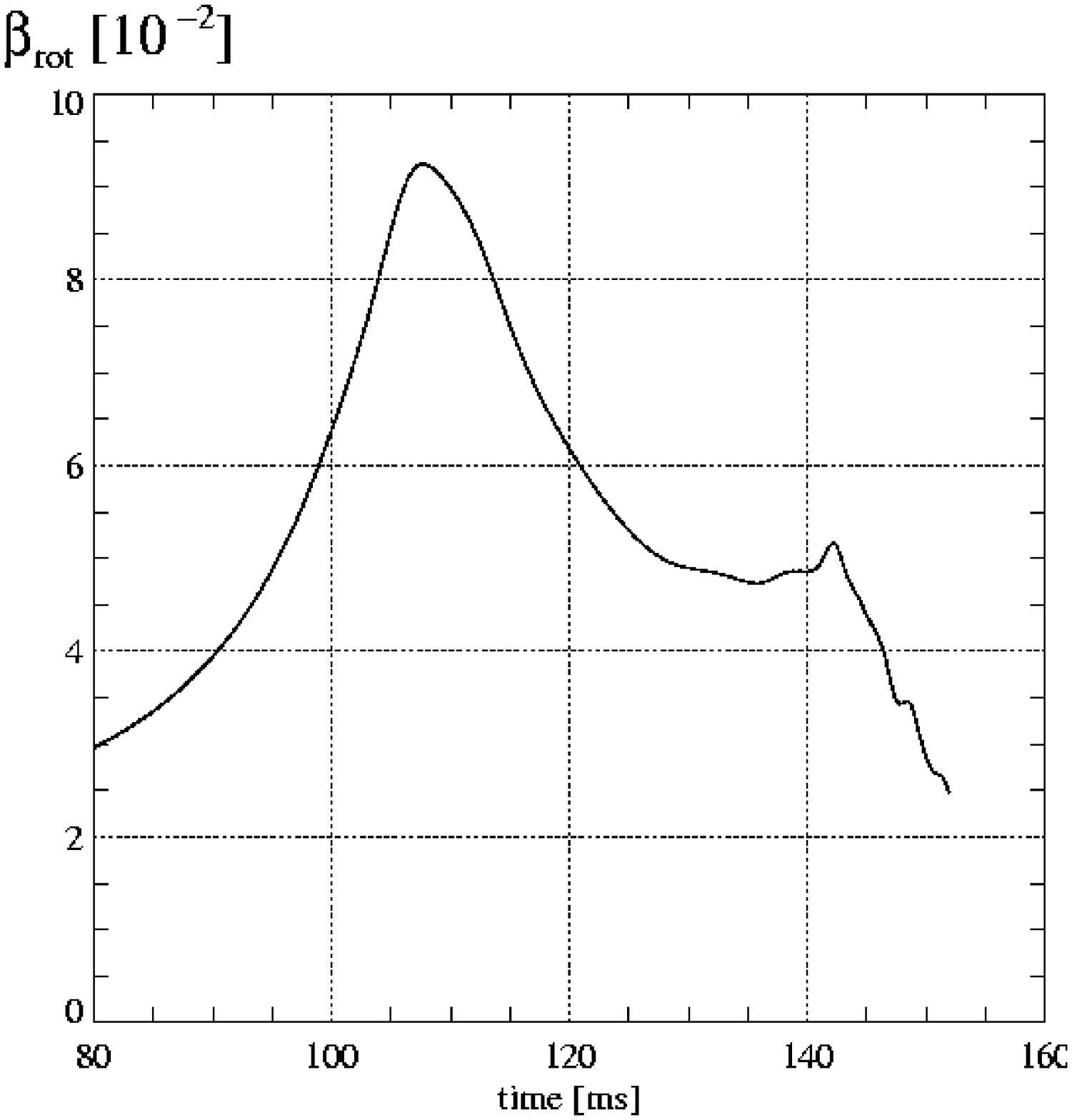}
  \includegraphics[width=5.6cm]{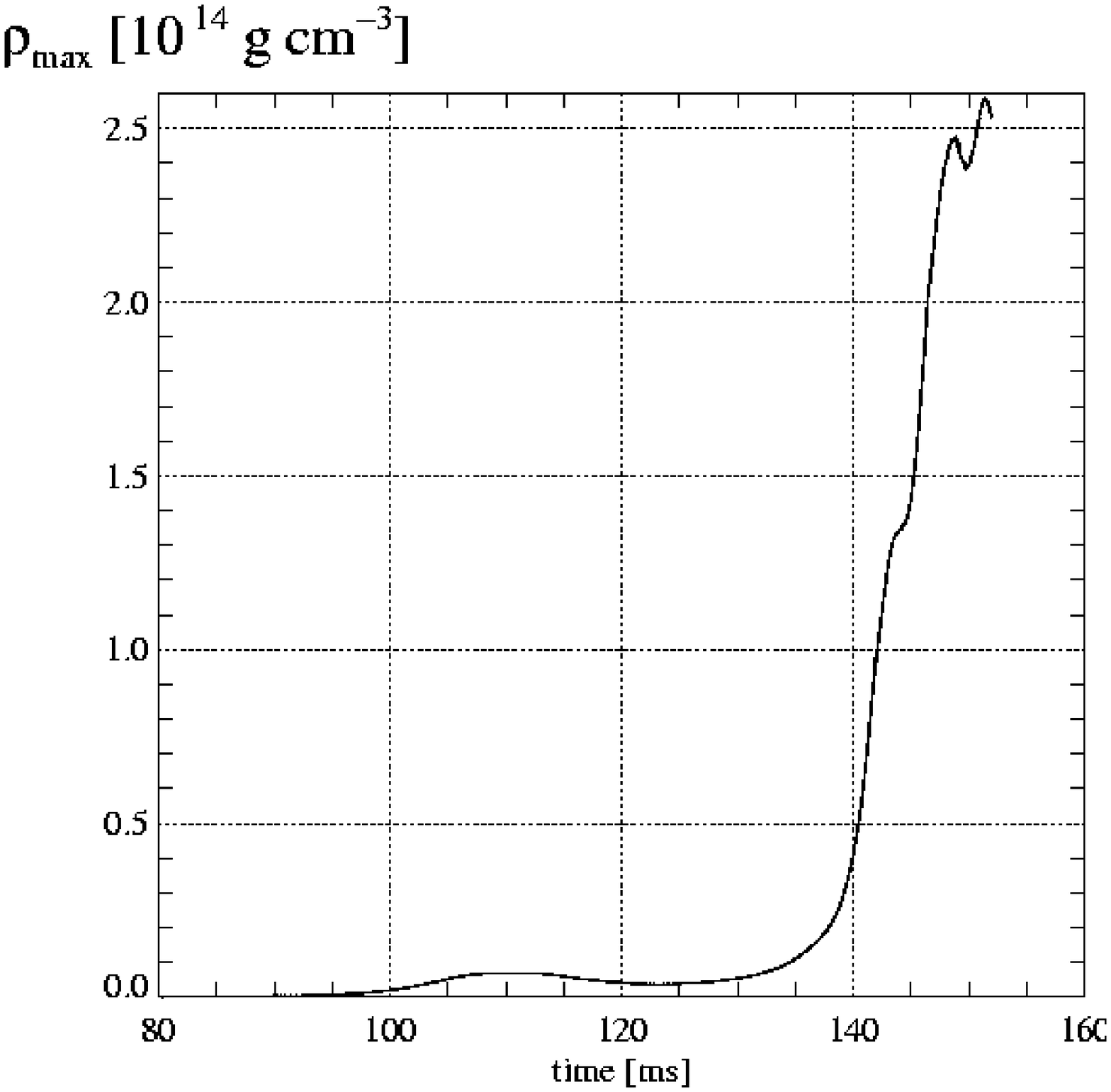}
  \includegraphics[width=5.6cm]{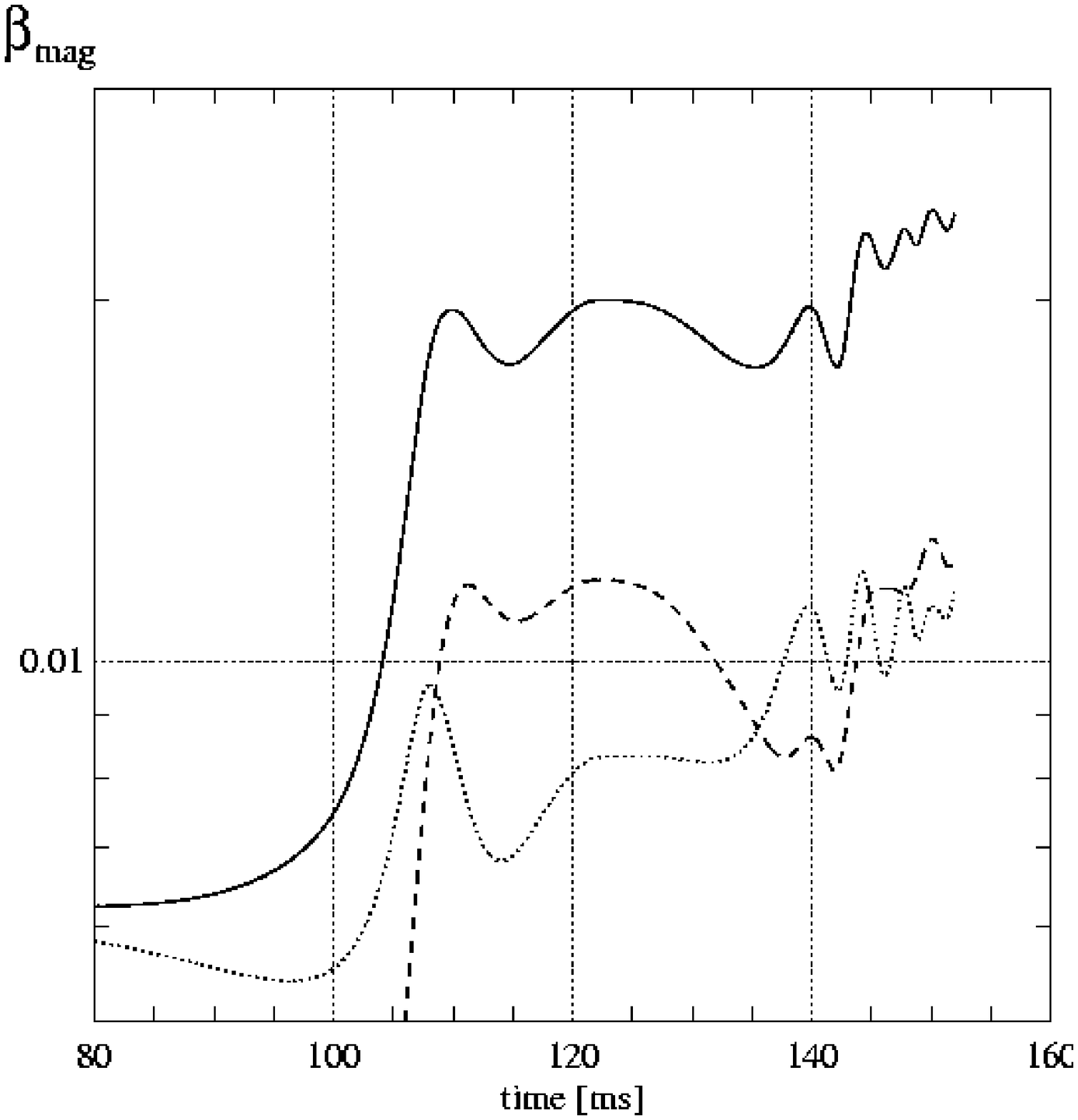}
  \includegraphics[width=5.6cm]{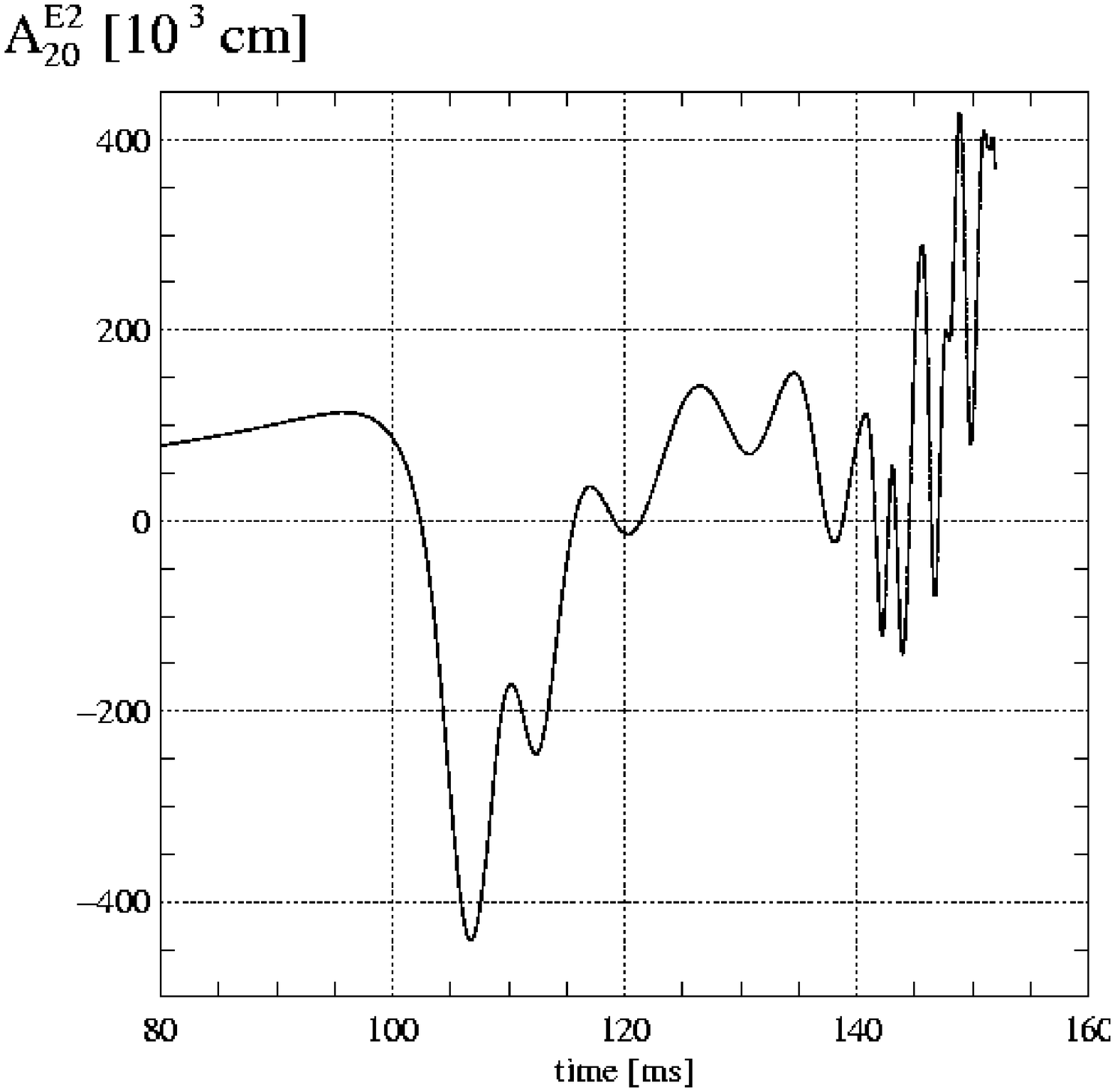}
  \includegraphics[width=5.6cm]{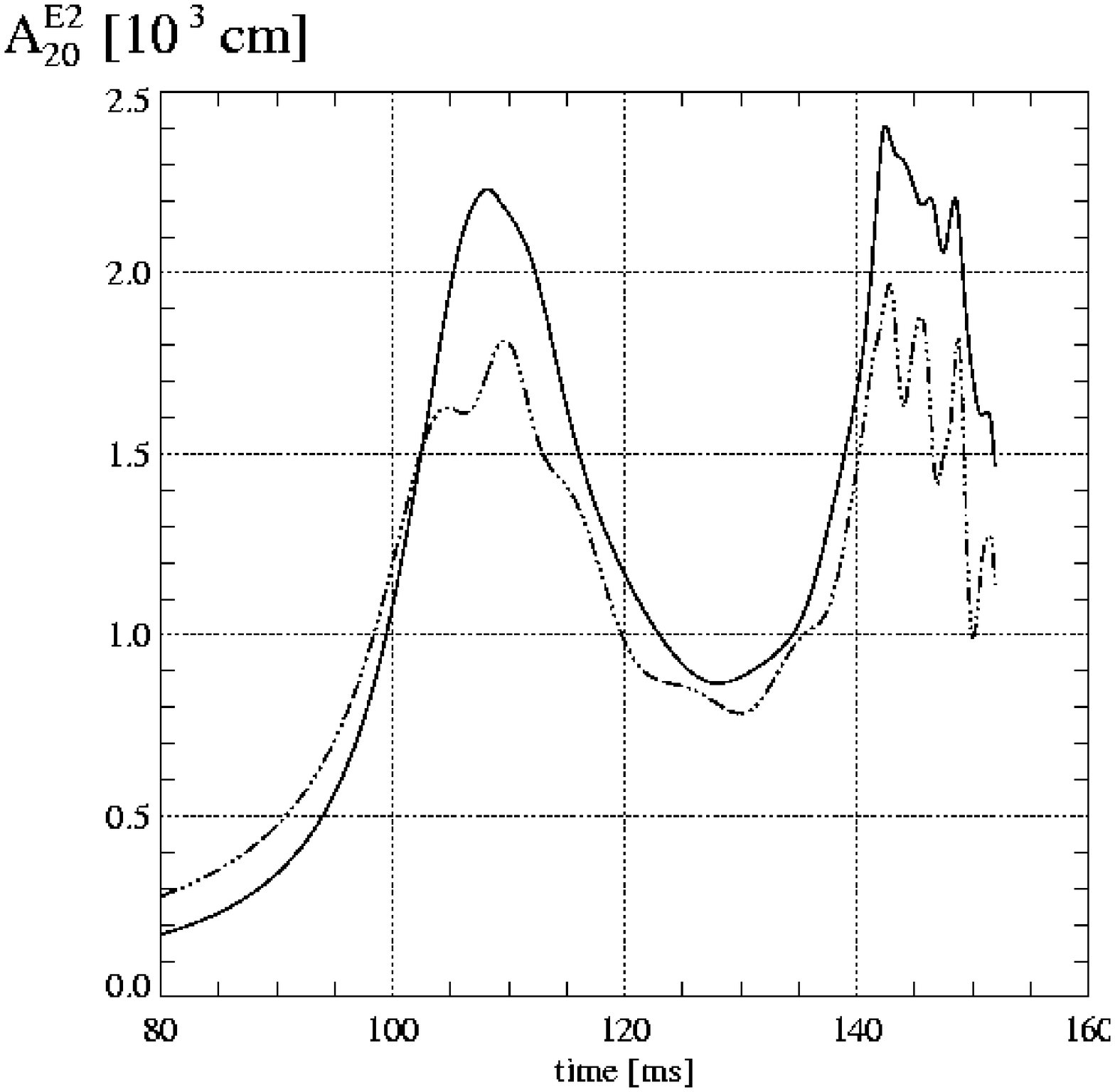}
  \includegraphics[width=5.6cm]{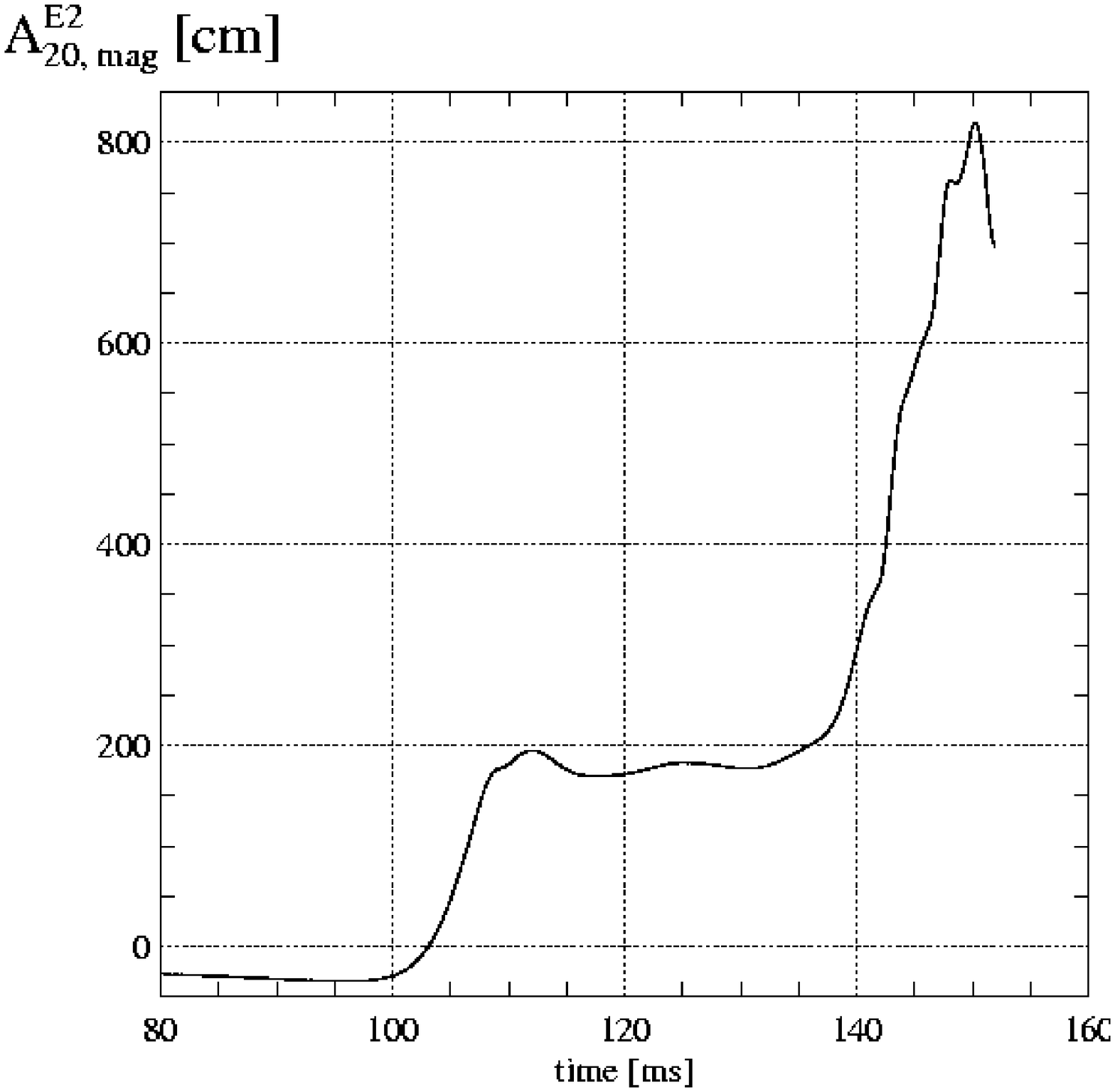}
  \caption[Evolution of model A2B4G1-D3M13]
  { 
    The dynamical evolution and the GW signal of model A2B4G1-D3M13.
    The upper panels display the temporal evolution of the rotational
    energy parameter $\beta_{\mathrm{rot}}$ (left), the maximum
    density $\rho_{\mathrm{max}}$ (middle), and the magnetic energy
    parameters (right; $\beta_{\mathrm{mag}}$ (solid line),
    $\beta_{\phi}$ (dashed line), and $\beta_{\mathrm{mag}} -
    \beta{\phi}$ (dotted line)), respectively.  The GW signal of the
    model is shown in the lower panels: total amplitude (solid line,
    left), $-A^{\mathrm{E}2}_{20;\mathrm{hyd}}$ (solid line,
    middle), $A^{\mathrm{E}2}_{20;\mathrm{grav}}$ (dashed line,
    middle), and $A^{\mathrm{E}2}_{20;\mathrm{mag}}$ (right).  
  }
  \label{Fig:241-313:Dyn}
\end{figure*}

\begin{figure}[!htbp]
  \resizebox{\hsize}{!}{\includegraphics{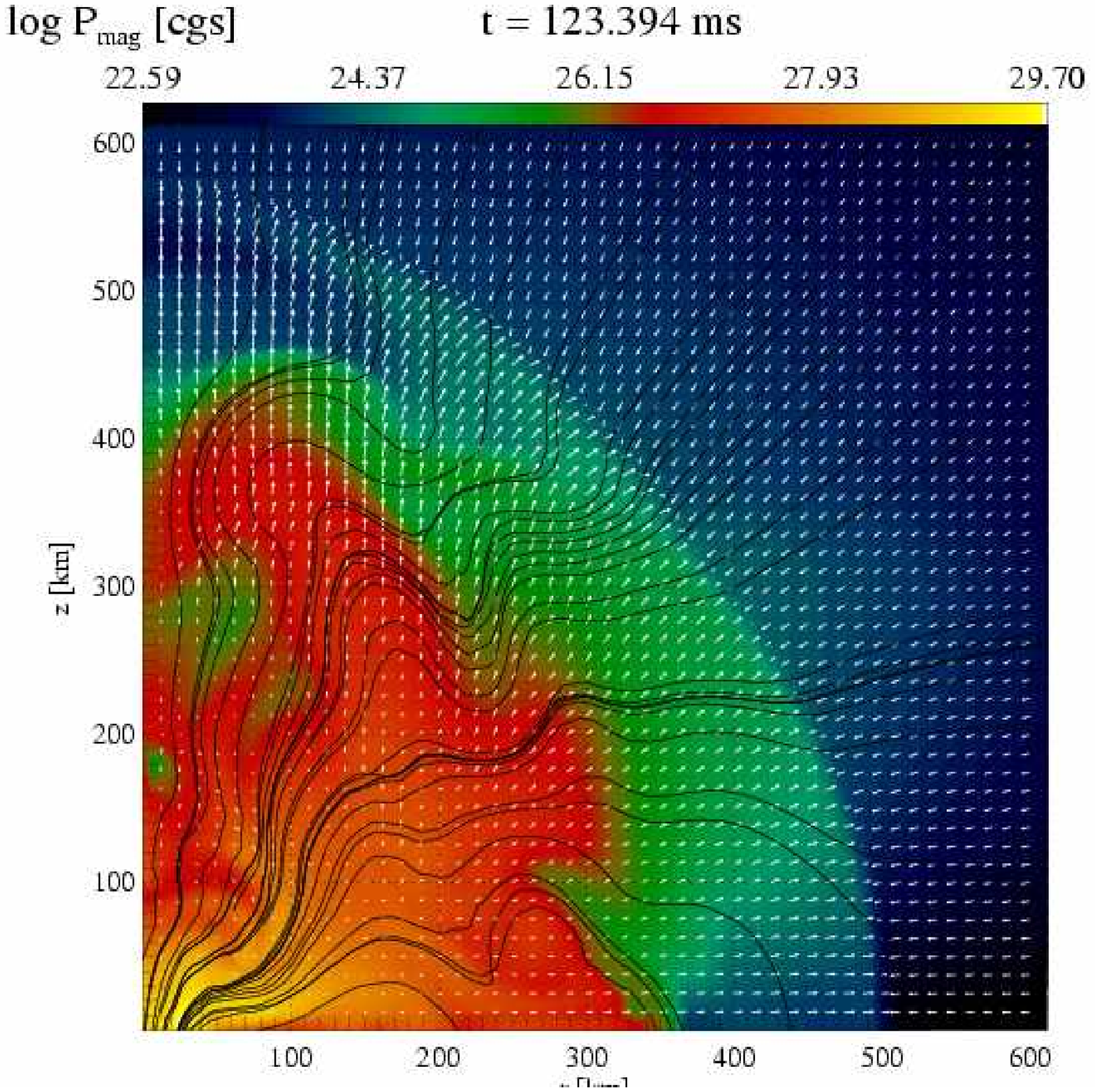}}
  \caption[Model A2B4G1-D3M13: Minimum Density]
  { 
    Model A2B4G1-D3M13 at $t=123.39\ \mathrm{ms}$ when the expansion of the core
    is maximum just prior to its second collapse. Besides the magnetic
    pressure (color shaded), the flow field (vectors), and the
    magnetic field lines are shown. 
  }
  \label{Fig:241-313:Minden}
\end{figure}

Unlike the other models, which collapse to higher densities when angular
momentum transport by the magnetic field becomes dynamically important,
model A2B4G1-D3M13 bounces at a lower density ($\rho_{\mathrm{b}} =
6.84\cdot 10^{12}\ \mathrm{cm \, s}^{-1}$ compared to $\rho_{\mathrm{b}}
= 1.14\cdot 10^{13}\ \mathrm{cm \, s}^{-1}$ in the weak--field case).
At bounce its rotational energy ($\beta_{\mathrm{max}} = 0.093$) is less
than that of model A2B4G1-D3M11 ($\beta_{\mathrm{max}} = 0.118$), which
has an initially 100 times weaker magnetic field.  After bounce the
rotational energy continues to decrease, and the core eventually suffers
a second collapse that is stopped at $\rho_{\mathrm{max}} = 2.5 \cdot
10^{14} \ \mathrm{cm \, s}^{-1}$ (Fig.\,\ref{Fig:241-313:Dyn}).  During
the second collapse $\beta_{\mathrm{rot}}$ rises temporarily slightly
again, but angular momentum redistribution prevents the shock formed by
centrifugal forces (at a similar density as the one resulting from the
first bounce) from stopping the infall.  Thus, a large fraction of the
inner core continues to fall towards the center, until the collapse is
halted by the stiffening of the equation of state, i.e.\,by a bounce due
to pressure forces.  A shock forms at the edge of the still very
extended inner core ($r_{\mathrm{shock}} = 70\ \mathrm{km}$ compared to
$r_{\mathrm{shock}} \approx 10\ \mathrm{km}$ in the previously discussed
models) at $t = 108\ \mathrm{ms}$ at the beginning of the ``plateau'' in
the evolution of $\rho_{\mathrm{max}}$ (see upper middle panel of
Fig.\,\ref{Fig:241-313:Dyn}), and it becomes slightly prolate.  About
15\,ms later ($t=123\ \mathrm{ms}$) this shock has reached a radius of
$r_{\mathrm{shock}} \approx 200\ \mathrm{km}$
(Fig.\,\ref{Fig:241-313:Minden}). The magnetic field of model
A2B4G1-D3M13 is less amplified and deformed than in the deeper
collapsing single--bounce models such as e.g.\, model A1B3G3-D3M13. The
largest magnetic to gas pressure ratios are reached well behind the
shock at about $0.75\, r_{\mathrm{shock}} \approx 400\ \mathrm{km}$.

The GW signal does not resemble any of the types used to classify
non--magnetic models. Hence, we suggest to introduce a new type-IV GW
signal (Fig.\,\ref{Fig:241-313:Dyn}) which it is weaker at bounce
($A^{\mathrm{E}2}_{20} = -440 \ \mathrm{cm}$) than in the non--magnetic
case ($A^{\mathrm{E}2}_{20} = -610 \ \mathrm{cm}$), but the amplitude
remains large for a longer time period due to the longer phase of
maximum contraction.  Until the second collapse the signal varies on
time scales ($\Delta t \approx 5 \ \mathrm{ms}$) that are short compared
to the pulsation periods of the non--magnetic case, but comparable to
the sound crossing time of the relatively extended inner core.  After
$t\approx 140\ \mathrm{ms}$ the oscillation frequency and the amplitude
of the GW signal strongly increases as the second collapse begins.
During the entire post--bounce evolution the large negative
gravitational and hydrodynamic (dominated by the centrifugal amplitude)
contributions to the GW signal, and the large positive magnetic
contribution add up to a total GW amplitude which is very small compared
to the individual contributions.

\begin{figure*}[!htbp]
  \centering
  \includegraphics[width=5.6cm]{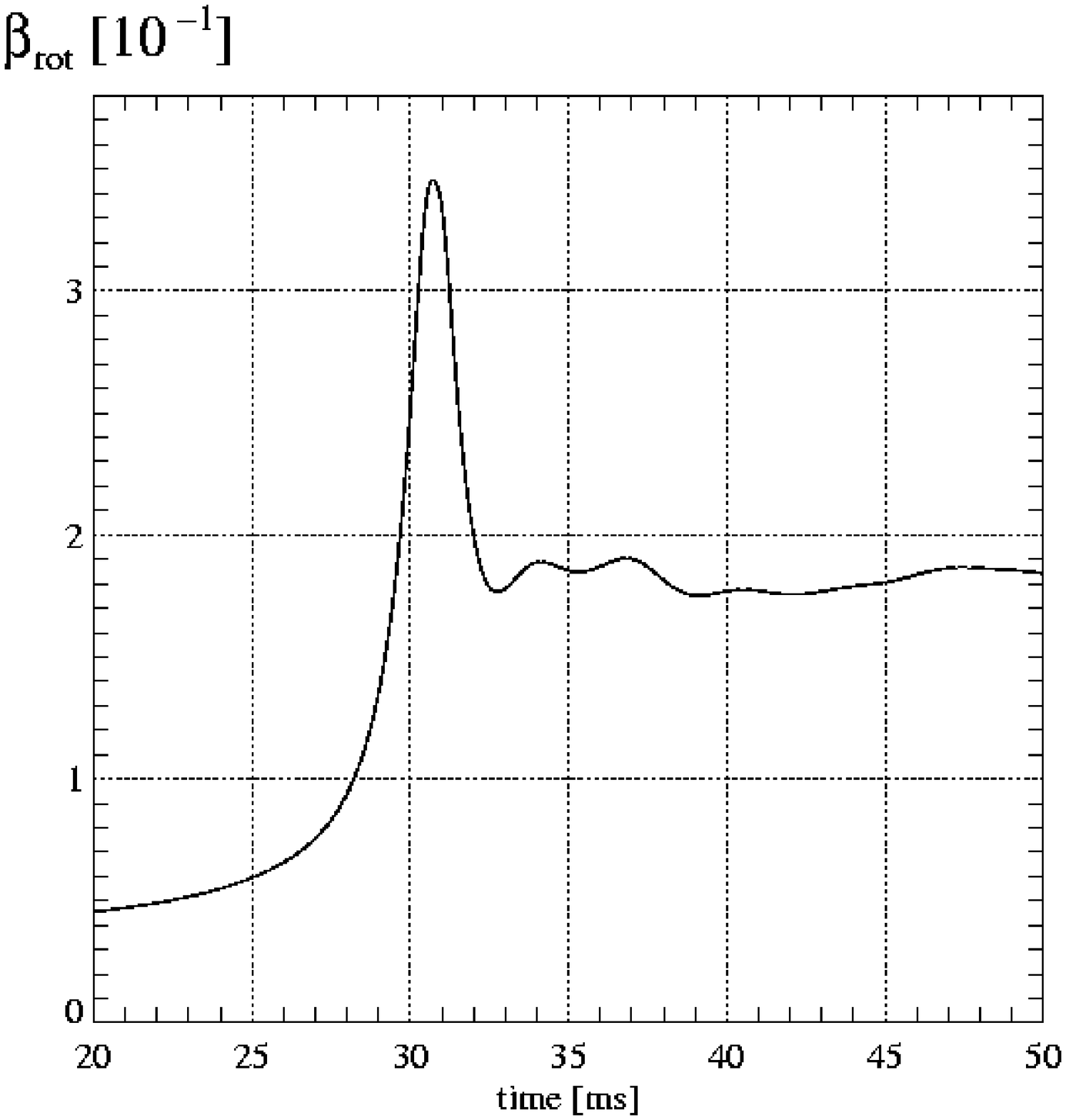}
  \includegraphics[width=5.6cm]{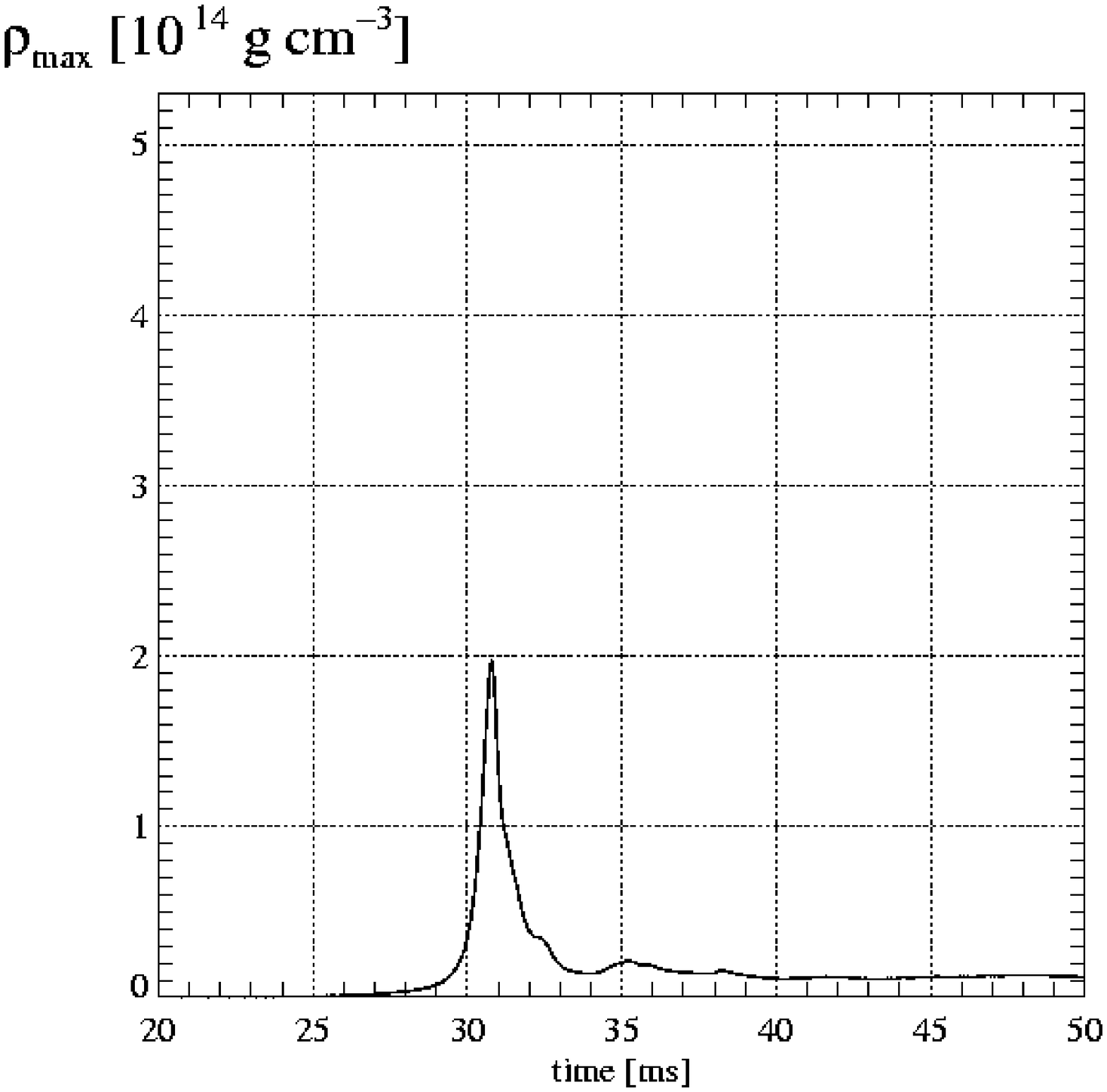}
  \includegraphics[width=5.6cm]{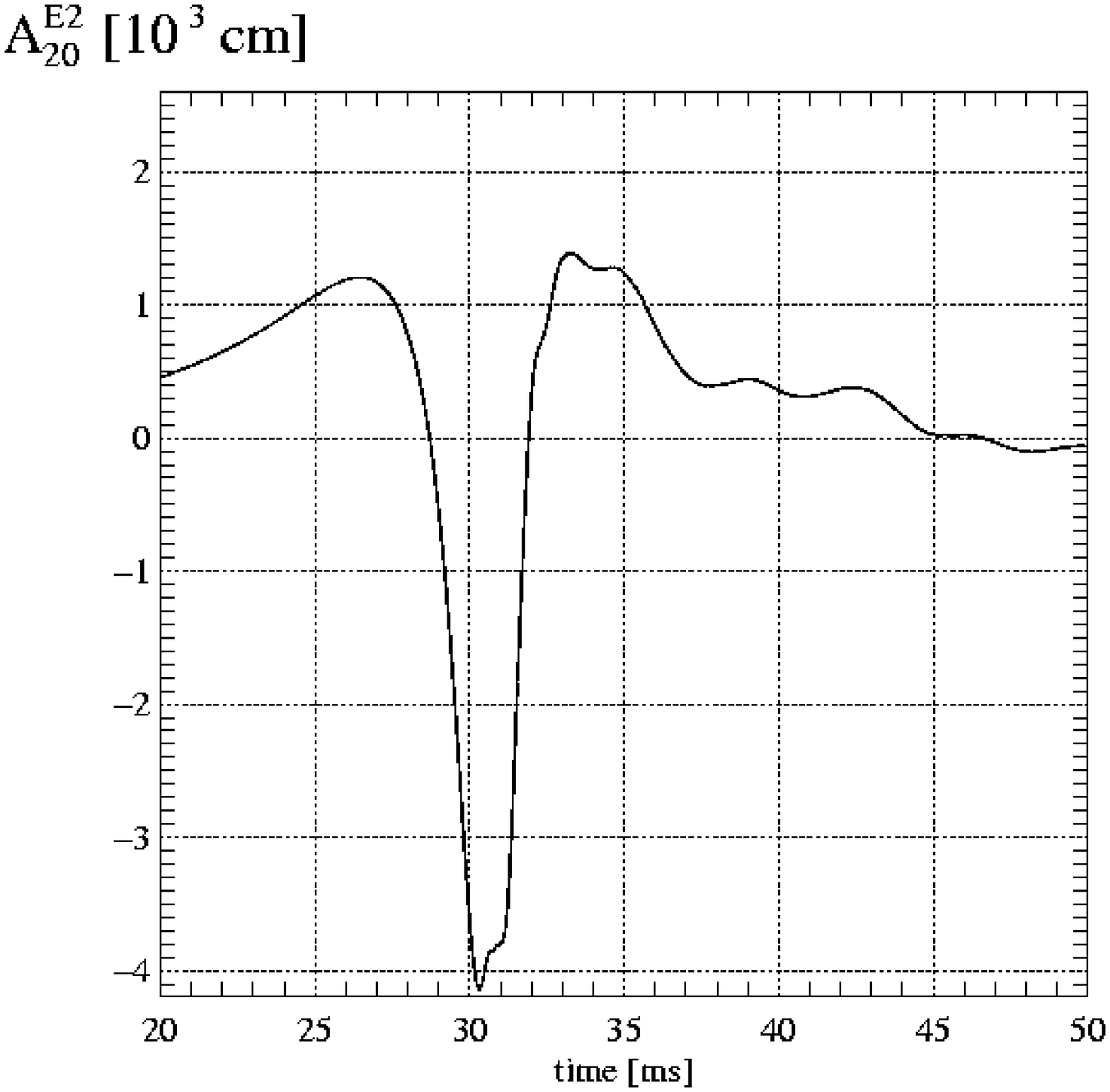}
  \includegraphics[width=5.6cm]{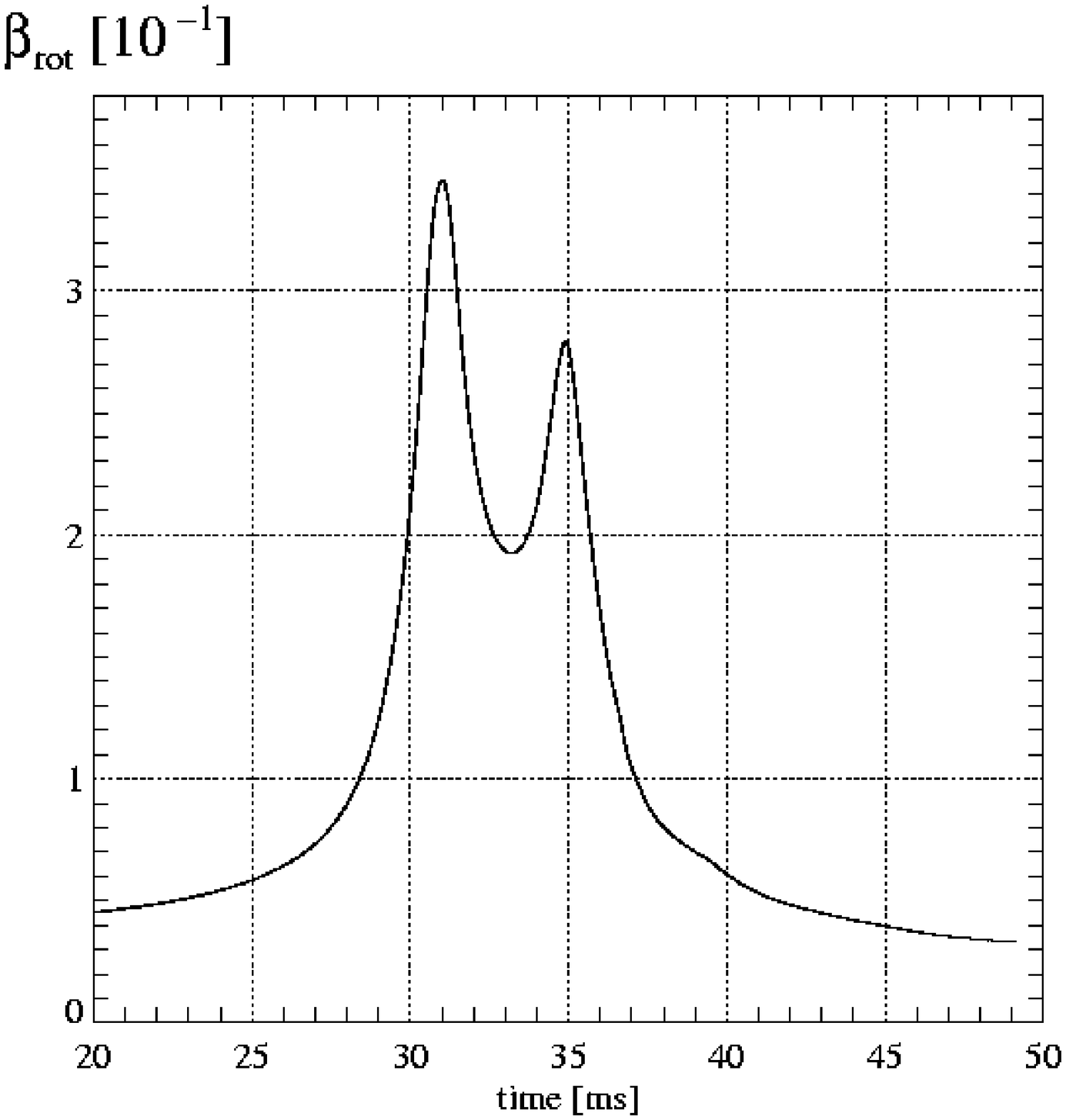}
  \includegraphics[width=5.6cm]{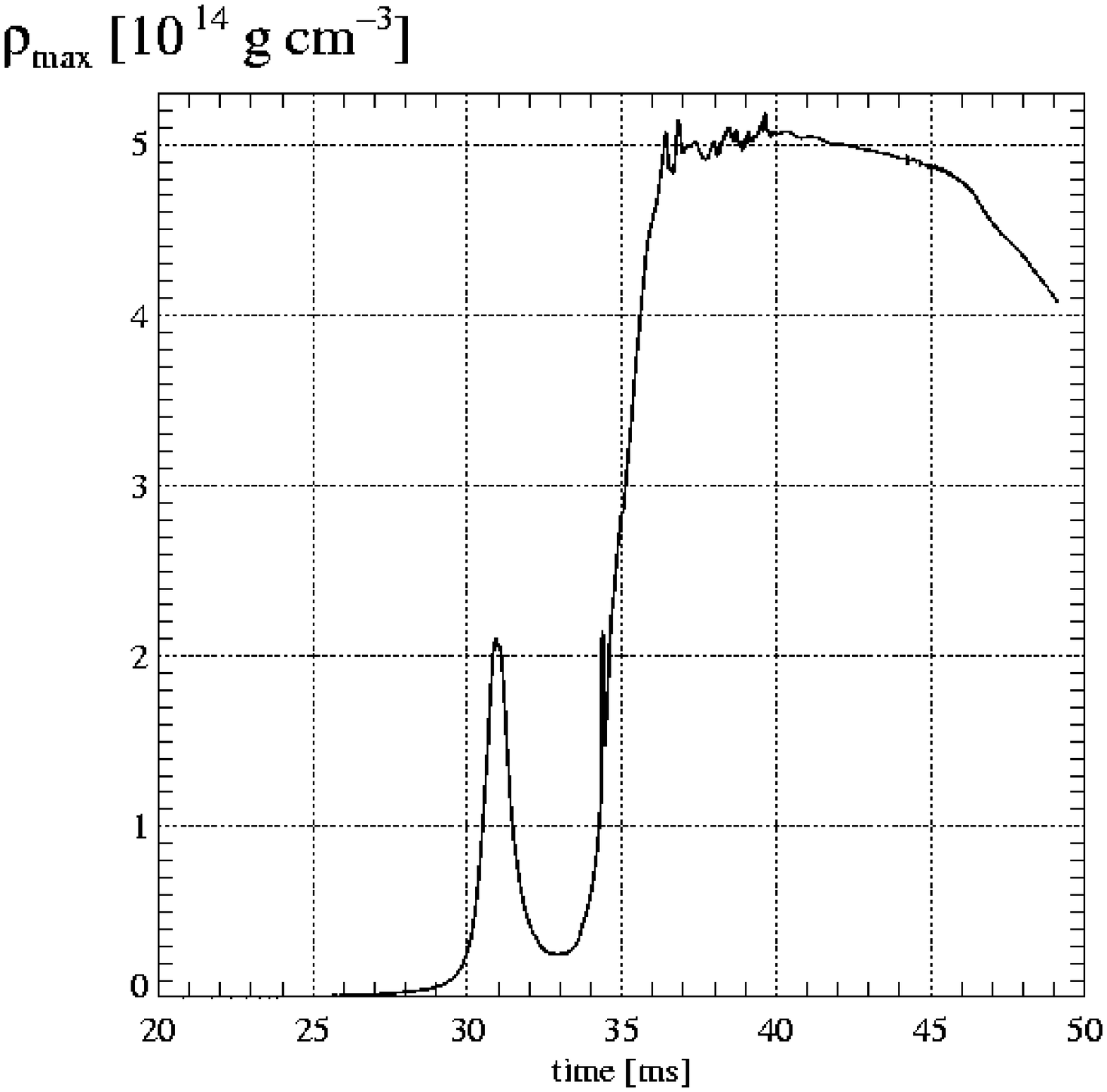}
  \includegraphics[width=5.6cm]{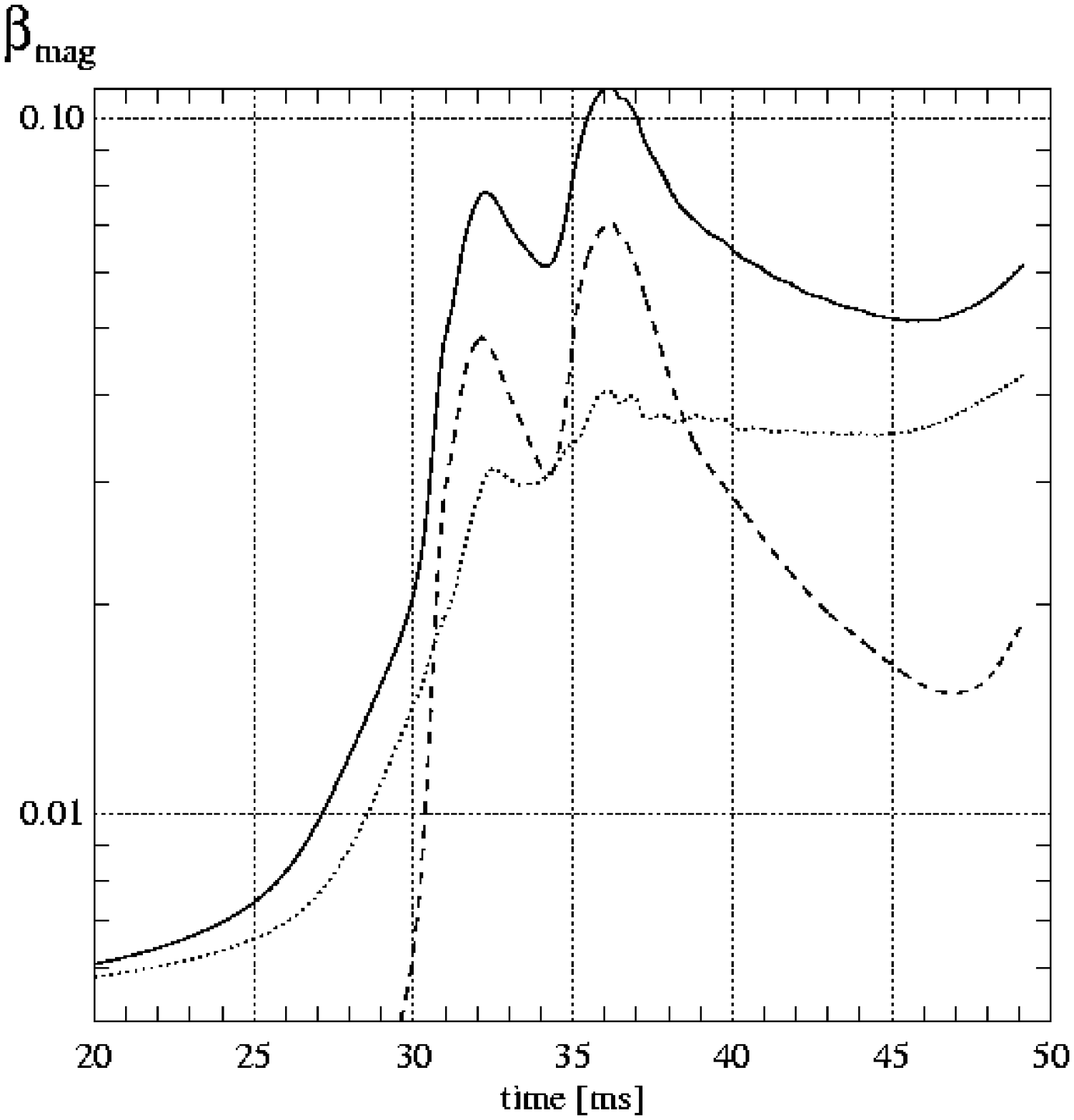}
  \includegraphics[width=5.6cm]{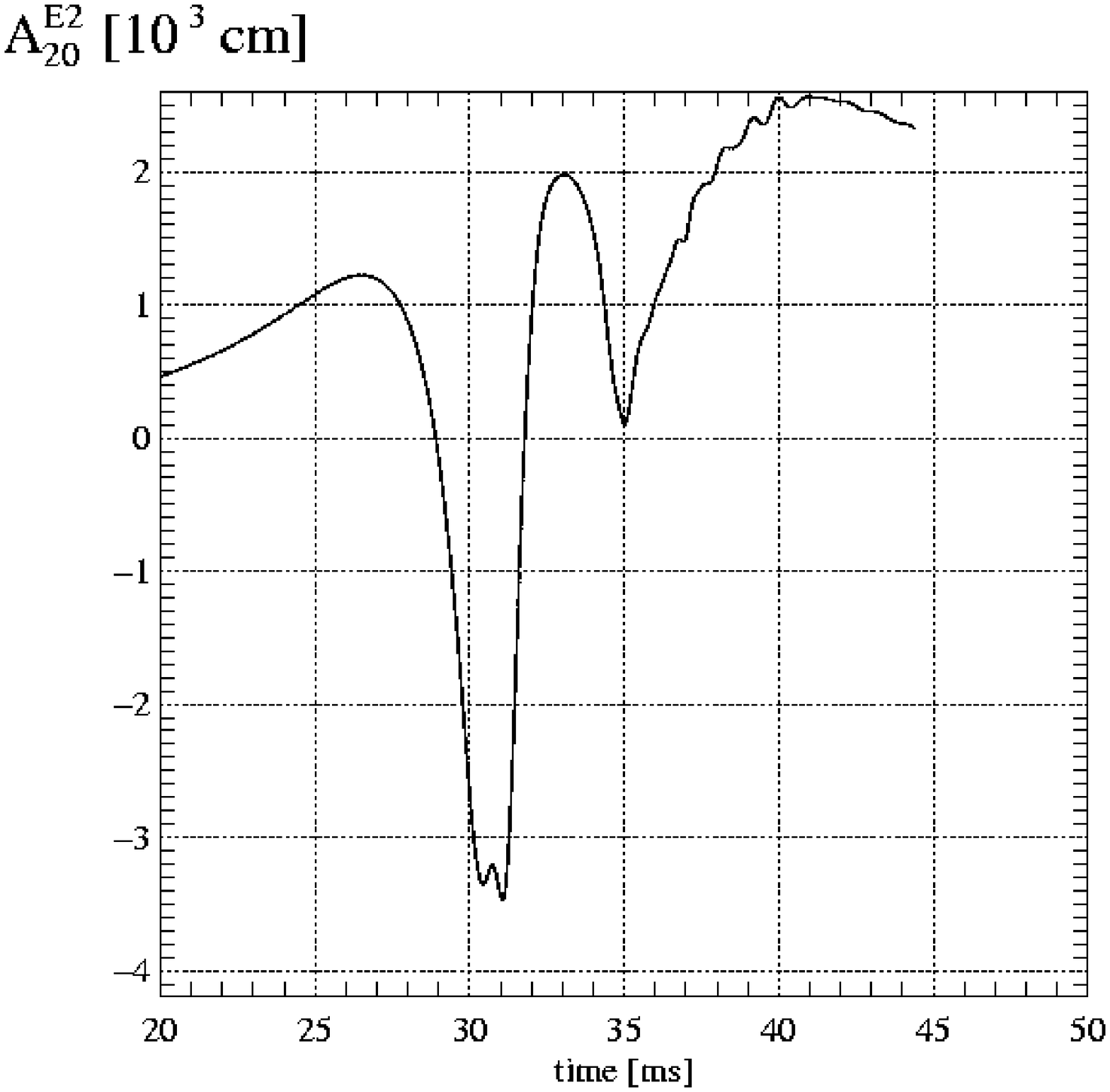}
  \includegraphics[width=5.6cm]{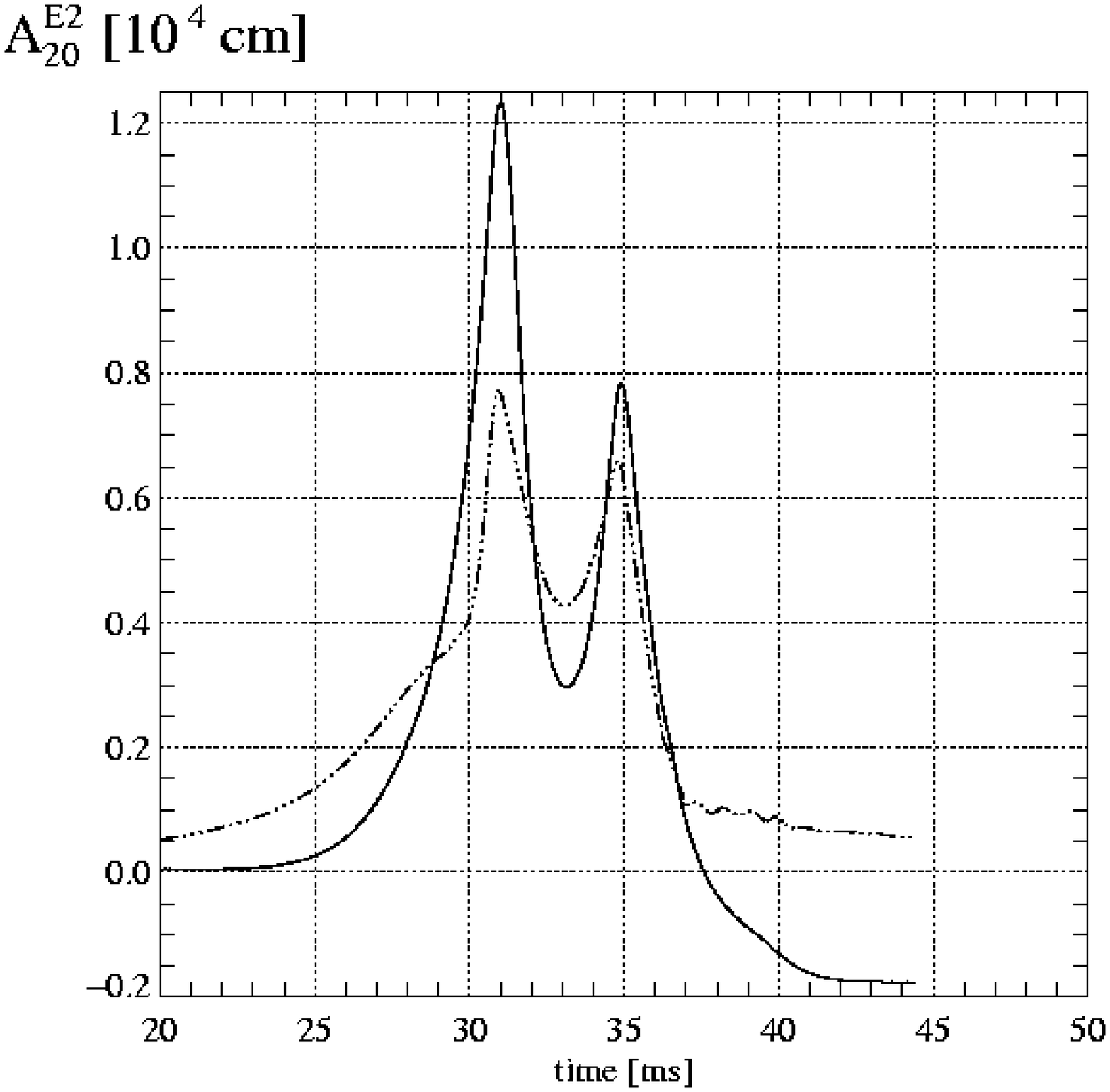}
  \includegraphics[width=5.6cm]{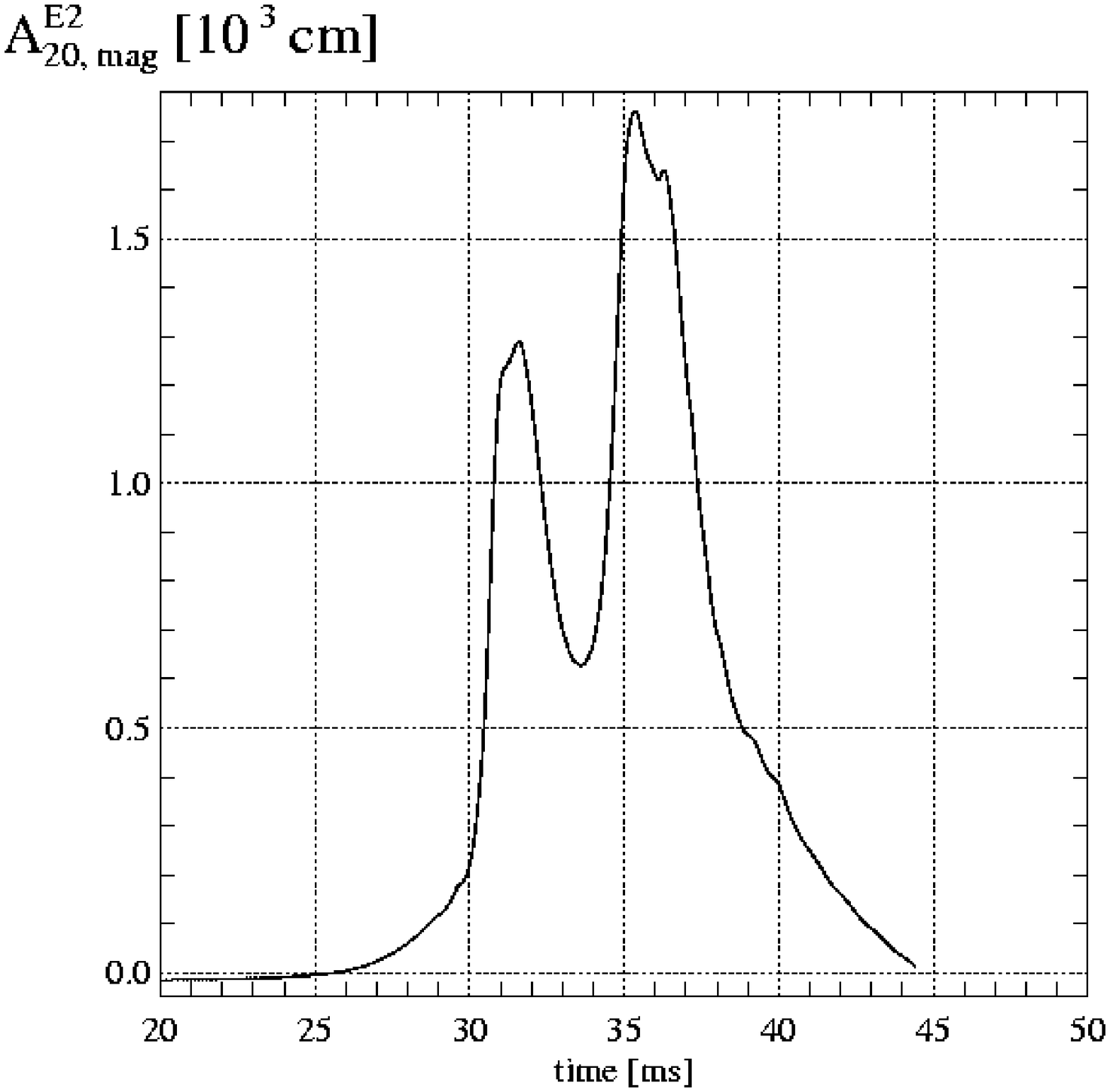}
  \caption[Dynamical Evolution of model A4B5G5-D3M13]
  { Evolution of the rotational energy parameter
    $\beta_{\mathrm{rot}}$ (left) and the maximum density (middle) of
    models A4B5G5-D3M10 (upper row) and A4B5G5-D3M13 (middle row),
    respectively. The total GW amplitude of model A4B5G5-D3M10 is
    shown in the upper right panel, and the evolution of the magnetic
    energy parameters, $\beta_{\mathrm{mag}}$ (solid), $\beta_{\phi}$
    (dashed), and $\beta_{\mathrm{mag}}-\beta_{\phi}$ in the middle
    right panel, respectively.
    The panels in the bottom row show different contributions to the
    GW amplitude of model A4B5G5-D3M13: total amplitude ( left),
    $-A^{\mathrm{E}2}_{20;\mathrm{hyd}}$ (solid line, middle),
    $A^{\mathrm{E}2}_{20;\mathrm{grav}}$ (dashed line, middle), and
    $A^{\mathrm{E}2}_{20;\mathrm{mag}}$ (right).  }
  \label{Fig:455:Dyn}
\end{figure*}

\begin{figure*}[!htbp]
  \centering
  \includegraphics[width=16cm]{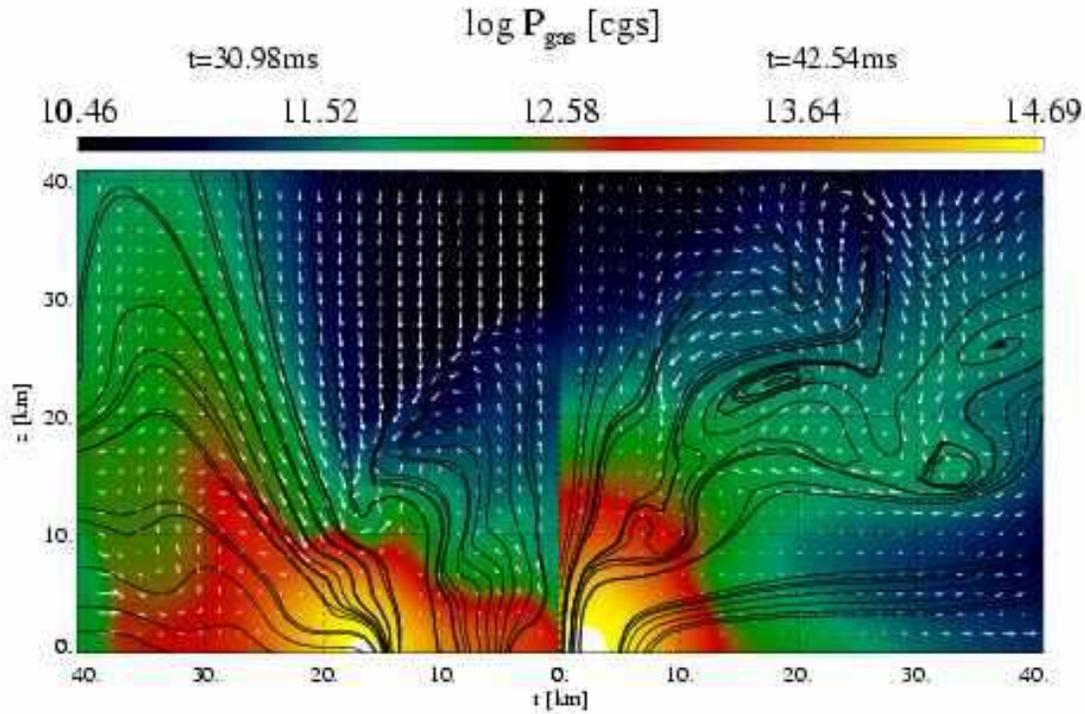}
  \caption[Model A4B5G5-D3M13: Density of the Central Core]
  { Comparison of two snapshots of model A4B5G5-D3M13.  The innermost
    region of the core is displayed at $t=30.98\ \mathrm{ms}$ (at core bounce;
    left), and at $t=42.54\ \mathrm{ms}$ (right).  The logarithm of the density
    (color--coded), the velocity (arrow), and the magnetic field lines
    are displayed, respectively.  }
  \label{Fig:455-313:Rho}
\end{figure*}

Model A4B5G5-D3M13 bounces at a slightly higher maximum density
($\rho_{\mathrm{b}} = 2.1\cdot 10^{14}\ \mathrm{cm \, s}^{-1}$) than its
weak--field variant A4B5G5-D3M10 ($\rho_{\mathrm{b}} = 1.97\cdot
10^{14}\ \mathrm{cm \, s}^{-1}$; Fig.\,\ref{Fig:455:Dyn}).  The maximum
rotational energy at bounce is barely affected by the presence of the
strong magnetic field, but it changes the density structure of the core
completely in the subsequent evolution.  Models A4B5G5-D3Mm ($m \leq
12$) maintain a toroidal density structure ($r_{\mathrm{tor}} \approx
40\ \mathrm{km}$) during their entire evolution
(Sect.\,\ref{Suk:Dis:SchwachFeld}), but in model A4B5G5-D3M13 the
density structure is changed by magnetic forces. Up to a few
milliseconds after bounce ($t\la 32\ \mathrm{ms}$), the core of
model A4B5G5-D3M13 evolves only marginally differently than its
weak--field variants.  Later, however, magnetic stresses transfer
angular momentum from the still toroidal density distribution into the
surrounding gas, whereby the torus transforms into a flattened
configuration, which has its density maximum only slightly off--center
(Fig.\,\ref{Fig:455-313:Rho}).  During this phase, the core develops a
retrograde rotating region near the equatorial plane.  The core's
transformation process is associated with a large increase of its
maximum density comparable with that occurring during the first collapse
(Fig.\,\ref{Fig:455:Dyn}).

During the second collapse, both the magnetic and the rotational energy
of the core increase.  The magnetic energy ($\beta_{\mathrm{mag}} =
0.11$) exceeds that of the primary collapse ($\beta_{\mathrm{mag}} =
0.078$), and the rotational energy exceeds slightly the limit for the
dynamic instability of MacLaurin ellipsoids ($\beta_{\mathrm{rot}}
\approx 0.275$), but only for about 1\,ms.  We note that the rotational
energy reaches its maximum value well before the density does, which is
different from all other models. Later on, the rotation parameter
rapidly decreases well below the corresponding value of the
non--magnetic case.

\begin{figure*}[!htbp]
  \centering
  \includegraphics[width=16cm]{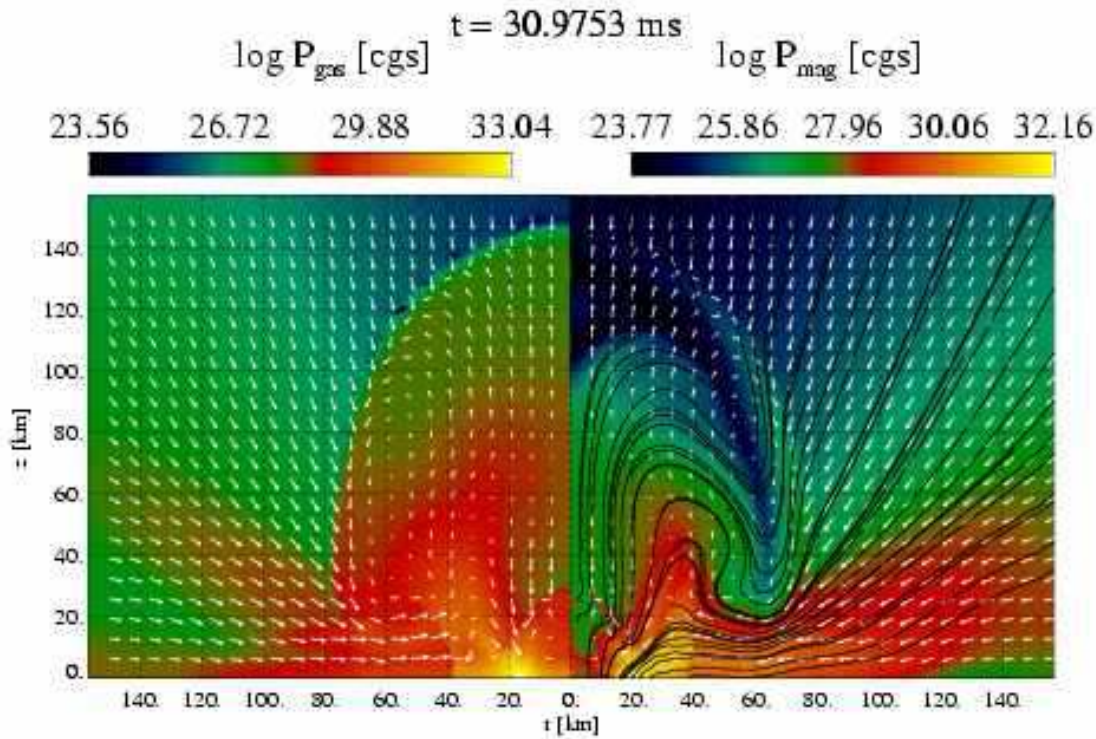}
  \caption[Model A4B5G5-D3M13: Bounce] 
  { Model A4B5G5-D3M13 at $t=30.98\ \mathrm{ms}$ (core bounce).  The flow field
    (vectors) and the magnetic field lines are displayed together with
    the gas pressure (color--coded, left) and the magnetic pressure
    (color--coded, right), respectively. }
\label{Fig:455-313:Bounce}
\end{figure*}

The shock has a very prolate shape at core bounce, its axis ratio being
$\ga 2:1$.  At large latitudes, the ratio of magnetic and gas
pressure is largest well behind the shock ($r < 0.75 \ 
r_{\mathrm{shock}} \approx 110\ \mathrm{km}$;
Fig.\,\ref{Fig:455-313:Bounce}).  Due to the rather extreme rotation of
the model, the high latitudes of the core are strongly rarefied, and the
shock propagates into a very thin medium near the axis, its structure
remaining unchanged.

The GW amplitude of the weakly magnetized ($\approx$ non--magnetic)
models A4B5G5-D3Mm ($m \leq 12$) is characterized by a very large
negative amplitude at bounce ($A^{\mathrm{E}2}_{20} \approx -4100\ 
\mathrm{cm}$; Fig.\,\ref{Fig:455:Dyn} and Table\,\ref{Tab:SynopseI}).
The signal is positive for several milliseconds after bounce due to the
aspherical shock wave, and approaches zero rapidly after the violent
re--expansion of the core.  The bounce amplitude is significantly
lowered ($A^{\mathrm{E}2}_{20} \approx -3470\ \mathrm{cm}$) in the
strongly magnetized model A4B5G5-D3M13 (Fig.\,\ref{Fig:455:Dyn} and
Table\,\ref{Tab:SynopseI}), immediately after bounce the GW amplitude is
strongly positive and eventually becomes even more positive due to the
growing asphericity of the shock wave.  The secondary collapse of this
models shows up in the GW amplitude in the form of a local minimum.
Afterwards the amplitude rises to values $A^{\mathrm{E}2}_{20} > 2000\ 
\mathrm{cm}$ and shows superimposed very weak oscillations with a period
of slightly less than 1\,ms.  In this model, the magnetic and the
combined hydrodynamic plus gravitational amplitude contributions show a
clear phase shift resulting from the opposing actions of hydrodynamic
(mainly centrifugal) and magnetic forces.

In all our models bouncing due to centrifugal forces the peak value of
the GW signal at bounce decreases with increasing magnetic field
strength, and for most models the ratio of the amplitudes of the
post--bounce to the pre--bounce signal peaks decreases.  Note that the
latter statement must be considered carefully since -- as discussed
above -- the GW signal may be subject to large modifications for models
with very strong fields.

\subsection{Magneto--rotational instability}
\label{Suk:MRI}

The magneto--rotational instability (MRI) is a shear instability
occurring in differentially rotating magnetized plasma, which generates
turbulence that amplifies the magnetic field and transfers angular
momentum \citep{Balbus_Hawley_ApJ_1991__MRI,
  Balbus_Hawley_ApJ_1992__MRI_Oort, Balbus_Hawley_RMP_1998__MRI}.  If
effects due to buoyancy are neglected, a linear stability analysis shows
that the condition for the instability of a mode with wavenumber $\vec
k$ is
\begin{equation}
  \label{Gl:MRI-Krit}
  \frac{\mathrm{d} \Omega^2}{\mathrm{d} \ln \varpi} +  
  \left( \vec c_{\mathrm{A}} \vec k \right)^2  < 0 \,,
\end{equation}
where $\Omega$ is the angular velocity, $\varpi$ the distance from the
rotation axis, and $\vec c_{\mathrm{A}} \equiv {\vec b}/{\sqrt{\rho}}$
the Alfv\'en velocity. When the magnetic field is very small (i.e.\,the
Alfv\'en velocity is very small compared to both the local sound speed
as well as the local rotation velocity) and/or the wavelength is very
long, $\left( \vec c_{\mathrm{A}} \vec k \right)$ is negligible,
i.e.\,the MRI occurs simply when the angular velocity gradient is
negative \citep{Balbus_Hawley_ApJ_1991__MRI, Balbus_Hawley_RMP_1998__MRI}:
\begin{equation}
  \label{Gl:MRI-Krits}
  \frac{\mathrm{d} \Omega^2}{\mathrm{d} \ln \varpi} < 0 \,.
\end{equation}
The time scale of the fastest growing unstable mode is
\begin{equation}
  \label{Gl:Scher-Wachstumsrate}
  \tau_{\mathrm{max}} = 4\pi \left| \frac{\mathrm{d} \Omega}{\mathrm{d} \ln \varpi} 
                             \right|^{-1} \,,
\end{equation}
which depends neither on the strength nor on the geometry of the
magnetic field.

General theoretical considerations and non--linear simulations show that
the magnetic energy achievable by the MRI field amplification process is
of the order of the rotational energy, which is comparable to the
saturation field expected from the process of winding--up field lines by
differential rotation \citep{Akiyama_etal_ApJ_2003__MRI_SN}. However, in
axisymmetric (non--linear) hydrodynamic simulations of accretion disks
the field built up by the MRI was found to decay with a rate depending
on the dissipation properties of the numerical scheme, and particularly
on the grid resolution \citep{Balbus_Hawley_ApJ_1991__MRI,
  Balbus_Hawley_RMP_1998__MRI}.  This is due to Cowling's anti--dynamo
theorem \citep{Shercliff_Book_1965__MHD} according to which an
axisymmetric dynamo cannot work in a dissipative system,
i.e.\,three--dimensional simulations are required.

In a star a convectively stable stratification will tend to stabilize
the MRI, while convective instability will strengthen the MRI.
Including the effects of buoyancy a mode with wave vector $\vec k$ is
unstable with respect to the MRI in a weakly magnetized rotating system,
if the instability criterion
\begin{equation}
  \label{Gl:BV_MRI-Wellenzahl-Krit}
  \left( \vec c_{\mathrm{A}} \vec k \right)^2 
  < 
  \left( \vec c_{\mathrm{A}} \vec k \right)^2_{\mathrm{crit}} 
  \equiv
  - N^2 - \frac{\mathrm{d} \Omega^2}{\mathrm{d} \ln \varpi}
\end{equation}
is fulfilled \citep{Balbus_Hawley_ApJ_1991__MRI,
  Balbus_ApJ_1995__stratified_MRI}, where $N$ is the
Brunt--V\"ais\"al\"a or buoyancy frequency.  The wave number of the
fastest growing mode is given by
\begin{equation}
  \label{Gl:BV_MRI-MaxWachs}
  \left( \vec c_{\mathrm{A}} \vec k \right)_{\mathrm{max}}^2 = \Omega^2 
  \left[
    1 - \frac{(-N^2 - \kappa^2)^2}{16 \Omega^4}
  \right] \,,
\end{equation}
where $\kappa^2 = 4\Omega^2 + \mathrm{d} \Omega^2 / \mathrm{d} \ln
\varpi$ is the epicyclic frequency. This mode grows exponentially with
the time scale
\begin{equation}
  \label{Gl:BV_MRI-Wachsrate}
  \tau_{\mathrm{max}} = 4\pi \left| \frac{N^2}{2\Omega} + 
                                    \frac{\mathrm{d} \Omega}{\mathrm{d} \ln \varpi} 
                             \right|^{-1} \, ,
\end{equation}
which is a generalization of Eq.\,\ref{Gl:Scher-Wachstumsrate}.

For decreasing wavenumber $k$ (provided that $k c_{\mathrm{A}} < (k
c_{\mathrm{A}})_{\mathrm{max}}$) the maximum growth rate ($= 2\pi /
\tau_{\mathrm{max}}$) decreases approaching the value
\begin{equation}
  \label{Gl:MRI-Anwachsrate_langWelle}
  \omega_{\mathrm{long}}^2 = -N^2 - \kappa^2
\end{equation}
in the limit of very long wavelengths ($k \to 0$) independently of the
Alfv\'en velocity.  If the MRI occurs in a buoyantly stable
stratification (where $N^2 > 0$), long modes will grow very slowly
compared to the maximum growth rate. In the opposite case modes with a
very small wave number grow very fast, and the growth rate of long modes
will be of the order of the maximum growth rate.  In the following
discussion we use the names ``magneto--shear'' and
``magneto--convective'' limit to discriminate the two limiting cases
without implying any deeper physical significance, because it is a
matter of ongoing debate whether there exists a fundamental physical
distinction between the two cases in the context of angular momentum
transport in accretion disks (see,
e.g.\,\citep{Balbus_Hawley_ApJ_2002__MRI_CDAF,
  Narayan_etal_ApJ_2002__MHD_CDAF}).

For a rotating magnetized configuration with axial symmetry the growth
time scale of the maximum growing MRI mode in the presence of entropy
gradients is given in spherical coordinates $(r, \theta)$ by
\citet{Akiyama_etal_ApJ_2003__MRI_SN}
\begin{eqnarray}
  \label{Gl:MRI-Zeit-BV}
  \tau_{\mathrm{MRI,en}} = &2& \pi\,
  \left| (\eta^2 - 2\eta+1) \Omega^2 + \frac{\eta-1}{2}
          \left( \xi N^2 + \eta \frac{\mathrm{d} \Omega^2}{\mathrm{d} \ln r}
          \right)
  \right.
  \nonumber \\
   &+&  
  \left. \frac{1}{16 \Omega^2}
    \left( \xi N^2 + \eta \frac{\mathrm{d} \Omega^2}{\mathrm{d} \ln r}
    \right)^2
  \right|^{-1/2} \,,
\end{eqnarray}
where

\begin{eqnarray}
  \label{Gl:eta_Akiyama}
  \eta^2 & = & \left(1-\sin 2\theta\right)^2 \, , \\
  \xi^2  & = & \sin^2\theta \, \left(1-\sin 2\theta\right) \,.
\end{eqnarray}
The additional subscript ``en'' refers to the entropy gradient included
into Eq.\,\ref{Gl:MRI-Zeit-BV} via the Brunt--V\"ais\"al\"a frequency.
In the equatorial plane, where the polar angle $\theta = \pi/2$, the
time scale given by Eq.\,\ref{Gl:MRI-Zeit-BV} is equivalent to that
obtained from Eq.\,\ref{Gl:BV_MRI-Wachsrate}.

In order to see the MRI in our simulations, we must have sufficient
spatial resolution to resolve at least the modes with the longest
wavelengths which grow fastest.  However, as these modes have the
smallest wavenumbers, the Alfv\'en velocity and thus the magnetic field
(parallel to the direction of $\vec k$) must be sufficiently large for
the product $\vec c_{\mathrm{A}}\cdot \vec k$ to be in the range of
maximum growth.  This is unproblematic in the magneto--convective case
due to the large growth rate in the limit $\vec c_{\mathrm{A}} \vec k
\to 0$.  In the magneto--shear limit only modes near
$\left(c_{\mathrm{A}} k\right)_{\mathrm{max}}$ grow fast, which are,
however, difficult to resolve.  Further numerical complications arise
from the fact that we have to identify the MRI in a very inhomogeneous
and highly dynamical background flow.

Apart from the direct solution of the MHD equations, there are
alternative ways to investigate the MRI and the turbulence driven by the
instability, such as the inclusion of well-suited models for turbulent
transport coefficients into the momentum and energy equations.  Various
closure models for magnetorotational turbulence exist. The ones by
\cite{Ogilvie_MNRAS_2001__MRI_excentric_disks,Ogilvie_MNRAS_2003__MRI_stress}
and
\cite{Williams_MNRAS_2005__Jet_collimation,Williams_NA_2004__Elasticity}
which make use of the analogy of MHD turbulence with viscoelastic flows
as observed, e.\,g.\,in polymer fluids in the laboratory, seem
particularly interesting. Such methods provide a way for including
turbulence effects into a numerical simulation that cannot treat these
effects due to, e.\,g.\,resolution or symmetry constraints. However, in
the simulations we report here, we did not consider any of these models.
Instead, we focussed on the possibility of directly simulating the
development of the instabilities.

In most of our models we find extended MRI unstable regions at various
epochs.  During collapse the cores are unstable in the magneto--shear
limit. However, the growth times are significantly larger than the
collapse time scale: even for the fastest growing mode they are of the
order of a second (in the initial models).  As the rotational energy and
the degree of differential rotation increase during collapse, the growth
times become smaller, but remains larger than the time until bounce,
i.e.\,the dynamical background evolves faster than any unstable mode.
Hence, we do not observe the growth of the MRI even for models where the
initial magnetic field is sufficiently strong for magneto--shear modes
to be numerically resolved.

During and shortly after bounce the cores possess a convectively stable
stratification, i.e.\,the MRI is of magneto--shear character.  In model
A1B3G3-D3M10 the immediate post--shock region is unstable shortly after
shock formation due to a large negative angular velocity gradient (in
$\varpi$--direction).  The growth times are in the sub--millisecond to
millisecond range, i.e.\,comparable to the dynamic time scales.  The
fastest growing modes correspond to spatial scales of less than
$100\,$m, which are significantly smaller than the (local) grid
resolution of $\sim 300\,$m.  For models with a stronger magnetic field
(e.g.\,A1B3G3-D3M12) the spatial scales of the fastest growing modes are
larger due to the larger Alfv\'en velocity, and they can be resolved.
However, they cannot be discriminated from the dynamic background flow
at the time of bounce which is dominated by numerous pressure waves
propagating through the inner core and launching the shock wave.
Moreover, obscuring the action of the MRI, the strong shear flow
surrounding the "surface" of the inner core produces a strong toroidal
magnetic field component. The positive entropy gradient behind the shock
wave does not allow the MRI to grow in most of the post--shock region,
while large parts of the inner core are MRI unstable throughout the
entire post--bounce evolution due to the very flat entropy profile of
the core.  However, as the instability is of magneto--shear nature, one
encounters the previously discussed spatial resolution problem unless
the magnetic field is already quite strong.  The maximum growth rates
are moderate ($t_{\mathrm{grow}} \la 10\,$ms).

\begin{figure}[htbp]
  \resizebox{\hsize}{!}{\includegraphics{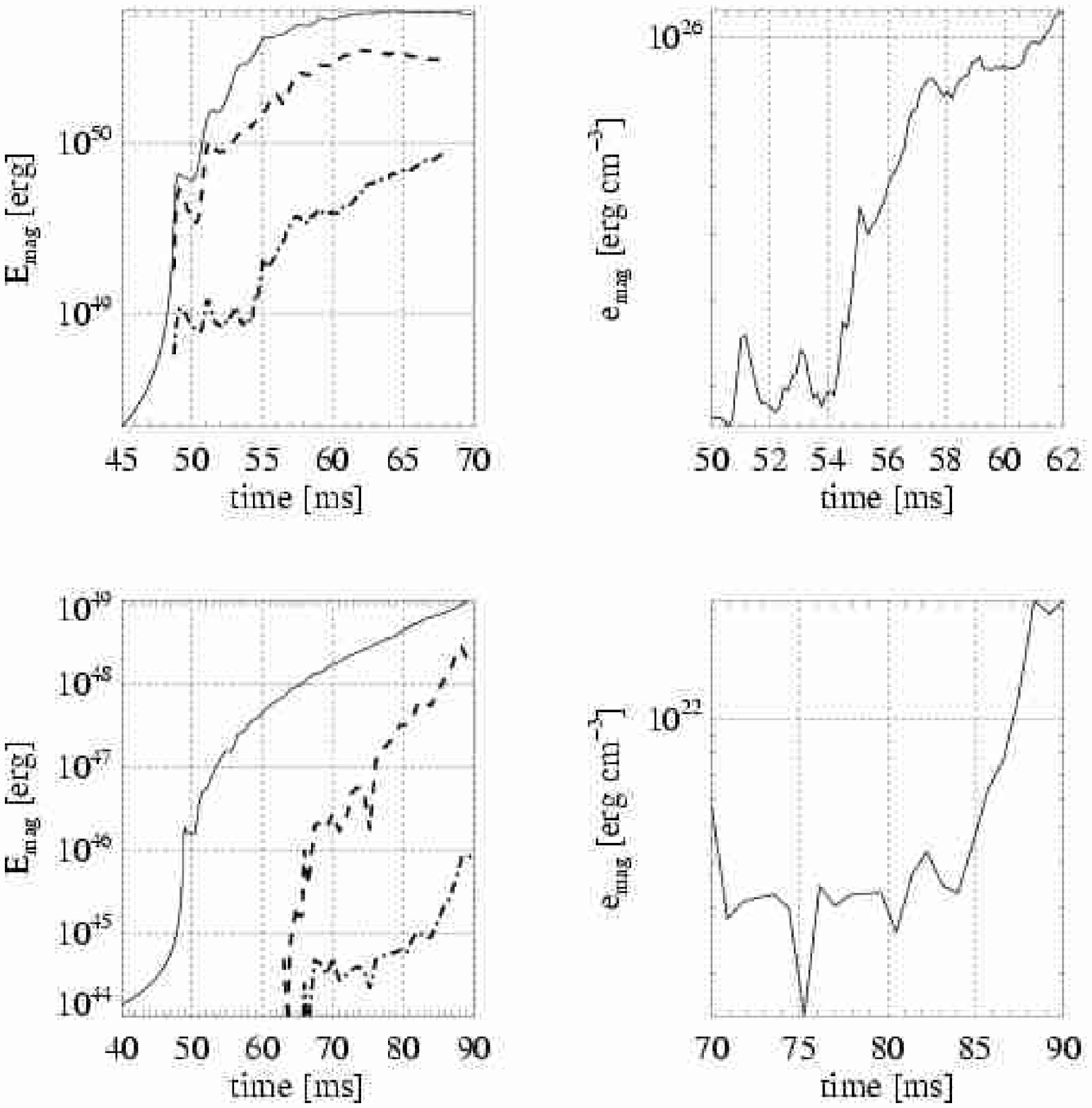}}
  \caption
  { Comparison of the evolution of models A1B3G3-D3M12 (upper panels)
    and A1B3G3-D3M10 (lower panels).  The left panels display the
    total magnetic energy (solid lines) of the core, and the total
    (dashed lines) and poloidal field energy (dash-dotted lines) of
    the MRI--unstable regions, which are defined according to two
    different criteria: For the strong--field model A1B3G3-D3M12 all
    zones are considered where the growth time of the fastest growing
    mode is less than $10\,$ms, while for the weak--field model
    A1B3G3-D3M10 the corresponding magneto--convective limit is used.
    The energy density $e_{\mathrm{mag}} = E_{\mathrm{mag}} / V$ of
    the poloidal field in the MRI--unstable regions is shown in the
    right panels.  The poloidal field increases rapidly when the MRI
    grows, which happens in the strong--field model (top right)
    shortly after bounce, and in the weak--field model (bottom right)
    by magneto--convective modes at later epochs.  }
  \label{Fig:Z133-310_12_mri_mag}
\end{figure}

Well after bounce when the shock wave begins to weaken and thus
(locally) negative entropy gradients develop, the MRI growth rates
become large even for long wavelengths (magneto--convective limit).
Thus, we observe the emergence of the corresponding modes, and a growth
of the vorticity $\vec \omega = \vec \nabla \times \vec v$ of the flow.
However, this occurs late after bounce, when our assumptions concerning
the microphysics do no longer hold.  For model A1B3G3-D3M10, we can
study the onset of magneto--convective modes. In regions where the
growth times of the MRI modes (in the limit $k \to 0$) are less than
$10\,$ms (Fig.\,\ref{Fig:Z133-310_12_mri_mag}), we find a rapid increase
of the magnetic energy, the magnetic energy density, and of the specific
magnetic energy.  Their temporal variation can be described by
exponential laws in the interval $57\,\mathrm{ms} \le t \le
96\,\mathrm{ms}$.  In phases when the field is amplified by differential
rotation only, the toroidal field increases but the poloidal field and
the corresponding magnetic field energy stay approximately constant
(Fig.\,\ref{Fig:Z133-310_12_mri_mag}).  However, when the MRI sets in,
the poloidal component grows too due to the stretching of the poloidal
field lines by meridional motions.  Because of this behavior the two
field amplification mechanisms can be distinguished.

For initially very strong magnetic fields ($ \ga 10^{12}\ 
\mathrm{G}$), the spatial scales of interest for the MRI can easily be
resolved even in the magneto--shear limit, and the MRI criterion is
fulfilled the growth rates being similar to those of the weak--field
case.  The proto--typical model A1B3G3-D3M12 shows the creation of a
large amount of vorticity both in the post--shock gas and in the MRI
unstable regions of the inner core (Fig.\,\ref{Fig:133-312:Outflow}).
Its vorticity is considerably larger than in the corresponding weak
field model A1B3G3-D3M10, where we cannot resolve the growth of the
unstable modes due to their very small spatial scales.  The flow is
organized in sheet--like structures which are typically a few kilometers
wide in $\varpi$--direction and $\sim 10\,$km long in z--direction.
Another prominent flow structure forms near the equatorial plane in the
outer layers of the inner core, where a violent meridional flow leads to
the largest amplification of the poloidal field and to angular momentum
transport out of the inner core.  The latter, which implies that the
loss of the rotational support of the inner core is caused by the MRI,
holds for all models where we can resolve the unstable modes, and where
the field is not too strong initially.  For the extreme models
AaBbGg-D3M13 the stress tensor is initially already large enough for
significant angular momentum transport to occur, i.e.\,only a very small
amount of field amplification is required.

Initially weak magnetic fields should lead to a similar dynamical
behavior, as the MRI does not depend on the initial field strength.  The
modes should grow on similar time scales, but smaller spatial scales.
In practice, however, structures resolvable on our grid are much larger
than the wavelengths of the fastest growing modes, i.e.\,they have
relatively small growth rates. Hence, the development of the MRI is
artificially delayed.  In addition, the smaller scales of the more
unstable modes may affect the global dynamics.  Indeed, if we simulate
the evolution of model A1B3G3-D3M11 for a longer time, the poloidal
field energy starts to rise after $t \approx 65\,$ms. This rise is
accompanied by the formation of a flow pattern consisting of sheets of
large vorticity, and the extraction of rotational energy from the core,
quite similar to our findings for model A1B3G3-D3M12, though less
pronounced.  Thus, the distinction between weak- and strong--field
models may be irrelevant, because even small initial fields give rise to
the same kind of instability and can become ``strong'' according to our
criterion very rapidly.

The poloidal field strength oscillates similarly as the density or the
rotational energy during the post--bounce phase for the models bouncing
due to centrifugal forces.  The field is subject to compression and
expansion, but apart from the oscillations it does not grow unless an
MRI--like mechanism sets in.

The magneto--convective motions found during the later stages of weak
field models are much less pronounced in the cores of strong field ones.
There are two reasons for this behavior: Firstly, due to the extraction
of angular momentum from the core one of the driving forces of the
instability is reduced, and secondly a strong magnetic field is known to
suppress convection.  The latter also holds for the MRI which is
hampered by too strong fields, as the seed field cannot be amplified
beyond the MRI saturation value.  This can be seen from
Eq.\,(\ref{Gl:BV_MRI-Wellenzahl-Krit}): For large Alfv\'en velocities
only modes with very long wavelengths (or small wavenumbers) are
unstable.  If the critical wavelength exceeds the intrinsic scales of
the problem, e.g.\,the size of the inner core or the size of potentially
unstable regions, the unstable longer modes cannot develop, and the
modes of smaller wavelengths are stable.

\subsection{Influence of the magnetic-field configuration}
\label{Suk:Dis:FC}

The dynamics and gravitational wave signature of the models A3B3G5-DdM12
differing only in the location of the field--generating current loop
(i.e.\,in their Dd parameter; see Sect.\,\ref{Suk:Modelle:IniMag}) are
qualitatively quite similar, except for model A3B3G5-D0M12 whose current
loop is located at infinity (see the discussion below).  The time scale
for the slow--down of the core exhibits some variation, but otherwise
the evolution and the GW signal agree qualitatively quite well with
those of the corresponding reference models A3B3G5-D3Mm (see above).
This may be surprising because the initial total magnetic energies of
models having uniform fields may be more than 100 times greater than
that of current--loop models (table \ref{Tab:MinitPars}).  However, the
outer parts of the very extended field structure of the more strongly
magnetized models, where most of the additional magnetic energy is
stored, are dynamically less important than the central regions, and in
particular the region around the field--generating current loop where
the magnetic field is strongest.

\begin{figure*}[!htbp]
  \centering
  \includegraphics[width=16cm]{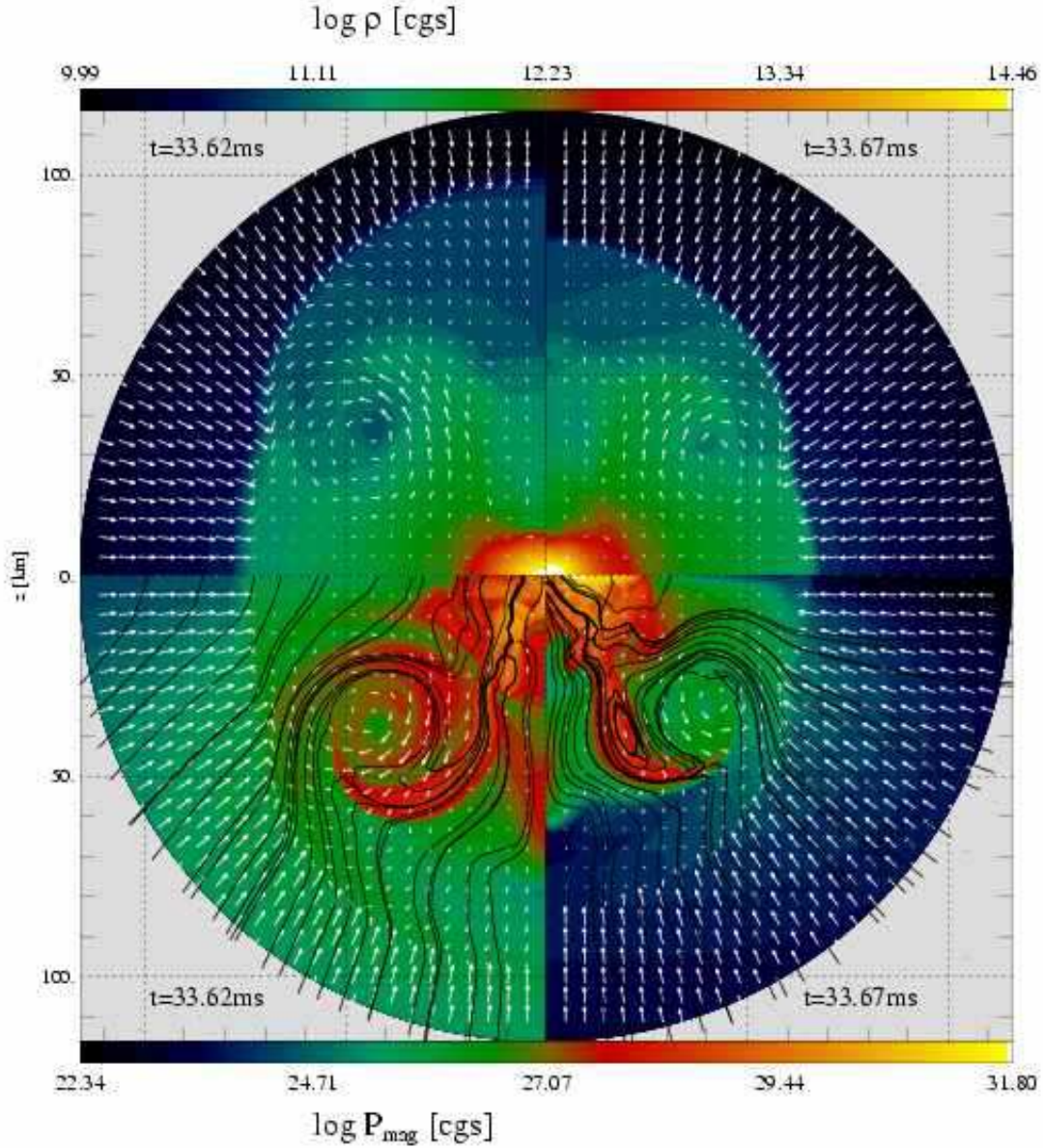}
  \caption[Model A3B3G5-D2M12: $t=33\ \mathrm{ms}$]
  { The post-shock region of models A3B3G5-D0M12 (left quadrants) and
    A3B3G5-D2M12 (right quadrants) at $t\approx 33.6\ \mathrm{ms}$ about 3\,ms
    after core bounce.  Density and magnetic pressure (color--coded)
    are displayed in the top and bottom quadrants, respectively.
    Additionally, the flow field (vectors) and the poloidal field
    lines of the magnetic field are displayed.  }
  \label{Fig:335-012_212:33ms}
\end{figure*}

\begin{figure*}[!htbp]
  \centering
  \includegraphics[width=5.6cm]{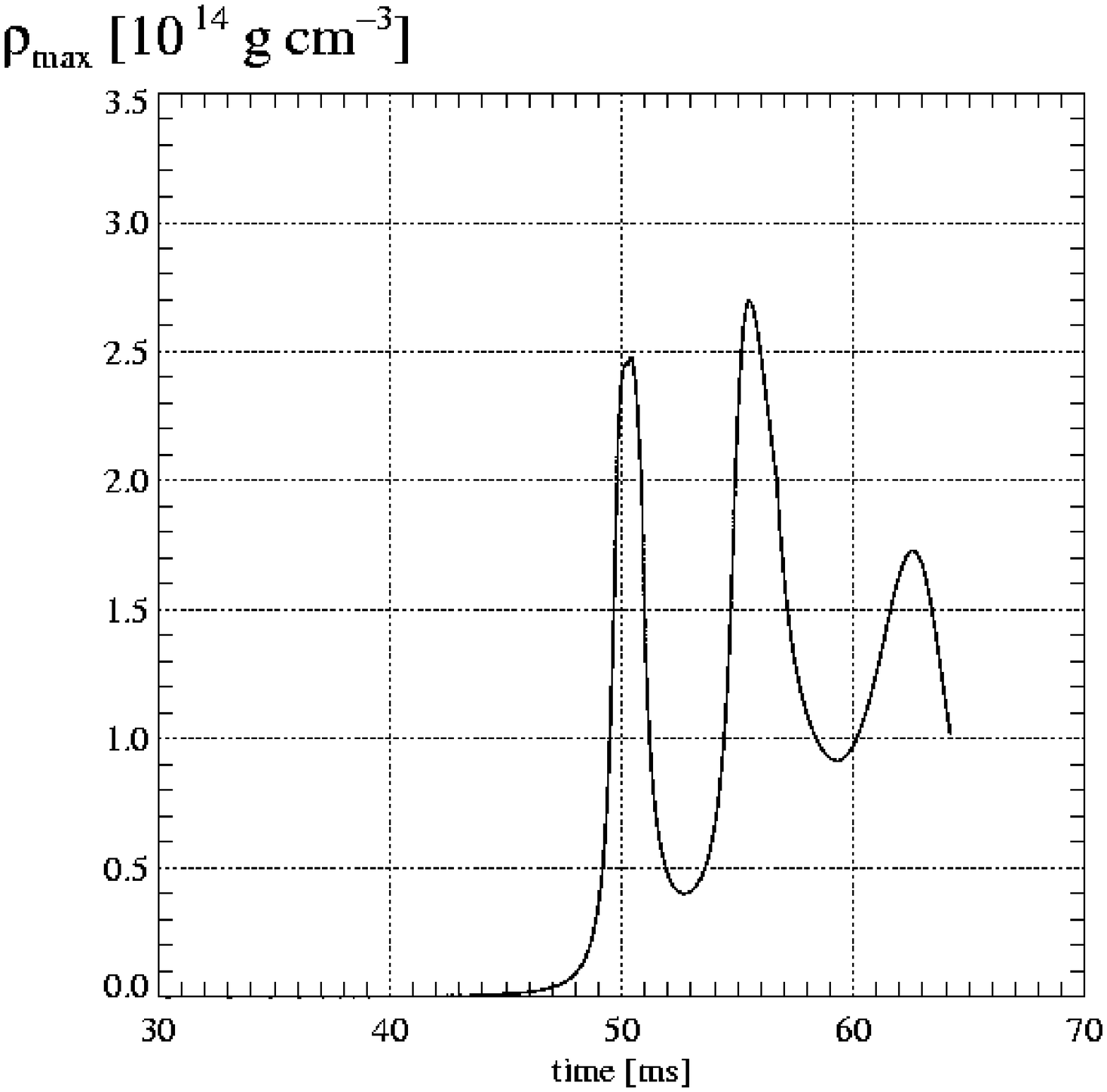}
  \includegraphics[width=5.6cm]{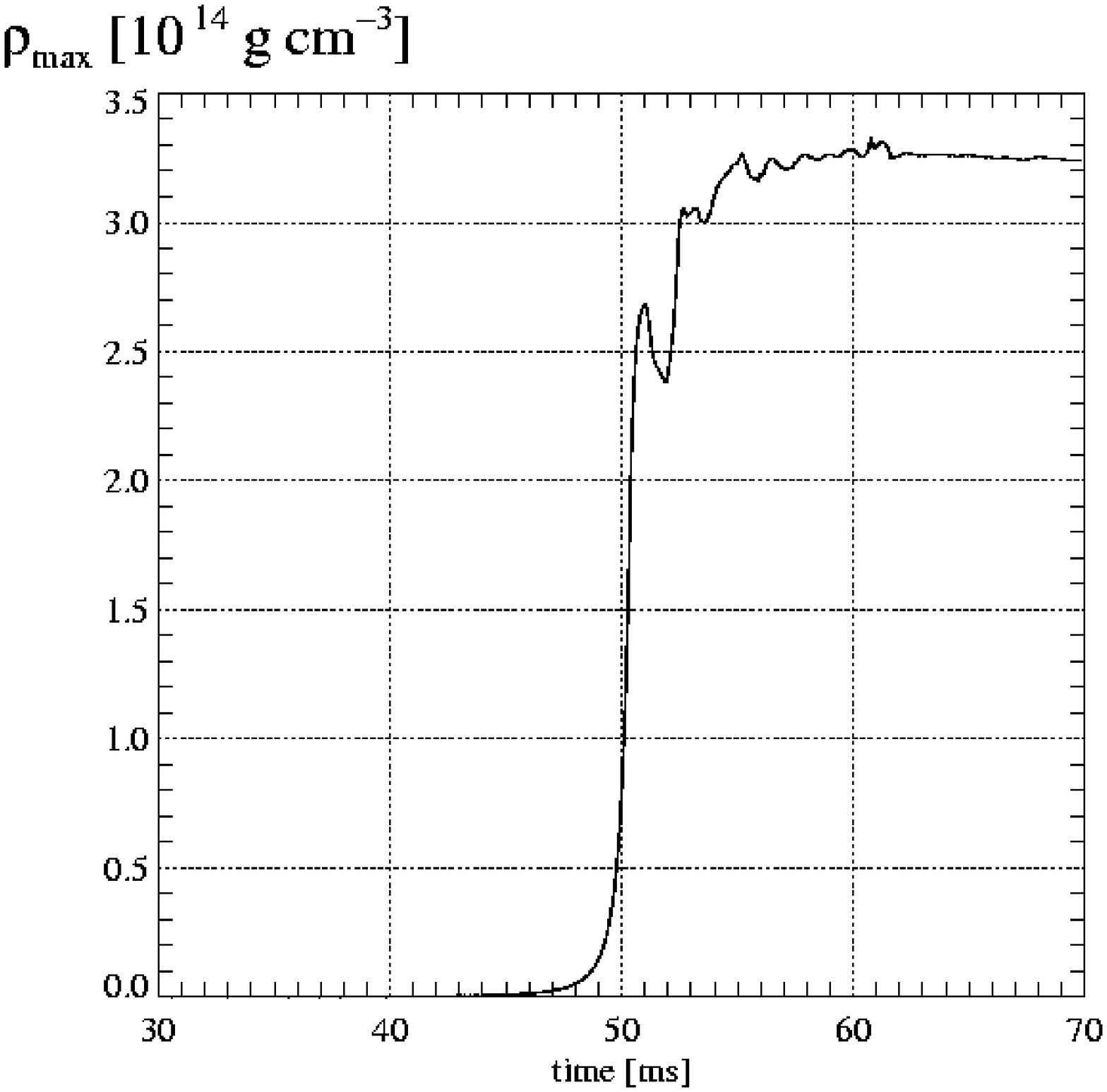}
  \includegraphics[width=5.6cm]{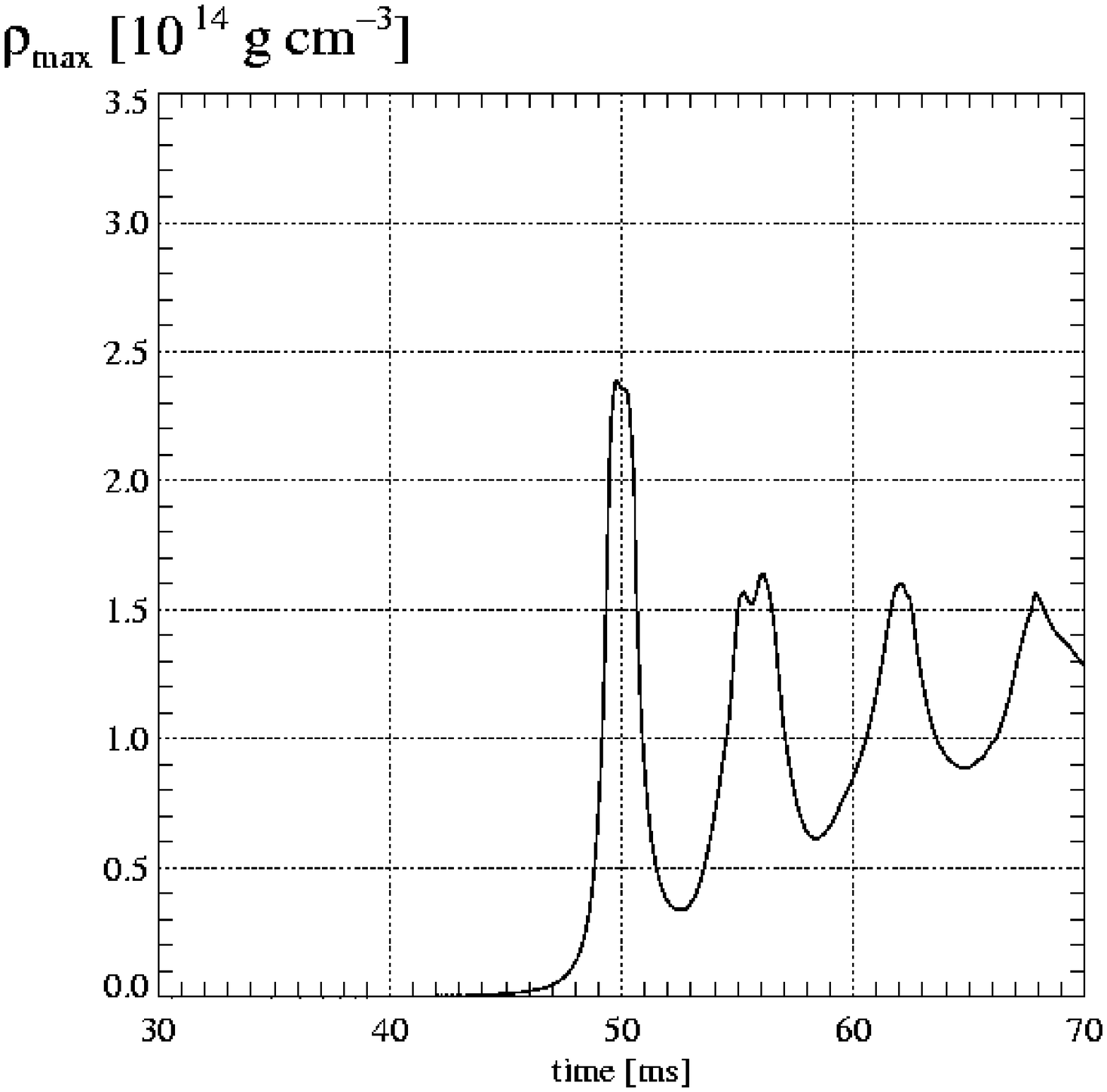}
  \includegraphics[width=5.6cm]{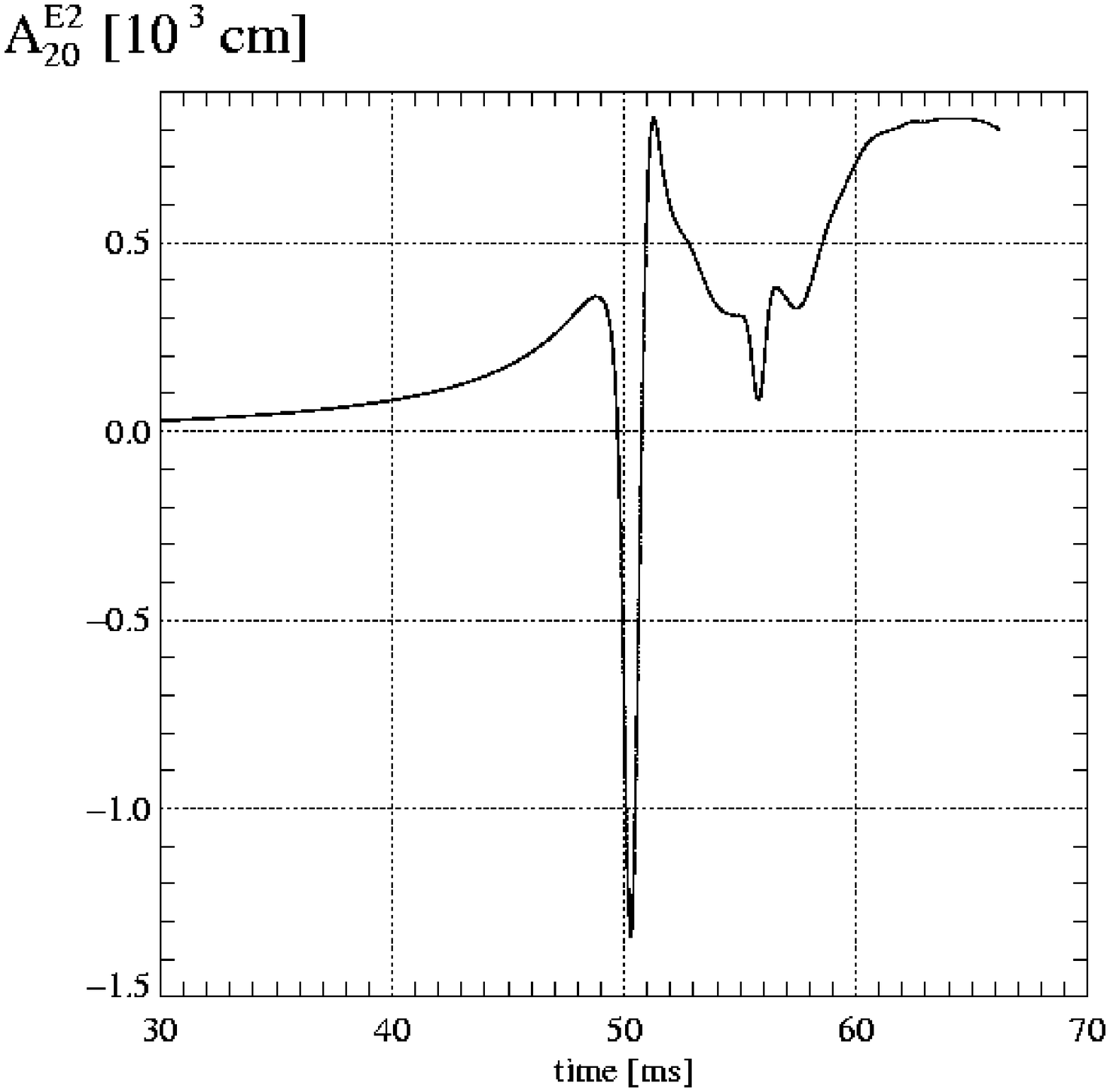}
  \includegraphics[width=5.6cm]{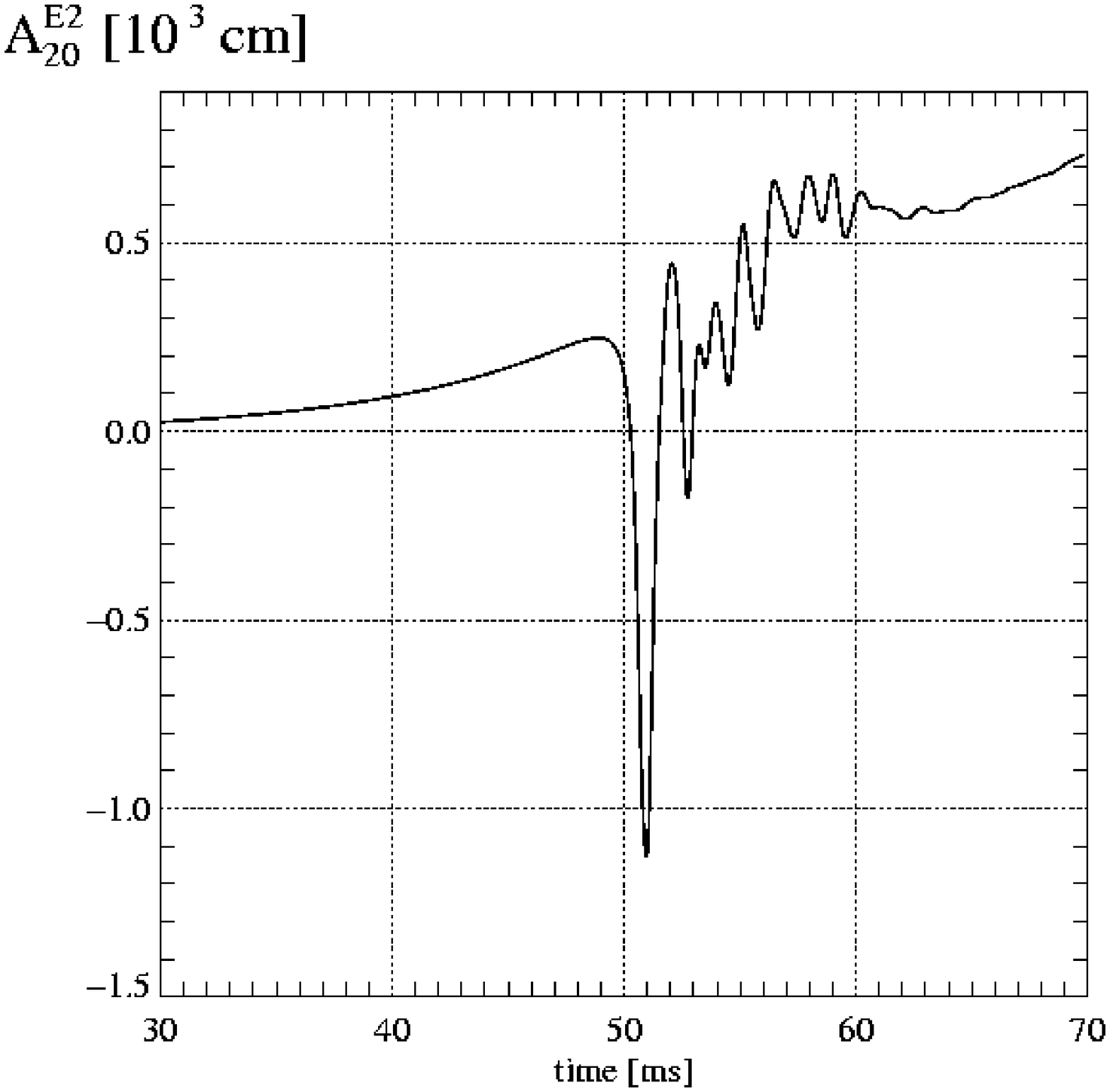}
  \includegraphics[width=5.6cm]{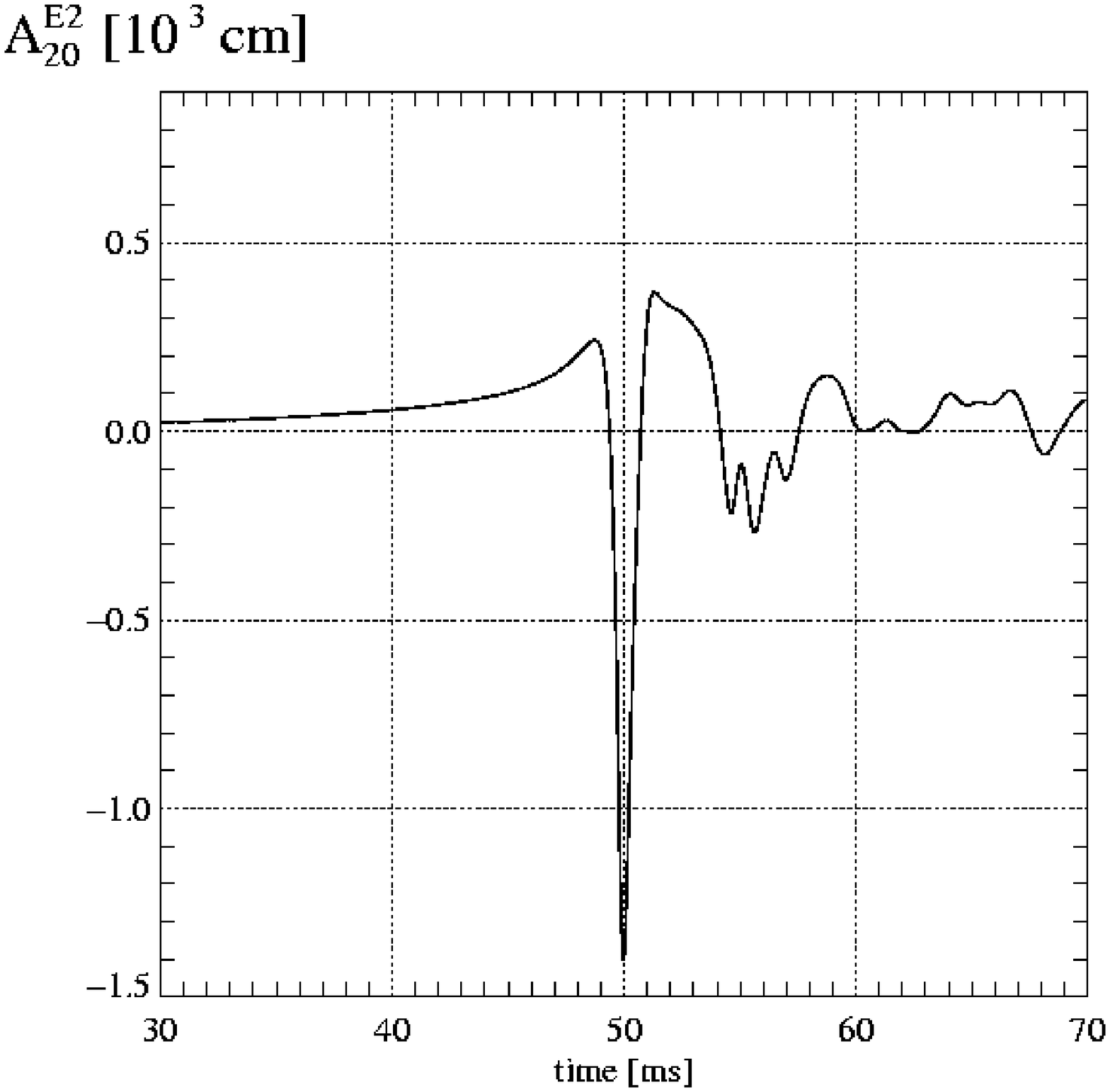}
  \caption[Density Evolution of models A3B3G3-DdM13]
  { The evolution of the maximum density (top panels) and the GW
    amplitude (bottom panels) of models A3B3G3-D0M13 (left panels),
    A3B3G3-D3M13 (middle panels), and A3B3G3-D1M13 (right panels),
    respectively.  The GW signal of model A3B3G3-D0M13 shows a
    constant offset from zero, which is subtracted here.  }
  \label{Fig:333D}
\end{figure*}

For the initially homogeneously magnetized model A3B3G5-D0M12 a less
efficient braking of the core's rotation is observed than for the
corresponding current-loop models.  Instead of a contracting secularly,
the core expands to sub--nuclear densities on a comparable time scale.
The rotation energy of the immediate post--bounce core of model
A3B3G5-D0M12 is slightly smaller ($\beta_{\mathrm{rot}} \approx 0.076$
at $t \approx 32\ \mathrm{ms}$) than for the corresponding non--magnetic
core ($\beta_{\mathrm{rot}} \approx 0.084$ at $t \approx 32\ 
\mathrm{ms}$), and it decreases sightly during the subsequent evolution.
Although the magnetic field gets amplified a little during collapse,
$\beta_{\mathrm{mag}}$ decreases until bounce because most efficient
amplification takes place in the central regions of the core whereas the
total magnetic energy is dominated by the outer regions where the field
stays roughly constant during collapse. Afterwards, both
$E_{\mathrm{mag}}$ and $\beta_{\mathrm{mag}}$ show a comparable growth
to the current--loop models, and the ratio of toroidal field energy and
total field energy grows steadily, but slowly. The magnetic field of the
uniformly magnetized core has a simpler structure at late times, while
the current--loop fields become more and more twisted. The gravitational
wave amplitude of model A3B3G5-D0M12 is strongly dominated by the fluid
contribution in spite of its large initial magnetic energy, and
resembles that of the corresponding current-loop cores in many respects.
The small magnetic contribution to the signal never exceeds that of
model A3B3G5-D3M12, although this model has a 15 times smaller magnetic
energy (but the same central field strength).

Model A3B3G5-D2M12 has an initial magnetic field energy which is a
factor of 150 greater than that of model A3B3G5-D0M12 (table
\ref{Tab:MinitPars}).  Nevertheless, at $t \approx 39\ \mathrm{ms}$, its
magnetic field is almost three times as energetic as that of model
A3B3G5-D0M12 ($\beta_{\mathrm{mag}} = 0.0075$ compared to
$\beta_{\mathrm{mag}} = 0.0028$) due to the different amount of field
amplification occurring in both cases.  At bounce both models have
developed a region of high magnetization near the edge of the inner core
close to the rotation axis.  In model A3B3G5-D2M12 this region also
extends towards larger radii along the equatorial plane forming a highly
magnetized ``sheet'' which is a relic of the initially highly magnetized
region around the field generating current--loop.  The shape of the
shock waves is similar in both models, but the regions of large magnetic
pressure are more extended in the current--loop model filling the whole
post--shock region, whereas they are more concentrated towards the very
center of the core for the uniform field model
(Fig.\,\ref{Fig:335-012_212:33ms}).

Similar trends as for model series A3B3G5-DdM12 can be observed in the
case of model series A3B3G3-DdM13 (Fig.\,\ref{Fig:333D}).  The dynamics
of the current--loop model A3B3G3-D1M13 is qualitatively comparable to
that of a A3B3G3-D3Mm model with a much weaker magnetic field. Hence,
its GW signal does not differ much from the non--magnetic case in spite
of its strong initial magnetic field.  However, we find large
differences between the uniform field model A3B3G3-D0M13
(Fig.\,\ref{Fig:333D}, left panels) and the current--loop models, as
e.g.\,A3B3G3-D3M13 (Fig.\,\ref{Fig:333D}, middle panels).  The maximum
density of the former model exhibits large--scale pulsations during
which the bounce density is exceeded by $9\,\%$, and
$\rho_{\mathrm{max}}$ is smaller than $\rho_{\mathrm{nuc}}$ for most of
the time.  The core has a very high magnetic energy initially,
$\beta_{\mathrm{mag}} \approx 20\,\%$, which decreases during the
collapse until it levels off at $\approx 7\,\%$.  This value is still
much larger than the largest one reached by the current--loop models
during their whole evolution.  The GW signature of model A3B3G3-D0M13
reflects the density oscillations of the core, and like in the case of
model A3B3G3-D3M13, it is considerably shifted towards positive
amplitudes (Fig.\,\ref{Fig:333D}) due to the very aspherical shape of
the shock wave.

\section{Summary and conclusions}
\label{Sek:Schluss}

Past and present realistic studies of core collapse supernovae neglect
magnetic fields.  Instead, they focus on an elaborate and accurate
description of the microphysics and (neutrino) transport physics.  Since
this is a formidable task requiring extensive computational resources,
these studies are necessarily limited to a very small number of
different initial models.  Bearing in mind our lack of knowledge of the
exact conditions in the pre--collapse star, we pursued a complimentary
approach. Instead of simulating the evolution of a few stars in great
detail, we performed a comprehensive parameter study covering the space
of possible progenitors with a large number of initial models and
exploring a relatively unknown territory, namely the influence of
magnetic fields.  Consequently, we are restricted to some simplified and
approximate treatment of the core's physics, and employ only simplified
progenitor models (i.e.\,polytropes).

Our parameter study focuses on the investigation of two main questions:
\begin{itemize}
\item 
How do magnetic fields affect the dynamics of the collapse and bounce,
and how is their presence reflected in the gravitational wave signal
emitted by the core?
\item 
Which pre--collapse magnetic fields (strength and geometry) and
angular momentum (amount and distribution) will cause important
dynamic effects for core collapse?
\end{itemize}

\noindent
To address these questions we have performed the most comprehensive
parameter study of magneto--rotational core collapse up to now. We
simulated the gravitational collapse and the immediate post--bounce
evolution of a series of differentially rotating, magnetized,
axisymmetric stellar core models using a newly developed 2D Newtonian
MHD code based on the algorithm of
\citet{Pen_Arras_Wong_APJS_2003__rTVD_MHD_Code} that employs the
relaxing TVD scheme of \citet{Jin_Xin_CPAM_1995__rTVD} and the
constraint--transport method \citep{Evans_Hawley_ApJ_1988__CTM} to keep
the magnetic field divergence--free.  Our code, which has been
comprehensively tested, incorporates an analytic equation of state
describing in an approximate way the thermodynamic properties of core
matter \citep{Janka_Zwerger_Moenchmeyer_AAP_1993__Art_vis_SN}.  We also
computed the gravitational wave signal produced by the core with the
Einstein quadrupole formula taking into account both the contributions
of the hydrodynamic and gravitational forces \citep{Moenchmeyer_1991},
and of the Lorentz forces \citep{Kotake_etal_PRD_2004__MagCollapse_GW}.
The initial models used in our study are identical to those used by
\citet{ZM97} and \citet{DFM1,DFM2}.  The initial magnetic field is
purely poloidal.

Our simulations show significant modifications of the dynamics and the
GW signal of strongly magnetized cores compared to non--magnetic ones.
Magnetic fields are amplified in some of our models to very large field
strengths which are in the magnetar range ($|\vec B| \sim 10^{15} \ 
\mathrm{G}$). Such strong magnetic fields efficiently extract rotational
energy from the central core, thereby triggering a secular post--bounce
contraction phase.  They also cause the formation of collimated
outflows, which give rise to distinctive features in the GW signal.
Hence, it is in principle possible to extract information about the
degree to which an explosion has a jet--like character from the GW
signal.  Further specific results are:
\begin{enumerate}
  
\item From the initially purely poloidal magnetic field, large toroidal
  field components are created by the action of the differentially
  rotating core in a so--called $\Omega$ dynamo, the energy of the
  saturation field being of the order of the rotational energy of the
  core.  The efficiency of this field winding process depends on the
  angular momentum distribution of the progenitor and the amount of
  differential rotation arising during collapse. The latter is
  determined by the degree of non--homologous collapse which is
  sensitively influenced by the equation of state and the rotation rate.
  The saturation fields estimated from the rotational energy of our
  collapsed cores are of the order of $10^{15} \ \mathrm{G}$, which
  are field strengths expected for magnetars, but not for typical
  neutron stars.  The field amplification process is found to be less
  efficient if some magnetic energy is consumed for the acceleration of
  matter by dynamically important fields.
  
\item As MHD instabilities acting on the toroidal field component are
  suppressed due to the assumed axisymmetry, we are unable to simulate
  the corresponding field amplification processes, and thus may
  underestimate the saturation value and the growth rate of the magnetic
  field. However, if the entire dynamo process of transforming poloidal
  and toroidal fields into one another and their amplification is mainly
  driven by differential rotation, the saturation fields and the growth
  rates are expected not to exceed those found in our simulations.  We
  point out that (i) without any symmetry restriction fields of similar
  strengths may arise when one evolves initially less magnetized cores,
  and (ii) that the strong initial fields of our progenitors (which are
  stronger than those expected in actual cores) may compensate for a
  possible additional dynamo action unconsidered in our models.  The
  field amplification is fastest for rapidly and very differentially
  rotating models.  In models producing a type-I GW signal the magnetic
  field is amplified more efficiently than both in the faster collapsing
  models of type-III and the centrifugally bouncing type-II models.
  
\item The influence of a magnetic field on the dynamics and the GW
  signature of the core depends mainly on the initial magnetic field
  strength, and to a lesser extent on the field topology.  For the
  weakest of our initial fields ($B\sim10^{10}\ \mathrm{G}$) we do not
  find any significant impact either on the dynamics or on the GW signal
  on the time scales covered by our simulations ($\la 100 \ 
  \mathrm{ms}$ after core bounce). However, sufficiently strong fields
  have a large impact on the dynamics of the core and on the
  gravitational wave signal.  Compared to the corresponding
  non--magnetized models, the core bounce is somewhat delayed for the
  majority of our very strongly magnetized models ($B\ga 10^{12}\ 
  \mathrm{G}$), and the bounce density is slightly enhanced.  Extremely
  strong initial magnetic fields ($B\ga 10^{13}\ \mathrm{G}$)
  efficiently slow down the rotation rate of the central core already
  during collapse, and thus reduce its.  centrifugal support by a
  considerable amount. Due to the loss of angular momentum from its
  central regions the core eventually begins to contract or even enters
  a second collapse phase after bounce.  Hence, an initially extremely
  magnetized and rapidly rotating core, which bounces due to centrifugal
  forces at low densities, evolves into a very compact configuration
  that is almost entirely pressure supported, its initially toroidal
  density stratification changing into a flattened centrally
  concentrated one.
  
\item The magnetic stresses of the toroidal field component (created by
  the differential rotation) can cause a strong pinch effect, which
  gives rise to a mildly relativistic, jet--like outflow along the
  rotation axis, and consequently to a very prolate shock wave. The
  aspherical outflow is mainly driven by the very high magnetic pressure
  in the post--shock plasma, as first pointed out by
  \citet{Yamada_Sawai_ApJ_2004__MHDCollapse}, and it is observed both in
  cores that bounce due to the stiffening of the EOS and (mainly) due to
  centrifugal forces.  In purely hydrodynamic calculations highly
  aspherical outflows are only encountered for very extreme rotators,
  whereas they are quite common for models with an initial magnetic
  field strength of $B \ga 10^{13} \mathrm{G}$.  On somewhat longer time
  scales, similar but weaker outflows are observed for models having
  only a tenth of this field strength.  This finding might be of
  relevance for the study of Gamma--Ray Bursts.  According to our
  results extreme initial conditions are required for the formation of
  jet--like outflows on the time scale of a few ten milliseconds.
  However, our results may also apply on longer time scales to the
  smaller rotation rates and the weaker magnetic field strengths of more
  realistic stellar progenitors.
  
\item All simulated cores develop MRI unstable regions.  However, in
  most cases, a fast growth of the MRI is limited to short--wavelength
  modes which we cannot resolve due to insufficient numerical
  resolution.  For initial fields $\ga 10^{12}\ \mathrm{G}$, the
  fastest growing modes have sufficiently long wavelengths to be
  resolved numerically.  A meridional flow pattern characterized by
  large vorticity arises the circulation pattern being organized in
  sheet--like structures which is also the topology of the poloidal
  field.  In contrast to the weak--field case, where we cannot resolve
  possibly unstable modes and where the poloidal field stays roughly
  constant after bounce, the poloidal field energy of strongly
  magnetized cores grows approximately exponentially.  The field
  strengths found in our simulations are of the order of the saturation
  field estimates derived for the MRI, i.e.\,$\sim 10^{15}\ \mathrm{G}$ at
  the surface of the inner core.  The growth of the field comes along
  with the transport of angular momentum out of the inner core, and thus
  a loss of rotational support.  Consequently, the maximum density of
  the core rises steadily during the post--bounce evolution.
  
\item The gravitational wave signature of the core is altered
  significantly in the presence of initially strong ($B\ga10^{12}\ 
  \mathrm{G}$) magnetic fields compared to the corresponding
  non--magnetic case.  Since pressure--bouncing type-I and type-III
  models are affected only quantitatively by strong magnetic fields (in
  contrast to the centrifugally bouncing models where magnetic fields
  change the collapse dynamics considerably), the GW amplitude of a
  strongly magnetized type-I or type-III model does not deviate strongly
  from the purely hydrodynamic case.  Depending on the actual rotational
  profile and the EOS, the GW bounce amplitude may be enhanced or
  weakened by a few $10 \,\%$. This confirms the findings of
  \citet{Yamada_Sawai_ApJ_2004__MHDCollapse} obtained for a much smaller
  set of models.
  
\item For most strongly magnetized models we observe a shift of the
  bounce signal towards more positive amplitudes.  A major part of this
  shift is due to magnetic forces: around the time of core bounce the
  magnetic part of the GW amplitude is positive and large indicating
  that the magnetic stresses act to diminish the core's oblateness.
  This reduces the absolute value of the pronounced (negative) bounce
  amplitude for most type-I and type-II models, and enhances the
  (positive) maximum GW amplitude of the type-III models reached shortly
  before or after bounce. In the latter models strong magnetic fields
  also decrease the size of the much less pronounced (negative) bounce
  minimum of the GW amplitude. Rigidly and modestly fast rotating type-I
  models behave differently, as for these models the size of the
  (negative) bounce peak is enhanced significantly for initially
  extremely strong magnetic fields ($\sim 10^{13}\mathrm{G}$).  After
  bounce, both the gravitational and the hydrodynamic parts of the GW
  amplitude decrease as the rotational flattening of the central core
  diminishes due to the extraction of angular momentum.  Later in the
  evolution the magnetic part of the GW amplitude, now becoming
  increasingly negative, dominates the hydrodynamic one for cores which
  are slowed down very strongly.  As in the case of purely hydrodynamic
  models, the GW signal of a strongly magnetized core bouncing due to
  pressure forces exhibits oscillations with periods that are comparable
  to the local dynamic time scale.
  
\item The dynamics of type-II models bouncing due to centrifugal forces
  is changed completely by the presence of strong magnetic fields.  This
  is reflected in a rather drastic change of the GW signal emitted by
  these cores.  As the large--scale pulsations of the core fade away
  during the secular contraction phase, the long--period modes of the GW
  signal become less important.  Instead high frequency oscillations,
  similar to the ones emitted by type-I models, dominate the GW signal.
  The corresponding periods are given by the local dynamic time scales
  of the core which is no longer supported by rotation, but -- at much
  higher densities -- by pressure forces.  This shift in the frequency
  of the GW amplitude provides observational evidence of the change of
  the core's dynamics and structure by a strong magnetic field.  We
  introduce the new GW signal type IV to refer to this phenomenon.
  
\item Aspherical outflows and shock waves get imprinted on the GW signal
  the signature being a large, slowly varying positive GW amplitude.  A
  few ($\sim 5$) milliseconds after bounce, we find amplitudes of the
  order of $A^{\mathrm{E}2}_{20} \sim 10^2 \ \mathrm{cm}$ which remain at roughly
  this level until the end of our calculations.  As bipolar outflows
  occur preferentially in strongly magnetized cores, the GW signal may
  be used to distinguish strongly magnetized cores from non--magnetic or
  weakly magnetized ones.

\end{enumerate}

The applicability of our findings to the supernova problem is limited by
our simplified treatment of the microphysics (approximate equation of
state), by the neglect of neutrino transport, and by our simplified and
probably too extremely rotating and magnetized initial models.
Nevertheless, after having shown that our code is suited for
multi-dimensional MHD simulations of core collapse, and after having
highlighted some general trends of magneto--rotational core collapse, we
are ready for a more sophisticated investigation of the phenomenon.  In
particular, the evolution of our (more realistic) weak--field models on
longer time scales will be the topic of further studies, as on longer
time scales even initially weak magnetic fields can be amplified,
e.g.\,by means of the magneto--rotational instability, to a strength
relevant both for the dynamics and the GW signature of the core.
Furthermore, the interplay of magnetic fields and various hydrodynamic
instabilities occurring in a supernova explosion is a challenging topic
to be addressed.  In particular, the convective region between the
proto--neutron star and the stalled hydrodynamic shock wave might be an
arena where generic MHD effects such as the magneto--rotational
instability do operate.

Imposing axial (and equatorial) symmetry is a serious restriction when
simulating MHD flows. It prevents us from investigating the fate of
cores whose magnetic field and rotational axis are misaligned as in
pulsars, and from simulating the intrinsically 3D magneto--rotational
instability to the full extent. We think that the first shortcoming will
influence our results only slightly, as the initial magnetic field is
soon completely dominated in our models by the field generated by
differential rotation. Thus, provided there is a poloidal seed field of
similar strength no significant changes are to be expected. The second
restriction is probably uncritical for the relatively short time scales
covered by our simulations, as the dominant amplification process for
the magnetic field during these early stages is the rapid winding--up of
the initial poloidal field by means of differential rotation.
Additional field growth by the MRI or other MHD instabilities is
probably less important.  In any case, limited by the present
computational resources, a 3D MHD parameter study of core collapse is
still some years ahead.

A further limitation arises from our Newtonian approach, as we observe
outflow velocities in some of our models that exceed $\sim 30\,\%$ of
the speed of light and that are still increasing when we had to stop our
simulations.  Although being only mildly relativistic in our models, a
study of the further evolution of these outflows may require a special
relativistic MHD code, if the jets continue to accelerate.  The effects
of general relativistic corrections of the gravitational potential on
magneto--rotational core collapse arising due to the high compactness of
the collapsed cores have been investigated by us in some detail. The
results of this complementary study will be presented in a separate
publication.

\begin{acknowledgements}
  MO thanks H.\,Dimmelmeier for help in preparing, performing and
  understanding these simulations, particularly for providing the
  routines to compute the hydrodynamic initial models.  Part of the
  simulations were performed at the \emph{Rechenzentrum Garching} (RZG)
  of the Max--Planck--Gesellschaft. MAA is a Ram\'on y Cajal Fellow of
  the Spanish Ministry of Education and Science.  MAA acknowledges the
  partial support of the Spanish Ministerio de Ciencia y Tecnolog\'{\i}a
  (AYA2004-08067-C03-C01) and of the Sonderforschungsbereich-Transregio
  7 ``Gravitationswellenastronomie'' of the German Ministry of Science.
\end{acknowledgements}


\appendix

\section{Relaxing TVD}
\label{Sek:rTVD}

In this section we summarize the basics of the relaxing TVD method.  For
further information the reader is referred to
\citet{Pen_Arras_Wong_APJS_2003__rTVD_MHD_Code,Jin_Xin_CPAM_1995__rTVD,Trac_Pen_PASP_2003__Primer_rTVD}.

A major drawback of MUSCL--type and Riemann solver schemes for systems
of non--linear conservation laws is the need for calculating the
eigenvalues $a_m$, and the eigenvectors of the Jacobian of the flux
vector (Hebrew indices $(\aleph = 1, \ldots, m)$ enumerate the
components of the system of $m$ equations)
\begin{equation}
  J^{\aleph\beth} = \frac{\partial F^{\aleph}}{\partial U^{\beth}} \,, 
\end{equation}
which usually is a computationally quite demanding step.  This
difficulty can be overcome by employing a \emph{relaxing TVD} scheme
\citep{Jin_Xin_CPAM_1995__rTVD,Trac_Pen_PASP_2003__Primer_rTVD} that does not require the
explicit calculation of the eigenvectors of the system, and in some
cases of most of the eigenvalues. The numerical method we have applied
turns out to be both robust and accurate, and is well suited for
simulations of core collapse. This is good news, as the method involves
several considerable simplifications compared to more elaborate methods
based on Riemann solvers, which make it computationally very cheap, and
hence attractive.  One replaces the one--dimensional system of
conservation laws
\begin{equation}
  \label{Gl:1dKonla}
  \partial_t U^{\aleph} + \nabla F^{\aleph} = 0
\end{equation}
with the initial condition $U^{\aleph}(\xi,0) = U^{\aleph}_0(\xi)$ by
the relaxation system
\begin{equation}
  \partial_t U^{\aleph} + \nabla V^{\aleph} = 0 \,,
  \label{Gl:RelSys I}
\end{equation}
\begin{equation}
  \partial_t V^{\aleph} + A^2 \nabla U^{\aleph} 
     = -\frac{1}{\tau}(V^{\aleph} - F^{\aleph}) \,,
  \label{Gl:RelSys II}
\end{equation}
with the initial conditions $U^{\aleph}(\xi,0) = U^{\aleph}_0(\xi)$
and $V^{\aleph}(\xi,0) = F^{\aleph}(U^{\beth}_0(\xi))$.  The constant
$\tau$ is called the relaxation rate.  The constant matrix $A =
\textrm{diag}(a_1,...,a_m)$ satisfies the sub-characteristic condition
\begin{equation}
  A^2 - J^{2} \ge 0 \,.
\label{Gl:subchara}
\end{equation}
Thereby, the non--linear system (\ref{Gl:1dKonla}) gets replaced by a
system of linear equations that can be solved without the need of
Riemann solvers.  In the zero relaxation limit $\tau \to \infty$, one
arrives at the relaxed system
\begin{equation}
  \label{Gl:Relax 1}
  \partial_t U^{\aleph} + \nabla V^{\aleph} = 0 \,,
\end{equation}
\begin{equation}
  \label{Gl:Relax 2}
  \partial_t V^{\aleph} + A^2 \nabla U^{\aleph} = 0 \,,
\end{equation}
where $V^{\aleph}=F^{\aleph}(U^{\beth})$.  Introducing right (R) and
left (L) moving variables,
\begin{equation}
  \label{Gl:ReLi}
  U^{\aleph; R,L} = \frac{1}{2} \cdot (U^{\aleph} \pm A^{-1}F^{\aleph})
\end{equation}
the system decouples yielding
\begin{equation}
  \label{Gl:Relax 3,4}
  \partial_t U^{\aleph;R,L} \pm \nabla ( AU^{\aleph;R,L}) = 0 \,.
\end{equation}
These two systems (R, L) of equations have the structure of linear
advection equations for rightward and leftward directed advection,
respectively.  The constant advection velocities in both systems of
equations are given by
\begin{equation}
  \label{Gl:rTVD-Geschw}
  c_{\aleph}^{R,L} = \pm a_{\aleph} \,.
\end{equation}
Being of a quite simple structure and having quite simple
characteristics with constant speeds $c_{\aleph}^{R,L}$, the system
(\ref{Gl:Relax 3,4}) can easily be solved using a first- or
second--order accurate upwind scheme.  The upwind fluxes at the zone
interface $\mathcal{S}_{i+1/2}$ will be computed from variables of zone
$\mathcal{Z}_{i}$ for the right--moving variable $U^{\aleph;R}$, and
from zone $\mathcal{Z}_{i+1}$ for $U^{\aleph;L}$, respectively.  The
value of $U^{\aleph}$ can then be recovered from $U^{\aleph;R,L}$ by
\begin{equation}
  \label{Gl:rTVD-Reko}
  U^{\aleph} = \frac{1}{2} (U^{\aleph;R}+U^{\aleph;L}) \,.
\end{equation}

As \citet{Jin_Xin_CPAM_1995__rTVD} showed, the resulting scheme will be TVD, if the
sub-characteristic condition (\ref{Gl:subchara}), and the CFL
condition arising form the advection terms are fulfilled.

\section{Convergence tests}
\label{Sek:Convergence}

To ensure the numerical convergence of our results, we performed a
number of simulations of the same model using different grid spacings.
Based on these simulations, we selected a grid of 380 zones in radial
and 60 zones in angular direction as our standard grid. The grid is
logarithmically spaced in radius, with a central resolution of
$(\Delta_r)_{\mathrm{c}} \approx 300 \ \mathrm{m}$ and a relative
increase of the cell size from zone to the zone of $\approx 1.1 \,\%$.
The angular grid is uniform.

\begin{figure*}[!htbp]
  \centering
  \includegraphics[width=5.6cm]{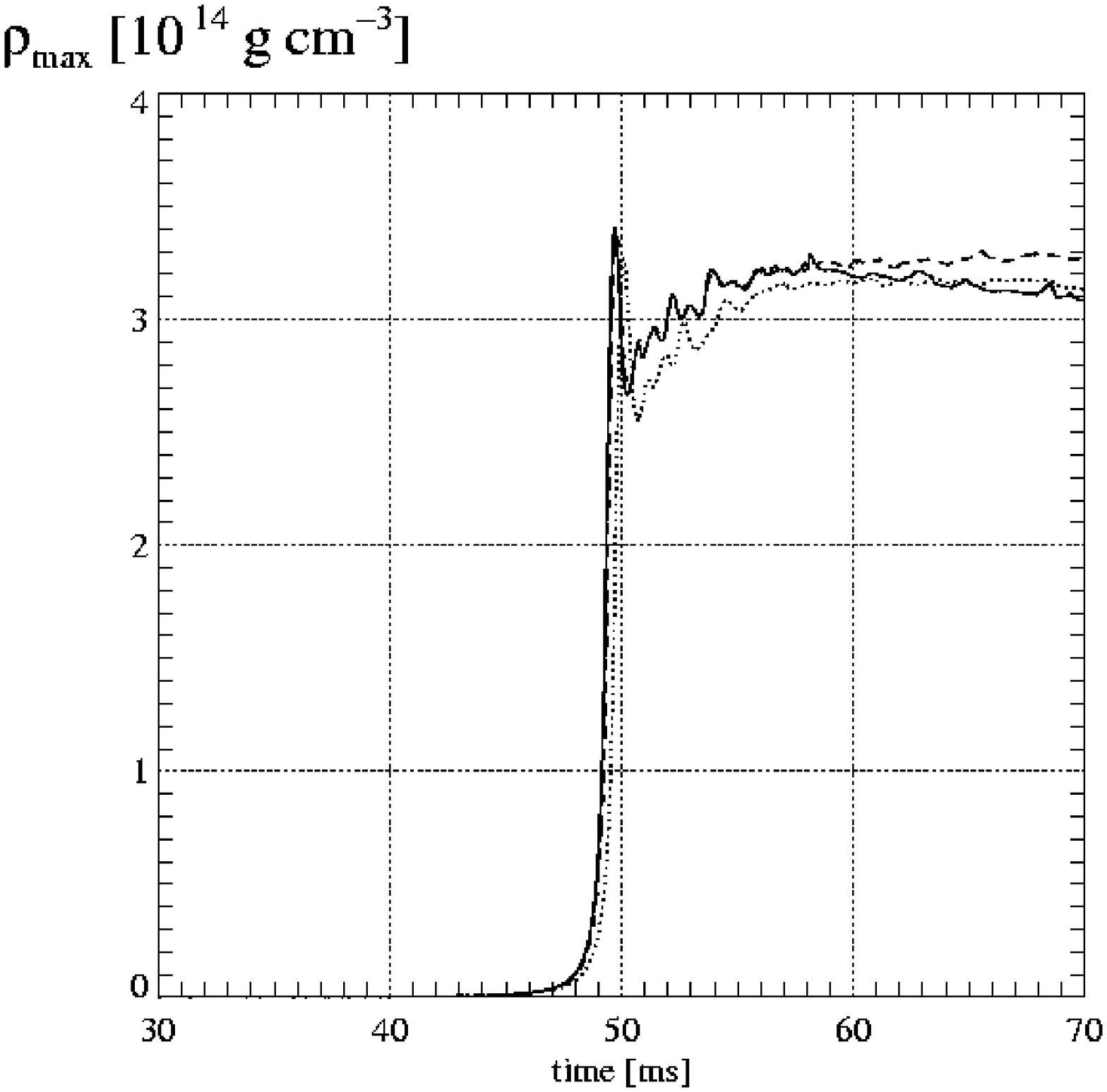}
  \includegraphics[width=5.6cm]{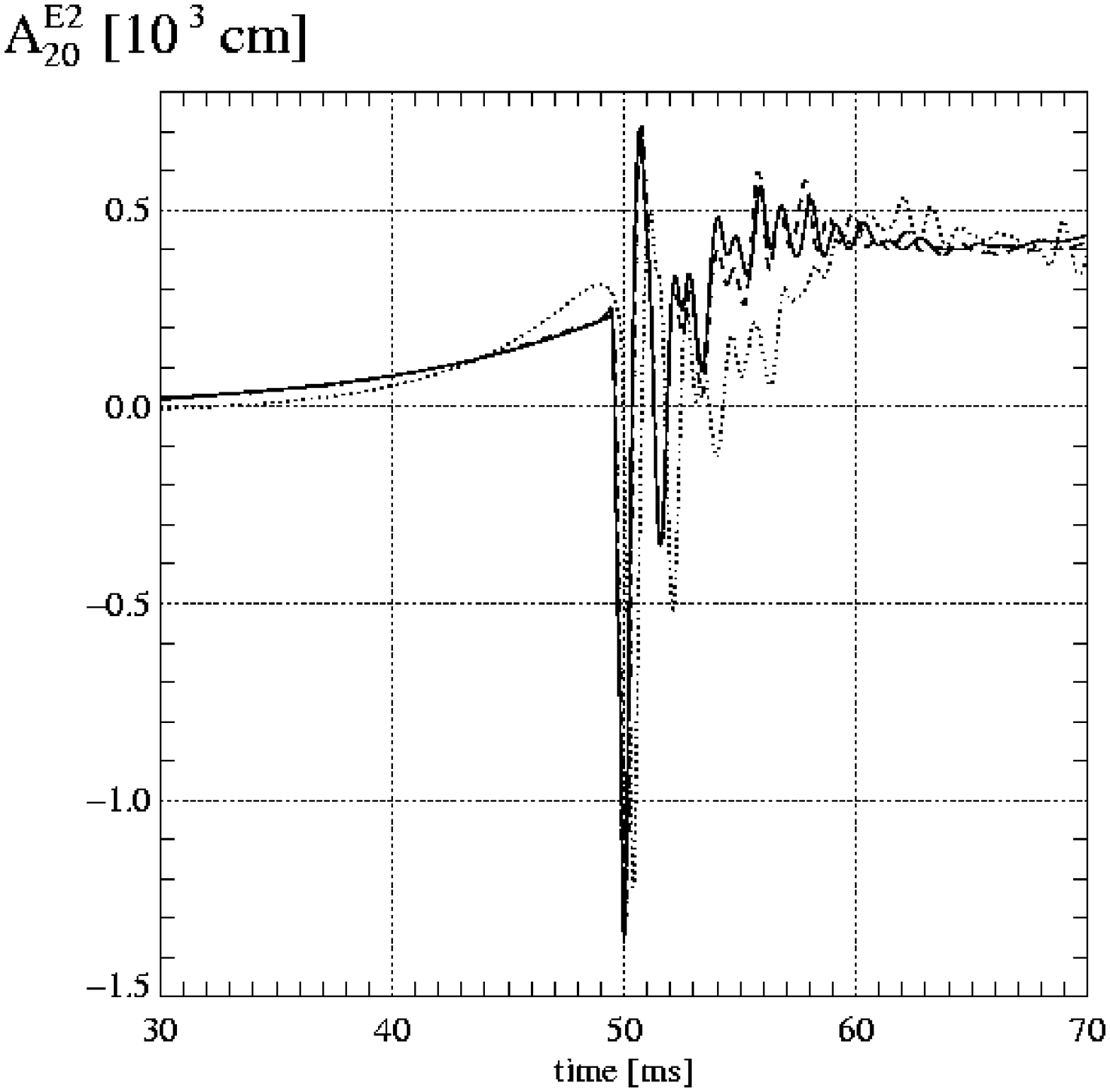}
  \includegraphics[width=5.6cm]{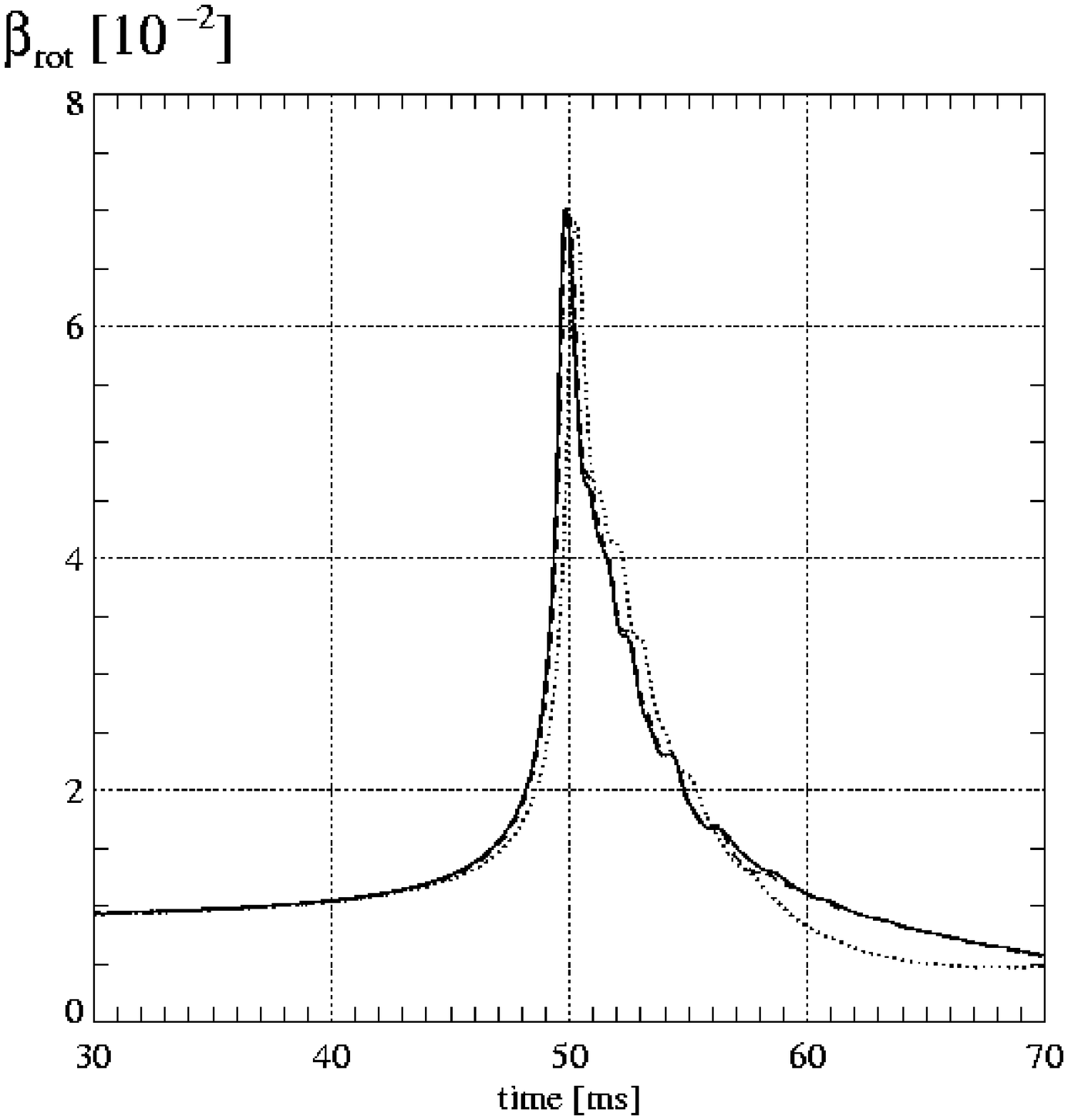}
  \caption[Convergence runs] 
  { The temporal evolution of the maximum density, the GW signal and the
    rotational energy parameter $\beta_{\mathrm{rot}}$ of model
    A1B3G3-D3M13 comparing simulations with grids of different
    resolution. Results from a run with the standard grid of 380
    logarithmically spaced zones in radius (central resolution of 300 m)
    and 60 zones uniformly distributed over $0 \leq \theta \leq \pi/2$
    are shown by the solid lines, and runs using a coarser grid of 280
    zones in radius (central resolution of 600 m) and 30 zones in angle
    and a finer grid of 580 zones (central resolution of 150 m) and 90
    zones in angle are plotted with dotted and dashed lines,
    respectively. 
  }
\label{Fig:convergence}
\end{figure*}

Results from a sample of the convergence runs are displayed in
Fig.\,\ref{Fig:convergence}. We show the temporal evolution of the
maximum density, the GW amplitude, and $\beta_{\mathrm{rot}}$ for model
A1B3G3-D3M13, computed with grids of
\begin{itemize}
\item 280 zones in radius and $(\Delta_r)_{\mathrm{c}} \approx 600 \ 
  \mathrm{m}$, and 30 zones in angle (dotted lines),
\item standard resolution as described above
  (solid lines), and
\item 580 zones in radius and $(\Delta_r)_{\mathrm{c}} \approx 150 \ 
  \mathrm{m}$, and 90 zones in angle (dashed lines).
\end{itemize}
We compare both local properties of the models such as the maximum
density of the core, and global ones such as the GW signal and the
evolution of $\beta_{\mathrm{rot}}$.  The results of these simulations
indicate that our standard resolution is adequate to follow the
evolution of our models.

\section{Gravitational-wave extraction}
\label{Sek:GW}

Since we do not evolve the consistent set of general--relativistic MHD
equations, we are not able to extract the GW signal from the metric.
But, in the post--Newtonian limit, it is still possible to extract the
lowest--order terms of the GW signal from the hydrodynamic variables.
For this purpose we have used the reformulation of the quadrupole
formula by \citet{Moenchmeyer_1991}, where temporal derivatives are
replaced by spatial ones using the continuity and the Euler equations.
Thus, the hydrodynamic fluxes and source terms are introduced into the
quadrupole formula.

As we use spherical coordinates $(r,\theta,\phi)$ in our simulations,
the metric perturbation associated with a gravitational wave is most
conveniently expanded in terms of pure spin tensor harmonics
$T^{\mathrm{E}2,lm}_{ab}$ and $T^{\mathrm{B}2,lm}_{ab}$ with the
corresponding ``electric'' and ``magnetic'' amplitudes
$A^{\mathrm{E}2}_{20}$ and $A^{\mathrm{B}2}_{20}$, respectively:
\begin{eqnarray}
  h_{ab} = 
  \frac{1}{R} 
  \sum_{l=2}^{\infty} \sum_{m=-l}^{+l}
  & & \left[  
    A^{\mathrm{E}2}_{lm}\left(t-\frac{R}{c}\right) 
    T^{\mathrm{E}2,lm}_{ab}(\theta,\phi) + 
  \right. 
  \nonumber \\
  & & \left. 
    + A^{\mathrm{B}2}_{lm}\left(t-\frac{R}{c}\right) 
      T^{\mathrm{B}2,lm}_{ab}(\theta,\phi) 
  \right].
\end{eqnarray}
Under the assumption of axisymmetry, the only non--vanishing amplitude
is $A^{\mathrm{E}2}_{20}$, which is given by \citep{Moenchmeyer_1991}:
\begin{equation}
  A^{E2}_{20} = \frac{\mathrm{d}^2}{\mathrm{d} t^2} M^{\mathrm{E}2}_{20} 
              = \frac{\mathrm{d}}{\mathrm{d} t} N^{\mathrm{E}2}_{20},
\end{equation}
where the radiative quadrupole moment $M^{\mathrm{E}2}_{20}$ is ($z =
\cos\theta$) defined as
\begin{equation} 
  \label{Gl:ME220}
  M^{\mathrm{E}2}_{20} 
  = \frac{G}{c^4}  \frac{32 \pi^{3/2}}{\sqrt{15}} 
  \int_0^1 \mathrm{d} z \int_0^{\infty} \mathrm{d} \frac{r^3}{3} \rho(r,z,t) r^2 
  \left(\frac{3}{2} z^2-\frac{1}{2} \right).
\end{equation}
Using the continuity equation, the first time derivative of
$M^{\mathrm{E}2}_{20}$ is given by
\begin{equation} 
  \label{Gl:NE220}
  N^{\mathrm{E}2}_{20} 
  = \frac{G}{c^4}  \frac{32 \pi^{3/2}}{\sqrt{15}} 
  \int_0^1 \mathrm{d} z \int_0^{\infty} \mathrm{d} \frac{r^3}{3} 
  \rho r \left[ v_r \left(\frac{3}{2} z^2-1 \right) - 
              3 v_{\theta} z \sqrt{1-z^2} \right] \, .
\end{equation}
The time derivative of $N^{\mathrm{E}2}_{20}$ is computed using the MHD
momentum equation which allows ones to eliminate
$\frac{\mathrm{d}}{\mathrm{d} t} (\rho v_r,\rho v_{\theta})$ with the
help of the hydro-magnetic and gravitational force terms given by
\begin{eqnarray}
  F^r_{\mathrm{MHD}} & = &
  - \frac{\partial r^2 F_{rr}}{\partial r^3/3 } 
  - \frac{1}{r} \frac{\partial \sin\theta F_{r\theta}}{
                      \partial(-\cos\theta)} + {} \nonumber \\
  {} & + & \frac{1}{r} ( 2P_{\star} + \rho v_{\theta}^2 - 
   b_{\theta}^2 + \rho v_{\phi}^2 - b_{\phi}^2 ), \\
  F^{\theta}_{\mathrm{MHD}} & = &
  - \frac{\partial r^2 F_{\theta r}}{\partial (r^3/3)}  
  - \frac{1}{r} \frac{\partial \sin\theta F_{\theta \theta}}{
                      \partial(-\cos\theta)} + {} \nonumber \\
  {} & + & \frac{\cot \theta}{r} (P_{\star} + \rho v_{\phi}^2 - 
   b_{\phi}^{2}) -  \frac{(\rho v_{r}v_{\theta} - b_{r}b_{\theta} )}{r}, \\
  F^{r}_{\mathrm{grav}} & = &
  - \rho \partial_r \Phi, \\
  F^{\theta}_{\mathrm{grav}} & = &
  - \frac{1}{r} \rho \partial_{\theta} \Phi       
  \, .
\end{eqnarray}
The components of the flux tensor in the above expressions are given
by
\begin{equation}
  F_{ab} = \rho v_{a} v_{b} + P_{\star} \delta_{ab} - b_{a} b_{b}.
\end{equation}  
Following the derivation of the quadrupole formula by \citep{Moenchmeyer_1991},
one then finds
\begin{eqnarray}
  \label{Gl:AE220}
  A^{\mathrm{E}2}_{20} & = &
  \frac{G}{c^4} \frac{32\pi^{\frac{3}{2}}}{\sqrt{15}}
  \int_{0}^{1} \mathrm{d} z \int_{0}^{\infty} \mathrm{d} \frac{r^3}{3} \nonumber \\
  && 
  \left[
    f_{rr} (3z^2-1) + f_{\theta\theta}(2-3z^2)  - f_{\phi\phi}
    - 6 f_{r\theta} z\sqrt{1-z^2} - \right. \nonumber \\
  && \left. - r \partial_r\Phi (3z^2-1) + 
            3 \partial_{\theta}\Phi z\sqrt{1-z^2}
  \right] \,,
\end{eqnarray}
where the components of $f_{ij}$ are given by
\begin{equation}
  f_{ij}  = \rho v_i v_j - b_i b_j \,.
\end{equation}
The total isotropic pressure $P_{\star}$ (sum of the gas pressure and
the magnetic pressure) cancels out, i.e.\,only the velocities and the
magnetic stresses appear in the hydro-magnetic part of the amplitude.

Eq.\,\ref{Gl:AE220} corresponds to the hydrodynamic and the Lorentz
force parts of the quadrupole formula given by
\citet{Kotake_etal_PRD_2004__MagCollapse_GW}, who derived the following
expression for the GW amplitude $A^{\mathrm{E}2}_{20} =
A^{\mathrm{E}2}_{20;\mathrm{hyd}} +
A^{\mathrm{E}2}_{20;\mathrm{Lorentz}} +
A^{\mathrm{E}2}_{20;\mathrm{mag}}$:
\begin{eqnarray}
  A^{\mathrm{E}2}_{20; \mathrm{hyd}}(t) & =  & 
  \frac{G}{c^4} \frac{16\pi^{\frac{3}{2}}}{\sqrt{15}}
  \int_{-1}^{1} \mathrm{d} z \int_{0}^{\infty} r^2 \mathrm{d} r \rho(r,z;t) {} 
  \nonumber \\
  & & {} 
  \left[
    v_rv_r(3z^2-1) + v_{\theta}v_{\theta}(2-3z^2) - \right. \nonumber \\
  & & \left. - v_{\phi}v_{\phi} - 6 v_r v_{\theta} z\sqrt{1-z^2}
  \right. {} \nonumber \\
  & & \left.  - r \partial_r\Phi (3z^2-1) + 3 \partial_{\theta}\Phi 
              z\sqrt{1-z^2} \right],
  \label{Gl:AE220hyd}
  \\
  A^{\mathrm{E}2}_{20; \mathrm{Lorentz}}(t) & = & 
  \frac{G}{c^4} \frac{16\pi^{\frac{3}{2}}}{\sqrt{15}}
  \int_{-1}^{1} \mathrm{d} z \int_{0}^{\infty} r^3 \mathrm{d} r 
  \frac{1}{c} \left[ (3z^2-1) (\vec j \times \vec B)_r-\right.{}\nonumber \\
  & & {}  \left. - 3 z\sqrt{1-z^2}(\vec j \times \vec B)_{\theta}
  \right],
  \label{Gl:AE220Lorentz}
  \\
  A^{\mathrm{E}2}_{20;\mathrm{mag}}(t) & = &
  \frac{G}{c^4} \frac{16\pi^{\frac{3}{2}}}{\sqrt{15}}
  \int_{-1}^{1} \mathrm{d} z \int_{0}^{\infty} \mathrm{d} r \frac{1}{8\pi c} \frac{\mathrm{d}}{\mathrm{d} t} \nonumber \\
  & & \left[ 
    \partial_{\theta}\left(B_rr^3(3z^2-1)\right) E_{\phi} - \right.\nonumber \\
  & & \left. \partial_r \left(B_{\theta} r^3(3z^2-1)\right)rE_{\phi} + 
      \right.{} \nonumber \\
  & & {} \left.
    + \partial_r\left( B_{\phi}r^3(3z^2-1)\right) rE_{\theta} -
  \right. {} \nonumber \\
  & & {} \left.
    - \frac{1}{\sin\theta} \partial_{\theta} \left(B_{\phi} 
               \sin\theta r^3 (3z^2-1) \right)E_r
  \right].
  \label{Gl:AE220mag}
\end{eqnarray}
To evaluate these expressions, the current density $\vec j$ and the
electric field $\vec E$ have to be calculated from the magnetic field
and the velocity.  The last term (Eq.\,\ref{Gl:AE220mag}) describes the
contribution resulting from the energy density of the magnetic field.
\citet{Kotake_etal_PRD_2004__MagCollapse_GW} found that the hydrodynamic
and the Lorentz force contributions are at least two orders of magnitude
larger than the magnetic energy one.  Therefore, and due to the time
derivatives still involved in its calculation, we neglect this
contribution and consider only the GW amplitude resulting from the
quadrupole moment of the matter.

\section{Synopsis of our results}
\label{Sek:Syn}

Tables \ref{Tab:SynopseI} through \ref{Tab:SynopseIII} provide an
overview of the dynamic evolution of the flow and the magnetic field,
and about the resulting gravitational wave signal of all our models.

\begin{table*}
  \caption[Global parameters of the models I] 
  { 
    Some characteristic
    model quantities: the first two columns give the model name and the
    classification of the GW signal (for the corresponding
    non--magnetized model). Columns\,3 and 4 give the time of bounce
    $t_{\mathrm{b}}$ (in milliseconds) and the maximum density at bounce
    $\rho_{\mathrm{b}}$ (in units of $10^{14}\ \mathrm{cm \,
      s}^{-1}$). An exclamation 
    mark behind the density value signifies that the maximum density of
    the model exceeds the bounce density during the later evolution.
    $A^{\mathrm{E}2}_{20}$ (column 5) and $A^{\mathrm{E}2}_{20;
      \mathrm{mag}}$ (column 6) are the maximum GW amplitude (in cm) and
    the corresponding magnetic contribution.  $A^{\mathrm{E}2}_{20;
      \infty}$ (Column\,7) is a \emph{rough} mean value of the wave
    amplitude (in cm) at some late epoch; no value is provided when the
    GW amplitude does not approach a quasi--constant asymptotic value.
    If the absolute value of this amplitude is large, the presence of an
    aspheric outflow at late epochs can be inferred.  The following
    columns give the maximum value of the rotational (column 8) and the
    magnetic beta parameter (column 9), the time when
    $\beta_{\mathrm{mag}}$ reaches its maximum (column 10), and the
    corresponding beta of the toroidal field (column 11).  If the
    magnetic field is still amplifying at the end of the simulation, an
    exclamation mark is added behind the table entry, and if the
    magnetic field is decreasing at this time, we give its final
    value $\beta_{\mathrm{mag}}^{\mathrm{fin}}$ in parentheses. 
  }
  \label{Tab:SynopseI}
  \centering
  \begin{tabular}{ccccccccccc}
    \hline\hline
    Model & type & $t_{\mathrm{b}}$ & $\rho_{\mathrm{b}}$ & 
    $A^{\mathrm{E}2}_{20}$
    & $A^{\mathrm{E}2}_{20; \mathrm{mag}}$ & $A^{\mathrm{E}2}_{20; \infty}$ &  
    $\beta_{\mathrm{rot}}^{\mathrm{max}}$  & 
    $\beta_{\mathrm{mag}}^{\mathrm{max}} \
    ( \beta_{\mathrm{mag}}^{\mathrm{fin}} ) $  &  
    $t_{\mathrm{m}}$ & 
    $\beta_{\mathrm{mag},\phi}^{\mathrm{max}}$ \\ 
    &                                      & $[\mathrm{ms}]$  & 
    $[10^{14}\ \mathrm{\frac{g} {cm^3}}]$    & 
    $[\mathrm{cm}]$ &    $[\mathrm{cm}]$   &    
    $[\mathrm{cm}]$ & $\%$ & $\%$ & $[\mathrm{ms}]$ & $\%$ \\
    \hline
    A1B1G3-D3M10 & I & 
    $49.29$ & $3.79$ & $-308.8$ & $0.0363$ & $10$ & $2.9$ &
    $0.036$! & $117.5$ &  $0.036$ \\  
    A1B1G3-D3M11 & I & 
    $49.29$ & $3.79$ & $-307.9$ & $0.33$ & $10$ & $2.9$ &
    $0.39$! & $96.1$ &  $0.38$ \\  
    A1B1G3-D3M12 & I & 
    $49.29$ & $3.79$ & $-282.3$ & $9.57$ & $10$ & $2.9$ & $0.74 \ (0.37)$ &
    $64.7$ &  $0.62$ \\  
    A1B1G3-D3M13 & I & 
    $50.46$ & $3.74$ & $-572.2$ & $-111.2$ & $130$ & $2.6$ & $1.9 \ (1.4)$
    & $55.4$ &  $0.71$ \\  
    \hline
    A1B3G1-D3M10 & II & 
    $95.16$ & $2.11$ & $-1305$ & $0.0028$ & $90$ & $10.8$ &
    $4.3\cdot 10^{-5}$! & $129.5$ &  $4.2\cdot 10^{-5}$ \\  
    A1B3G1-D3M11 & II & 
    $95.16$ & $2.11$ & $-1305$ & $0.26$ & $90$ &   $10.8$ &
    $4.7\cdot 10^{-3}$! & $130.3$ & $4.5\cdot 10^{-3}$ \\  
    A1B3G1-D3M12 & II & 
    $95.23$ & $2.11$ & $-1297$ & $15.3$ & $$ &   $10.7$ & $0.42$! &
    $148.7$ & $0.36$ \\  
    A1B3G1-D3M13 & II & $102.1$ & $1.10$! & $-900.6$ & $59.6$  & $$ &
    $7.6$ & $2.2$ & $115.0$ & $1.1$ \\  
    \hline
    A1B3G3-D3M10 & I & 
    $48.62$ & $3.40$ & $-1037$ & $0.0093$ & $40$ & $8.1$ &
    $2.0\cdot 10^{-3}$! & $66.16$ &  $2.0\cdot 10^{-3}$ \\  
    A1B3G3-D3M11 & I & 
    $48.62$ & $3.40$ & $-1037$ & $0.85$ & $40$ &   $8.1$ &
    $0.048$! & $58.5$ & $0.048$ \\  
    A1B3G3-D3M12 & I & 
    $48.64$ & $3.40$ & $-1016$ & $31.3$ & $80$ &   $8.1$ & $1.1 \ (1.0)$ & 
    $71.3$ & $0.85$ \\  
    A1B3G3-D3M13 & I & $49.68$ & $3.41$ & $-1344$ & $191$  & $420$ &
    $7.0$ & $3.1 \ (2.1) $ & $53.69$ & $1.7$ \\  
    \hline
    A1B3G5-D3M10 & III & $29.94$ & $4.21$ & $133.7$ & $-2.2\cdot 10^{-5}$ & $16$ 
    & $3.5$ & $1.8\cdot 10^{-4}$! & $48.33$ & $1.8\cdot 10^{-4}$ \\ 
    A1B3G5-D3M11 & III & $29.94$ & $4.21$ & $133.8$ & $-2.3\cdot 10^{-3}$ & $16$ 
    & $3.5$ & $0.017$! & $48.11$ & $0.017$ \\ 
    A1B3G5-D3M12 & III & $29.94$ & $4.21$ & $136.4$ & $-0.38$ & $5.5$ &
    $3.5$ & $6.3$! & $61.5$ & $0.51$ \\  
    A1B3G5-D3M13 & III & $30.08$ & $4.21$ & $259$ & $-52$ & $18$      &
    $3.2$ & $2.8 \ ( 1.8 )$ & $33.84$ & $1.1$ \\  
    \hline
    A2B4G1-D3M10 & II & $99.87$ & $0.114$ & $-608.6$ & $0.0014$ & $10$ & 
    $11.8$ & $1.08\cdot 10^{-4}$! & $215.6$ & $1.07\cdot 10^{-4}$ \\ 
    A2B4G1-D3M11 & II & 
    $99.87$ & $0.114$ & $-608.6$ & $0.13$ & $10$ & $11.8$ &
    $3.6\cdot 10^{-3}$! & $153.6$ & $3.4\cdot 10^{-3}$ \\  
    A2B4G1-D3M12 & II & $99.96$ & $0.114$ & $-606.3$ & $7.1$ & $10$ &
    $11.8$ & $0.19$! & $159.8$ & $0.17$ \\  
    A2B4G1-D3M13 & II & $112.5$ & $0.0685$! & $-441.0$ & $111.1$ & $80$
    & $9.3$ & $2.4!$ & $147.1$ & $1.4$ \\  
    \hline
    A2B4G4-D3M10 & I & $39.77$ & $2.79$ & $-743.8$ & $0.0017$ & $0$ &  
    $15.3$ & $7.9\cdot 10^{-4}$! & $74.8$ & $7.8\cdot 10^{-4}$ \\ 
    A2B4G4-D3M11 & I & $39.77$ & $2.79$ & $-743.7$ & $0.17$ & $0$ & 
    $15.3$ & $0.062$! & $70.7$ & $0.062$ \\ 
    A2B4G4-D3M12 & I & $39.77$ & $2.80$ & $-742.3$ & $14.83$ & $100$ & 
    $15.3$ & $0.79$! & $48.98$ & $0.74$ \\ 
    A2B4G4-D3M13 & I & $40.31$ & $2.88$! & $-720.1$ & $370.1$ & $400$ & 
    $13.9$ & $4.9 \ (4.4)$ & $43.37$ & $3.1$ \\ 
    \hline
    A2B4G5-D3M10 & III & $30.37$ & $3.53$ & $331.6$ & $2.0\cdot 10^{-4}$ &
    $-30$ &  $9.6$ & $1.6\cdot 10^{-4}$! & $45.6$ & $1.6\cdot 10^{-4}$ \\ 
    A2B4G5-D3M11 & III & $30.37$ & $3.53$ & $331.7$ & $2.0\cdot 10^{-2}$ &
    $-30$ & $9.6$ & $0.03.6$! & $52.9$ & $0.036$ \\ 
    A2B4G5-D3M12 & III & $30.37$ & $3.53$! & $331.3$ & $1.7$ & $25$ &
    $9.6$ & $1.4 \ ( 1.3) $ & $65.8$ & $1.1$ \\  
    A2B4G5-D3M13 & III & $30.45$ & $3.53$! & $338.1$ & $-29.4$ & $140$ &
    $9.0$ & $5.2 \ (3.2)$ & $37.6$ & $2.6$ \\ 
    \hline
    A3B2G4-D3M10 & I & $39.15$ & $3.45$ & $-734.1$ & $0.0011$ & $10$ &
    $9.1$ & $5.5\cdot 10^{-4}$! & $58.4$ & $5.5\cdot 10^{-4}$ \\
    A3B2G4-D3M11 & I & $39.15$ & $3.45$ & $-734.1$ & $0.11$ & $10$ &
    $9.1$ & $0.051$! & $57.9$ & $0.050$ \\
    A3B2G4-D3M12 & I & $39.16$ & $3.45$ & $-726.2$ & $9.6$ & $16$ &
    $9.1$ & $1.0$! & $49.7$ & $0.92$ \\
    A3B2G4-D3M13 & I & $39.70$ & $3.48$! & $-626$ & $39.94$ & $250$ &
    $8.0$ & $3.6 \ (3.0)$ & $44.23$ & $1.8$ \\
    \hline
    A3B3G3-D3M10 & II/I & 
    $49.70$ & $2.41$ & $-1400$ & $0.017$ & $30$ & $16$ &
    $1.3\cdot 10^{-3}$! & $75.0$ & $1.2\cdot 10^{-3}$ \\  
    A3B3G3-D3M11 & II/I & $49.70$ & $2.41$ & $-1401$ & $1.5$ & $30$ &
    $16$ & $0.18$! & $73.4$ & $0.18$ \\  
    A3B3G3-D3M12 & II/I & $49.71$ & $2.42$! & $-1379$ & $56$ & $40$ &
    $16$ & $14$! & $71.1$ & $11$ \\  
    A3B3G3-D1M13 & II/I & $49.79$ & $2.39$ & $-1400$ & $2.2$ & $40$ &
    $16$ & $0.49$! & $77.1$ & $0.46$ \\  
    A3B3G3-D3M13 & II/I & $51.00$ & $2.68$! & $-1128$ & $705$ & $600$ &
    $13.5$ & $5.3 \ (2.6)$ & $52.9$ & $3.1$ \\  
    A3B3G3-D0M13 & II/I & $50.41$ & $2.48$! & $-1339$ & $204$ & $600$ &
    $15$ & $21 \ (7.4)$ & $0.93$ & $2.4\cdot 10^{-4}$ \\  
    \hline
    A3B3G4-D3M10 & I & 
    $39.64$ & $2.86$ & $-895.7$ & $0.0021$ & $-25$ & $14.6$ &
    $0.017$! & $145.8$ & $0.017$ \\  
    A3B3G4-D3M11 & I & 
    $39.64$ & $2.86$ & $-894.9$ & $0.21$ & $15$ & $14.6$ &
    $0.092$! & $69.7$ & $0.092$ \\  
    A3B3G4-D3M12 & I & 
    $39.65$ & $2.86$ & $-888.4$ & $18.1$ & $65$ & $14.5$ & $1.3 \ (1.2)$ &
    $52.4$ & $1.1$ \\  
    A3B3G4-D3M13 & I & 
    $40.14$ & $2.96$! & $-871.8$ & $412.4$ & $450$ & $13.1$ & $4.8 \ (4.8)$ &
    $43.0$ & $2.9$ \\  
    \hline
    A3B3G5-D3M10 & III & $30.35$ & $3.47$ & $262.5$ & $1.3\cdot 10^{-5}$ &
    $-30$ &  $9.6$ & $1.6\cdot 10^{-4}$! & $44.3$ & $1.5\cdot 10^{-4}$ \\  
    A3B3G5-D3M11 & III & $30.35$ & $3.47$ & $262.5$ & $1.3\cdot 10^{-3}$ &
    $-30$ & $9.5$ & $0.016$! & $44.29$ & $0.01.6$ \\  
    A3B3G5-D0M12 & III & $30.34$ & $3.58$ & $332.1$ & $-5.5$ & $-50$ &
    $9.5$ & $0.70$! & $52.7$ & $0.57$ \\  
    A3B3G5-D1M12 & III & $30.35$ & $3.47$ & $262.4$ & $0.034$ & $-10$ &
    $9.5$ & $0.12$! & $56.3$ & $0.12$ \\  
    A3B3G5-D2M12 & III & $30.35$ & $3.47$! & $261.6$ & $0.47$ & $-25$ &
    $9.5$ & $1.0$ & $45.01$ & $0.93$ \\  
    A3B3G5-D3M12 & III & $30.35$ & $3.47$! & $264.3$ & $-.079$ & $-7.5$
    & $9.5$ & $1.6$! & $61.9$ & $1.4$ \\  
    A3B3G5-D4M12 & III & 
    $30.35$ & $3.47$! & $262.0$ & $-1.6$ & $-3$ & $9.5$ &
    $1.0$! & $64.0$ & $0.85$ \\  
    A3B3G5-D3M13 & III & $30.56$ & $3.47$! & $343$ & $-31.9$ & $130$ &
    $8.9$ & $5.1 \ (3.1)$ & $37.4$ & $2.5$ \\  
    \hline\hline
  \end{tabular}
\end{table*}

\begin{table*}[htbp]
  \caption{Continuation of Table \ref{Tab:SynopseI}}
  
  \label{Tab:SynopseIa}
  \centering
  
  \begin{tabular}{ccccccccccc}
    \hline\hline
    Model & type & $t_{\mathrm{b}}$ & $\rho_{\mathrm{b}}$ & 
    $A^{\mathrm{E}2}_{20}$
    & $A^{\mathrm{E}2}_{20; \mathrm{mag}}$ & $A^{\mathrm{E}2}_{20; \infty}$ &  
    $\beta_{\mathrm{rot}}^{\mathrm{max}}$  & 
    $\beta_{\mathrm{mag}}^{\mathrm{max}} \
    ( \beta_{\mathrm{mag}}^{\mathrm{fin}} ) $  &  
    $t_{\mathrm{m}}$ & 
    $\beta_{\mathrm{mag},\phi}^{\mathrm{max}}$ \\ 
    &                                      & $[\mathrm{ms}]$  & 
    $[10^{14}\ \mathrm{\frac{g} {cm^3}}]$    & 
    $[\mathrm{cm}]$ &    $[\mathrm{cm}]$   &    
    $[\mathrm{cm}]$ & $\%$ & $\%$ & $[\mathrm{ms}]$ & $\%$ \\
    \hline
    A4B5G5-D3M10 & I/II 
    & $30.80$ & $1.97$ & $-4141$ & $0.0070$ & $140$ & $34.6$ &
    $3.3\cdot 10^{-4}$! & $64.0$ & $3.2\cdot 10^{-4}$ \\ 
    A4B5G5-D3M11 & I/II
    & $30.80$ & $1.97$ & $-4140$ & $0.69$ & $140$ & $34.6$ &
    $0.022$! & $57.6$ & $0.021$\\  
    A4B5G5-D3M12 & I/II 
    & $30.79$ & $2.00$ & $-4101$ & $43.6$ & $140$ & $34.5$ & $0.60$!
    & $32.1$ & $0.52$ \\  
    A4B5G5-D3M13 & I/II
    & $30.93$ & $2.10$! & $-3473$ & $1225$ & $1900$ & $34.5$ & $11.1 \ (6.2)$
    & $36.2$ & $7.0$ \\  
    \hline\hline
  \end{tabular}
\end{table*}

\begin{table*}
  \caption[Global parameters of the models II] { Some characteristic
  model quantities (name of model given in column 1) at time $t$ (in
  msec; column 2) when the core has reached a quasi--equilibrium
  state.  For models which do not reach a quasi--equilibrium state
  until the end of the simulation (e.g.\,type--II models with large
  scale core pulsations) we provide upper (top value) and lower
  (bottom value) bounds estimated from the values at maximum and
  minimum contraction.  Columns\,3 and 4 give the surface radius
  $r_{\mathrm{c}}$ (in km) and the mass $M_{\mathrm{c}}$ (in solar
  masses) of the quasi--equilibrium configuration, respectively.
  Since it is still surrounded by an (expanding) envelope of high
  density matter, the definition of its surface radius
  $r_{\mathrm{c}}$ is somewhat uncertain.  As the rotation rate
  $2\pi/\Omega$ (in msec), where $\Omega$ is the angular velocity
  averaged over the angle $\theta$, as well as the total magnetic
  field $|\vec b|$ and (the absolute value of) its toroidal component
  $b_{\mathrm{\phi}}$ (both in Gauss) vary strongly near the surface
  and on short time scales, the corresponding values in columns\,5, 6
  and 7 should be used with care.  Negative values of the rotation
  rate signify counter--rotating cores. Finally, in columns\,8 and 9
  we give the radii of the shock at the polar axis,
  $r_{\mathrm{sh}}^{\mathrm{p}}$, and at the equator,
  $r_{\mathrm{sh}}^{\mathrm{e}}$ (both in cm), respectively.  No entry
  in these columns implies that the shock has already left the
  computational grid.  }
  \label{Tab:SynopseII} \centering
  
  \begin{tabular}{ccccccccc}
    \hline\hline
    Model & $t$ & $r_{\mathrm{c}}$ & $M_{\mathrm{c}}$ 
    & $2\pi/\Omega$ &
    $|\vec b|$ & $|b_{\phi}|$ & 
    $r_{\mathrm{sh}}^{\mathrm{p}}$ & $r_{\mathrm{sh}}^{\mathrm{e}}$ \\
    & $[\mathrm{ms}]$ & $[\mathrm{km}]$ & $[\mathrm{M_{\odot}]}$ 
    & $[\mathrm{ms}]$ &
    $[\mathrm{G}]$ & $[\mathrm{G}]$ & 
    $[\mathrm{km}]$ & $[\mathrm{km}]$ \\
    \hline
    A1B1G3-D3M10 & $75$ & $22.5$ & $0.59$ & $9.6$ & $4.9\cdot 10^{13}$ &
    $4.7\cdot 10^{13}$ & $725$ & $692$ \\ 
    A1B1G3-D3M11 & $75$ & $22.3$ & $0.59$ & $8.2$ & $5.7\cdot 10^{14}$ &
    $5.6\cdot 10^{13}$ & $725$ & $692$ \\ 
    A1B1G3-D3M12 & $75$ & $21.5$ & $0.59$ & $11.9$ & $8.7\cdot 10^{14}$ &
    $4.8\cdot 10^{14}$ & $725$ & $692$ \\ 
    A1B1G3-D3M13 & $75$ & $23.4$ & $0.61$ & $-104.9$ & $8.5\cdot 10^{14}$ &
    $4.6\cdot 10^{13}$ & $768$ & $669$ \\ 
    \hline
    A1B3G1-D3M10 & $116$ & $49.8$ & $1.2$ & $8.0$ & $1.9\cdot 10^{12}$ &
    $1.9\cdot 10^{12}$ & $920$ & $745$ \\ 
                 & $126$ & $129$ & $1.2$ & $52.4$ & $3.6\cdot 10^{11}$ &
                 $3.6\cdot 10^{11}$ & & \\ 
    A1B3G1-D3M11 & $116$ & $49.7$ & $1.2$ & $8.0$ & $1.9\cdot 10^{13}$ &
    $1.9\cdot 10^{13}$ & $920$ & $745$ \\ 
                 & $126$ & $129$ & $1.2$ & $46.7$ & $3.2\cdot 10^{12}$ &
                 $3.2\cdot 10^{12}$ & & \\ 
    A1B3G1-D3M12 & $116$ & $49.0$ & $1.2$  & $7.9$ & $2.0\cdot 10^{14}$ &
    $1.9\cdot 10^{14}$ & $920$ & $745$ \\ 
                 & $125$ & $129$ & $1.2$ & $45.5$ & $3.2\cdot 10^{13}$ &
                 $3.2\cdot 10^{13}$ & & \\ 
    A1B3G1-D3M13 & $117$ & $35.3$ & $1.1$  & $6.3$ & $1.4\cdot 10^{15}$ &
    $4.7\cdot 10^{14}$ & $813$ & $568$ \\ 
    \hline
    A1B3G3-D3M10 & $70$ & $25.5$ & $0.68$ & $6.6$ & $5.4\cdot 10^{13}$ &
    $5.4\cdot 10^{13}$ & $617$ & $544$ \\ 
    A1B3G3-D3M11 & $58.5$ & $29.1$ & $0.68$ & $8.3$ & $2.2\cdot 10^{14}$ &
    $2.2\cdot 10^{14}$ & $330$ & $2.77$ \\ 
    A1B3G3-D3M12 & $70$ & $23.6$ & $0.70$ & $4.6$ & $1.4\cdot 10^{15}$ &
    $1.2\cdot 10^{15}$ & $625$ & $557$ \\ 
    A1B3G3-D3M13 & $70$ & $27.1$ & $0.73$ & $140.7$ & $9.9\cdot 10^{14}$ &
    $1.2\cdot 10^{14}$ & $1068$ & $568$ \\ 
    \hline
    A1B3G5-D3M10 & $48$ & $13.6$ & $0.21$ & $4.9$ & $4.4\cdot 10^{13}$ &
    $4.4\cdot 10^{13}$ & $282$ & $278$ \\ 
    A1B3G5-D3M11 & $48$ & $13.6$ & $0.21$ & $4.8$ & $4.9\cdot 10^{14}$ &
    $4.9\cdot 10^{14}$ & $282$ & $278$ \\ 
    A1B3G5-D3M12 & $48$ & $14.2$ & $0.22$ & $4.9$ & $8.8\cdot 10^{14}$ &
    $7.2\cdot 10^{14}$ & $282$ & $278$ \\ 
    A1B3G5-D3M13 & $48$ & $13.6$ & $0.24$ & $-215$ & $1.9\cdot 10^{15}$ &
    $4.8\cdot 10^{13}$ & $285$ & $278$ \\ 
    \hline
    A2B4G1-D3M10 & $114$ & $119$ & $0.86$ & $58.1$ & $3.9\cdot 10^{11}$ &
    $3.9\cdot 10^{11}$ & $518$ & $438$ \\ 
                 & $151$ & $110$ & $1.24$ & $29.4$ & $3.6\cdot 10^{11}$ &
                 $3.6\cdot 10^{11}$ &   &  \\ 
                 & $179$ & $118$ & $0.91$ & $57.0$ & $6.9\cdot 10^{11}$ &
                 $6.9\cdot 10^{11}$ &   &  \\ 
    A2B4G1-D3M11 & $115$ & $119$ & $0.85$ & $58.0$ & $4.0\cdot 10^{12}$ &
    $3.9\cdot 10^{12}$ & $518$ & $438$ \\ 
                 & $151$ & $111$ & $1.24$ & $29.4$ & $3.6\cdot 10^{12}$ &
                 $3.6\cdot 10^{12}$ &   &  \\ 
    A2B4G1-D3M12 & $124$ & $120$ & $0.76$ & $72.2$ & $2.6\cdot 10^{13}$ &
    $2.1\cdot 10^{13}$ & $881$ & $634$ \\ 
    A2B4G1-D3M13 & $123$ & $120$ & $1.1$ &  $40.3$ & $1.5\cdot 10^{14}$ &
    $1.1\cdot 10^{14}$ & $570$ & $506$ \\ 
                 & $143$ & $57.4$ & $1.1$ & $8.3$ & $5.4\cdot 10^{14}$ &
                 $3.3\cdot 10^{14}$ & &  \\ 
                 & $146$ & $60.6$ & $1.1$ & $8.2$ & $5.8\cdot 10^{14}$ &
                 $3.1\cdot 10^{14}$ & & \\ 
    \hline
    A2B4G4-D3M10 & $48$ & $28$ & $0.47$ & $6.4$ & $8.8\cdot 10^{12}$ &
    $8.6\cdot 10^{12}$ & $246$ & $186$ \\ 
                 & $60$ & $29.5$ & $0.48$ & $7.1$ & $1.1\cdot 10^{13}$ &
                 $1.1\cdot 10^{13}$ & $482$ & $370$ \\ 
    A2B4G4-D3M11 & $48$ & $28$ & $0.47$ & $6.4$ & $8.7\cdot 10^{13}$ &
    $8.6\cdot 10^{13}$ & $246$ & $184$ \\ 
                 & $60$ & $29.4$ & $0.48$ & $7.0$ & $9.3\cdot 10^{13}$ &
                 $9.0\cdot 10^{13}$ & $482$ & $370$ \\ 
    A2B4G4-D3M12 & $48$ & $28.3$ & $0.47$ & $6.0$ & $8.0\cdot 10^{14}$ &
    $7.8\cdot 10^{14}$ & $266$ & $201$ \\ 
    A2B4G4-D3M13 & $48$ & $17.6$ & $0.48$ & $10.6$ & $1.9\cdot 10^{15}$ &
    $4.5\cdot 10^{14}$ & $332$ & $249$ \\ 
    \hline
    A2B4G5-D3M10 & $50$ & $13.9$ & $0.21$ & $3.1$ & $2.7\cdot 10^{13}$ &
    $2.7\cdot 10^{13}$ & $308$ & $290$ \\ 
    A2B4G5-D3M11 & $50$ & $13.9$ & $0.22$ & $3.1$ & $3.4\cdot 10^{14}$ &
    $3.4\cdot 10^{14}$ & $308$ & $290$ \\ 
    A2B4G5-D3M12 & $50$ & $12.0$ & $0.23$ & $3.2$ & $3.2\cdot 10^{15}$ &
    $2.9\cdot 10^{15}$ & $316$ & $293$ \\ 
    A2B4G5-D3M13 & $50$ & $15.2$ & $0.28$ & $-70.7$ & $2.4\cdot 10^{15}$ &
    $1.1\cdot 10^{14}$ & $891$ & $320$ \\ 
    \hline\hline
  \end{tabular}
\end{table*}

\begin{table*}
  \caption[Global parameters of the models III]
  {
    Continuation of Table\,\ref{Tab:SynopseII}.
  }
  \label{Tab:SynopseIII}
  \centering
  
  \begin{tabular}{ccccccccc}
    \hline\hline
    Model & $t$ & $r_{\mathrm{c}}$ & $M_{\mathrm{c}}$ 
    & $2\pi/\Omega$ &
    $|\vec b|$ & $|b_{\phi}|$ & 
    $r_{\mathrm{sh}}^{\mathrm{p}}$ & $r_{\mathrm{sh}}^{\mathrm{e}}$ \\
    & $[\mathrm{ms}]$ & $[\mathrm{km}]$ & $[\mathrm{M_{\odot}]}$ 
    & $[\mathrm{ms}]$ &
    $[\mathrm{G}]$ & $[\mathrm{G}]$ & 
    $[\mathrm{km}]$ & $[\mathrm{km}]$ \\
    \hline
    A3B2G4-D3M10 & $57$ & $19.6$ & $0.44$ & $4.6$ & $4.7\cdot 10^{13}$ &
    $4.7\cdot 10^{13}$ & $378$ & $340$ \\ 
    A3B2G4-D3M11 & $57$ & $20.0$ & $0.44$ & $4.8$ & $4.0\cdot 10^{14}$ &
    $4.0\cdot 10^{14}$ & $382$ & $340$ \\ 
    A3B2G4-D3M12 & $57$ & $28.2$ & $0.44$ & $8.7$ & $8.7\cdot 10^{14}$ &
    $8.4\cdot 10^{14}$ & $382$ & $348$ \\ 
    A3B2G4-D3M13 & $57$ & $17.1$ & $0.45$ & $41.2$ & $2.4\cdot 10^{15}$ &
    $2.1\cdot 10^{14}$ & $906$ & $405$ \\ 
    \hline
    A3B3G3-D3M10 & $64.7$ & $54.9$ & $0.73$ & $20.3$ & $2.0\cdot 10^{13}$ &
    $1.9\cdot 10^{13}$ & $491$ & $398$ \\ 
                 & $67.8$ & $37.4$ & $0.71$ & $10.0$ & $1.1\cdot 10^{13}$ &
                 $1.1\cdot 10^{13}$ & $564$ & $463$ \\ 
    A3B3G3-D3M11 & $64.7$ & $54.9$ & $0.73$ & $20.3$ & $2.0\cdot 10^{13}$ &
    $1.9\cdot 10^{13}$ & $485$ & $398$ \\ 
                 & $67.7$ & $37.5$ & $0.71$ & $9.9$ & $1.2\cdot 10^{14}$ &
                 $1.2\cdot 10^{14}$ & $558$ & $458$ \\ 
    A3B3G3-D3M12 & $68$ & $39.7$ & $0.65$ & $6.2$ & $7.0\cdot 10^{14}$ &
    $6.9\cdot 10^{14}$ & $564$ & $469$ \\ 
    A3B3G3-D0M13 & $64.2$ & $63.1$ & $0.68$ & $12.8$ & $4.2\cdot 10^{14}$ & 
    $2.5\cdot 10^{14}$ & $911.5$ & $508.4$ \\  
    A3B3G3-D1M13 & $68$ & $39.7$ & $0.71$ & $14.8$ & $4.9\cdot 10^{13}$ &
    $4.8\cdot 10^{13}$ & $571$ & $469$ \\ 
    A3B3G3-D3M13 & $68$ & $24.7$ & $0.63$ & $11.1$ & $1.1\cdot 10^{15}$ &
    $3.4\cdot 10^{14}$ & $901$ & $571$ \\ 
    \hline
    A3B3G4-D3M10 & $61.7$ & $25.2$ & $0.47$ & $6.3$ & $2.7\cdot 10^{13}$ &
    $2.7\cdot 10^{13}$ & $497$ & $412$ \\ 
    A3B3G4-D3M11 & $62$ & $24.4$ & $0.46$ & $5.7$ & $4.3\cdot 10^{14}$ &
    $4.3\cdot 10^{14}$ & $503$ & $417$ \\ 
    A3B3G4-D3M12 & $62$ & $35.9$ & $0.44$ &  $5.5$ & $9.4\cdot 10^{14}$ &
    $7.0\cdot 10^{14}$ & $520$ & $417$ \\ 
    A3B3G4-D3M13 & $43$ & $27.72$ & $0.48$ & $34.2$ & $2.3\cdot 10^{14}$ &
    $1.4\cdot 10^{14}$ & $131$ & $111$ \\ 
    \hline
    A3B3G5-D3M10 & $44.3$ & $13.4$ & $0.22$ & $3.3$ & $2.8\cdot 10^{13}$ &
    $2.7\cdot 10^{13}$ & $224$ & $221$ \\ 
    A3B3G5-D3M11 & $44.3$ & $13.7$ & $0.22$ & $3.6$ & $3.3\cdot 10^{14}$ &
    $3.3\cdot 10^{14}$ & $226$ & $224$ \\ 
    A3B3G5-D0M12 & $52.7$ & $29.2$ & $0.23$ & $7.7$ & $5.3\cdot 10^{14}$ &
    $5.2\cdot 10^{14}$ & $375$ & $358$ \\ 
    A3B3G5-D1M12 & $54.6$ & $16.4$ & $0.24$ & $5.6$ & $1.3\cdot 10^{14}$ &
    $1.3\cdot 10^{14}$ & $371$ & $371$ \\ 
    A3B3G5-D2M12 & $54.6$ & $13.7$ & $0.25$ & $3.4$ & $2.0\cdot 10^{15}$ &
    $1.9\cdot 10^{15}$ & $384$ & $388$ \\ 
    A3B3G5-D3M12 & $56.9$ & $12.6$ & $0.24$ & $2.3$ & $3.1\cdot 10^{15}$ &
    $2.8\cdot 10^{15}$ & $417$ & $417$ \\ 
    A3B3G5-D4M12 & $57.6$ & $14.8$ & $0.24$ & $3.4$ & $5.3\cdot 10^{14}$ &
    $5.2\cdot 10^{14}$ & $375$ & $358$ \\ 
    A3B3G5-D3M13 & $59.6$ & $15.1$ & $0.27$ & $-38.0$ & $2.3\cdot 10^{15}$ &
    $8.9\cdot 10^{13}$ & $1009$ & $354$ \\ 
    \hline
    A4B5G5-D3M10 & $52.7$ & $86.0$ & $0.74$ & $21.6$ & $2.3\cdot 10^{12}$ &
    $1.2\cdot 10^{12}$ &   &   \\ 
    A4B5G5-D3M11 & $52.7$ & $71.3$ & $0.68$ & $17.1$ & $2.3\cdot 10^{13}$ &
    $2.22\cdot 10^{13}$ &   &   \\ 
    A4B5G5-D3M12 & $36.1$ & $74.4$ & $0.70$ & $10.9$ & $1.4\cdot 10^{14}$ &
    $1.0\cdot 10^{14}$ &   &   \\ 
    A4B5G5-D3M13 & $46.8$ & $13.9$ & $0.36$ & $34.8$ & $3.4\cdot 10^{15}$ &
    $2.0\cdot 10^{14}$ &   &   \\ 
    \hline\hline
  \end{tabular}
\end{table*}


\bibliographystyle{aa}

\bibliography{4306}

\end{document}